\documentclass{book}[12pt]

\usepackage[usenames,dvipsnames]{xcolor}
\usepackage{etex}
\usepackage[thinlines]{easytable}
\usepackage[a4paper]{geometry} 
\usepackage{natbib}
\usepackage{float}
\usepackage{morefloats}
\usepackage{subcaption}
\usepackage{tikz}
\usepackage{mathtools}
\usepackage{multirow}
\usepackage{algorithmicx}
\usepackage{algpseudocode}
\usepackage[toc,page]{appendix}
\usepackage{amssymb,amsthm,amsmath,graphicx}
\usepackage{epsfig}
\usepackage{pstricks,pst-node,pst-tree}
\usepackage{listings}
\usepackage{longtable}
\usepackage{booktabs}
\usepackage{setspace}
\usepackage[nottoc,numbib]{tocbibind}
\usepackage{soul}
\usepackage{graphics}
\usepackage{braket}
\usepackage{verbatim}
\usepackage{underscore}
\usepackage{stackengine}
\usepackage{afterpage}

\graphicspath{ {images/} }
\allowdisplaybreaks

\newcommand\blankpage{%
    \null
    \thispagestyle{empty}%
    \addtocounter{page}{-1}%
    \newpage}

\usetikzlibrary{positioning,arrows}
\usetikzlibrary{shapes.geometric}
\allowdisplaybreaks

\newcases{<case name>}{<space>}{%
  <left col align>}{<right col align>}{<left delim>}{<right delim>}

\makeatletter
\newcases{mycases}{\quad}{%
  \hfil$\m@th\displaystyle{##}$}{$\m@th\displaystyle{##}$\hfil}{\lbrace}{.}
\makeatother

\newtheorem{theorem}{Theorem}[section]
\newtheorem{definition}[theorem]{Definition}

\newtheorem{lemma}[theorem]{Lemma}
\newtheorem{proposition}[theorem]{Proposition}
\newtheorem{corollary}[theorem]{Corollary}

\theoremstyle{definition}

\makeatletter
 \def\@textbottom{\vskip \z@ \@plus 200pt}
 \let\@texttop\relax
\makeatother

\begin{document}

\onehalfspacing

\begin{titlepage}

\begin{center}

\includegraphics{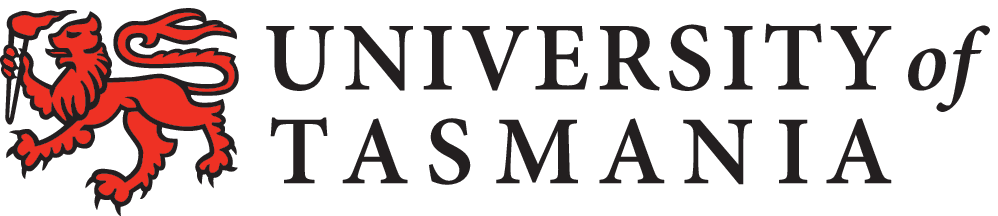}

\vfill

\Huge
\textsc{Distinguishing Convergence on Phylogenetic Networks}\\

\large

\vfill

{Submitted in fulfilment of the requirements of the degree\\[0.5\baselineskip]
of\\[0.5\baselineskip]
Doctor of Philosophy\\[0.5\baselineskip]
in\\[0.5\baselineskip]
Mathematics\\[0.5\baselineskip]
at\\[0.5\baselineskip]
University of Tasmania\\[0.5\baselineskip]
Faculty of Science, Engineering \& Technology\\[0.5\baselineskip]
School of Physical Sciences}\\[0.5\baselineskip]
Discipline of Maths

\vspace{1cm}
Jonathan Mitchell, BSc (Hons) \emph{UTAS}\\[0.5\baselineskip]
June $2016$

\end{center}

\end{titlepage}

\pagenumbering{alph}
\frontmatter

\afterpage{\blankpage}

\newpage{}

\chapter*{\centering\Large\textbf{Declaration of Originality}}
\addcontentsline{toc}{chapter}{Declaration of Originality}

This thesis contains no material which has been accepted for a degree or diploma by the University or any other institution, except by way of background information and duly acknowledged in the thesis, and to the best of my knowledge and belief no material previously published or written by another person except where due acknowledgement is made in the text of the thesis, nor does the thesis contain any material that infringes copyright.

\vfill

Signature\hspace{0.5cm}
\includegraphics[scale=0.9]{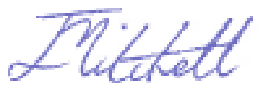}
Date\hspace{0.5cm} {\LARGE \textbf{21/06/2016}}

\newpage{}

\chapter*{\centering\Large\textbf{Authority of Access}}
\addcontentsline{toc}{chapter}{Authority of Access}

This thesis may be made available for loan and limited copying and communication in accordance with the Copyright Act $1968$.

\vfill

Signature\hspace{0.5cm}
\includegraphics[scale=0.9]{Signature.eps}
Date\hspace{0.5cm} {\LARGE \textbf{21/06/2016}}

\newpage{}

\chapter*{\centering\Large\textbf{Abstract}}
\addcontentsline{toc}{chapter}{Abstract}

In phylogenetics, the evolutionary history of a group of taxa, for example, groups of species, genera or subspecies, can be modelled using a phylogenetic tree. Alternatively, we can model evolutionary history with a phylogenetic network. On phylogenetic networks, edges that have previously evolved independently from a common ancestor may subsequently converge for a period of time. Examples of processes in biology that are better represented by networks than trees are hybridisation, horizontal gene transfer and recombination.

Molecular phylogenetics uses information in biological sequences, for example, sequences of DNA nucleotides, to infer a phylogenetic tree or network. This requires models of character substitution. A group of these models is called the \emph{Abelian group-based models}. The rate matrices of the Abelian group-based models can be diagonalised in a process often referred to as \emph{Hadamard conjugation} in the literature. The time dependent probability distributions representing the probabilities of each combination of states across all taxa at any site in the sequence are referred to as \emph{phylogenetic tensors}. The phylogenetic tensors representing a given tree or network can be expressed in the diagonalised basis that may allow them to be analysed more easily. We look at the diagonalising matrices of various Abelian group-based models in this thesis.

We compare the phylogenetic tensors for various trees and networks for two, three and four taxa. If the probability spaces between one tree or network and another are not identical then there will be phylogenetic tensors that could have arisen on one but not the other. We call these two trees or networks \emph{distinguishable} from each other. We show that for the binary symmetric model there are no two-taxon trees and networks that are distinguishable from each other, however there are three-taxon trees and networks that are distinguishable from each other.

We compare the time parameters for the phylogenetic tensors for various taxon label permutations on a given tree or network. If the time parameters on one taxon label permutation in terms of the other taxon label permutation are all non-negative then we say that the two taxon label permutations are not \emph{network identifiable} from each other. We show that some taxon label permutations are network identifiable from each other.

We show that some four-taxon networks do not satisfy the four-point condition, while others do. There are two ``structures'' of four-taxon rooted trees. One of these structures is defined by the cluster, b,c,d, where the taxa are labelled alphabetically from left to right, starting with a. The network with this structure and convergence between the two taxa with the root as their most recent common ancestor satisfies the four-point condition.

The phylogenetic tensors contain polynomial equations that cannot be easily solved for four-taxon or higher trees or networks. We show how methods from algebraic geometry, such as Gr\"obner bases, can be used to solve the polynomial equations. We show that some four-taxon trees and networks can be distinguished from each other.

\newpage{}

\chapter*{\centering\Large\textbf{Acknowledgements}}
\addcontentsline{toc}{chapter}{Acknowledgements}

I would like to thank my supervisor, Barbara Holland, and my co-supervisor, Jeremy Sumner, for their wonderful support over the course of my PhD. I am very appreciative of everything they have done for me. They have taught me almost everything I know about phylogenetics. Regular meetings often contained insightful discussions that lead me to make discoveries that I may not have made otherwise. I would not have been able to pursue a PhD in the topic I chose if it weren't for the previous work of Jeremy, Barbara and their collaborators, in particular Peter Jarvis. The help they have given me on writing my thesis has been extremely constructive. I am also very grateful for their support of my wish to take some time off and suspend my candidature. If it were not for this then I may not have completed my PhD.

Barbara's knowledge of phylogenetics is incredibly broad. She is a leader in the field of phylogenetics and brings together researchers from various disciplines, which is vital in phylogenetics. This is reflected in her many collaborators in the biological fields, for which she provides statistical and mathematical expertise. Barbara almost invariably had the answer when I had a general question about phylogenetics or a question about statistics. It has been Barbara's reassurance and knowledge of what constitutes a PhD thesis that has given me confidence that my work is of the standard of a PhD.

Despite being officially a co-supervisor, Jeremy has provided me help on the level of a main supervisor. Most of our meetings were in Jeremy's office with both Jeremy and Barbara. Jeremy's skills in algebra are excellent. He played a major part in developing the convergence-divergence networks for which my thesis relies on. Jeremy was of great assistance when I had a question about algebra and how it related to the convergence-divergence networks.

I would also like to thank Amelia Taylor, who specialises in computational commutative algebra at Colorado College. I had the idea of using techniques from algebraic geometry, eg. using Gr\"obner bases, to solve the polynomial equations that I was dealing with. While visiting UTAS, Amelia helped me to understand the necessary algebraic geometric concepts and helped me to install and use Macaulay$2$, a computer algebra system. Macaulay$2$ was required for a large part of Chapter $5$.

I would like to thank Michael Brideson, the Graduate Research Coordinator in Mathematics, and Karen Bradford, the Executive Officer of the Maths and Physics department, for their general help during my PhD.

I would like to thank my family, particularly my parents, Bron and Chris, and my sister, Emma, for their continued love and belief in me.

I would like to thank my wonderful partner, Ella, who has been an amazing addition to my life.

I would like to thank the University of Tasmania, in particular the Mathematics and Physics department.

Finally, I would like to thank the Australian Federal Government for providing me with an Australian Postgraduate Award.

\tableofcontents

\mainmatter

\chapter{Introduction \& Literature Review}
\label{chapter1}

\section{Introduction}

\subsection{Overview}

In this thesis we will discuss a range of topics in phylogenetics and extend upon the current knowledge in the field. The thesis focuses on the analysis of phylogenetic networks, structures that differ from more simple phylogenetic trees and can allow for convergence of multiple edges. For instance, introgression is an example of convergence, where two taxa (eg. species) become more genetically similar with time. This convergence cannot be modelled on regular phylogenetic trees that allow only for divergence.

Chapter~\ref{chapter1} introduces phylogenetics, beginning with an explanation of what phylogenetics is and its importance in biology. This section will be followed by brief introductory sections to the chapters to follow. Finally, we will conclude with a review of the literature in some of the most important current topics in phylogenetics.

Chapter~\ref{chapter2} discusses the importance of Abelian group-based models in phylogenetics and their relationships to group theory and representation theory. We will introduce the reader to Markov chains and their application to transition matrices in phylogenetics. We will then discuss the resulting properties of the transition matrices and their corresponding rate matrices and argue the importance of Abelian group-based models. We will describe how the Abelian group-based models can be diagonalised and why diagonalisation is a valuable tool for finding expressions for probability distributions in phylogenetics. We will expand the work by \citet{hendy1989framework} and \citet{hendy1989relationship} on Hadamard conjugation to include a number of Abelian group-based models, some that have had their diagonalisation matrices described already by \citet{sumner2014tensorial} and others that have not. We conclude the chapter with a collation of the diagonalising matrices for the Abelian group-based model examples.

Chapter~\ref{chapter3} focusses on phylogenetic tensors, tensor representations of probability distributions. We show how the splitting operator, introduced by \citet{bashford2004u} and used to split an edge on a network into two, can be pushed back to the root of the network, with all bifurcations then occurring at the root \citep{sumner2012algebra}. We will show that this process results in correlated changes, where multiple identical edges are forced to remain identical. We then argue that this result can be used to not only maintain identical edges, but also to model the convergence of diverged edges. We will call these networks \emph{convergence-divergence networks}. This is a fairly new and unique method of modelling networks in phylogenetics. Finally, we will develop some new explicit expressions, which will be represented as polynomials expressions in Chapter~\ref{chapter4} and Chapter~\ref{chapter6}, for the phylogenetic tensor elements of trees and networks on the binary symmetry model. These expressions will be represented in the transformed, or Hadamard, basis. This polynomial representation of the transformed phylogenetic tensor elements will be necessary for the comparison of trees and networks in Chapter~\ref{chapter4} and Chapter~\ref{chapter6}.

Chapter~\ref{chapter4} will be a detailed analysis of two-taxon and three-taxon trees and convergence-divergence networks. We will address the issue of identifiability of models and introduce some new terminology. We will raise the question of whether different network ``structures'' can be \emph{distinguished} from one another. For example, if the set of all pattern frequencies that could have arisen on one tree or network is identical to the set of all pattern frequencies that could have arisen on the other tree or network then we say that they are \emph{indistinguishable}. If the sets of all pattern frequencies are not identical then we say that the two trees or networks are \emph{distinguishable}.

Chapter~\ref{chapter5} will examine the permuting of taxon labels on the three-taxon convergence-divergence network with convergence between ``non-sister'' taxa, taxa that do not share a cluster other than the cluster containing all taxa. We will look at whether appropriate time parameters can still be found on the network if we permute the taxon labels. We will compare maximum likelihoods for the taxon label permutations on the network.

Chapter~\ref{chapter6} will discuss methods of extending the work on distinguishability in Chapter~\ref{chapter4} to higher taxon trees and networks. We will discuss how concepts from algebraic geometry, namely ideals and Gr\"{o}bner bases, can be utilised to extend our work in the two-taxon and three-taxon cases to trees and networks involving a higher number of taxa. We will refer back to our three-taxon examples to illustrate the use of Gr\"{o}bner bases, before arguing that Gr\"{o}bner bases provide a more feasible way of analysing more complicated, higher taxon networks. Finally, we will examine a number of four-taxon trees and networks and determine whether these trees and networks are distinguishable from each other. We will conclude the chapter by showing that our algebraic geometric techniques can be applied to a very complicated convergence-divergence network with multiple convergence periods.

In the final chapter, Chapter~\ref{chapter7}, we will discuss the major findings from the previous chapters and outline the future work that could follow on from the work done in this thesis.

\subsection{Phylogenetics}

Phylogenetics is the study and categorisation of the evolutionary history of groups of organisms. These groups, called taxa, are genetically distinct from one another, such as different species, genera or subspecies. Taxa are classified according to a hierarchy. The hierarchy reflects the assumption that distinct taxa evolved from a common ancestor in the hierarchy and have shared traits. With progression from the time of the common ancestor, each taxon gradually acquired different traits from each other taxon, eventually resulting in the diverse range of life on Earth today.

\begin{figure}[H]
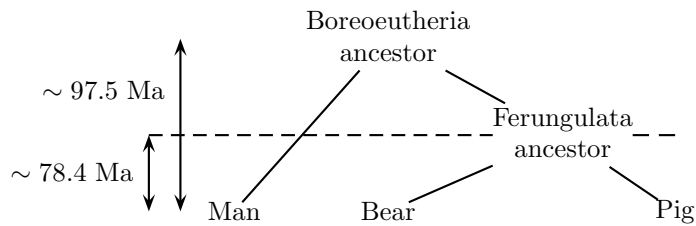

	\centering
		\psmatrix[colsep=.3cm,rowsep=.4cm,mnode=r]
		~ & ~ & && \begin{tabular}[t]{@{}c@{}}Boreoeutheria \\ancestor\end{tabular} \\
		~ & ~ && && \begin{tabular}[t]{@{}c@{}}Ferungulata \\ancestor\end{tabular} & ~ \\
		~ & ~ & Man && Bear && Pig
		\ncline{1,5}{3,3}
		\ncline{1,5}{2,6}
		\ncline{2,6}{3,5}
		\ncline{2,6}{3,7}
		\ncline{<->,arrowscale=1.5}{1,2}{3,2}
		\tlput[tpos=.3]{$\sim{}97.5$ Ma}
		\ncline{<->,arrowscale=1.5}{2,1}{3,1}
		\tlput{$\sim{}78.4$ Ma}
		\psset{linestyle=dashed}\ncline{2,1}{2,6}
		\psset{linestyle=dashed}\ncline{2,6}{2,7}
		\endpsmatrix
		\caption{A simple phylogenetic tree representing the genetic relationships between three species: man (\emph{Homo sapiens}), bear (\emph{Ursus arctos}) and pig (\emph{Sus scrofa}), using divergence time estimates from \citet{hedges2015tree}.}
		\label{exampletree}
\end{figure}

Most commonly, a rooted tree is used to display the evolutionary history of a number of taxa. Rooted trees start with a common ancestor of all taxa on the tree, called the root. Edges coming from the root represent groups of taxa which have evolved independently of, or diverged from, each other after splitting apart at the root. Further splits of edges can occur further down the tree at positions called nodes, with further divergence of edges below these nodes. The bottom of the tree represents the taxa present at the final stage in the tree, the leaves. This is often taken to be the present time, but more generally can be any cross-section in time. Times between nodes, including the root, the leaves of the tree and all internal nodes, are often included. A common task in phylogenetics is to find the ``structure'' of a phylogenetic tree and the lengths of its edges, representing the divergence times. In Figure~\ref{exampletree} below is a simple example of a phylogenetic tree.

Time progresses down the page from the ancestral Boreoeutheria species at the root to the three present day species: man, bear and pig. The most recent common ancestor of the Boreoeutheria clade, which includes man, bears and pigs, lived approximately $97.5$ million years ago, while the most recent common ancestor of the Ferungulata clade, which includes bears and pigs, lived approximately $78.4$ million years ago. It is usual for the horizontal axis to hold no meaning. However, in this thesis we will allow the horizontal axis to represent how much divergence from a common ancestor has occurred and how much subsequent convergence has occurred. The closer two taxa are on the horizontal axis the less they have diverged from a common ancestor or the more convergence has occurred. Swapping the order of two edges below a node has no consequence. For example, we could have swapped the labels for bear and pig and the tree would have the same meaning.

\citet{sumner2012algebra} argued that convergence-divergence networks could be used to model convergence. In this thesis, our convergence-divergence networks can be thought of as regular phylogenetic trees with some added convergence periods. When two or more taxa are no longer diverging from each other, but are becoming more similar over time, we say they are converging. Examples of biological processes that convergence-divergence networks could be used to model are introgression and horizontal gene transfer. Alternatively, convergence-divergence networks could be used to model convergence of morphological characteristics.

\citet{anderson1938hybridization} and \citet{anderson1949introgressive} define introgression as the exchange of alleles from one genome to another through hybridisation or backcrossing. Introgression may result in a more complicated mixture of genomes than a single occurrence of hybridisation between two organisms.

\citet{keeling2008horizontal} describe horizontal gene transfer as the exchange of genetic material between genomes not related to reproduction. Genetic material from one genome can be added to another or exchanged between the genomes. Horizontal gene transfer is well known in bacteria, however it also occurs in eukaryotes. \citet{andersson2003origin} argue that mitochondria, found in most eukaryotes, came from alpha-proteobacteria. Furthermore, they argue that despite the mitochondria and nuclear genomes being distinct, horizontal gene transfer has occurred between them.

Analogous morphological characteristics can arise in taxa that are not related due to natural selection \citep{parker2013genome}. An example of convergence of analogous morphological traits is wings in bats and birds. \citet{theissen2002orthology} argues that although bats and birds inherited forelimbs from a tetrapod common ancestor, they are not closely related and have evolved wings for flight separately due to similar selective pressures.

When attempting to make comparisons between different types of organisms we must first separate these organisms into different taxa, with each taxon being a genetically distinct group of organisms. An example of taxon separations is the commonly used taxonomic hierarchy, which places species at the bottom and domain (\emph{Eukaryota}, \emph{Bacteria} and \emph{Archaea}) below life at the top. An example of taxonomic distinctions at the species level is the \emph{Macropus} genus. This genus includes many different Kangaroo \emph{species} and also contains wallaroo species and some wallaby species \citep{wilson2005mammal}. Taxa can be distinguished further beyond the taxonomic hierarchy, for example at the subspecies level or at the subgenus level between genus and species.

In recent decades, organisms have often been assigned to a particular taxon based on genetic evidence. Sometimes, however, distinctions are made from other types of evidence, such as morphological differences found in fossils for which the DNA has degraded over time. An example of morphological differences is the presence or absence of vertebrae in animals. The \emph{Vertebrata} subphylum is an example of equivalent genetic and morphological taxonomic distinctions.

A question that often arises in molecular phylogenetics is where a set of taxa fit on a phylogenetic tree or network. To answer this question, we usually start with a character sequence for each taxon. A character sequence is a string of characters, with all characters being the same type of object. The state that each character takes must be from the same state space. For example, the state space for nucleic acid sequences is the four DNA nucleobases described earlier, C, G, A and T. The four nucleobases are the same type of object. We cannot include, for example, a state for the presence or absence of vertebrae into this state space.

\subsection{Markov Models}

Markov models are often used in phylogenetics to model the rates of changes between states in the state space. If these changes in state obey the Markov property then we can use a Markov model. For example, if the state space is the four nucleobases, C, G, A and T, then the substitution rates between the nucleobases can be modelled as a Markov chain.

A random walk is a process in which some object, which takes a state, changes its state randomly. These state changes can happen at fixed time intervals, known as discrete time intervals, or at random times intervals, known as continuous time intervals. In phylogenetics, the object we are dealing with may be a single character in a character sequence. For example, the state of our character may be one of the DNA nucleobases (cytosine (C), guanine (G), adenine (A) and thymine (T)). Suppose the initial state of the character at a position, or site, in the genome of a species is G. After some random period of time there may be a character substitution into the state T in the genome. At some further random time, another substitution may change the state back to G. Furthermore, after another random period of time, the character may change state to T. This is an example of a random walk in phylogenetics.

The outcome of a step in the random walk may or may not depend on the previous state that the object took. From our example, the final substitution was from state G to state T. This substitution may or may not depend on the history of the random walk, that is the previous substitutions from G to T to G. If we assume that all state changes depend only on the current state that the object takes and not on any previous states then we say that the object obeys the Markov property and we call it a Markov chain. We often assume processes in phylogenetics to be Markov in nature. If the joint probability distribution is assumed to obey a Markov process then the probabilities for transitions between states can be organised into stochastic transition matrices. For ergodic random walks, where every state in the state space can be reached from every other state in some finite number of steps, $s$, it is shown that after a certain number of substitutions, further substitutions depend very little on the initial state \citep{diaconis1988group}. We can therefore assume that if enough substitutions have occurred then an objects' state changes display Markov behaviour.

We call the position in the sequence that a particular character is its site. Generally each site is assumed to behave independently of every other site, and with an identical process. This is called the assumption of independent and identically distributed random variables (i.i.d. assumption), a very commonly used assumption in phylogenetics. The consequence of the independence property of the i.i.d. assumption is that the behaviour of each site can be modelled by a Markov chain, which is independent of every other site. The consequence of the identical distributions property is that the probability of a substitution from one state in the state space to another state in the state space over a given period of time must be equal for every site in the sequence and independent of every other site in the sequence. What it does not mean is that every character must be the same at each site in the sequence. In Table~\ref{DNAsequences} below is an example of three aligned nucleic acid sequences, each with ten sites.
\begin{table}[H]
\begin{align*}
\left.\begin{array}{c|c|c|c|c|c|c|c|c|c} $C$ & $\bf{T}$ & $\bf{C}$ & $G$ & $\bf{T}$ & $A$ & $G$ & $T$ & $G$ & $C$ \\
$C$ & $A$ & $G$ & $G$ & $\bf{G}$ & $A$ & $G$ & $\bf{A}$ & $G$ & $C$ \\
$C$ & $A$ & $G$ & $G$ & $\bf{C}$ & $\bf{T}$ & $G$ & $T$ & $G$ & $C$ \\
\end{array}\right.
\end{align*}
\caption{An example of three aligned nucleic acid sequences, each with ten sites. Each site is designated by its own column. Sites which differ between sequences are highlighted in boldface.}
\label{DNAsequences}
\end{table}

We can now use a Markov model to describe the behaviour of the entire set of taxa. Markov models are square stochastic matrices, with elements representing either the probabilities or rates over time of substitutions between states in the state space. If the rate matrices can be represented in terms of the elements of an Abelian group then the model is \emph{Abelian group-based} or simply \emph{group-based}. Abelian group-based models have some desirable mathematical properties that will be discussed in more detail in Chapter~\ref{chapter2}. Group-based models can be diagonalised by a diagonalising matrix. When the diagonalising matrix is a Hadamard matrix, first described in a form by \citet{silvester1867thoughts} and later expanded upon by \citet{hadamard1893resolution}, we refer to the diagonalisation process as \emph{Hadamard conjugation} or a \emph{Hadamard transformation}. Hadamard conjugation was first introduced by \citet{hendy1989framework} and \citet{hendy1989relationship}. Exponentiation of the resulting diagonalised rate matrix is then straightforward. We will explain in Chapter~\ref{chapter2} the desirable effects that diagonalisation has on the probability distribution and finding the diagonalising matrices for a number of Abelian group-based models.

The simplest Markov model of evolution for sequences of four states was described by \citet{jukes1969evolution} and assumes the rates of substitutions to be equal for every substitution. In other words, the rate matrix representing the substitution rates has only one parameter. If we assume some initial probability distribution of character states at the root of the tree or network, the earliest position in time, then we can use a Markov model to model substitutions between character states leading from the root to a node, where one taxon splits into multiple, usually two, descendant taxa. The Markov model must assume constant rates of substitutions between characters from the root to the node. We can model every edge between two nodes in this fashion, with nodes including the root and leaves of the tree or network. If two sequences are known, each for a different taxon, with no other information available, it cannot be determined if the sequences are ancestral and descendant to each other nor which sequence is ancestral and which is descendant. For this reason time-reversible Markov models are often preferred when the root position is unknown, as the root of the tree can be placed in an arbitrary position. Almost always, the model parameters are fixed across a tree or network. When all parameters are equal on the tree or network we say that the process is homogeneous. This raises the question of what happens when a Markov model on an edge is partitioned into several successive Markov models comprising the length of the edge. \citet{sumner2012general} showed that for some Markov models the combination of many successive Markov processes along the edge cannot be modelled by a single process along the edge under that same model. The general time-reversible model was shown to be one of these models lacking closure.

If we have a set of aligned sequences, each representing a different taxon, we can compare each taxon by comparing the states that each sequence takes at each site independently. The more sites taking the same state between two sequences, the closer those two taxa will be on a phylogenetic tree or network. The limit is where two sequences are identical, as would be the case with identical twins, assuming no new substitutions have occurred in their DNA. The fewer sites taking the same state between two sequences, the further apart those two taxa will be. With a set of more than two taxa we can determine where each taxon should be placed on the tree or network.

\subsection{Phylogenetic Tensors on Convergence-Divergence Networks}

Suppose we are given a phylogenetic tree or network. We want to know what the theoretical probability distribution representing the tree or network in the diagonalised, or transformed, basis will be. In Chapter~\ref{chapter3} we will discuss the relevant algebra and notation needed to express the theoretical probability distribution of a given tree or network. We call the probability distribution tensor representing the probability distribution the phylogenetic tensor. If there are $m$ states in the state space and each of the $n$ taxa can take any of these $m$ states at a particular site, then there will be $m^{n}$ elements in the phylogenetic tensor.

To derive explicit expressions for the phylogenetic tensor we start with the initial probability distribution, the probabilities of each state at the root, where there is just one taxon, the taxon ancestral to all other taxa. In practice, the initial probability distribution is rarely known and the probability distribution is only known at the leaves, the most recent time. We must therefore make a choice for our initial probability distribution. The choice made is often the stationary distribution. The stationary distribution represents the long run probability distribution for each individual taxon.

The next task after deciding on the initial probability distribution is to model the splitting of the single root taxon into the multiple descendant edges. In this thesis we will only be examining bifurcating trees and networks, where all nodes, including the root, split into two descendant edges. \citet{bashford2004u} introduced the \emph{splitting operator}, an operator which takes a single edge at the root or a node and instantaneously outputs two descendant edges.

An Abelian group-based Markov model is then utilised as described in the previous section on Markov models. The Abelian group-based Markov model is applied to every edge of the tree. From here, the phylogenetic tensor for any tree formed by an Abelian group-based Markov model can be found. \citet{sumner2012algebra} showed that the splitting operator can be used on an Abelian group-based Markov model to model convergence, allowing us to find the phylogenetic tensor for networks allowing convergence. We will call these networks \emph{convergence-divergence networks}.

Convergence-divergence networks are not equivalent to any previous methods of generalising the structure of trees. Unlike methods based on splits it is directed in time. It is also different from the approaches for implementing maximum likelihood on networks that are described in \citet{nakhleh2011evolutionary}'s review. In \citet{nakhleh2011evolutionary}'s review a network is thought of as encoding a set of trees (those displayed by the network). The likelihood is then either a mixture model over these trees \citep{jin2006maximum}, or each site is allowed to pick the tree that suits it best. Convergence-divergence networks will clearly have different limiting properties to either of these implementations. If the convergence process is run for a long enough period of time then the taxa that are converging can get arbitrarily close - this is not the case in either the mixture-model setting or for the $n$-taxon process as described by \citet{bryant2009hadamard}.

The convergence-divergence models provide a lot of flexibility, perhaps too much flexibility. Before suggesting that they are a practical tool for phylogenetics we need to address two points that were well put by \citet{steel2005should}. Steel's two key points to keep in mind when developing models are:
\begin{enumerate}
\item Are they capturing a process that is important biologically?
\item Do they overfit the data?
\end{enumerate}

As soon as we leave the safe waters of tree-based inference where the number of ``structural'' parameters we need is determined by the number of species under consideration (e.g. in a rooted binary tree with $n$ taxa, we know we need $2n-2$ edges) we have to think much more carefully about variable selection.

Addressing the first point, it seems that the convergence-divergence model might be appropriate for modelling introgression, in the extreme case leading to despeciation. In this sense, the model can be thought of as a species-level analogue to the population-level isolation/migration model of \citet{hey2010isolation}. \citet{seehausen2008speciation} argued that a loss of diversity can break down ecological boundaries, allowing more opportunities for the exchange of genetic material among previously independent populations, which can in turn lead to convergence. \citet{taylor2006speciation} described a case where environmental changes may be resulting in the convergence of three-spined sticklebacks (\textit{G. aculeatus}) in Enos Lake, Vancouver Island. \citet{sheppard2008convergence} and \citet{sheppard2011introgression} identified a case in which two species of bacteria, \textit{C. jejuni} and \textit{C. coli}, appear to be in the process of undergoing convergence through horizontal gene transfer. A further scenario where we might consider applying convergence-divergence networks is for morphological data where selection acts similarly on taxa on different parts of the tree, causing some of the morphological characters to converge.

We will address the second point from \citet{steel2005should} in this thesis. Can we put limits on what structures are sensible to consider with these models? We examine some cases on two, three and four taxa and determine in what circumstances parameters are recoverable. In other words, given some particular time periods, is there a one-to-one map between the time parameters and the probability distribution on the pattern frequencies? This is a question of \emph{identifiability}. Given a two-taxon network where taxon $1$ and taxon $2$ diverge for time $t_{1}$ and then converge for time $t_{2}$, is there an equivalent time parameter where they diverge for time $t'$? This is a question of \emph{network identifiability}. Given a pattern frequency, could that pattern frequency have arisen on either of the trees or networks? If the set of pattern frequencies that could have arisen on one tree or network is identical to the set of pattern frequencies that could have arisen on the other tree or network then we say that the two trees and networks are \emph{indistinguishable}. If the sets of pattern frequencies are not identical then we say that the two trees and networks are \emph{distinguishable}. Generally if two trees or networks are not distinguishable we will choose the tree or network which is least parameter rich or is most biologically feasible.

A framework will be developed to find the phylogenetic tensors and the individual phylogenetic tensor elements for a given Abelian group-based Markov model and a given tree or network. We will show that we can use the algebraic properties of a Markov model to find explicit expressions for the phylogenetic tensor in the transformed basis. Simple substitutions for the exponential terms can turn the phylogenetic tensor elements into non-linear polynomial equations. In the transformed basis some of the elements of the phylogenetic tensor will always be constant. These constraints are similar to phylogenetic invariants, introduced by \citet{cavender1987invariants} and \citet{lake1987rate}. We will also show that some of the constraints are variable, similar to the Fourier transform inequalities described by \citet{matsen2009fourier}.

Arbitrarily many convergence-divergence networks could be constructed under the framework we will describe. For example, we could have a network where a process involving divergence of a group of taxa followed by convergence of the same group is then repeated arbitrarily many times. We will put a limit on convergence and divergence periods and restrictions on when and how they can occur.

In Chapter~\ref{chapter4} and Chapter~\ref{chapter6} we will examine various two-taxon, three-taxon and four-taxon trees and networks and determine which are distinguishable from each other. We will determine the sets of equality and inequality constraints on the transformed phylogenetic tensor for each tree and network. We will determine whether the trees and networks are distinguishable from each other by comparing their sets of constraints. If the two sets of constraints are identical then the two trees or networks are indistinguishable. If the two sets of constraints are different then the two trees or networks are distinguishable. In the case that the two trees and networks are distinguishable we will determine the intersection between these two sets of constraints.

As well as comparing different trees and networks, we will compare the time parameters on different taxon label permutations on the convergence-divergence network with convergence between non-sister taxa in Chapter~\ref{chapter5}. Given a set of time parameters on one taxon label permutation we will determine whether an equivalent set of time parameters can be found on another taxon label permutation that satisfies the same constraints and satisfies the condition that time parameters must be non-negative. Finally, we will show how the maximum likelihood on one taxon label permutation can be compared to the maximum likelihood on another taxon label permutation.

\subsection{Ideals and Gr\"{o}bner Bases}

In Chapter~\ref{chapter4} we determine the distinguishability of many two-taxon and three-taxon convergence-divergence trees and networks on the binary symmetric model. These examples are simple enough that we can compare the phylogenetic tensor elements of each tree and network using elementary algebra. By making simple substitutions, we can turn exponential equations into non-linear polynomial equations, as shown in earlier chapters, which are relatively straightforward to solve.

For more complicated Abelian group-based models than the binary symmetric model or for trees and networks with more than three taxa, it will quickly become infeasible to continue using elementary algebraic methods. In Chapter~\ref{chapter6} we will discuss a more general method for dealing with our phylogenetic tensor equations than was used in Chapter~\ref{chapter4}. As was the case with the three-taxon trees and convergence-divergence networks, all phylogenetic tensor elements for four-taxon trees and networks will be able to be expressed as non-linear polynomial equations by making simple substitutions to convert the exponential expressions into polynomial expressions. Once we have the polynomial equations, we can rearrange them to be identically zero and express the polynomials from the non-zero sides of these equations as generators on an ideal. We can then find a different set of generators that generates the same ideal. The polynomials from this set of generators will have the same solutions as the polynomials from the original set of generators when expressed as polynomial equations identical to zero. We will show that if the basis of the ideal is the \emph{Gr\"{o}bner basis} then we can solve the polynomial equations from the polynomial generators and find all of the inequality and equality constraints on the tree or network. We will compare various four-taxon trees and networks and determine which trees and networks are distinguishable from each other.

\section{Literature Review}

\subsection{Networks in Phylogenetics}

Networks play a role in phylogenetics as an alternative to or extension of phylogenetic trees. A network may be appropriate when the best representation of the evolutionary history of a set of taxa is multiple trees or a generalisation of a single tree to no longer require independent evolution of all edges. Networks may be desirable when convergent evolution, or despeciation, has occurred. Introgression, horizontal gene transfer and recombination can all lead to convergent evolution. Networks can also be used to represent model heterogeneity, sampling error and uncertainty caused by parallel evolution, where unrelated taxa evolve similar traits independently through similar ecological pressures. A network may be preferred even when a single tree may be the best model for the evolutionary history of a set of taxa. If this tree is unknown and several candidate trees are roughly equally good fits, then this information can be represented in a network.

Much of the work on networks in phylogenetics up until now has dealt with \emph{splits}, where taxa are bipartitioned into non-intersecting sets based on their evolutionary history. The structure of a tree can be represented by a series of splits, with each split representing an edge on the tree. However, these splits are not necessarily required to fit on a tree, but must fit on a more general \emph{split network}. These split networks may not necessarily have splits which are all \emph{compatible} with each other. In other words, a taxon may appear on different sides of two different splits, grouped with one set of taxa on one split and away from the same set of taxa on the other split. In a tree, a taxon must group with a set of taxa on every split or away from that set of taxa on every split.

\citet{bandelt1992split} introduced the split decomposition method. The split decomposition method takes a set of distances and returns a weakly compatible split system. It is a generalisation of the four-point condition. Suppose there are $n$ taxa, with distances, $d\left(a,b\right)$, between each pair of taxa, $\left(a,b\right)$. For every set of four taxa, $i$, $j$, $k$ and $l$, the condition, $d\left(i,j\right)+d\left(k,l\right)<max\left(d\left(i,k\right)+d\left(j,l\right),d\left(i,l\right)+d\left(j,k\right)\right)$, must be met.

\citet{posada2001intraspecific} provided a summary of many of the network methods available in phylogenetics. A thorough and comprehensive textbook for networks in phylogenetics is \citet{huson2010phylogenetic}. Reticulograms \citep{legendre2002reconstruction} and ancestral recombination graphs \citep{song2004minimum} are some examples of network approaches. \citet{legendre2002reconstruction} heuristically created a reticulation network from an underlying tree, adding reticulations if they improved the fit. \citet{baroni2005framework} used a directed graph (digraph) approach, called ``hybrid phylogenies''. \citet{lapointe2000account} developed graphs and used four methods: pyramids, weak hierarchies, split graphs and reticulograms. \citet{hein1990reconstructing,hein1993heuristic,wang2001perfect,gusfield2003efficient,gusfield2004fine,nakhleh2004reconstructing} explored how to recreate a reticulate network if it is assumed that a set of taxa evolved under a reticulate network. \citet{hudson1985statistical,myers2003bounds,bafna2004number,gusfield2004fine} looked at finding bounds on the number of reticulations needed to fit a set of data.

\citet{maddison1997gene} used a ``separate analysis'' technique. A reticulation network with a single reticulation corresponds to two gene trees inside the network. Each gene evolves according to one of the gene trees and the two gene trees differ by one rooted subtree prune and regraft (rSPR) step. If both gene trees are given, a reticulation network can be found for them. If there are multiple reticulations, then the number of rSPR steps connecting the two gene trees is no greater than the number of reticulations.

\citet{nakhleh2004reconstructing} developed two polynomial time algorithms from the work of \citet{maddison1997gene}. The first algorithm takes two gene trees inside a network and outputs a reticulation network. Although the algorithm allows for any number of reticulations, cycles in the reticulation network must be ``galled'' (node-disjoint). The second algorithm computes a reticulation network with a single reticulation and can be used when there are errors in the inferred gene trees.

\citet{huson2005reconstruction} provided an algorithm for computing a most parsimonious reticulate network. The algorithm depends on ``tangles'', reticulation cycles that have at least one edge in common. The algorithm takes polynomial time if the reticulations in any tangle obey an overlapping property and if the maximum number of reticulations in a tangle is a constant. They also provided an algorithm for determining the structure of reticulate networks and a statistical procedure to determine whether reticulations occurred due to hybridisation or other means, for example lineage sorting or tree estimation error.

\subsubsection{Split Networks}

The following section on split networks follows the notation used by \citet{huson2010phylogenetic}.

Every edge on an unrooted tree can be interpreted as a bipartion of the set of all taxa on the tree. As every taxon is represented exactly once on the tree, each taxon must be represented on exactly one side of each edge. This idea of separating taxa based on edges gives rise to the concept of \emph{splits}. We will see that splits are not only useful on unrooted trees, but can be applied to more general \emph{split networks}.

\begin{definition}[\textbf{Split}]
A set of taxa, $\mathbb{N}$, can be bipartitioned into a \textbf{split} of two sets, $A$ and $B$, written as $A|B$, which are both non-empty, do not intersect and have the set of all taxa as their union,
\begin{align*}
\begin{mycases}
A,B&\neq{}\emptyset{},\\
A\cap{}B&=\emptyset{},\\
A\cup{}B&=\mathbb{N}.
\end{mycases}
\end{align*}
\end{definition}

The two sets of the split, $A$ and $B$, are called the \emph{split parts}. The ordering of a split is inconsequential. Each taxon can always be separated from all other taxa in what is called a \emph{trivial split}. If a set of splits can be displayed as an unrooted tree, the set of splits are said to be \emph{compatible} with each other. To read more about splits see Definition 5.2.1 in \citet{huson2010phylogenetic}.

\begin{definition}[\textbf{Compatibility}]
Suppose we have the two splits, $A|B$ and $C|D$, where $A$, $B$, $C$ and $D$ are sets of taxa. The two splits are \textbf{compatible} if one of the intersections of the split parts, $A\cap{}C$, $A\cap{}D$, $B\cap{}C$ or $B\cap{}D$, is the empty set. If none of the intersections are the empty set then the two splits are \emph{incompatible}. If every pair of splits in a group of splits is compatible then the group of splits is compatible.
\end{definition}
If we have a set of compatible splits for every possible combination of the set of taxa, these splits can always be represented on an unrooted tree. To read more about compatibility see Definition 5.3.1 in \citet{huson2010phylogenetic}.

\begin{theorem}[\textbf{Compatibility Theorem}]
For an $n$-taxon unrooted binary tree, there will be $2n-3$ edges, of which $n$ will be edges representing taxa at the tips and $n-3$ will be internal edges. A unique $n$-taxon unrooted binary tree will exist if and only if a compatible split system exists, with $2n-3$ splits, of which $n$ will be trivial splits corresponding to the tips of the tree and $n-3$ will be non-trivial splits corresponding to the internal edges of the tree.
\begin{proof}
Proof can be found in \citet{buneman1971recovery}.
\end{proof}
\end{theorem}

From this point onwards it can be assumed that all trees we are dealing with are binary, unless specifically stated otherwise. From a combinatorial point of view it is sufficient to only consider binary trees since multifurcating trees can be constructed from binary trees by setting the appropriate edge lengths to zero.

A set of all trivial splits is a set of compatible splits, however unless $n=3$ it is not enough to define an unrooted tree. In general, as a consequence of the split equivalence theorem, a set of compatible splits must define at least one unrooted tree and possibly multiple unrooted trees. For example, suppose we have a four-taxon unrooted tree. There are four trivial splits, each corresponding to one taxon. However, one more compatible split, out of three possibilities, is required to define an unrooted tree. To read more about the compatibility theorem see Theorem 5.3.2 in \citet{huson2010phylogenetic}.

A similar concept to compatibility is \emph{weak compatibility}.

\begin{definition}[\textbf{Weakly Compatible}]
Three sets of splits, $A|B$, $C|D$ and $E|F$, where $A$, $B$, $C$, $D$, $E$ and $F$ are sets of taxa, are \textbf{weakly compatible} if at least one of the intersections of the split parts, $A\cap{}C\cap{}E$, $A\cap{}D\cap{}F$, $B\cap{}C\cap{}F$, or $B\cap{}D\cap{}E$, is empty and at least one of the intersections of the split parts, $B\cap{}D\cap{}F$, $B\cap{}C\cap{}E$, $A\cap{}D\cap{}E$, or $A\cap{}C\cap{}F$, is also empty.
\end{definition}

Weakly compatible splits can be formed from algorithms that attempt to create sets of incompatible splits from data sets that do not allow for compatible split systems. To read more about weakly compatible splits see Definition 5.8.1 in \citet{huson2010phylogenetic}.

\begin{figure}[H]
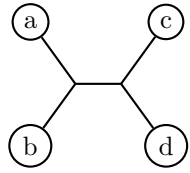

	\centering
		\psmatrix[colsep=.3cm,rowsep=.4cm,mnode=r]
		[mnode=circle] a && && [mnode=circle] c \\
		& {} && {} \\
		[mnode=circle] b && && [mnode=circle] d \\
		\ncline{1,1}{2,2}
		\ncline{2,2}{2,4}
		\ncline{2,2}{3,1}
		\ncline{2,4}{1,5}
		\ncline{2,4}{3,5}
		\endpsmatrix
\caption{A four-taxon unrooted tree.}
\label{fourtaxonunrooted}
\end{figure}

We have so far only discussed unrooted trees. While splits only apply to unrooted trees, a similar concept, called \emph{clusters}, corresponds to rooted trees. Clusters define groups of similar taxa, while splits separate taxa according to their differences. We can transform a compatible set of splits into a compatible set of clusters if we wish to generate a rooted tree. Firstly, the root is placed on an arbitrary position on the tree. Every set of taxa below a node, including the root, is then assigned to a cluster. For example, if there is a bifurcation below a node then all of the taxa below the node will form a cluster. All of the taxa below the root, that is all of the taxa on the rooted tree, will also form a cluster. We will look at the four-taxon unrooted tree in Figure~\ref{fourtaxonunrooted} below as an example.

The splits for this tree are
\begin{align}
\begin{mycases}
\left\{a\right\}&|\left\{b,c,d\right\},\\
\left\{b\right\}&|\left\{a,c,d\right\},\\
\left\{c\right\}&|\left\{a,b,d\right\},\\
\left\{d\right\}&|\left\{a,b,c\right\},\\
\left\{a,b\right\}&|\left\{c,d\right\}.
\end{mycases}
\end{align}

Suppose we now root the tree by placing a root on an arbitrary edge. Let's put the root on the edge representing the split $\left\{a,b\right\}|\left\{c,d\right\}$ and label it $r$. The rooted tree is displayed in Figure~\ref{fourtaxonrootedtree} below.
\begin{figure}[H]
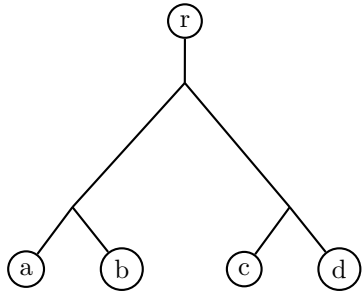

	\centering
		\psmatrix[colsep=.3cm,rowsep=.4cm,mnode=r]
		& && [mnode=circle] r \\
		& && ~ \\
		\\
		& ~ && && ~ \\		
		[mnode=circle] a && [mnode=circle] b && [mnode=circle] c && [mnode=circle] d \\
		\ncline{1,4}{2,4}
		\ncline{2,4}{4,2}
		\ncline{4,2}{5,1}
		\ncline{4,2}{5,3}
		\ncline{2,4}{4,6}
		\ncline{4,6}{5,5}
		\ncline{4,6}{5,7}
		\endpsmatrix
\caption{A four-taxon rooted tree.}
\label{fourtaxonrootedtree}
\end{figure}

The clusters for this tree are the sets of taxa under each node, including the leaves,
\begin{align}
\begin{mycases}
a,b,c,d, \\
a,b, \\
c,d, \\
a, \\
b, \\
c, \\
d.
\end{mycases}
\end{align}

For the inverse problem, we wish to go from a set of clusters to a set of splits. Supposing the set of clusters is compatible, they must correspond to a rooted tree. The set of splits can be found by simply unrooting the tree and finding the set of splits by dividing the unrooted tree along every edge. With incompatible clusters the task of finding a set of splits becomes more challenging. Suppose we have three taxa: $a$, $b$ and $c$. Suppose we have the two clusters: $a,b$ and $b,c$. These two clusters are incompatible since they cannot be represented on the same rooted tree. The corresponding set of splits, however, is compatible since they can all be represented on the same unrooted tree, shown below in Figure~\ref{threetaxonunrootedtree}.
\begin{figure}[H]
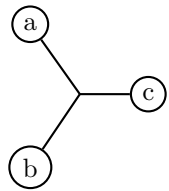

	\centering
		\psmatrix[colsep=.3cm,rowsep=.4cm,mnode=r]
		[mnode=circle] a \\
		& ~ && [mnode=circle] c \\
		[mnode=circle] b \\
		\ncline{1,1}{2,2}
		\ncline{2,2}{2,4}
		\ncline{2,2}{3,1}
		\endpsmatrix
\caption{A three-taxon unrooted tree.}
\label{threetaxonunrootedtree}
\end{figure}

To turn the set of clusters into a set of splits we first introduce a dummy taxon set, $r$. The dummy taxon set is placed with the split part that does not contain the cluster. In our example, the splits become $\left\{a,b\right\}|\left\{c,r\right\}$ and $\left\{b,c\right\}|\left\{a,r\right\}$, which can be represented on the two incompatible unrooted trees shown in Figure~\ref{twoincompatibleunrootedtrees} below, each corresponding to one of the splits.
\begin{figure}[H]
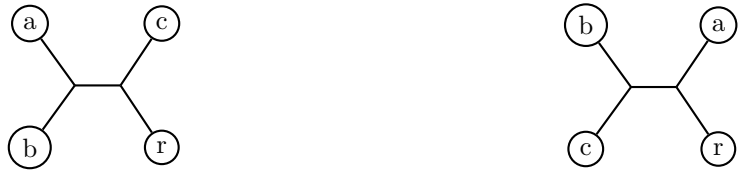

	\centering
		\begin{subfigure}[h]{0.49\textwidth}
		\centering
		\psmatrix[colsep=.3cm,rowsep=.4cm,mnode=r]
		[mnode=circle] a && && [mnode=circle] c \\
		& {} && {} \\
		[mnode=circle] b && && [mnode=circle] r \\
		\ncline{1,1}{2,2}
		\ncline{2,2}{2,4}
		\ncline{2,2}{3,1}
		\ncline{2,4}{1,5}
		\ncline{2,4}{3,5}
		\endpsmatrix
		\caption{A four-taxon unrooted tree representing the non-trivial split, $\left\{a,b\right\}|\left\{c,r\right\}$.}
		\end{subfigure}
	\centering
		\begin{subfigure}[h]{0.49\textwidth}
		\centering
		\psmatrix[colsep=.3cm,rowsep=.4cm,mnode=r]
		[mnode=circle] b && && [mnode=circle] a \\
		& {} && {} \\
		[mnode=circle] c && && [mnode=circle] r \\
		\ncline{1,1}{2,2}
		\ncline{2,2}{2,4}
		\ncline{2,2}{3,1}
		\ncline{2,4}{1,5}
		\ncline{2,4}{3,5}
		\endpsmatrix
		\caption{A four-taxon unrooted tree representing the non-trivial split, $\left\{b,c\right\}|\left\{a,r\right\}$.}
		\end{subfigure}
		\caption{Two incompatible four-taxon unrooted trees, each corresponding to a different split.}
		\label{twoincompatibleunrootedtrees}
\end{figure}

If a set of splits are incompatible, such as the two trees above, and we wish to obtain a tree, some of the splits or some of the taxa can be removed until the splits are compatible. Alternatively, a \emph{split network} may be preferred instead.

A split network can be derived from a \emph{split graph}, a finite, connected graph. To get from a split graph to a split network we apply taxon labels to the leaves and split labels to the edges. Single edges represent splits which are compatible with every split on the graph. Sets of parallel edges represent splits which are incompatible with at least one split on the graph. Removing every edge corresponding to a particular split will result in two new split graphs, with each split graph representing one split part.

If the split network corresponds to an unrooted tree then every split will be represented by a single edge and removing one of these edges will result in the two new split graphs, both corresponding to unrooted trees. Each unrooted tree will contain the set of taxa from one partition of the split, with the two unrooted trees not sharing any taxa.

Split networks are not unique. There may be multiple split networks all representing the same set of splits. A split network for a compatible set of splits may not be an unrooted tree, however an unrooted tree will always exist for a compatible set of splits.

To see how incompatible splits on split networks work, let's look at an example. Suppose we have the six-taxon split graph displayed in Figure~\ref{sixtaxonsplitgraph} below.
\begin{figure}[H]
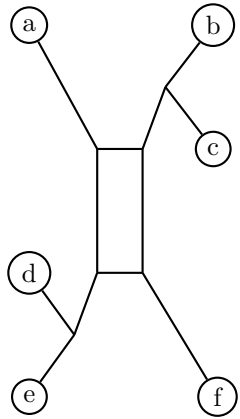

	\centering
		\psmatrix[colsep=.3cm,rowsep=.4cm,mnode=r]
		[mnode=circle] a && && && [mnode=circle] b \\
		&& && & {} \\
		&& {} && {} && [mnode=circle] c \\
		\\
		[mnode=circle] d && {} && {} \\
		& {} \\
		[mnode=circle] e && && && [mnode=circle] f \\
		\ncline{1,1}{3,3}
		\ncline{3,3}{3,5}
		\ncline{3,5}{2,6}
		\ncline{2,6}{1,7}
		\ncline{2,6}{3,7}
		\ncline{3,3}{5,3}
		\ncline{3,5}{5,5}
		\ncline{5,3}{6,2}
		\ncline{6,2}{7,1}
		\ncline{6,2}{5,1}
		\ncline{5,3}{5,5}
		\ncline{5,5}{7,7}
		\endpsmatrix
\caption{A six-taxon split graph.}
\label{sixtaxonsplitgraph}
\end{figure}

There are two sets of parallel edges, each representing a split which is incompatible with at least one other split on the graph. The two parallel vertical edges represent the split, $\left\{a,b,c\right\}|\left\{d,e,f\right\}$. The two parallel horizontal edges represent the split, $\left\{a,d,e\right\}|\left\{b,c,f\right\}$. Suppose we now removed the two edges corresponding to the split, $\left\{a,b,c\right\}|\left\{d,e,f\right\}$. We now have two split graphs, each representing one of the split parts, shown in Figure~\ref{twosplitgraphs} below.
\begin{figure}[H]
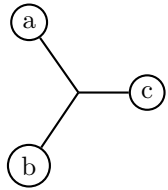
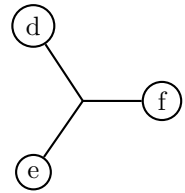

	\centering
		\begin{subfigure}[h]{0.49\textwidth}
		\centering
		\psmatrix[colsep=.3cm,rowsep=.4cm,mnode=r]
		[mnode=circle] a \\
		& ~ && [mnode=circle] c \\
		[mnode=circle] b
		\ncline{1,1}{2,2}
		\ncline{2,2}{3,1}
		\ncline{2,2}{2,4}
		\endpsmatrix
		\caption{The remaining three-taxon split graph \\*
representing the split part $\left\{a,b,c\right\}$ after the \\*
edges representing the split $\left\{a,b,c\right\}|\left\{d,e,f\right\}$ \\*
are removed.}
		\end{subfigure}
		\begin{subfigure}[h]{0.49\textwidth}
		\centering
		\psmatrix[colsep=.3cm,rowsep=.4cm,mnode=r]
		[mnode=circle] d \\
		& ~ && [mnode=circle] f \\
		[mnode=circle] e
		\ncline{1,1}{2,2}
		\ncline{2,2}{3,1}
		\ncline{2,2}{2,4}
		\endpsmatrix
		\caption{The remaining three-taxon split graph representing the split part $\left\{d,e,f\right\}$ after the edges representing the split $\left\{a,b,c\right\}|\left\{d,e,f\right\}$ are removed.}
		\end{subfigure}
		\caption{Two split graphs, each representing one of the split parts.}
		\label{twosplitgraphs}
\end{figure}

We could also remove the two edges representing the split $\left\{a,d,e\right\}|\left\{b,c,f\right\}$, shown in Figure~\ref{twosplitgraphssecond} below.
\begin{figure}[H]
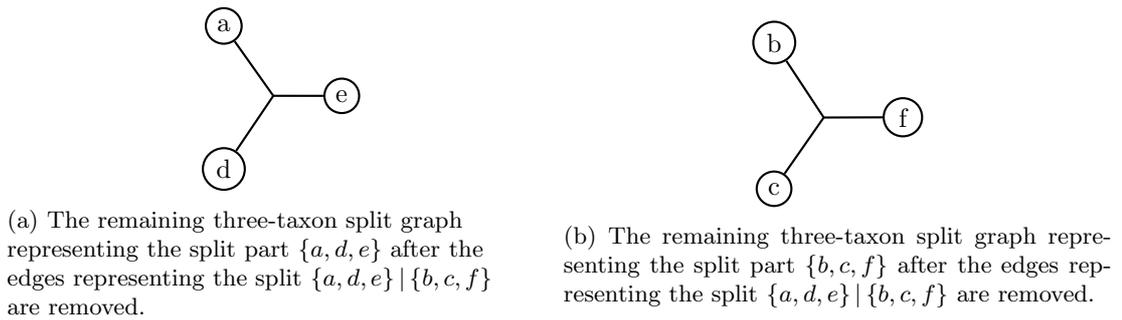

	\centering
		\begin{subfigure}[h]{0.49\textwidth}
		\centering
		\psmatrix[colsep=.3cm,rowsep=.4cm,mnode=r]
		[mnode=circle] a \\
		& ~ && [mnode=circle] e \\
		[mnode=circle] d
		\ncline{1,1}{2,2}
		\ncline{2,2}{3,1}
		\ncline{2,2}{2,4}
		\endpsmatrix
		\caption{The remaining three-taxon split graph \\*
representing the split part $\left\{a,d,e\right\}$ after the \\*
edges representing the split $\left\{a,d,e\right\}|\left\{b,c,f\right\}$ \\*
are removed.}
		\end{subfigure}
		\begin{subfigure}[h]{0.49\textwidth}
		\centering
		\psmatrix[colsep=.3cm,rowsep=.4cm,mnode=r]
		[mnode=circle] b \\
		& ~ && [mnode=circle] f \\
		[mnode=circle] c
		\ncline{1,1}{2,2}
		\ncline{2,2}{3,1}
		\ncline{2,2}{2,4}
		\endpsmatrix
		\caption{The remaining three-taxon split graph representing the split part $\left\{b,c,f\right\}$ after the edges representing the split $\left\{a,d,e\right\}|\left\{b,c,f\right\}$ are removed.}
		\end{subfigure}
		\caption{Two split graphs, each representing one of the split parts.}
		\label{twosplitgraphssecond}
\end{figure}

To see why the two splits are incompatible with each other, we need to compare the split parts from the two splits. The intersections of the split parts are
\begin{align}
\begin{mycases}
\left\{a,b,c\right\}\cap{}\left\{a,d,e\right\}&=\left\{a\right\}\neq{}\emptyset{},\\
\left\{a,b,c\right\}\cap{}\left\{b,c,f\right\}&=\left\{b,c\right\}\neq{}\emptyset{},\\
\left\{d,e,f\right\}\cap{}\left\{a,d,e\right\}&=\left\{d,e\right\}\neq{}\emptyset{},\\
\left\{d,e,f\right\}\cap{}\left\{b,c,f\right\}&=\left\{f\right\}\neq{}\emptyset{}.
\end{mycases}
\end{align}

Since none of these intersections are the empty set, the two splits are incompatible with each other.

When dealing with a data set, we rarely have the option of describing the data with a compatible split system. We may need to resort to an algorithm which returns an incompatible split system instead. The first attempts at obtaining sets of incompatible splits from data sets were made using the split decomposition method, introduced by \citet{bandelt1992split}. The split decomposition method uses a distance matrix and returns a set of weakly compatible weighted splits.

Split networks describe data sets, but do not necessarily have obvious biological meanings. For example, it is not clear what the biological meaning of a node connecting two incompatible splits is, such as on Figure~\ref{sixtaxonsplitgraph}~on~page~\pageref{sixtaxonsplitgraph}.

Directed acyclic graphs (DAGs) provide an alternative to split networks when a direction in time is known. Figure~\ref{fig:rootedDAG}~on~page~\pageref{fig:rootedDAG} shows a rooted DAG, where two edges join at the same node.

Generally, split systems and cluster systems are used to describe unrooted trees and rooted trees, respectively. We can, however, also use cluster systems to describe our convergence-divergence models.

\begin{figure}[H]
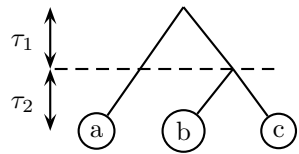

	\centering
		\psmatrix[colsep=.3cm,rowsep=.4cm,mnode=r]
		~ & && ~ \\
		~ & && & ~ & ~ \\
		~ & [mnode=circle] a && [mnode=circle] b && [mnode=circle] c
		\ncline{1,4}{3,2}
		\ncline{1,4}{2,5}
		\ncline{2,5}{3,4}
		\ncline{2,5}{3,6}
		\ncline{<->,arrowscale=1.5}{1,1}{2,1}
		\tlput{$\tau_{1}$}
		\ncline{<->,arrowscale=1.5}{2,1}{3,1}
		\tlput{$\tau_{2}$}
		\psset{linestyle=dashed}\ncline{2,1}{2,6}
		\endpsmatrix
		\caption{Three-taxon clock-like tree.}
		\label{threetaxonclocktwo}
\end{figure}

Let's look at the three-taxon clock-like tree in Figure~\ref{threetaxonclocktwo} below as an example.

The clusters for this tree are
\begin{align}
\begin{mycases}
a,b,c, \\
b,c, \\
a, \\
b, \\
c.
\end{mycases}
\end{align}

\begin{figure}[H]
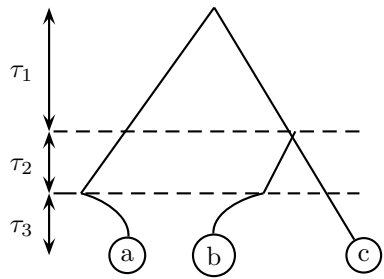

	\centering
		\psmatrix[colsep=.3cm,rowsep=.4cm,mnode=r]
		~ && && ~ \\
		~ \\
		~ && && && ~ && ~ \\
		~ & ~ && && ~ & && ~ \\
		~ && [mnode=circle] a && [mnode=circle] b && && [mnode=circle] c
		\ncline{1,5}{4,2}
		\ncline{1,5}{5,9}
		\ncline{3,7}{4,6}
		\ncarc[arcangle=40]{4,2}{5,3}
		\ncarc[arcangle=-40]{4,6}{5,5}
		\ncline{<->,arrowscale=1.5}{1,1}{3,1}
		\tlput{$\tau_{1}$}
		\ncline{<->,arrowscale=1.5}{3,1}{4,1}
		\tlput{$\tau_{2}$}
		\ncline{<->,arrowscale=1.5}{4,1}{5,1}
		\tlput{$\tau_{3}$}
		\psset{linestyle=dashed}\ncline{3,1}{3,7}
		\psset{linestyle=dashed}\ncline{3,7}{3,9}
		\psset{linestyle=dashed}\ncline{4,1}{4,2}
		\psset{linestyle=dashed}\ncline{4,2}{4,6}
		\psset{linestyle=dashed}\ncline{4,6}{4,9}
		\endpsmatrix
		\caption{Three-taxon clock-like convergence-divergence network with a convergence period involving the first and second taxa in the third time period. Convergence is represented by curved lines.}
		\label{threetaxoncondiv}
\end{figure}

Now let's look at the three-taxon convergence-divergence network in Figure~\ref{threetaxoncondiv} below as another example.

It is not clear what the cluster system should be for this convergence-divergence network. If there is a ``small'' amount of convergence then the convergence-divergence network will be ``similar'' to the three-taxon clock-like tree that we just discussed. Therefore, taxa $b$ and $c$ will still form a cluster and the cluster system will remain the same. However, if a ``large'' amount of convergence is allowed to happen, then taxa $a$ and $b$ will have progressively converged towards each other. Taxa $a$ and $b$ will now cluster together. The new cluster system would then be
\begin{align}
\begin{mycases}
a,b,c, \\
a,b, \\
a, \\
b, \\
c.
\end{mycases}
\end{align}

This is the same cluster system as described by the three-taxon clock-like tree in Figure~\ref{threetaxonclockthree} below.
\begin{figure}[H]
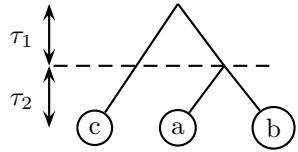

	\centering
		\psmatrix[colsep=.3cm,rowsep=.4cm,mnode=r]
		~ & && ~ \\
		~ & && & ~ & ~ \\
		~ & [mnode=circle] c && [mnode=circle] a && [mnode=circle] b
		\ncline{1,4}{3,2}
		\ncline{1,4}{2,5}
		\ncline{2,5}{3,4}
		\ncline{2,5}{3,6}
		\ncline{<->,arrowscale=1.5}{1,1}{2,1}
		\tlput{$\tau_{1}$}
		\ncline{<->,arrowscale=1.5}{2,1}{3,1}
		\tlput{$\tau_{2}$}
		\psset{linestyle=dashed}\ncline{2,1}{2,6}
		\endpsmatrix
		\caption{Three-taxon clock-like tree.}
		\label{threetaxonclockthree}
\end{figure}

Our convergence-divergence networks can be interpreted as representing a particular cluster system if the ``amount'' of convergence is known. If it unknown how much convergence has happened, then the convergence-divergence networks may correspond to multiple incompatible cluster systems. If sister taxa converge, then the cluster system describing the convergence-divergence network will be the same as the cluster system describing the tree with no convergence.

\subsubsection{Directed Acyclic Graphs (DAGs)}

Split networks generalise unrooted trees to allow for non-tree like behaviour. \emph{Directed acyclic graphs (DAGs)} allow us to generalise rooted trees to allow for non-tree like behaviour.

\begin{definition}[\textbf{Rooted DAG}]
A \textbf{DAG} is a graph which is both directed, such as in time, and has no directed cycles. A \textbf{rooted DAG} contains a single root, a node which has no edges leading to it in the specified direction.
\end{definition}

\begin{figure}[H]
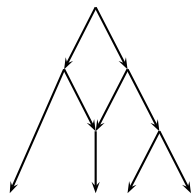

	\centering
		\psmatrix[colsep=.3cm,rowsep=.4cm,mnode=r]
		& && ~ \\
		&& ~ && ~ \\
		& && ~ && ~ \\
		~ & && ~ & ~ && ~ \\
		\psset{arrows=->}
		\ncline{1,4}{2,3}
		\ncline{2,3}{4,1}
		\ncline{2,3}{3,4}
		\ncline{3,4}{4,4}
		\ncline{1,4}{2,5}
		\ncline{2,5}{3,4}
		\ncline{2,5}{3,6}
		\ncline{3,6}{4,5}
		\ncline{3,6}{4,7}
		\endpsmatrix
\caption{A rooted DAG. Arrows refer to the direction of time.}
\label{fig:rootedDAG}
\end{figure}

Rooted DAGs are described in Definition 1.4.1 in \citet{huson2010phylogenetic}.

The definition of a DAG does not exclude cycles, it only excludes \emph{directed} cycles. Suppose the direction associated with the DAG is removed and we allow for any direction along an edge. There are some DAGs, such as the one above, for which a node can be returned to without the path requiring any edge being traversed twice.

Suppose we have two nodes, $a$ and $d$, on a DAG. We define $a$ as an \emph{ancestor} of $d$ and $d$ as a \emph{descendant} of $a$ if $d$ can be arrived at from $a$ in the specified direction of the DAG. \emph{Parents} and \emph{children} are specific types of ancestors and descendants, respectively, which are separated by a single edge on the DAG. The term \emph{lower} is used to define a descendant node in comparison to an ancestor node. We can compare edges in a similar way. If the ancestor of the first edge is lower than the descendant of the second edge then the first edge is lower than the second edge. If neither of a pair of nodes or edges is lower than the other then we call the pair of nodes or edges \emph{incomparable}.

The root of a DAG is the only node that does not have any ancestor node. Every other node in the DAG must be a descendant of the root, with those descendant nodes which are separated by a single edge also being child nodes. Conversely, the leaves of a DAG are the only nodes that do not have any descendant nodes. However, generally only some nodes will be ancestor nodes of an individual leaf. Every internal node, that is every node that is not the root or a leaf, must have the root as an ancestor node, at least one leaf as a descendant node and potentially some other internal nodes as ancestor or descendant nodes. DAGs may also contain \emph{sub-DAGs}. A sub-DAG fits the definition of a DAG, but is found within the DAG itself.

If the only node with no ancestor node is the root and every other node has exactly one ancestor node then the DAG is a rooted tree. If each node also has two descendant nodes, including leaves, then the tree is a rooted bifurcating tree. If we wish to obtain a rooted tree from a DAG we can selectively remove some of the edges from the DAG. For every node with multiple ancestor nodes, we must remove all but one of the edges connecting the node to an ancestor node. Let's look at Figure~\ref{fig:rootedDAG} as an example. There is only one node with multiple ancestor nodes. This node has two ancestor nodes. To convert the DAG into a rooted tree we must remove one of the edges connecting this node to an ancestor nodes. The two rooted trees which result are those in Figure~\ref{tworootedtrees} below.
\begin{figure}[H]
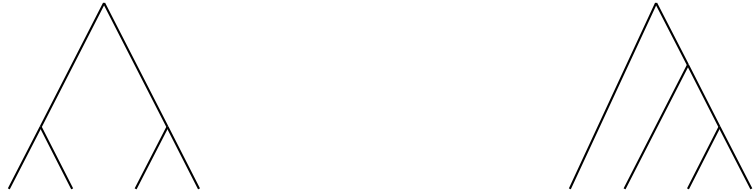

	\centering
		\begin{subfigure}[h]{0.49\textwidth}
		\centering
		\psmatrix[colsep=.3cm,rowsep=.4cm,mnode=r]
		& && ~ \\
		\\
		& ~ && && ~ \\
		~ && ~ && ~ && ~ \\
		\ncline{1,4}{3,2}
		\ncline{3,2}{4,1}
		\ncline{3,2}{4,3}
		\ncline{1,4}{3,6}
		\ncline{3,6}{4,5}
		\ncline{3,6}{4,7}
		\endpsmatrix
		\caption{The rooted tree which remains when the right \\*
edge above the node of the DAG with two \\*
ancestor nodes is removed.}
		\end{subfigure}
		\begin{subfigure}[h]{0.49\textwidth}
		\centering
		\psmatrix[colsep=.3cm,rowsep=.4cm,mnode=r]
		& && ~ \\
		&& && ~ \\
		& && && ~ \\
		~ && ~ && ~ && ~ \\
		\ncline{1,4}{4,1}
		\ncline{1,4}{2,5}
		\ncline{2,5}{4,3}
		\ncline{2,5}{3,6}
		\ncline{3,6}{4,5}
		\ncline{3,6}{4,7}
		\endpsmatrix
		\caption{The rooted tree which remains when the left edge above the node of the DAG with two ancestor nodes is removed.}
		\end{subfigure}
		\caption{The two rooted trees which remain when one of the edges is removed.}
		\label{tworootedtrees}
\end{figure}

We can see that a DAG can be used in circumstances where multiple rooted trees may be of interest. Situations where this may arise are when different sections of character sequences have evolved on different trees. Clearly DAGs are preferable over split networks in situations where a root is known, a direction in time is known or pairs of ancestor and descendant nodes or edges are known. DAGs can model processes success as hybridisation, horizontal gene transfer and recombination. However, they cannot model convergence, where diverged edges no longer undergo independent evolution and instead gradually become more similar over time. To model convergence we will need our convergence-divergence model.

\subsection{Identifiability}

An issue that often arises in model inference in phylogenetics is the issue of inferring the appropriate tree or network and its parameters from the phylogenetic tensor. Suppose we have chosen a model of evolution and we wish to compare the phylogenetic tensors for multiple tree and network structures. Recall that the phylogenetic tensor is the probability distribution for a specified tree or network. The phylogenetic tensor is dependent on the tree or network structure, the rate parameters of the model and the edge parameters. For the binary symmetric model, for each edge parameter we are often only interested in the product of the rate and edge parameters as the two parameters cannot be resolved independently.

\citet{allman2008identifying} defined \emph{identifiability} with respect to both numerical parameters and models. Some other recent articles addressing the issue of identifiability from Allman and Rhodes are \citet{allman2006identifiability,allman2009identifiability,allman2008identifiability,allman2009identifiability2,allman2011identifiability,allman2011parameter,allman2015parameter}. We will discuss some of the important concepts in their articles and adapt them to our needs, in the process introducing some new definitions to help us compare tree and network structures.

\begin{definition}[\textbf{Numerical Parameter Identifiability}]
Numerical parameters are \textbf{identifiable} if there is an injective map between the parameter space and the pattern frequencies.
\end{definition}

\begin{definition}[\textbf{Model Identifiability}]
A model is \textbf{identifiable} if there is a unique tree for every possible pattern frequency from the trees that we are considering.
\end{definition}

\citet{allman2008identifying} recognised that there are some issues with these definitions. The first problem they stated was that the numerical parameters may not be identifiable if the numerical parameter space is unrestricted. They used the four-taxon star tree as an example. They argued that any phylogenetic tensor arising from this tree could have arisen from any other four-taxon tree with the appropriate edge lengths. The reason for this is that any four-taxon tree, or network, can be made to be equivalent to the star tree by letting the appropriate edge lengths tend towards zero. Consequently, the space of the phylogenetic tensor for the star tree contains the spaces of the phylogenetic tensors of every other tree, as well as network.

\begin{figure}[H]
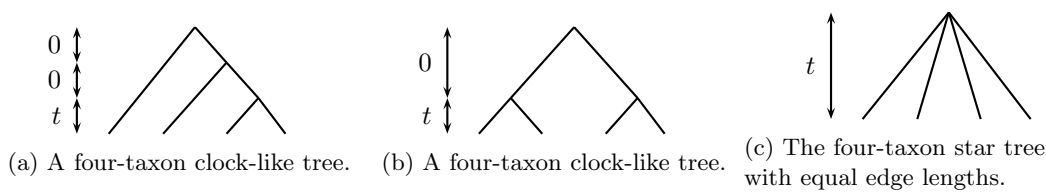

	\begin{subfigure}[h]{0.33\textwidth}
	\centering
		\psmatrix[colsep=.3cm,rowsep=.05cm,mnode=r]
		~ && && ~ \\
		~ & && && ~ \\
		~ && && && ~ \\
		~ & ~ && ~ && ~ && ~
		\ncline{<->}{1,1}{2,1}
		\tlput{$0$}
		\ncline{<->}{2,1}{3,1}
		\tlput{$0$}
		\ncline{<->}{3,1}{4,1}
		\tlput{$t$}
		\ncline{1,5}{4,2}
		\ncline{1,5}{2,6}
		\ncline{2,6}{4,4}
		\ncline{2,6}{3,7}
		\ncline{3,7}{4,6}
		\ncline{3,7}{4,8}
		\endpsmatrix
		\caption{A four-taxon clock-like tree.}
	\end{subfigure}
	\begin{subfigure}[h]{0.33\textwidth}
	\centering
		\psmatrix[colsep=.3cm,rowsep=.05cm,mnode=r]
		~ && && ~ \\
		\\
		~ && ~ && && ~ \\
		~ & ~ && ~ && ~ && ~
		\ncline{<->}{1,1}{3,1}
		\tlput{$0$}
		\ncline{<->}{3,1}{4,1}
		\tlput{$t$}
		\ncline{1,5}{3,3}
		\ncline{3,3}{4,2}
		\ncline{3,3}{4,4}
		\ncline{1,5}{3,7}
		\ncline{3,7}{4,6}
		\ncline{3,7}{4,8}
		\endpsmatrix
		\caption{A four-taxon clock-like tree.}
	\end{subfigure}
	\begin{subfigure}[h]{0.33\textwidth}
	\centering
		\psmatrix[colsep=.3cm,rowsep=.05cm,mnode=r]
		~ && && ~ \\
		\\
		\\
		~ & ~ && ~ && ~ && ~
		\ncline{<->}{1,1}{4,1}
		\tlput{$t$}
		\ncline{1,5}{4,2}
		\ncline{1,5}{4,4}
		\ncline{1,5}{4,6}
		\ncline{1,5}{4,8}
		\endpsmatrix
		\caption{The four-taxon star tree \\* 
		with equal edge lengths.}
	\end{subfigure}
	\caption{Four-taxon tree structures.}
	\label{fourtaxontreestructures}
\end{figure}

Let's look at some four-taxon tree structures in Figure~\ref{fourtaxontreestructures} below as an example.

The three tree structures are clearly identical when some of the edge lengths are set to zero. Likewise, we could have identical tree structures for any other four-taxon tree, or even a convergence-divergence network, with some of the edges, including any with convergence, set to zero. We will address this issue later with a definition that we will call \emph{distinguishability}.

We will use the definitions of \citet{allman2008identifying}, with some slight name changes. \emph{Numerical parameter identifiability} will simply be called \emph{identifiability}. \emph{Model identifiability} will be called \emph{network identifiability} to reflect our generalisation from tree structures to network structures.

Can we put limits on what structures are sensible to consider with these models? In this thesis we will examine some simple cases on two, three and four taxa and determine in what circumstances parameters are recoverable. In other words, given a particular set of time parameters, is there a one-to-one map between time parameters and the phylogenetic tensor? This is a question of {\bf identifiability}. Now suppose we have a two-taxon network where taxon $1$ and taxon $2$ diverge for time $t_1$ and then converge for time $t_2$. Is there an equivalent set of time parameters where they diverge for time $t'$ with no later convergence? This is a question of {\bf network identifiability}. Thirdly, given a set of pattern frequencies, could they have arisen on two trees or networks or only on one of the trees or networks? This is a question of {\bf distinguishability}.

\begin{definition}[\textbf{Distinguishable}]
Two networks are said to be \textbf{distinguishable} if the spaces of their phylogenetic tensors, the family of probability distributions, are not identical. That is, there are some phylogenetic tensors, or pattern frequencies, that can arise on one network, but not the other.
\end{definition}

To answer questions about identifiability, network identifiability or distinguishability we need to be able to describe what the space of the phylogenetic tensors looks like. What constraints are there under different sets of time parameters? By making use of the Hadamard basis for comparisons we are able to fully describe the situation for the two and three-taxon cases, as well as some of the four-taxon cases, under the binary symmetric model.

\subsection{Group-Based Models}

In a series of papers from the late 80s and 90s, Hendy, Penny, Steel and co-authors showed that for Abelian group-based models, the Kimura $3$ parameter model and its submodels, there was an invertible transformation between edge weights and site pattern probabilities \citep{hendy1989framework,hendy1993spectral,hendy1989relationship,hendy1994discrete,steel1992spectral}. They called this transformation ``Hadamard conjugation''. \citet{von1993network} were the first to note that Hadamard conjugation provided one way to think about the likelihood of more general split systems than trees. \citet{bryant2009hadamard} followed up on this viewpoint, showing how to extend likelihood to general split systems for the Abelian models, the so called ``$n$-taxon process''. \citet{bashford2004u} introduced the ``splitting operator'', which can be used to represent bifurcations at nodes on a tree. In their follow up paper, \citet{sumner2012algebra} discussed in more detail the properties of the splitting operator, arguing that it can be applied to the general Markov model on trees and suggested that it could also be used to model convergence on a network.

A question that often arises in molecular phylogenetics is how similar sequences are and over how long those sequences have diverged from each other. With a molecular clock assumption, we can assume that the total number of substitutions in a character sequence is proportional to the time that those substitutions have occurred over. We call the total number of substitutions that have occurred from the root to all of the leaves the ``evolutionary distance''. The early work of Hendy and Penny showed how Hadamard conjugation can be used to compare the evolutionary distance of a tree and the pairwise distances on a tree to the probability distribution of the tree.

To illustrate how this is done, we will use the K$3$ST model as an example. The K$3$ST model was introduced by \citet{kimura1981estimation} to model three different types of nucleotide substitutions, one type of ``transition'' and two types of ``transversions''. Transitions between the DNA nucleotides, C and T, or A and G, were all given the same time independent rate, $\alpha{}$. Transversions between the DNA nucleotides, A and T, or C and G, were given the time independent rate, $\beta{}$. The other type of transversions between the DNA nucleotides, A and C, or G and T, were given the time independent rate, $\gamma{}$. Diagrammatically, these substitutions are represented in Figure~\ref{K3STmodel} below.
\begin{figure}[H]
	\centering
		\begin{tikzpicture}[->,auto,node distance=2cm,thick,main node/.style={circle,draw}]
			\node[main node] (1) {T};
			\node[main node] (2) [below of=1] {A};
			\node[main node] (3) [right of=1] {C};
			\node[main node] (4) [right of=2] {G};
			\path[every node/.style={}]
      (1) edge [<->] node {$\beta{}$} (2)
			(1) edge [<->] node {$\alpha{}$} (3)
			(1) edge [<->] node [pos=0.6] {$\gamma{}$} (4)
			(2) edge [<->] node [pos=0.4] {$\gamma{}$} (3)
			(2) edge [<->] node {$\alpha{}$} (4)
			(3) edge [<->] node {$\beta{}$} (4);
		\end{tikzpicture}
		\caption{The K$3$ST model.}
		\label{K3STmodel}
\end{figure}

Suppose that we have two taxa that have evolved independently on a clock-like tree from a common ancestor. We wish to estimate the number of substitutions of each type that have occurred and the evolutionary distance. We cannot simply look at the two sequences and find the number of substitutions of each type. The problem with this process is that intermediate substitutions will always be ignored and the estimated number of substitutions will always be less than or equal to the actual number of substitutions. For example, suppose we can only ``see'' the site at the ``start'' and at the ``end'' of the process and that substitutions on a site are from C to G to A. We will only be able to see a substitution from C at the start to A at the end of the process. Assuming a Poisson process, Kimura found expressions for the expected number of each type of substitution which took into account intermediate substitutions. \citet{hendy2005hadamard} denoted these numbers of substitutions $r$, with the subscript referring to the type of substitution. The expressions they gave for the numbers of substitutions were
\begin{align}
\begin{mycases}
r_{\emptyset{}}&=\frac{1}{4}\left[\ln\left(1-2p_{\alpha{}}-2p_{\gamma{}}\right)+\ln\left(1-2p_{\beta{}}-2p_{\gamma{}}\right)+\ln\left(1-2p_{\alpha{}}-2p_{\beta{}}\right)\right], \\
r_{\alpha{}}&=\frac{1}{4}\left[-\ln\left(1-2p_{\alpha{}}-2p_{\gamma{}}\right)+\ln\left(1-2p_{\beta{}}-2p_{\gamma{}}\right)-\ln\left(1-2p_{\alpha{}}-2p_{\beta{}}\right)\right], \\
r_{\beta{}}&=\frac{1}{4}\left[\ln\left(1-2p_{\alpha{}}-2p_{\gamma{}}\right)-\ln\left(1-2p_{\beta{}}-2p_{\gamma{}}\right)-\ln\left(1-2p_{\alpha{}}-2p_{\beta{}}\right)\right], \\
r_{\gamma{}}&=\frac{1}{4}\left[-\ln\left(1-2p_{\alpha{}}-2p_{\gamma{}}\right)-\ln\left(1-2p_{\beta{}}-2p_{\gamma{}}\right)+\ln\left(1-2p_{\alpha{}}-2p_{\beta{}}\right)\right],
\end{mycases}
\end{align}
where $p$ is the time dependent probability of a substitution, with the subscript referring to the type of substitution and $-r_{\emptyset{}}=r_{\alpha{}}+r_{\beta{}}+r_{\gamma{}}$ being the evolutionary distance.

$p_{\emptyset{}}=1-p_{\alpha{}}-p_{\beta{}}-p_{\gamma{}}$ is the probability that the two sequences have the same nucleotide state at any given site. The equations then become
\begin{align}
\begin{mycases}
r_{\emptyset{}}&=\frac{1}{4}\left[\ln\left(p_{\emptyset{}}-p_{\alpha{}}+p_{\beta{}}-p_{\gamma{}}\right)+\ln\left(p_{\emptyset{}}+p_{\alpha{}}-p_{\beta{}}-p_{\gamma{}}\right)+\ln\left(p_{\emptyset{}}-p_{\alpha{}}-p_{\beta{}}+p_{\gamma{}}\right)\right], \\
r_{\alpha{}}&=\frac{1}{4}\left[-\ln\left(p_{\emptyset{}}-p_{\alpha{}}+p_{\beta{}}-p_{\gamma{}}\right)+\ln\left(p_{\emptyset{}}+p_{\alpha{}}-p_{\beta{}}-p_{\gamma{}}\right)-\ln\left(p_{\emptyset{}}-p_{\alpha{}}-p_{\beta{}}+p_{\gamma{}}\right)\right], \\
r_{\beta{}}&=\frac{1}{4}\left[\ln\left(p_{\emptyset{}}-p_{\alpha{}}+p_{\beta{}}-p_{\gamma{}}\right)-\ln\left(p_{\emptyset{}}+p_{\alpha{}}-p_{\beta{}}-p_{\gamma{}}\right)-\ln\left(p_{\emptyset{}}-p_{\alpha{}}-p_{\beta{}}+p_{\gamma{}}\right)\right], \\
r_{\gamma{}}&=\frac{1}{4}\left[-\ln\left(p_{\emptyset{}}-p_{\alpha{}}+p_{\beta{}}-p_{\gamma{}}\right)-\ln\left(p_{\emptyset{}}+p_{\alpha{}}-p_{\beta{}}-p_{\gamma{}}\right)+\ln\left(p_{\emptyset{}}-p_{\alpha{}}-p_{\beta{}}+p_{\gamma{}}\right)\right].
\end{mycases}
\end{align}

It can be immediately seen that these expressions are all sums of the same three logarithmic terms, with only the signs of the terms differing between the expressions. In matrix form,
\begin{align}
\begin{split}
\left[\begin{array}{c} r_{\emptyset{}} \\
r_{\alpha{}} \\
r_{\beta{}} \\
r_{\gamma{}} \\
\end{array}\right]&=\frac{1}{4}\left[\begin{array}{cccc} 1 & 1 & 1 & 1  \\
1 & -1 & 1 & -1 \\
1 & 1 & -1 & -1 \\
1 & -1 & -1 & 1 \\
\end{array}\right]\cdot{}\left[\begin{array}{c} 0  \\
\ln\left(p_{\emptyset{}}-p_{\alpha{}}+p_{\beta{}}-p_{\gamma{}}\right) \\
\ln\left(p_{\emptyset{}}+p_{\alpha{}}-p_{\beta{}}-p_{\gamma{}}\right) \\
\ln\left(p_{\emptyset{}}-p_{\alpha{}}-p_{\beta{}}+p_{\gamma{}}\right)
\end{array}\right].
\end{split}
\end{align}
Note that $\ln\left(p_{\emptyset{}}+p_{\alpha{}}+p_{\beta{}}+p_{\gamma{}}\right)=\ln\left(1\right)=0$.

Recognising the Sylvester representation of the Hadamard matrix,
\begin{align}
\begin{split}
H_{4}=\left[\begin{array}{cccc} 1 & 1 & 1 & 1  \\
1 & -1 & 1 & -1 \\
1 & 1 & -1 & -1 \\
1 & -1 & -1 & 1 \\
\end{array}\right],
\end{split}
\end{align}
and defining the vector logarithm as the component-wise logarithm of every element of the vector, this is the equation
\begin{align}
\begin{split}
r&=H_{4}^{-1}\cdot{}\ln\left(H_{4}\cdot{}p\right),
\end{split}
\end{align}
where
\begin{align}
\begin{split}
r=\left[\begin{array}{c} r_{\emptyset{}} \\
r_{\alpha{}} \\
r_{\beta{}} \\
r_{\gamma{}} \\
\end{array}\right],\quad{}p=\left[\begin{array}{c} p_{\emptyset{}} \\
p_{\alpha{}} \\
p_{\beta{}} \\
p_{\gamma{}} \\
\end{array}\right].
\end{split}
\end{align}

Inverting the equation,
\begin{align}
\begin{split}
p&=H_{4}^{-1}\cdot{}\exp\left(H_{4}\cdot{}r\right).
\end{split}
\end{align}

This process is referred to as ``Hadamard conjugation''. We will see in the second chapter why the Hadamard matrix is chosen for the K$3$ST model. We will show that a more general matrix, for which the Hadamard matrix is contained within, will ``transform'' the K$3$ST model. We will expand on the Hadamard conjugation for the K$3$ST model and show that analogous conjugations, or transformations, can be found for any Abelian group-based model.

In this thesis we will generally be working in the transformed basis. For the K$3$ST model our transformed basis will be
\begin{align}
\begin{split}
q&=H_{4}\cdot{}p=\left[\begin{array}{cccc} 1 & 1 & 1 & 1  \\
1 & -1 & 1 & -1 \\
1 & 1 & -1 & -1 \\
1 & -1 & -1 & 1 \\
\end{array}\right]\cdot{}\left[\begin{array}{c} p_{\emptyset{}} \\
p_{\alpha{}} \\
p_{\beta{}} \\
p_{\gamma{}} \\
\end{array}\right]=\left[\begin{array}{c} p_{\emptyset{}}+p_{\alpha{}}+p_{\beta{}}+p_{\gamma{}} \\
p_{\emptyset{}}-p_{\alpha{}}+p_{\beta{}}-p_{\gamma{}} \\
p_{\emptyset{}}+p_{\alpha{}}-p_{\beta{}}-p_{\gamma{}} \\
p_{\emptyset{}}-p_{\alpha{}}-p_{\beta{}}+p_{\gamma{}} \\
\end{array}\right].
\end{split}
\end{align}

For $n$ taxa, we simply need the corresponding Hadamard matrix of dimension $4^{n}$. The transformed basis will be
\begin{align}
\begin{split}
q&=H_{4}^{\otimes{n}}\cdot{}p.
\end{split}
\end{align}

Hadamard conjugation has the capacity to generalise model based inference beyond trees. We will later show how we can use Hadamard conjugation on our convergence-divergence models.

\subsection{Tensor Representation of Models}

In \citet{bashford2004u} the splitting operator was introduced. The splitting operator represents the ``branching process'' and is a linear operator which takes a vector space and outputs the tensor product of the space with itself. In other words, if we have unit vectors, each representing a single character state for a single taxon, the tensor product of each unit vector with itself is another unit vector, now representing the same character on two descendant taxa. Initially after the branching process, the two taxa must be identical. Since the splitting operator is a linear operator, the action of the splitting operator on the phylogenetic tensor for one taxon is a phylogenetic tensor for the two descendant taxa, which must initially be identical.

The phylogenetic tensor for a single taxon can be represented as a sum of the probabilities over all of the possible states. For example, for a binary character state space, $\{0,1\}$, the phylogenetic tensor can be represented as
\begin{align}
\begin{split}
P^{\left(1\right)}\left(t\right)&=p_{0}\left(t\right)e_{0}+p_{1}\left(t\right)e_{1},
\end{split}
\end{align}
where the $p_{i}\left(t\right)$ are the probabilities for the respective states after time $t$ has elapsed from the root and the $e_{0}$ are the respective unit vectors.

\begin{figure}[H]
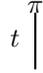

	\centering
		\psmatrix[colsep=.3cm,rowsep=.4cm,mnode=r]
		$\pi{}$ \\
		~
		\ncline{1,1}{2,1}
		\tlput{$t$}
		\endpsmatrix
\caption{A single taxon phylogenetic tree, where $\pi{}$ is the probability distribution at the root and $P^{\left(1\right)}\left(t\right)$ is the phylogenetic tensor at the single leaf.}
\label{singletaxontree}
\end{figure}

In Figure~\ref{singletaxontree} below is a graphical representation of a single taxon phylogenetic tree, with time $t$ elapsed from the root.

The splitting operator, denoted $\delta{}$, acts on the unit vectors as follows,
\begin{align}
\begin{mycases}
\delta{}\cdot{}e_{0}=e_{0}\otimes{}e_{0}=\left[\begin{array}{c} 1 \\
0 \\
\end{array}\right]\otimes{}\left[\begin{array}{c} 1 \\
0 \\
\end{array}\right]=\left[\begin{array}{c} 1 \\
0 \\
0 \\
0 \\
\end{array}\right]=e_{00}, \\
\delta{}\cdot{}e_{1}=e_{1}\otimes{}e_{1}=\left[\begin{array}{c} 0 \\
1 \\
\end{array}\right]\otimes{}\left[\begin{array}{c} 0 \\
1 \\
\end{array}\right]=\left[\begin{array}{c} 0 \\
0 \\
0 \\
1 \\
\end{array}\right]=e_{11},
\end{mycases}
\end{align}
where $\otimes{}$ is the Kronecker product.

The phylogenetic tensor for two taxa initially after the branching process is then
\begin{align}
\begin{split}
P^{\left(2\right)}\left(t\right)&=\delta{}\cdot{}\left(p_{0}\left(t\right)e_{0}+p_{1}\left(t\right)e_{1}\right) \\
&=\delta{}\cdot{}\left(p_{0}\left(t\right)e_{0}\right)+\delta{}\cdot{}\left(p_{1}\left(t\right)e_{1}\right) \\
&=p_{00}\left(t\right)\delta{}\cdot{}e_{0}+p_{11}\left(t\right)\delta{}\cdot{}e_{1} \\
&=p_{00}\left(t\right)e_{00}+p_{11}\left(t\right)e_{11},
\end{split}
\end{align}
where $p_{00}\left(t\right)=p_{0}\left(t\right)$ and $p_{11}\left(t\right)=p_{1}\left(t\right)$ are scalar probability functions of time, $t$.

The two-taxon phylogenetic tree immediately after the branching process is shown in Figure~\ref{twotaxonbranching} below.
\begin{figure}[H]
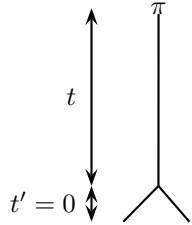

	\centering
		\psmatrix[colsep=.3cm,rowsep=0.05cm,mnode=r]
		~ && $\pi{}$ \\
		&& \\
		&& \\
		&& \\
		&& \\
		~ && ~ \\
		~ & ~ && ~
		\ncline{1,3}{6,3}
		\ncline{<->,arrowscale=1.5}{1,1}{6,1}
		\tlput{$t$}
		\ncline{6,3}{7,2}
		\ncline{6,3}{7,4}
		\ncline{<->,arrowscale=1.5}{6,1}{7,1}
		\tlput{$t'=0$}
		\endpsmatrix
\caption{A two-taxon phylogenetic tree initially after the branching process, where $\pi{}$ is the probability distribution at the root and $t$ and $t'$ are the time parameters.}
\label{twotaxonbranching}
\end{figure}

Supposing the two edges then diverge over some time period, $t'$, according to a Markov model, with transition matrix, $M\left(t'\right)$, the phylogenetic tensor then becomes
\begin{align}
\begin{split}
P^{\left(2\right)}\left(t+t'\right)&=\left(M\left(t'\right)\otimes{}M\left(t'\right)\right)\cdot{}P^{\left(2\right)}\left(t\right) \\
&=p_{00}\left(t+t'\right)e_{00}+p_{01}\left(t+t'\right)e_{01}+p_{10}\left(t+t'\right)e_{10}+p_{11}\left(t+t'\right)e_{11}.
\end{split}
\end{align}

Additional branching processes and diverging edges can be applied to the tree to generate any $n$-taxon tree that is desired. Non-clock-like trees and star trees can also be generated by applying different time periods to adjacent edges and by applying consecutive branching processes with no divergence between them. For example, suppose we applied the splitting operator to the root of a tree. We would now have a two-taxon tree with the phylogenetic tensor being the initial probability distribution at the root on both taxa. Suppose now that we applied the splitting operator again to one of the edges descending from the root. We would now have a three-taxon tree, with all three taxa splitting from the root and with the phylogenetic tensor being the initial probability distribution at the root on all three taxa. Finally, suppose we now applied transition matrices from a Markov model to each of the three edges, but over different time periods. We would then essentially have a three-taxon non-clock-like tree or a three-taxon star tree with different edge lengths, shown in Figure~\ref{threetaxonstar} below.
\begin{figure}[H]
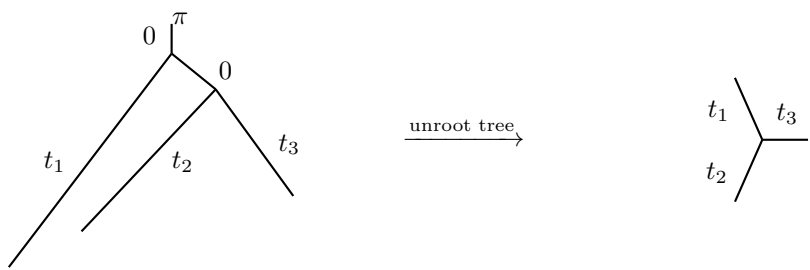

	\centering
		\begin{subfigure}[h]{0.43\textwidth}
			\centering
				\psmatrix[colsep=.3cm,rowsep=.05cm,mnode=r]
				&& && && ~ $\pi{}$ \\
				&& && && ~ \\
				&& && && & ~ \\
				&& && && && \\
				&& && && && & \\
				&& && && && && ~ \\
				&& & ~ \\
				~
				\ncline{1,7}{2,7}
				\tlput{$0$}
				\ncline{2,7}{8,1}
				\tlput{$t_{1}$}
				\ncline{2,7}{3,8}
				\trput{$0$}
				\ncline{3,8}{7,4}
				\trput{$t_{2}$}
				\ncline{3,8}{6,11}
				\trput{$t_{3}$}
				\endpsmatrix
		\end{subfigure}
		$\xrightarrow{\text{unroot tree}}$
		\begin{subfigure}[h]{0.43\textwidth}
			\centering
				\psmatrix[colsep=.3cm,rowsep=.4cm,mnode=r]
				~ \\
				& {} && ~ \\
				~
				\ncline{1,1}{2,2}
				\tlput{$t_{1}$}
				\ncline{2,2}{3,1}
				\tlput{$t_{2}$}
				\ncline{2,2}{2,4}
				\taput{$t_{3}$}
				\endpsmatrix
		\end{subfigure}
		\caption{A three-taxon non-clock-like tree or star tree with uneven edges.}
		\label{threetaxonstar}
		\end{figure}

Recently, \citet{sumner2012algebra} showed how the $n$-taxon process could be extended to the general Markov model. The convergence-divergence networks that they introduced can be thought of as applying combinations of two different processes to a set of $n$ taxa. There is a convergent process which acts to make taxa more similar (or in the case of taxa that have yet to diverge it keeps them identical), and a divergent process which implements the standard idea of conditionally independently evolving edges. For models where taxa that have diverged never later converge, this is completely equivalent to the standard tree approach. We will consider what might happen if the convergent process is applied to a subset of the taxa that have previously been diverging. In its most general form the convergence-divergence model places no restriction on the number of time parameters, each with associated partitions of the taxa defining which subsets of the taxa are converging with each other and which are diverging from every other taxon.

Suppose now that we applied the action of the splitting operator to a transition matrix. We will explain the motivation for applying the splitting operator to a transition matrix later in the thesis. Using the binary symmetric model as an example, the splitting operator performs the following action on a transition matrix,
\begin{align}
\begin{split}
\delta{}\cdot{}M\left(t\right)&=e^{\lambda{}R_{11}t}\cdot{}\delta{},
\end{split}
\end{align}
where
\begin{align}
\begin{split}
\lambda{}R_{11}=\left[\begin{array}{c|cccc}
& 00 & 01 & 10 & 11 \\
\hline
00 & -\lambda{} & \lambda{} & \lambda{} & \lambda{} \\
01 & 0 & -2\lambda{} & 0 & 0 \\
10 & 0 & 0 & -2\lambda{} & 0 \\
11 & \lambda{} & \lambda{} & \lambda{} & -\lambda{} \\
\end{array}\right].
\end{split}
\end{align}

By inspecting $\lambda{}R_{11}$, it can be seen that this rate matrix can be used to model convergence of two or more diverged edges. The rate matrix can be represented graphically, as shown below in Figure~\ref{ratematrix}.
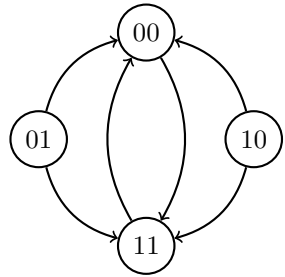
\begin{figure}[H]
	\centering
		\begin{tikzpicture}[->,node distance=2cm,thick,main node/.style={circle,draw}]
			\node[main node] (1) {00};
			\node[main node] (2) [below left of=1] {01};
			\node[main node] (3) [below right of=2] {11};
			\node[main node] (4) [below right of=1] {10};
			\path[every node/.style={}]
				(1) edge [bend left] node[above] {} (3)
				(2) edge [bend left] node[left] {} (1)
				(2) edge [bend right] node[left] {} (3)
				(3) edge [bend left] node[below] {} (1)
				(4) edge [bend right] node[right] {} (1)
				(4) edge [bend left] node[right] {} (3);
		\end{tikzpicture}
		\caption{$\lambda{}R_{11}$ on two taxa.}
		\label{ratematrix}
\end{figure}
Each character state transition has the same rate, $\lambda{}$, from the binary symmetrical model.

If we want to model convergence, we must apply the convergence to two edges that have diverged away from each other. It may appear that the presence of the splitting operator immediately after the transition matrix for the convergence of two edges is problematic. Originally we had the splitting operator acting on a transition matrix, however it is no longer apparent what the splitting operator is acting on as it now appears after the transition matrix for the convergence of two edges. We call this process ``pushing back'' the splitting operator. We will show that every splitting operator on a tree or convergence-divergence network can be ``pushed back'' as many times as necessary to act on the root of the tree or network. The splitting operator is therefore not only useful for modelling bifurcations at nodes, but also the convergence of two adjacent edges. We will show later in the thesis that this convergence can be generalised to allow for convergence of any subset of $n$ adjacent taxa.

For more detail on the algebra of the splitting operator see \citet{bashford2004u} and \citet{sumner2012algebra}.

\chapter{Diagonalisation of Markov Models}
\label{chapter2}

\section{Markov Chains in Phylogenetics}

Markov chains are stochastic processes acting on a state space. The defining property of a Markov chain is that the probability, or rate over time, of a transition from one state in the state space to another depends only on the current state and not on the path that was taken to arrive at that state. Markov chains are utilised in many diverse areas of mathematics involving stochastic processes, including phylogenetics. Phylogenetic methods, such as likelihood and Bayesian distance correction, rely on stochastic models of nucleotide substitutions. It is common to make some simplifying assumptions, such as the i.i.d. assumption. The i.i.d. assumption is the assumption that every character in a sequence is an independent and identically distributed random variable. Each character behaves independently of every other character and obeys the same Markov process. Each character is defined by a Markov chain, with the same Markov model of evolution, including the parameters defining the rates of mutations.

The state space is the set of all possible objects in a system. States depend on the type of characters to be considered. Examples of state spaces for biological sequences in phylogenetics are the DNA nucleobases (cytosine (C), guanine (G), adenine (A) and thymine (T)), the amino acids or a set of morphological traits. The DNA nucleobases can be partitioned into the purines (A and G) and the pyrimidines (C and T), reducing the states in the state space from four to two. Other than in this chapter, we will only examine the binary symmetric model, a state space with two states. We might choose these two states to be the purines and pyrimidines or two distinct morphological traits. Alternatively, we could use a two state model for single nucleotide polymorphisms (SNPs). SNPs are variations at a single site in nucleotide sequences within a species. \citet{hodgkinson2010human} state that there are usually two alleles in a SNP. Hence, a two state model is commonly used. 

Mendelian traits are characterised by the presence of a dominant or recessive phenotypic trait. If an offspring receives two recessive alleles, one from each parent, on two paired chromosomes, then they will display the recessive phenotypic trait. Alternatively, if an offspring receives at least one dominant allele then they will display the dominant phenotypic trait. The state space for Mendelian traits will contain two states, one for each allele.

In phylogenetics, mutations passed on to offspring follow a homogeneous discrete time Markov chain, since the separation between generations is a discrete period of time. However, the evolution of a population follows a homogeneous continuous time Markov chain as mutations to the genome of a population can happen continuously through time.

\section{The Markov Property on Transition Matrices}

The Markov property is the expression that the behaviour of a Markov chain is independent of the path taken from time $t$ to time $u$ and is only dependent on the states at times $t$ and $u$. For any set of non-negative times, $s_{1}\leq{}s_{2}\leq{}\ldots{}\leq{}s_{k}\leq{}t$, the Markov property can be expressed on a probability distribution as
\begin{align}\begin{split}
\mathbb{P}\left[X\left(t\right)=i|X\left(s_{1}\right)=i_{1},X\left(s_{2}\right)=i_{2},\ldots{},X\left(s_{k}\right)=i_{k}\right]=\mathbb{P}\left[X\left(t\right)=i|X\left(s_{k}\right)=i_{k}\right],
\end{split}\end{align}
where $i_{1},i_{2},\ldots{},i_{k},i$ are the states the Markov chain, $X$, takes at the respective times.

Assuming we have some process, which is not necessarily Markov, to find the probability of taking a particular final state given an initial state we must sum over all of the possible states between the initial and final states. The probability can be expressed as
\begin{align}\begin{split}
\mathbb{P}\left[X\left(t\right)=i|X\left(s_{1}\right)=i_{1}\right]=&\sum\limits_{i_{2},i_{3},\ldots{},i_{k}}^{}\mathbb{P}\left[X\left(s_{2}\right)=i_{2}|X\left(s_{1}\right)=i_{1}\right] \\
&\mathbb{P}\left[X\left(s_{3}\right)=i_{3}|X\left(s_{1}\right)=i_{1},X\left(s_{2}\right)=i_{2}\right] \\
&\ldots{}\mathbb{P}\left[X\left(t\right)=i|X\left(s_{1}\right)=i_{1},X\left(s_{2}\right)=i_{2},\ldots{},X\left(s_{k}\right)=i_{k}\right].
\end{split}\end{align}

In practice we may not have any knowledge of the behaviour of the process over a period of time. We may only know the initial state of the Markov chain and the final or current state. If the process is Markov then we can utilise the Markov property as follows,
\begin{align}\begin{split}
\mathbb{P}\left[X\left(t\right)=i|X\left(s_{1}\right)=i_{1}\right]=&\sum\limits_{i_{2},i_{3},\ldots{},i_{k}}^{}\mathbb{P}\left[X\left(s_{2}\right)=i_{2}|X\left(s_{1}\right)=i_{1}\right] \\
&\mathbb{P}\left[X\left(s_{3}\right)=i_{3}|X\left(s_{2}\right)=i_{2}\right]\ldots{}\mathbb{P}\left[X\left(t\right)=i|X\left(s_{k}\right)=i_{k}\right].
\end{split}\end{align}

We use a stochastic transition matrix, denoted $M\left(t\right)$, to represent the probabilities for transitions between states. The elements of the transition matrix, $\left[M\left(t\right)\right]_{ij}$, represent the probabilities of transitions from the states $j$ to the states $i$. The notation then becomes
\begin{align}\begin{split}
\left[M\left(s_{1},t\right)\right]_{ii_{1}}=&\sum\limits_{i_{2},i_{3},\ldots{},i_{k}}^{}\left[M\left(s_{1},s_{2}\right)\right]_{i_{2}i_{1}}\left[M\left(s_{2},s_{3}\right)\right]_{i_{3}i_{2}}\ldots{}\left[M\left(s_{k},t\right)\right]_{ii_{k}}.
\end{split}\end{align}

In terms of the transition matrices themselves,
\begin{align}\begin{split}
M\left(s_{1},t\right)=&M\left(s_{1},s_{2}\right)M\left(s_{2},s_{3}\right)\ldots{}M\left(s_{k},t\right).
\end{split}\end{align}

It is sufficient to consider only a single intermediate time, $u$, where $s\leq{}u\leq{}t$ are a set of non-negative times,
\begin{align}\begin{split}
M\left(s,t\right)&=M\left(s,u\right)M\left(u,t\right).
\end{split}\end{align}

If we assume homogeneity, the transition probabilities depend only on the differences in time between the initial and final states and not on the actual times themselves. This can be expressed as
\begin{align}\begin{split}
M\left(s,t\right)&=M\left(0,t-s\right).
\end{split}\end{align}

We then simplify the notation to specify only the time difference and not the initial and final times,
\begin{align}\begin{split}
M\left(t\right)&:=M\left(0,t\right).
\end{split}\end{align}

If no time has elapsed, the state must not change from the initial state. In other words, the probability of transitioning from the initial state to any other state must be zero and the probability of staying in the same state must be one. The transition matrix initially must therefore be the identity matrix,
\begin{align}\begin{split}
M\left(0\right)&=I.
\end{split}\end{align}

After some small amount of time has progressed there will be non-negative probabilities of transitions away from the initial state. The probability of staying in the same state must have decreased by the same amount as the increase in the sum of the probabilities of transitions, conserving probability.

\section{Expressing Transition Matrices in Terms of the Rate Matrices}

We will now refer back to the Markov property to attempt to find an expression for the transition matrix. If we start by setting $s=0$ in our original transition matrix expression for the Markov property,
\begin{align}\begin{split}
M\left(0,t\right)&=M\left(0,u\right)M\left(u,t\right) \\
M\left(t\right)&=M\left(u\right)M\left(t-u\right).
\end{split}\end{align}

Now re-parametrising,
\begin{align}\begin{split}
M\left(u+t\right)&=M\left(u\right)M\left(t\right).
\end{split}\end{align}

We derive the transition matrix from the definition of a derivative,
\begin{align}\begin{split}
M'\left(t\right)&=\lim_{s\rightarrow{}0}\frac{M\left(t+s\right)-M\left(t\right)}{s} \\
&=\lim_{s\rightarrow{}0}\frac{M\left(t\right)M\left(s\right)-M\left(t\right)}{s} \\
&=M\left(t\right)\lim_{s\rightarrow{}0}\frac{M\left(s\right)-M\left(0\right)}{s} \\
&=M\left(t\right)M'\left(0\right).
\end{split}\end{align}

We define the initial rate matrix to be the initial time derivative of the transition matrix,
\begin{align}\begin{split}
Q:=M'\left(0\right).
\end{split}\end{align}

A solution to the differential equation is then
\begin{align}\begin{split}
M\left(t\right)&=Ae^{Qt},
\end{split}\end{align}
where $A$ is some unknown $n\times{}n$ matrix.

Initially,
\begin{align}\begin{split}
M\left(0\right)&=Ae^{Q\cdot{}0}=A.
\end{split}\end{align}

Since $M\left(0\right)=I$,
\begin{align}\begin{split}
M\left(t\right)&=e^{Qt}.
\end{split}\end{align}

Since $Q$ is the time derivative of the transition matrix initially, it will be the matrix of the initial rates of transitions. Rates are fixed in time due to the homogeneity of the Markov chain. We can derive the transition matrix and utilise the conservation of probability to show that a similar property holds for the rate matrix.
\begin{align}\begin{split}
\sum\limits_{i=1}^{n}\left[M\left(t\right)\right]_{ij}&=1\textrm{ for all }j \\
\frac{d}{dt}\sum\limits_{i=1}^{n}\left[M\left(t\right)\right]_{ij}&=0 \\
\sum\limits_{i=1}^{n}\left[M'\left(t\right)\right]_{ij}&=0.
\end{split}\end{align}

Since this property holds for all $t$, it must hold initially,
\begin{align}\begin{split}
\sum\limits_{i=1}^{n}\left[M'\left(0\right)\right]_{ij}&=0\textrm{ for all }j \\
\sum\limits_{i=1}^{n}Q_{ij}&=0.
\end{split}\end{align}

From the conservation of probability in the transition matrix, we can conclude that the sum of transition rates in the rate matrix must always be zero. Recalling that the off-diagonal elements of the transition matrix must increase over a short period of time after the initial time, we can show with a bit of manipulation that the off-diagonal elements of the rate matrix must be non-negative.
\begin{align}\begin{split}
\left[M\left(s\right)\right]_{ij}&\geq{}\left[M\left(0\right)\right]_{ij}\textrm{ for }i\neq{}j \\
\lim_{s\rightarrow{}0}\frac{\left[M\left(s\right)\right]_{ij}-\left[M\left(0\right)\right]_{ij}}{s}&\geq{}0 \\
Q_{ij}&\geq{}0.
\end{split}\end{align}

From a similar argument we find that the diagonal elements of the rate matrix must be non-positive, ensuring that the sum of the transition rates is zero. In summary,
\begin{align}\begin{split}
\begin{mycases}
\sum\limits_{i=1}^{n}Q_{ij}&=0, \\
Q_{ij}&\geq{}0\quad{}\textrm{for }i\neq{}j, \\
Q_{ii}&\leq{}0.
\end{mycases}
\end{split}\end{align}

\section{Transition Matrices as Matrix Exponentials}

Since the number of current and final states is equal, the rate matrix must be square. This ensures that the transition matrix can be evaluated by taking a Taylor series expansion,
\begin{align}\begin{split}
M\left(t\right)=e^{Qt}=I+Qt+\frac{Q^{2}t^{2}}{2!}+\ldots{}.
\end{split}\end{align}

In general, matrix exponentials are not easy to compute. If the matrix is diagonal, however, then matrix exponentiation is easy. The rate matrix, $Q$, is said to be \emph{diagonalisable} if there exists some matrix, $h$, such that
\begin{align}\begin{split}
\widehat{Q}&=h^{-1}Qh=diag\left(\lambda_{1},\lambda_{2},\ldots{},\lambda_{n}\right)
\end{split}\end{align}
is a diagonal matrix with diagonal elements of $\lambda_{1}$, $\lambda_{2}$, to $\lambda_{n}$.

The matrix exponentiation then simply becomes
\begin{align}\begin{split}
e^{Qt}&=e^{h\widehat{Q}th^{-1}} \\
&=I+h\left(\widehat{Q}t\right)h^{-1}+\frac{h\left(\widehat{Q}t\right)^{2}h^{-1}}{2!}+\ldots{} \\
&=hIh^{-1}+h\left(\widehat{Q}t\right)h^{-1}+h\frac{\left(\widehat{Q}t\right)^{2}}{2!}h^{-1}+\ldots{} \\
&=he^{\widehat{Q}t}h^{-1} \\
&=he^{diag\left(\lambda_{1},\lambda_{2},\ldots{},\lambda_{n}\right)t}h^{-1} \\
&=h\cdot{}diag\left(e^{\lambda_{1}t},e^{\lambda_{2}t},\ldots{},e^{\lambda_{n}t}\right)\cdot{}h^{-1},
\end{split}\end{align}
where $diag\left(e^{\lambda_{1}t},e^{\lambda_{2}t},\ldots{},e^{\lambda_{n}t}\right)$ is a diagonal matrix with the diagonal elements being $e^{\lambda_{1}t}$, $e^{\lambda_{2}t}$, to $e^{\lambda_{n}t}$.

We will now discuss some models that are diagonalisable.

\section{The Abelian Group-Based Models}

Suppose the permutation matrices, $\left\{I,K_{1},K_{2},\ldots{},K_{m}\right\}$ under matrix multiplication are isomorphic to an Abelian group. We will then refer to the model as being an \emph{Abelian group-based} model or simply a \emph{group-based} model, as defined by \citet{semple2003phylogenetics}. If a model is Abelian group-based then it has a one dimensional irreducible representation. This means that the transformed rate matrix, $\widehat{Q}$, can be expressed as a block matrix, with the off-diagonal blocks being the zero matrices (one dimensional) and the diagonal blocks being the scalar (one dimensional) eigenvalues of the rate matrix. The transformed rate matrix can then be expressed as
\begin{align}\begin{split}
\widehat{Q}&=h^{-1}Qh=diag\left(\lambda_{1},\lambda_{2},\ldots{},\lambda_{n}\right).
\end{split}\end{align}

In other words, if a model is Abelian group-based then it is diagonalisable and the matrix exponential is easy to find. We will discuss how to diagonalise Abelian group-based models and provide some examples. Some of these examples can be found in \citet{sumner2014tensorial}.

\subsection{Diagonalising the Abelian Group-Based Models}

Suppose we have an Abelian group-based model that we wish to diagonalise. To diagonalise our model we need to find its eigenvectors. We start by expressing the rate matrix as a linear combination of Markov generators,
\begin{align}\begin{split}
Q&=\sum\limits_{i=1}^{k}\alpha_{i}L_{i},
\end{split}\end{align}
where $L_{i}$ are the $k$ Markov generators, each with a rate parameter, $\alpha_{i}$.

The Markov generators are the set of Markov matrices that span the space of the Markov model. For example, for the binary symmetric model there is only one rate parameter and consequently only one Markov generator is required.

Furthermore, we can express the Markov generators as linear combinations of the elements of the Abelian group. This then allows us to express the rate matrix as a linear combination of the group elements,
\begin{align}\begin{split}
Q&=\sum\limits_{i=1}^{n-1}\alpha_{i}\left(-I+K_{i}\right),
\end{split}\end{align}
where $n-1\geq{}k$ since rate parameters for different group elements may be equal.

We now use the Cayley-Hamilton theorem to find the eigenvectors and eigenvalues of an element of the Abelian group by finding the minimal polynomial of the group element from its characteristic polynomial. Each factor of the characteristic polynomial is an eigenvalue of the group element subtracted from the group element itself. The degree of each factor of the characteristic polynomial is the number of eigenvectors that share than eigenvalue. 

From the Cayley-Hamilton theorem, the characteristic polynomial of each group element, including $I$, must be identically zero,
\begin{align}\begin{split}
p\left(K_{i}\right)=K_{i}^{n}+c_{n-1}K_{i}^{n-1}+\ldots{}+c_{1}K_{i}+c_{0}I&=0 \\
\left(K_{i}-\lambda_{1}I\right)^{r_{1}}\left(K_{i}-\lambda_{2}I\right)^{r_{2}}\ldots{}\left(K_{i}-\lambda_{m}I\right)^{r_{m}}&=0,
\end{split}\end{align}

So $\lambda_{1},\lambda_{2},\ldots{},\lambda_{m}$ are the $m$ distinct eigenvalues of $K_{i}$. We can conclude that the eigenvectors and eigenvalues of $K_{i}$ can be found by finding the minimal polynomial of $K_{i}$, the lowest degree polynomial such that $p\left(K_{i}\right)=0$. The factors in the minimal polynomial are the same as those in the characteristic polynomial. The degree of each factor in the characteristic polynomial is the number of eigenvectors with that eigenvalue, whereas the degree of every factor in the minimal polynomial is one. The minimal polynomial, therefore, does not tell us how many eigenvectors share a given eigenvalue.

Now suppose a group element, $K_{i}$, has a unique eigenvector, $v$, and an eigenvalue, $\lambda{}$. The equation relating the group element to the eigenvector and the eigenvalue is
\begin{align}\begin{split}
K_{i}v&=\lambda{}v.
\end{split}\end{align}

For an Abelian group, any two elements commute,
\begin{align}\begin{split}
\left[K_{i},K_{j}\right]&:=K_{i}K_{j}-K_{j}K_{i}=0.
\end{split}\end{align}

As a consequence, for two arbitrary group elements, $K_{i}$ and $K_{j}$,
\begin{align}\begin{split}
K_{i}K_{j}v&=K_{j}K_{i}v=K_{j}\lambda{}v=\lambda{}K_{j}v.
\end{split}\end{align}

We conclude that $K_{j}v$ is also an eigenvector of $K_{i}$, with the same eigenvalue, $\lambda{}$.

The degree of the characteristic polynomial is equal to the number of states in the state space of the Markov model, $n$. If the degree of the minimal polynomial for $K_{i}$ is equal to the degree of the characteristic polynomial for $K_{i}$, then the $n$ eigenvectors of $K_{i}$ must all have distinct eigenvalues. These $n$ eigenvectors must be eigenvectors of all of the elements of the group.

Suppose now that the degree of the minimal polynomial for $K_{i}$ is less than the degree of the characteristic polynomial. Some of the eigenvectors must belong to a multi-dimensional eigenspace. In other words, there are multiple eigenvectors with the same eigenvalue. We can use this fact to find the remaining eigenvectors of $K_{j}$. We continue this process if necessary to find the eigenvectors of all of the elements of the group. Alternatively, we can first find the minimal polynomials for all of the group elements. If at least one of the minimal polynomials of the group elements has degree $n$, the eigenvectors of these group elements must all be eigenvectors of every element of the group.

Supposing we have now found the set of eigenvectors for all of the group elements, we can multiply linear combinations of the group elements by each eigenvector and show that these eigenvectors are also eigenvectors of the entire model,
\begin{align}\begin{split}
\left[\sum\limits_{i=1}^{n-1}\alpha_{i}\left(-I+K_{i}\right)\right]v&=\left[\sum\limits_{i=1}^{n-1}\alpha_{i}\left(-1+\lambda_{i}\right)\right]v \\
Qv&=\left[\sum\limits_{i=1}^{n-1}\alpha_{i}\left(-1+\lambda_{i}\right)\right]v,
\end{split}\end{align}
where the eigenvalue of $I$ is one and the $\lambda_{i}$ are the eigenvalues of the $K_{i}$.

Now that we have the set of eigenvectors for the Markov model, the matrix formed by the set of these eigenvectors will diagonalise the model.

An appropriate exact method in our context of finding the minimal polynomial of the group element is to start with the lowest possible degree polynomial and increase the degree by one at a time until the polynomial becomes identically zero. This is a sufficient method for our needs, but it is only feasible without resorting to numerical methods because the degree of the minimal polynomials are at most the dimensions of the Markov models in question. For example, for an $n$ state Markov model, the degree of the minimal polynomial must be no greater than $n$. Apart from the identity element, since the group element will never be the zero matrix or a matrix proportional to the identity matrix, the degree of the minimal polynomial must be at least two. To find the minimal polynomial we start by squaring the group element. If the square of the group element can be expressed in terms of just the group element itself and the identity matrix, then we have found the minimal polynomial. If this cannot be done (i.e. the square of the group element must be expressed in terms of the other non-identity group elements) then the degree of the minimal polynomial must be at least three. We then proceed to find the cube of the group element and repeat the process until we find the minimal polynomial.

To illustrate how to find the eigenvectors and eigenvalues from the minimal polynomial, let's look at a minimal polynomial as an example. An arbitrary minimal polynomial can be expressed as
\begin{align}\begin{split}
\left(K_{i}-\lambda_{1}I\right)\left(K_{i}-\lambda_{2}I\right)\ldots{}\left(K_{i}-\lambda_{m}I\right)=0.
\end{split}\end{align}

Since this is the minimal polynomial, there exists a vector, $v\in{\mathbb{C}^{n}}$, such that
\begin{align}\begin{split}
u:&=\left[\left(K_{i}-\lambda_{2}I\right)\left(K_{i}-\lambda_{3}I\right)\ldots{}\left(K_{i}-\lambda_{m}I\right)\right]v\neq{}0.
\end{split}\end{align}

Therefore,
\begin{align}\begin{split}
K_{i}u&=K_{i}\left[\left(K_{i}-\lambda_{2}I\right)\left(K_{i}-\lambda_{3}I\right)\ldots{}\left(K_{i}-\lambda_{m}I\right)\right]v \\
&=\lambda_{i}\left[\left(K_{i}-\lambda_{2}I\right)\left(K_{i}-\lambda_{3}I\right)\ldots{}\left(K_{i}-\lambda_{m}I\right)\right]v,
\end{split}\end{align}
where
\begin{align}\begin{split}
\left[\left(K_{i}-\lambda_{2}I\right)\left(K_{i}-\lambda_{3}I\right)\ldots{}\left(K_{i}-\lambda_{m}I\right)\right]v
\end{split}\end{align}
is an eigenvector of $K_{i}$.

Since the factors of the minimal polynomial commute we can use the appropriate orderings on the factors of the minimal polynomial to find all of the eigenvectors of $K_{i}$.

\subsection{The Binary Symmetric Model}

We will now look at diagonalising the binary symmetric model, $\mathbb{Z}_{2}\cong{}\{e,\left(12\right)\}$, as an example. The binary symmetric model is also known in the literature as the Cavender-Farris-Neyman (CFN) model or the Neyman $2$-state model. More generally, the $r$ state model with the same substitution rate for every substitution is known as the $N_{r}$ model. The $N_{4}$ model is more commonly known as the Jukes-Cantor (JC) model.

The rate matrix for the binary symmetric model is
\begin{align}\begin{split}
Q&=\left[\begin{array}{cc} -\alpha{} & \alpha{} \\
\alpha{} & -\alpha{} \\
\end{array}\right],
\end{split}\end{align}
where $\alpha{}>0$ is the rate of mutations between the two states in the state space.

The rate matrix can be written as
\begin{align}\begin{split}
Q&=\alpha{}L=\alpha{}\left(-I+K\right),
\end{split}\end{align}
where
\begin{align}\begin{split}
K=\left[\begin{array}{cc} 0 & 1 \\
1 & 0 \\
\end{array}\right],
\textrm{ with }
K^{2}&=I.
\end{split}\end{align}

Clearly the group formed by the set $\{I,K\}$ under matrix multiplication is isomorphic to $\mathbb{Z}_{2}\cong{}\{e,\left(12\right)\}$, the cyclic group of order $2$, where $e$ is the identity permutation and $\left(12\right)$ is the permutation of the objects ``$1$'' and ``$2$''. This means that the binary symmetric model is an Abelian group-based model.

We will now proceed to diagonalise the binary symmetric model. We have already argued that $K^{2}=I$, therefore the minimal polynomial for $K$ must be
\begin{align}\begin{split}
K^{2}-I&=0.
\end{split}\end{align}

Factorising the minimal polynomial,
\begin{align}\begin{split}
K^{2}-I&=\left(K-I\right)\left(K+I\right)=0.
\end{split}\end{align}

The degree of the minimal polynomial is two and there are two distinct eigenvalues. We can conclude that the minimal polynomial must also be the characteristic polynomial and the eigenvectors of $K$ will diagonalise the model. Since $\left(K-I\right)\left(K+I\right)$ is the minimal polynomial, there exists a vector, $v\in{\mathbb{C}^{2}}$, such that
\begin{align}\begin{split}
u:=\left(K+I\right)v\neq{}0.
\end{split}\end{align}

Therefore
\begin{align}\begin{split}
Ku=K\left(K+I\right)v=\left(K+I\right)v=\left\{k_{1}\left[\begin{array}{c} 1 \\
1 \\
\end{array}\right],k_{1}\in{\mathbb{C}\setminus{}\left\{0\right\}}\right\}.
\end{split}\end{align}

Similarly, there exists a vector, $v'\in{\mathbb{C}^{2}}$, such that
\begin{align}\begin{split}
u':=\left(K-I\right)v'\neq{}0.
\end{split}\end{align}

Therefore
\begin{align}\begin{split}
Ku'=K\left(K-I\right)v'=-\left(K-I\right)v'=\left\{k_{2}\left[\begin{array}{c} 1 \\
-1 \\
\end{array}\right],k_{2}\in{\mathbb{C}\setminus{}\left\{0\right\}}\right\}.
\end{split}\end{align}

These eigenvectors, in union with the zero vector, then form an eigenspace for $K$, thus diagonalising it. The diagonalising matrix will then be
\begin{align}\begin{split}
h=\left[\begin{array}{cc} k_{1} & k_{2} \\
k_{1} & -k_{2} \\
\end{array}\right],\\
\end{split}\end{align}
where the determinant must be non-zero by definition. For this diagonalising matrix, the determinant will be $−2k_{1}k_{2}$, which implies that $k_{1},k_{2}\in{\mathbb{C}\setminus{}\left\{0\right\}}$.

The diagonalised rate matrix for the binary symmetric model will then be
\begin{align}\begin{split}
\widehat{Q}=\left[\begin{array}{cc} 0 & 0 \\
0 & -2\alpha{} \\
\end{array}\right].\\
\end{split}\end{align}

\subsection{The $\mathbb{Z}_{3}$ Model}

We will now look at the $\mathbb{Z}_{3}$ model, $\mathbb{Z}_{3}\cong{}\left\{e,\left(123\right),\left(132\right)\right\}$, as a slightly more complicated example. The rate matrix for the model is
\begin{align}\begin{split}
Q&=\left[\begin{array}{ccc} -(\alpha{}+\beta{}) & \alpha{} & \beta{} \\
\beta{} & -(\alpha{}+\beta{}) & \alpha{} \\
\alpha{} & \beta{} & -(\alpha{}+\beta{}) \\
\end{array}\right] \\
&=\alpha{}L_{\alpha{}}+\beta{}L_{\beta{}},
\end{split}\end{align}
where
\begin{align}\begin{split}
L_{\alpha{}}=\left[\begin{array}{ccc} -1 & 1 & 0 \\
0 & -1 & 1 \\
1 & 0 & -1 \\
\end{array}\right],
\quad{}
L_{\beta{}}=\left[\begin{array}{ccc} -1 & 0 & 1 \\
1 & -1 & 0 \\
0 & 1 & -1 \\
\end{array}\right]
\end{split}\end{align}
are the zero column sum Markov generators.

In terms of the elements of the Abelian group, $\mathbb{Z}_{3}$, we can express the Markov generators as
\begin{align}\begin{split}
L_{\alpha{}}=-I+K_{\alpha{}},
\quad{}
L_{\beta{}}=-I+K_{\beta{}},
\end{split}\end{align}
where
\begin{align}\begin{split}
K_{\alpha{}}=\left[\begin{array}{ccc} 0 & 1 & 0 \\
0 & 0 & 1 \\
1 & 0 & 0 \\
\end{array}\right],
\quad{}
K_{\beta{}}=\left[\begin{array}{ccc} 0 & 0 & 1 \\
1 & 0 & 0 \\
0 & 1 & 0 \\
\end{array}\right].
\end{split}\end{align}

It can easily be shown that
\begin{align}\begin{split}
\begin{mycases}
K_{\alpha{}}^{2}&=K_{\beta{}},\\
K_{\alpha{}}^{3}&=I,
\end{mycases}
\end{split}\end{align}
confirming the isomorphism to the $\mathbb{Z}_{3}$ group.

Since $K_{\alpha{}}^{2}=K_{\beta{}}$, the square of $K_{\alpha{}}$ must be written in terms of the other elements of the group. The degree of the minimal polynomial must therefore be at least three. Since $K_{\alpha{}}^{3}=I$, the minimal polynomial must be
\begin{align}\begin{split}
K_{\alpha{}}^{3}-I&=0.
\end{split}\end{align}

Factorising the minimal polynomial,
\begin{align}\begin{split}
K_{\alpha{}}^{3}-I&=\left(K_{\alpha{}}-I\right)\left(K_{\alpha{}}-\omega{}I\right)\left(K_{\alpha{}}-\omega^{2}I\right)=0,
\end{split}\end{align}
where
\begin{align}\begin{split}
\omega{}=e^{i\frac{2\pi{}}{3}}.
\end{split}\end{align}

Since the degree of the minimal polynomial for $K_{\alpha{}}$ is three, and each eigenvector has a distinct eigenvalue, the minimal polynomial must also be the characteristic polynomial. The eigenvector for $K_{\alpha{}}$ must therefore diagonalise the model and it is not necessary to find the minimal polynomial for the other non-trivial (non-identity) element of the group.

Following the same argument used for the binary symmetric model we find that there are some vectors, $v_{1}$, $v_{2}$ and $v_{3}$, such that the eigenvectors for $K_{\alpha{}}$ satisfy
\begin{align}\begin{split}
K_{\alpha{}}\left[\left(K_{\alpha{}}-\omega{}I\right)\left(K_{\alpha{}}-\omega^{2}I\right)\right]v_{1}&=\left[\left(K_{\alpha{}}-\omega{}I\right)\left(K_{\alpha{}}-\omega^{2}I\right)\right]v_{1}=\left\{k_{1}\left[\begin{array}{c} 1 \\
1 \\
1 \\
\end{array}\right],k_{1}\in{\mathbb{C}\setminus{}\left\{0\right\}}\right\},\\
K_{\alpha{}}\left[\left(K_{\alpha{}}-I\right)\left(K_{\alpha{}}-\omega^{2}I\right)\right]v_{2}&=\omega{}\left[\left(K_{\alpha{}}-I\right)\left(K_{\alpha{}}-\omega^{2}I\right)\right]v_{2}=\left\{k_{2}\left[\begin{array}{c} 1 \\
\omega{} \\
\omega{}^{2} \\
\end{array}\right],k_{2}\in{\mathbb{C}\setminus{}\left\{0\right\}}\right\}, \\
K_{\alpha{}}\left[\left(K_{\alpha{}}-I\right)\left(K_{\alpha{}}-\omega{}I\right)\right]v_{3}&=\omega^{2}\left[\left(K_{\alpha{}}-I\right)\left(K_{\alpha{}}-\omega{}I\right)\right]v_{3}=\left\{k_{3}\left[\begin{array}{c} 1 \\
\omega{}^{2} \\
\omega{} \\
\end{array}\right],k_{3}\in{\mathbb{C}\setminus{}\left\{0\right\}}\right\}.
\end{split}\end{align}

These eigenvectors, in union with the zero vector, then form an eigenspace for $K_{\alpha{}}$ and in turn for the model itself, thus diagonalising it. We then find the diagonalising matrix to be
\begin{align}\begin{split}
h=\left[\begin{array}{ccc} k_{1} & k_{2} & k_{3} \\
k_{1} & k_{2}e^{i\frac{2\pi{}}{3}} & k_{3}e^{-i\frac{2\pi{}}{3}} \\
k_{1} & k_{2}e^{-i\frac{2\pi{}}{3}} & k_{3}e^{i\frac{2\pi{}}{3}} \\
\end{array}\right],\\
\end{split}\end{align}
where the determinant must be non-zero by definition, which implies that $k_{1},k_{2},k_{3}\in{\mathbb{C}\setminus{}\left\{0\right\}}$.

The diagonalised rate matrix will then be
\begin{align}\begin{split}
\widehat{Q}=\left[\begin{array}{ccc} 0 & 0 & 0 \\
0 & \frac{3\left(\beta{}-\alpha{}\omega{}\right)}{-1+\omega{}} & 0 \\
0 & 0 & \frac{3\left(\alpha{}-\beta{}\omega{}\right)}{-1+\omega{}} \\
\end{array}\right].
\end{split}\end{align}

\subsection{The K$3$ST Model}

For a further example, we will now look at the K$3$ST model,
\begin{align}\begin{split}
\mathbb{Z}_{2}\times{}\mathbb{Z}_{2}\cong{}\left\{e,\left(12\right)\left(34\right),\left(13\right)\left(24\right),\left(14\right)\left(23\right)\right\}.
\end{split}\end{align}

The rate matrix for the model is
\begin{align}\begin{split}
Q&=\left[\begin{array}{cccc} -\left(\alpha{}+\beta{}+\gamma{}\right) & \alpha{} & \beta{} & \gamma{} \\
\alpha{} & -\left(\alpha{}+\beta{}+\gamma{}\right) & \gamma{} & \beta{} \\
\beta{} & \gamma{} & -\left(\alpha{}+\beta{}+\gamma{}\right) & \alpha{} \\
\gamma{} & \beta{} & \alpha{} & -\left(\alpha{}+\beta{}+\gamma{}\right) \\
\end{array}\right] \\
&=\alpha{}L_{\alpha{}}+\beta{}L_{\beta{}}+\gamma{}L_{\gamma{}},
\end{split}\end{align}
where
\begin{gather}
\begin{split}
L_{\alpha{}}=\left[\begin{array}{cccc} -1 & 1 & 0 & 0 \\
1 & -1 & 0 & 0 \\
0 & 0 & -1 & 1 \\
0 & 0 & 1 & -1 \\
\end{array}\right],
\quad{}
L_{\beta{}}=\left[\begin{array}{cccc} -1 & 0 & 1 & 0 \\
0 & -1 & 0 & 1 \\
1 & 0 & -1 & 0 \\
0 & 1 & 0 & -1 \\
\end{array}\right], \\
L_{\gamma{}}=\left[\begin{array}{cccc} -1 & 0 & 0 & 1 \\
0 & -1 & 1 & 0 \\
0 & 1 & -1 & 0 \\
1 & 0 & 0 & -1 \\
\end{array}\right]
\end{split}
\end{gather}
are the zero column sum Markov generators.

In terms of the elements of the Abelian group, $\mathbb{Z}_{2}\times{}\mathbb{Z}_{2}$, we can express the Markov generators as
\begin{align}\begin{split}
L_{\alpha{}}=-I+K_{\alpha{}},
\quad{}
L_{\beta{}}=-I+K_{\beta{}},
\quad{}
L_{\gamma{}}=-I+K_{\gamma{}}
\end{split}\end{align}
where
\begin{align}\begin{split}
K_{\alpha{}}=\left[\begin{array}{cccc} 0 & 1 & 0 & 0 \\
1 & 0 & 0 & 0 \\
0 & 0 & 0 & 1 \\
0 & 0 & 1 & 0 \\
\end{array}\right],
\quad{}
K_{\beta{}}=\left[\begin{array}{cccc} 0 & 0 & 1 & 0 \\
0 & 0 & 0 & 1 \\
1 & 0 & 0 & 0 \\
0 & 1 & 0 & 0 \\
\end{array}\right],
\quad{}
K_{\gamma{}}=\left[\begin{array}{cccc} 0 & 0 & 0 & 1 \\
0 & 0 & 1 & 0 \\
0 & 1 & 0 & 0 \\
1 & 0 & 0 & 0 \\
\end{array}\right].
\end{split}\end{align}

For all $i,j,k\in{\left\{\alpha{},\beta{},\gamma{}\right\}}$, with all subscripts being different elements, it can easily be shown that
\begin{align}\begin{split}
\begin{mycases}
K_{i}K_{j}&=K_{j}K_{i}=K_{k}, \\
K_{i}^{2}&=I,
\end{mycases}
\end{split}\end{align}
confirming the isomorphism to the $\mathbb{Z}_{2}\times{}\mathbb{Z}_{2}$ group.

We can immediately see the minimal polynomials from the group structure. Note the similarity of the minimal polynomials for the K$3$ST model, which is isomorphic to $\mathbb{Z}_{2}\times{}\mathbb{Z}_{2}$, to the minimal polynomial for the binary symmetric model, which is isomorphic to the $\mathbb{Z}_{2}$ group. These minimal polynomials will therefore have the same eigenvalues as each other and can be factorised as
\begin{align}\begin{split}
K_{i}^{2}-I&=\left(K_{i}-I\right)\left(K_{i}+I\right)=0.
\end{split}\end{align}

Following the same argument used for the previous examples we find that there are some vectors, $v_{1}$ and $v_{2}$, such that the eigenvectors for $K_{\alpha{}}$ satisfy
\begin{align}\begin{split}
\begin{mycases}
K_{\alpha{}}\left[\left(K_{\alpha{}}+I\right)\right]v_{1}&=\left[\left(K_{\alpha{}}+I\right)\right]v_{1}, \\
K_{\alpha{}}\left[\left(K_{\alpha{}}-I\right)\right]v_{2}&=-\left[\left(K_{\alpha{}}-I\right)\right]v_{2}.
\end{mycases}
\end{split}\end{align}

Recall that if some vector, $v$, is an eigenvector of some group element, $K_{i}$, with eigenvalue, $\lambda{}$, then $v$ is also an eigenvector of some other group element, $K_{j}$, with eigenvalue, $\mu{}$. If
\begin{align}\begin{split}
K_{\alpha{}}v&=\lambda{}v,
\end{split}\end{align}
then
\begin{align}\begin{split}
K_{\alpha{}}K_{\beta{}}v&=K_{\beta{}}K_{\alpha{}}v=K_{\beta{}}\lambda{}v=\lambda{}K_{\beta{}}v.
\end{split}\end{align}

Adding linear combinations of these equations
\begin{align}\begin{split}
K_{\alpha{}}\left(K_{\beta{}}-\mu{}I\right)v&=\lambda{}\left(K_{\beta{}}-\mu{}I\right)v.
\end{split}\end{align}

For $\left(K_{\beta{}}-\mu{}I\right)v$ to also be the eigenvectors of $K_{\beta{}}$, we require
\begin{align}\begin{split}
K_{\beta{}}\left(K_{\beta{}}-\mu{}I\right)v&=\nu{}\left(K_{\beta{}}-\mu{}I\right)v.
\end{split}\end{align}

From the same argument as that used for $K_{\alpha{}}$, these equations are satisfied when $\mu{}=\pm{}1$, giving
\begin{align}\begin{split}
\begin{mycases}
K_{\beta{}}\left(K_{\beta{}}+I\right)v&=\left(K_{\beta{}}+I\right)v, \\
K_{\beta{}}\left(K_{\beta{}}-I\right)v&=-\left(K_{\beta{}}-I\right)v.
\end{mycases}
\end{split}\end{align}

Applying this result to our original equations for the eigenvectors of $K_{\alpha{}}$,
\begin{align}\begin{split}
\begin{mycases}
K_{\alpha{}}\left[\left(K_{\beta{}}-\mu{}I\right)\left(K_{\alpha{}}+I\right)\right]v_{1}&=\left[\left(K_{\beta{}}-\mu{}I\right)\left(K_{\alpha{}}+I\right)\right]v_{1}, \\
K_{\alpha{}}\left[\left(K_{\beta{}}-\mu{}I\right)\left(K_{\alpha{}}-I\right)\right]v_{2}&=-\left[\left(K_{\beta{}}-\mu{}I\right)\left(K_{\alpha{}}-I\right)\right]v_{2}.
\end{mycases}
\end{split}\end{align}

The four eigenvectors are then
\begin{align}\begin{split}
K_{\alpha{}}\left[\left(K_{\beta{}}+I\right)\left(K_{\alpha{}}+I\right)\right]v_{1}&=\left[\left(K_{\beta{}}+I\right)\left(K_{\alpha{}}+I\right)\right]v_{1}=\left\{k_{1}\left[\begin{array}{c} 1 \\
1 \\
1 \\
1 \\
\end{array}\right],k_{1}\in{\mathbb{C}\setminus{}\left\{0\right\}}\right\}, \\
K_{\alpha{}}\left[\left(K_{\beta{}}+I\right)\left(K_{\alpha{}}-I\right)\right]v_{2}&=\left[\left(K_{\beta{}}+I\right)\left(K_{\alpha{}}-I\right)\right]v_{2}=\left\{k_{2}\left[\begin{array}{c} 1 \\
-1 \\
1 \\
-1 \\
\end{array}\right],k_{2}\in{\mathbb{C}\setminus{}\left\{0\right\}}\right\}, \\
K_{\alpha{}}\left[\left(K_{\beta{}}-I\right)\left(K_{\alpha{}}+I\right)\right]v_{1}&=-\left[\left(K_{\beta{}}-I\right)\left(K_{\alpha{}}+I\right)\right]v_{1}=\left\{k_{3}\left[\begin{array}{c} 1 \\
1 \\
-1 \\
-1 \\
\end{array}\right],k_{3}\in{\mathbb{C}\setminus{}\left\{0\right\}}\right\}, \\
K_{\alpha{}}\left[\left(K_{\beta{}}-I\right)\left(K_{\alpha{}}-I\right)\right]v_{2}&=-\left[\left(K_{\beta{}}-I\right)\left(K_{\alpha{}}-I\right)\right]v_{2}=\left\{k_{4}\left[\begin{array}{c} 1 \\
-1 \\
-1 \\
1 \\
\end{array}\right],k_{4}\in{\mathbb{C}\setminus{}\left\{0\right\}}\right\}.
\end{split}\end{align}

These eigenvectors, in union with the zero vector, then form an eigenspace for $K_{\alpha{}}$ and $K_{\beta{}}$. Since the eigenspace is of rank four it must diagonalise the model. We then find the diagonalising matrix to be
\begin{align}\begin{split}
h=\left[\begin{array}{cccc} k_{1} & k_{2} & k_{3} & k_{4} \\
k_{1} & -k_{2} & k_{3} & -k_{4} \\
k_{1} & k_{2} & -k_{3} & -k_{4} \\
k_{1} & -k_{2} & -k_{3} & k_{4} \\
\end{array}\right],\\
\end{split}\end{align}
where the determinant must be non-zero by definition, which implies that $k_{1},k_{2},k_{3},k_{4}\in{\mathbb{C}\setminus{}\left\{0\right\}}$.

The diagonalised rate matrix will then be
\begin{align}\begin{split}
\widehat{Q}=\left[\begin{array}{cccc} 0 & 0 & 0 & 0 \\
0 & -2(\alpha{}+\gamma{}) & 0 & 0 \\
0 & 0 & -2(\beta{}+\gamma{}) & 0 \\
0 & 0 & 0 & -2(\alpha{}+\beta{}) \\
\end{array}\right].
\end{split}\end{align}

\section{Finding Diagonalising Matrices in \emph{Mathematica}}
\label{mathematica}

\emph{Mathematica} has a function called ``Eigenvectors''. This function will output \emph{a} set of eigenvectors for our rate matrix, $Q$, of a given Abelian group-based model or, more generally, for any diagonalisable matrix. This set of eigenvectors can be used to diagonalise the rate matrix. Unfortunately the function does not output the most general set of eigenvectors for a given matrix. The code below can be used in \emph{Mathematica} to find the diagonalising matrix formed by the most general set of eigenvectors.
\begin{lstlisting}[breaklines=true]
(* The rate matrix, Q, of an Abelian group-based model. Qij is the element in the i^th row and j^th column of the rate matrix. *)
Q = {{Q11, Q12, ..., Q1n}, {Q21, Q22, ..., Q2n}, ..., {Qn1, Qn2, ..., Qnn}};

(* n is the number of columns and rows in Q and can be taken to be the dimension of any row or column in Q. Here n is the dimension of the first row of Q. *)
n = Dimensions[Q[[1]]][[1]];

(* An n*n table with arbitrary elements. Each column of the table will be an eigenvector of Q. *)
K = Table[k[i, j], {i, n}, {j, n}];

(* The eigenvalues of Q. *)
L = lambda /. FullSimplify[Solve[{Det[Q - lambda*IdentityMatrix[n]] == 0}, {lambda}]];

(* The diagonalising matrix formed by the eigenvectors of Q. *)
h = MatrixForm[Transpose[Table[K[[All, i]] /. FullSimplify[Solve[{Q.K[[All, i]] == L[[i]]*K[[All, i]]}, K[[All, i]]]], {i, n}]]]
\end{lstlisting}

For example, for the binary symmetric model, we would take $Q$ to be
\begin{lstlisting}[breaklines=true]
Q = {{-alpha, alpha}, {alpha, -alpha}};
\end{lstlisting}

The diagonalising matrix output for the binary symmetric model is shown below.
\begin{lstlisting}[breaklines=true, mathescape]
$\left(\left(\begin{array}{c} k[1, 1] \\
k[1, 1]
\end{array}\right)\left(\begin{array}{c} k[1, 2] \\
-k[1, 2]
\end{array}\right)\right)$
\end{lstlisting}

\section{The Four-State General Markov Model}

We have shown that Abelian group-based models are diagonalisable and provided some examples. This raises the question as to whether models that are not Abelian group-based can be diagonalised. Although having an isomorphism to an Abelian group is a sufficient condition for a Markov model to be diagonalisable, it is not a necessary condition. There are some models which are not Abelian group-based, which are nonetheless diagonalisable. The lowest order group which is not Abelian is the $\mathfrak{S}_{3}$ group, which has order six. Despite not being an Abelian group-based model, the $\mathfrak{S}_{3}$ group-based model can be diagonalised. For an example of a model which is not Abelian group-based, we will look at the general Markov model for four states. The rate matrix is
\begin{align}\begin{split}
Q=\sum\limits_{i=1,j\neq{}i}^{n}\alpha_{ij}L_{ij},
\end{split}\end{align}
where $n$ is the number of states in the state space, $\alpha_{ij}>0$ are the rate parameters and $L_{ij}$ are the generators. In \citet{sumner2012lie} it was shown that the commutators for the irreducible representations of the generators satisfy
\begin{align}\begin{split}
\left[L_{ij},L_{kl}\right]=\left(L_{il}-L_{jl}\right)\left(\delta_{jk}-\delta_{jl}\right)-\left(L_{kj}-L_{lj}\right)\left(\delta_{il}-\delta_{jl}\right),
\end{split}\end{align}
where $\delta_{ij}$ is the Kronecker delta.

Since the commutator is not generally equal to zero, we can conclude that the general Markov model is not Abelian group-based. Therefore it cannot be assumed that the general Markov model can be diagonalised. Indeed, it is not possible to diagonalise the general Markov model.

\section{Abelian Group-Based Models and Their Diagonalising Matrices}

We can follow a similar process to what was used for the binary symmetric model, the $\mathbb{Z}_{3}$ model and the K$3$ST model to diagonalise any Abelian group-based model. Refer to the \emph{Mathematica} code in Section~\ref{mathematica} to see how we found the diagonalising matrices for the remaining Abelian group-based models. Here we have swapped the order of some of the eigenvectors in the diagonalising matrices and relabelled the parameters. Below in Table~\ref{ratematrixcatalogue} is a list of the rate matrices and diagonalising matrices for Abelian group-based models.
\begin{center}
\small{}
\begin{longtable}{c|c|c}
\normalsize{Model} & \normalsize{Rate Matrix $\left(Q\right)$} & \normalsize{Diagonalising Matrix $\left(h\right)$} \\
\hline
Binary Symmetric $\left(\mathbb{Z}_{2}\right)$ & $\left[\begin{array}{cc} \ast{} & \alpha{} \\
\alpha{} & \ast{} \\
\end{array}\right]$ & $\left[\begin{array}{cc} k_{1} & k_{2} \\
k_{1} & -k_{2} \\
\end{array}\right]$ \\
\hline
$\mathbb{Z}_{3}$ & $\left[\begin{array}{ccc} \ast{} & \alpha{} & \beta{} \\
\beta{} & \ast{} & \alpha{} \\
\alpha{} & \beta{} & \ast{} \\
\end{array}\right]$ & $\left[\begin{array}{ccc} k_{1} & k_{2} & k_{3} \\
k_{1} & k_{2}e^{i\frac{2\pi{}}{3}} & k_{3}e^{-i\frac{2\pi{}}{3}} \\
k_{1} & k_{2}e^{-i\frac{2\pi{}}{3}} & k_{3}e^{i\frac{2\pi{}}{3}} \\
\end{array}\right]$ \\
\hline
Three State Neyman $\left(\alpha{}=\beta{}\right)$ $\left(\mathbb{Z}_{3}\right)$ & $\left[\begin{array}{ccc} \ast{} & \alpha{} & \alpha{} \\
\alpha{} & \ast{} & \alpha{} \\
\alpha{} & \alpha{} & \ast{} \\
\end{array}\right]$ & $\left[\begin{array}{ccc} k_{1} & k_{2} & k_{4} \\
k_{1} & k_{3} & k_{5} \\
k_{1} & -\left(k_{2}+k_{3}\right) & -\left(k_{4}+k_{5}\right) \\
\end{array}\right]$ \\
\hline
\begin{tabular}{c}
$\mathbb{Z}_{2}\times{}\mathbb{Z}_{2}\cong{}$ \\
$\left\{e,\left(12\right),\left(34\right),\left(12\right)\left(34\right)\}\right)$
\end{tabular}
& $\left[\begin{array}{cccc} \ast{} & \alpha{} & 0 & 0 \\
\alpha{} & \ast{} & 0 & 0 \\
0 & 0 & \ast{} & \beta{} \\
0 & 0 & \beta{} & \ast{} \\
\end{array}\right]$ & $\left[\begin{array}{cccc} k_{1} & k_{2} & k_{4} & 0 \\
-k_{1} & k_{2} & k_{4} & 0 \\
0 & k_{3} & k_{5} & k_{6} \\
0 & k_{3} & k_{5} & -k_{6} \\
\end{array}\right]$ \\
\hline{}
\begin{tabular}{c}
$\left(\alpha{}=\beta{}\right)$ $\left(\mathbb{Z}_{2}\times{}\mathbb{Z}_{2}\cong{}\right.$ \\
$\left\{e,\left(12\right),\left(34\right),\left(12\right)\left(34\right)\}\right)$
\end{tabular}
& $\left[\begin{array}{cccc} \ast{} & \alpha{} & 0 & 0 \\
\alpha{} & \ast{} & 0 & 0 \\
0 & 0 & \ast{} & \alpha{} \\
0 & 0 & \alpha{} & \ast{} \\
\end{array}\right]$ & $\left[\begin{array}{cccc} k_{1} & k_{3} & k_{5} & k_{7} \\
k_{1} & -k_{3} & k_{5} & -k_{7} \\
k_{2} & k_{4} & k_{6} & k_{8} \\
k_{2} & -k_{4} & k_{6} & -k_{8} \\
\end{array}\right]$ \\
\hline
\begin{tabular}{c}
K$3$ST $\left(\mathbb{Z}_{2}\times{}\mathbb{Z}_{2}\cong{}\right.$ \\
$\left\{e,\left(12\right)\left(34\right),\left(13\right)\left(24\right),\left(14\right)\left(23\right)\}\right)$
\end{tabular}
& $\left[\begin{array}{cccc} \ast{} & \alpha{} & \beta{} & \gamma{} \\
\alpha{} & \ast{} & \gamma{} & \beta{} \\
\beta{} & \gamma{} & \ast{} & \alpha{} \\
\gamma{} & \beta{} & \alpha{} & \ast{} \\
\end{array}\right]$ & $\left[\begin{array}{cccc} k_{1} & k_{2} & k_{3} & k_{4} \\
k_{1} & -k_{2} & k_{3} & -k_{4} \\
k_{1} & k_{2} & -k_{3} & -k_{4} \\
k_{1} & -k_{2} & -k_{3} & k_{4} \\
\end{array}\right]$ \\
\hline
\begin{tabular}{c}
K$2$ST $\left(\alpha{}=\beta{}\right)$ $\left(\mathbb{Z}_{2}\times{}\mathbb{Z}_{2}\cong{}\right.$ \\
$\left\{e,\left(12\right)\left(34\right),\left(13\right)\left(24\right),\left(14\right)\left(23\right)\}\right)$
\end{tabular}
& $\left[\begin{array}{cccc} \ast{} & \alpha{} & \alpha{} & \gamma{} \\
\alpha{} & \ast{} & \gamma{} & \alpha{} \\
\alpha{} & \gamma{} & \ast{} & \alpha{} \\
\gamma{} & \alpha{} & \alpha{} & \ast{} \\
\end{array}\right]$ & $\left[\begin{array}{cccc} k_{1} & k_{2} & k_{4} & k_{6} \\
k_{1} & -k_{3} & k_{5} & -k_{6} \\
k_{1} & k_{3} & -k_{5} & -k_{6} \\
k_{1} & -k_{2} & -k_{4} & k_{6} \\
\end{array}\right]$ \\
\hline
$\mathbb{Z}_{4}\cong{}\left\{e,\left(12\right)\left(34\right),\left(1324\right),\left(1423\right)\right\}$ & $\left[\begin{array}{cccc} \ast{} & \alpha{} & \beta{} & \gamma{} \\
\alpha{} & \ast{} & \gamma{} & \beta{} \\
\gamma{} & \beta{} & \ast{} & \alpha{} \\
\beta{} & \gamma{} & \alpha{} & \ast{} \\
\end{array}\right]$ & $\left[\begin{array}{cccc} k_{1} & k_{2} & k_{3} & k_{4} \\
k_{1} & -k_{2} & k_{3} & -k_{4} \\
k_{1} & ik_{2} & -k_{3} & -ik_{4} \\
k_{1} & -ik_{2} & -k_{3} & ik_{4} \\
\end{array}\right]$ \\
\hline
\begin{tabular}{c}
$2$ Parameter $\mathbb{Z}_{4}$ $\left(\alpha{}=\beta{}\right)$ $\left(\mathbb{Z}_{4}\cong{}\right.$ \\
$\left.\left\{e,\left(12\right)\left(34\right),\left(1324\right),\left(1423\right)\right\}\right)$
\end{tabular}
& $\left[\begin{array}{cccc} \ast{} & \alpha{} & \alpha{} & \gamma{} \\
\alpha{} & \ast{} & \gamma{} & \alpha{} \\
\gamma{} & \alpha{} & \ast{} & \alpha{} \\
\alpha{} & \gamma{} & \alpha{} & \ast{} \\
\end{array}\right]$ & $\left[\begin{array}{cccc} k_{1} & k_{2} & k_{3} & k_{4} \\
k_{1} & -k_{2} & k_{3} & -k_{4} \\
k_{1} & ik_{2} & -k_{3} & -ik_{4} \\
k_{1} & -ik_{2} & -k_{3} & ik_{4} \\
\end{array}\right]$ \\
\hline
\begin{tabular}{c}
$2$ Parameter $\mathbb{Z}_{4}$ $\left(\beta{}=\gamma{}\right)$ $\left(\mathbb{Z}_{4}\cong{}\right.$ \\
$\left.\left\{e,\left(12\right)\left(34\right),\left(1324\right),\left(1423\right)\right\}\right)$
\end{tabular}
& $\left[\begin{array}{cccc} \ast{} & \alpha{} & \beta{} & \beta{} \\
\alpha{} & \ast{} & \beta{} & \beta{} \\
\beta{} & \beta{} & \ast{} & \alpha{} \\
\beta{} & \beta{} & \alpha{} & \ast{} \\
\end{array}\right]$ & $\left[\begin{array}{cccc} k_{1} & k_{2} & k_{4} & k_{5} \\
k_{1} & -k_{2} & k_{4} & -k_{5} \\
k_{1} & k_{3} & -k_{4} & -k_{6} \\
k_{1} & -k_{3} & -k_{4} & k_{6} \\
\end{array}\right]$ \\
\hline
JC $\left(\mathbb{Z}_{2}\times{}\mathbb{Z}_{2}\right.$ or $\left.\mathbb{Z}_{4}\right)$ & $\left[\begin{array}{cccc} \ast{} & \alpha{} & \alpha{} & \alpha{} \\
\alpha{} & \ast{} & \alpha{} & \alpha{} \\
\alpha{} & \alpha{} & \ast{} & \alpha{} \\
\alpha{} & \alpha{} & \alpha{} & \ast{} \\
\end{array}\right]$ & $\left[\begin{array}{cccc} k_{1} & k_{2} & k_{5} & k_{8} \\
k_{1} & k_{3} & k_{6} & k_{9} \\
k_{1} & k_{4} & k_{7} & k_{10} \\
k_{1} & \ast{} & \ast{} & \ast{} \\
\end{array}\right]$ \\
\caption{The Abelian group-based models for two to four states and their diagonalising matrices. Each $\ast{}$ is the negative of the sum of all other elements in that column. Every parameter in the diagonalising matrices can be an arbitrary, non-zero complex number. The parameters of the diagonalising matrices must all be constrained to have non-zero determinants and are therefore invertible.}
\label{ratematrixcatalogue}
\end{longtable}
\end{center}

We have left out some models that are equivalent to the models shown up to permutation of the elements of the state space. For example, suppose we took the rate matrix for the $\mathbb{Z}_{4}\cong{}\left\{e,\left(12\right)\left(34\right),\left(1324\right),\left(1423\right)\right\}$ model and labelled the rows and columns as shown below,
\begin{align}\begin{split}
Q&=\left[\begin{array}{c|cccc} & A & G & C & T \\
\hline{}
A & \ast{} & \alpha{} & \beta{} & \gamma{} \\
G & \alpha{} & \ast{} & \gamma{} & \beta{} \\
C & \gamma{} & \beta{} & \ast{} & \alpha{} \\
T & \beta{} & \gamma{} & \alpha{} & \ast{} \\
\end{array}\right],
\end{split}\end{align}
where $Q_{ij}$ is the rate of substitutions from state $j$ to state $i$.

If we swapped the row labels for $G$ and $C$ then the rate matrix would become
\begin{align}\begin{split}
Q&=\left[\begin{array}{c|cccc} & A & C & G & T \\
\hline{}
A & \ast{} & \beta{} & \alpha{} & \gamma{} \\
C & \gamma{} & \ast{} & \beta{} & \alpha{} \\
G & \alpha{} & \gamma{} & \ast{} & \beta{} \\
T & \beta{} & \alpha{} & \gamma{} & \ast{} \\
\end{array}\right].
\end{split}\end{align}

Clearly the rate matrix is displaying the same substitution rates, however it is now for the $\mathbb{Z}_{4}\cong{}\left\{e,\left(13\right)\left(24\right),\left(1234\right),\left(1432\right)\right\}$ isomorphism. Hence we do not need to consider the $\mathbb{Z}_{4}\cong{}\left\{e,\left(13\right)\left(24\right),\left(1234\right),\left(1432\right)\right\}$ isomorphism. Similarly, we have not considered the $2$ Parameter $\mathbb{Z}_{4}$ $\left(\beta{}=\gamma{}\right)$ $\left(\mathbb{Z}_{4}\cong{}\left\{e,\left(12\right)\left(34\right),\left(1324\right),\left(1423\right)\right\}\right)$ model.

It should be noted that some of these models are submodels of other models. For example, the four state JC model is a submodel of the $\mathbb{Z}_{2}\times{}\mathbb{Z}_{2}\cong{}\left\{e,\left(12\right)\left(34\right),\left(13\right)\left(24\right),\left(14\right)\left(23\right)\right\}$ isomorphism of the K$3$ST model. If we set $\alpha{}=\beta{}=\gamma{}$ in this isomorphism of the K$3$ST model then we get the JC model. The diagonalising matrix for the isomorphism of the K$3$ST model diagonalises the model for \emph{all} parameter choices. Consequently, since the JC model is a parameter choice of the isomorphism of the K$3$ST model, the JC model will be diagonalised by the diagonalising matrix of the isomorphism of the K$3$ST model. There will, however, be other choices for the diagonalising matrix for the JC model that do not diagonalise the isomorphism of the K$3$ST model.

\citet{Bryant17extending} suggested a method of generating rate matrices from Abelian group-based models. If the process is ergodic then the stationary distribution will be the uniform distribution. Furthermore, if the process is time reversible then the rate matrix will be symmetric. Imposing the symmetry condition on the rate matrix for $\mathbb{Z}_{4}$ gives the K$2$ST model. Since the matrix representations of $\mathbb{Z}_{2}\times{}\mathbb{Z}_{2}$ must be symmetrical, $\mathbb{Z}_{2}\times{}\mathbb{Z}_{2}$ generates the K$3$ST model.

\section{Choosing the Diagonalising Matrix Parameters}

For every diagonalising matrix, we have some degrees of freedom. We will argue that some of these choices are more appropriate than others, particularly the choices that lead us to the statement of the conservation of probability. With the exception of the $\mathbb{Z}_{2}\times{}\mathbb{Z}_{2}\cong{}\left\{e,\left(12\right),\left(34\right),\left(12\right)\left(34\right)\right\}$ isomorphism of the K$3$ST model, setting every element of the first row of every diagonalising matrix to one sets the first element of the transformed phylogenetic tensor to be the statement of the conservation of probability. It should also be noted that this results in the diagonalising matrices for the binary symmetric model and the K$3$ST $\left(\mathbb{Z}_{2}\times{}\mathbb{Z}_{2}\cong{}\left\{e,\left(12\right)\left(34\right),\left(13\right)\left(24\right),\left(14\right)\left(23\right)\right\}\right)$ model being the Hadamard matrices of the appropriate orders. If we choose the appropriate parameters for the diagonalising matrices for the K2ST models, some of the $2$ Parameter $\mathbb{Z}_{4}$ models and the JC model, then we again get the Hadamard matrix. Clearly the Hadamard matrix is important in phylogenetic analysis. As stated earlier, the Hadamard matrix provides the statement of conservation of probability for many Abelian group-based models since every element of the first row is one. By definition, every element of every row of a Hadamard matrix must be $\pm{}1$ and each row must be orthogonal to every other row. Consequently, exactly half the elements of every row other than the first row must be $1$ and the other half must be $-1$. Other than the first element representing the conservation of probability, every other element of the transformed phylogenetic tensor must represent the difference in two sums of probabilities, with each sum containing the same number of elements.

\chapter[The Splitting Operator and Convergence]{Phylogenetic Tensors, the Splitting Operator and Convergence}
\label{chapter3}

The following chapters of this thesis will focus on the analysis of various tree and network structures under Markov models of evolution. A framework for phylogenetic tensors will be crucial to our comparisons of the trees and networks. Phylogenetic tensors are functions of the rate parameters from the model in question and the time parameters from the tree or network. Phylogenetic tensors express the theoretical probabilities for each combination of states in the character space across the set of taxa. For a given number of taxa, we can compare the phylogenetic tensors for various trees and networks. For a given tree or network structure, we can compare all of the elements of the phylogenetic tensor to determine a set of equality and inequality constraints on a phylogenetic tensor. This set of constraints will correspond to a probability space that the phylogenetic tensor lies in. We can then compare the probability spaces for various trees and networks.

\section{Character Sequence Alignments}

Suppose we have an $n$-taxon character sequence. The state space has $m$ states. These $m$ states are labelled $0,1,\ldots{},m-1$. At each site in the character sequence, each taxon takes one state from the state space. The combination of states, \textbf{i}, at a site in the sequence alignment can be identified with $\textrm{\textbf{i}}\equiv{}i_{1}i_{2}\ldots{}i_{n}$, where each subscript refers to the taxon in question.

\section{Tensor and Kronecker Products}

Before discussing phylogenetic tensors, we will make a slight diversion to discuss tensor products. While scalars can be represented as zero dimensional quantities, vectors can be represented as one dimensional arrays and matrices can be represented as two dimensional arrays. Tensors are the $n$-dimensional generalisations of these structures. Fortunately, we can express our tensors in vector representation by using the appropriate indices. For example, suppose we wish to express a three dimensional tensor in matrix form. For a two dimensional matrix, each element must have two indices, one representing the row and one representing the column. If we were to express a three dimensional tensor in matrix form, however, each element must have three indices.

Now suppose we have two vector spaces, $V$ and $W$. The tensor product of the two vector spaces is defined over some field, $F$, and denoted $V\otimes{}W$. It also forms a vector space.

The tensor product can be represented using the Kronecker product. For an example of the Kronecker product, suppose $A$ is an $m\times{}n$ matrix and $B$ is a $p\times{}q$ matrix. The Kronecker product of the two matrices, $A\otimes{}B$, is the $mp\times{}nq$ matrix,
\begin{align}\begin{split}
A\otimes{}B=\left[\begin{array}{cccc}
a_{11}B & a_{12}B & \ldots{} & a_{1n}B \\
a_{21}B & a_{22}B & \ldots{} & a_{2n}B \\
\vdots{} & \vdots{} & \ddots{} & \vdots{} \\
a_{m1}B & a_{m2}B & \ldots{} & a_{mn}B \\
\end{array}\right],
\end{split}\end{align}
where $a_{ij}B$ is the matrix $B$, multiplied by the scalar, $a_{ij}$.

\section{Phylogenetic Tensors on the General Markov Model}

We will start with a general discussion of phylogenetic tensors for the general Markov model. We will then derive the expressions for the two-taxon and three-taxon phylogenetic tensors for various trees and networks under the binary symmetric model.

Phylogenetic tensors are represented as column vectors of length $m^{n}$, where $m$ and $n$ refer to the number of states and taxa, respectively. They represent the probability distribution of the combinations of states at an arbitrary site in the character sequences and have $m^{n}$ elements, each denoted $p_{\textrm{\textbf{i}}}$. We can define a one-to-one function from the sequence alignment to the probability distribution. The probability distribution is then indexed as follows,
\begin{align}\begin{split}
p_{\textrm{\textbf{i}}}\equiv{}p_{i_{1}i_{2}\ldots{}i_{n}}.
\end{split}\end{align}

The index of each phylogenetic tensor element has an equivalent integer representation and $m$-ary representation. The function required to go from the $m$-ary representation to the integer representation is
\begin{align}\begin{split}
\textrm{\textbf{i}}=1+i_{1}m^{n-1}+i_{2}m^{n-2}+\ldots{}+i_{n-1}m+i_{n}=1+\sum\limits_{j=1}^{n}i_{j}m^{n-j}.
\end{split}\end{align}

For example, if the state space is binary ($m=2$), the subscript of the phylogenetic tensor will correspond to the combination of states with the equivalent binary representation.

We will see that this element labelling corresponds to the tensor space spanned by the tensor products of the unit vectors. For an $m$ state Markov model, the unit vectors will be
\begin{align}\begin{split}
e_{0}=\left[\begin{array}{c}
1 \\
0 \\
0 \\
\vdots{} \\
0 \\
\end{array}\right],\quad{}e_{1}=\left[\begin{array}{c}
0 \\
1 \\
0 \\
\vdots{} \\
0 \\
\end{array}\right],\quad{}\ldots{},\quad{}e_{m-1}=\left[\begin{array}{c}
0 \\
0 \\
\vdots{} \\
0 \\
1 \\
\end{array}\right],
\end{split}\end{align}
where each vector, $e_{j}$, has $m$ elements, such that
\begin{align}\begin{split}
\left[e_{j}\right]_{i}=\begin{cases} 1 & \text{ if $i=j+1$}, \\
0 & \text{ if $i\neq{}j+1$}.
\end{cases}
\end{split}\end{align}

The phylogenetic tensor space is then spanned by the $m^{n}$ dimensional space of the tensor products of the unit vectors. The tensor products of the units vectors in $\mathbb{R}_{m}$ are unit vectors in $\mathbb{R}_{m^{n}}$, where the ``1'' entry depends on choice \textbf{i}.

For the general Markov model, the phylogenetic tensor can now be represented as
\begin{align}\begin{split}
P=\sum\limits_{\textrm{\textbf{i}}=1}^{m^{n}}p_{\textrm{\textbf{i}}}e_{\textrm{\textbf{i}}}\equiv{}\sum\limits_{i_{1},i_{2},\ldots{},i_{n}=0}^{m-1}p_{i_{1}i_{2}\ldots{}i_{n}}e_{i_{1}}\otimes{}e_{i_{2}}\otimes{}\ldots{}\otimes{}e_{i_{n}},
\end{split}\end{align}
where $i_{k}$ is the state taxon $k$ takes at an arbitrary site in the character sequence, $p_{i_{1}i_{2}\ldots{}i_{n}}$ is the corresponding expression for the probability in the $\textrm{\textbf{i}}=i_{1}i_{2}\ldots{}i_{n}$ term of the phylogenetic tensor and $\otimes{}$ is the Kronecker product.

\subsection{The Phylogenetic Tensor on the Binary Symmetric Model}

For simplicity, the only Markov model we will examine is the binary symmetric model. For all two state models ($m=2$), such as the binary symmetric model, there are only two unit vectors to span $\mathbb{R}_{2}$ and the phylogenetic tensor will be
\begin{align}\begin{split}
P=\sum\limits_{\textrm{\textbf{i}}=1}^{2^{n}}p_{\textrm{\textbf{i}}}e_{\textrm{\textbf{i}}}\equiv{}\sum\limits_{i_{1},i_{2},\ldots{},i_{n}=0}^{1}p_{i_{1}i_{2}\ldots{}i_{n}}e_{i_{1}}\otimes{}e_{i_{2}}\otimes{}\ldots{}\otimes{}e_{i_{n}}.
\end{split}\end{align}

\section{The Splitting Operator}

We are now in a position to find expressions for phylogenetic tensors in terms of the rate parameters from the Markov model and the time parameters from the tree or network structure. To find these expressions we will utilise a tool called the \emph{splitting operator}, introduced by \citet{bashford2004u} and denoted $\delta{}$. The purpose of the splitting operator is to model the instantaneous ``splitting'' of one ancestral edge into two descendant edges. The splitting operator introduces a new taxon onto the tree or network. The dimension of the tensor space must therefore increase by a factor of $m$, the number of states in the state space, with each application of the splitting operator on the tree or network. Immediately after the splitting process has occurred, the two new taxa must have identical character sequences. In other words, the phylogenetic tensor elements representing the cases where the two new taxa are in different states must be identically zero immediately after the splitting process.

The splitting operator was first applied to phylogenetic networks by \citet{sumner2012algebra}. The splitting operator maps a vector space to the tensor product space of the vector space with itself. The splitting operator is only defined on unit vectors, and maps each unit vector to the tensor product of the unit vector with itself. In summary,
\begin{align}\begin{split}
\begin{cases}
\text{i) } & \delta{}: V\rightarrow{}V\otimes{}V, \\
\text{ii) } & \delta{}\cdot{}e_{i}:=e_{i}\otimes{}e_{i}, \\
\text{iii) } & \delta{}\left(u+v\right)=\delta\left(u\right)+\delta\left(v\right),
\end{cases}
\end{split}\end{align}
where $V$ is a vector space, $\otimes{}$ is the tensor product, $\cdot{}$ is the action of the splitting operator on a vector and is represented by the matrix product, $e_{i}$ is an arbitrary unit vector and $u$ and $v$ are arbitrary vectors in the vector space.

It is important to note that in general,
\begin{align}\begin{split}
\delta{}\left(u\right)\neq{}u\otimes{}u.
\end{split}\end{align}

This statement is only true when $u\equiv{}e_{i}$,
\begin{align}\begin{split}
\delta{}\left(u\right)={}u\otimes{}u\Leftrightarrow{}u\equiv{}e_{i}.
\end{split}\end{align}

For example, suppose that $u=e_{0}+e_{1}$. Then
\begin{align}\begin{split}
\delta{}\left(u\right)=\delta{}\left(e_{0}+e_{1}\right)=\delta{}\left(e_{0}\right)+\delta{}\left(e_{1}\right)=e_{0}\otimes{}e_{0}+e_{1}\otimes{}e_{1}.
\end{split}\end{align}

Conversely,
\begin{align}\begin{split}
u\otimes{}u&=\left(e_{0}+e_{1}\right)\otimes{}\left(e_{0}+e_{1}\right) \\
&=e_{0}\otimes{}e_{0}+e_{0}\otimes{}e_{1}+e_{1}\otimes{}e_{0}+e_{1}\otimes{}e_{1} \\
&\neq{}e_{0}\otimes{}e_{0}+e_{1}\otimes{}e_{1} \\
&=\delta{}\left(e_{0}+e_{1}\right) \\
&=\delta{}\left(u\right).
\end{split}\end{align}

\subsection{Pushing Back the Splitting Operator}

We will now look at how the splitting operator acts on the two state general Markov model as an example. The two state general Markov model has the rate matrix,
\begin{align}\begin{split}
Q=\left[\begin{array}{cc}
-\alpha{} & \beta{} \\
\alpha{} & -\beta{} \\
\end{array}\right]=\alpha{}L_{\alpha{}}+\beta{}L_{\beta{}},
\end{split}\end{align}
where columns sum to zero, the matrix element in the $\textrm{\textbf{i}}^{th}$ row and $\textrm{\textbf{j}}^{th}$ column, $Q_{\textrm{\textbf{i}}\textrm{\textbf{j}}}$, is the rate of substitutions from state $\textrm{\textbf{j}}-1$ to state $\textrm{\textbf{i}}-1$ and the generators are
\begin{align}\begin{split}
L_{\alpha{}}=\left[\begin{array}{cc}
-1 & 0 \\
1 & 0 \\
\end{array}\right],\quad{}L_{\beta{}}=\left[\begin{array}{cc}
0 & 1 \\
0 & -1 \\
\end{array}\right].
\end{split}\end{align}

In this thesis we will be examining only the binary symmetric model, obtained by taking $\alpha{}=\beta{}=\lambda{}$ in the general Markov matrix. The rate matrix for the binary symmetric model is
\begin{align}\begin{split}
Q=\left[\begin{array}{cc}
-\lambda{} & \lambda{} \\
\lambda{} & -\lambda{} \\
\end{array}\right].
\end{split}\end{align}

For two state models, the matrix representation of the splitting operator is
\begin{align}\begin{split}
\delta{}=\left[\begin{array}{cc}
1 & 0 \\
0 & 0 \\
0 & 0 \\
0 & 1 \\
\end{array}\right].
\end{split}\end{align}

The transition matrix, $M$, on an edge over time $t$ will be the matrix exponential, $M=e^{Qt}$. The matrix exponential can be expressed in the form
\begin{align}\begin{split}
M=e^{Qt}=I-\frac{1}{2}\left(e^{-2\lambda{}t}-1\right)\left(L_{\alpha{}}+L_{\beta{}}\right).
\end{split}\end{align}

As noted in \citet{sumner2012algebra}, the splitting operator performs the following action on the transition matrix,
\begin{align}\begin{split}
\delta{}\cdot{}M=\delta{}\cdot{}e^{Qt}&=\delta{}\cdot{}\left(I-\frac{1}{2}\left(e^{-2\lambda{}t}-1\right)\left(L_{\alpha{}}+L_{\beta{}}\right)\right), \\
&=\delta{}\cdot{}I-\frac{1}{2}\left(e^{-2\lambda{}t}-1\right)\left(\delta{}\cdot{}L_{\alpha{}}+\delta{}\cdot{}L_{\beta{}}\right).
\end{split}\end{align}

The splitting operator acts on the identity matrix and generating matrices as follows,
\begin{align}\begin{split}
\begin{cases} \delta{}\cdot{}I&=\left(I\otimes{}I\right)\cdot{}\delta{}, \\
\delta{}\cdot{}L_{\alpha{}}&=\left(L_{\alpha{}}\otimes{}L_{\alpha{}}+L_{\alpha{}}\otimes{}I+I\otimes{}L_{\alpha{}}\right)\cdot{}\delta{}, \\
\delta{}\cdot{}L_{\beta{}}&=\left(L_{\beta{}}\otimes{}L_{\beta{}}+L_{\beta{}}\otimes{}I+I\otimes{}L_{\beta{}}\right)\cdot{}\delta{}.
\end{cases}
\end{split}\end{align}

We can check that these expressions are correct by substituting in the appropriate matrix representations for the splitting operator, the identity matrix and each of the generators.

Here it should be noted that these expressions are not unique. To see why, consider the following actions of each of the generators on a unit vector,
\begin{align}\begin{split}
L_{\beta{}}\cdot{}e_{0}=\left[\begin{array}{cc}
0 & 1 \\
0 & -1 \\
\end{array}\right]\cdot{}\left[\begin{array}{c}
1 \\
0 \\
\end{array}\right]=\left[\begin{array}{c}
0 \\
0 \\
\end{array}\right],
\end{split}\end{align}
\begin{align}\begin{split}
L_{\alpha{}}\cdot{}e_{1}=\left[\begin{array}{cc}
-1 & 0 \\
1 & 0 \\
\end{array}\right]\cdot{}\left[\begin{array}{c}
0 \\
1 \\
\end{array}\right]=\left[\begin{array}{c}
0 \\
0 \\
\end{array}\right].
\end{split}\end{align}

Now consider the following expression,
\begin{align}\begin{split}
L_{\alpha{}}\otimes{}L_{\beta{}}\cdot{}\delta{}\cdot{}e_{0}=L_{\alpha{}}\otimes{}L_{\beta{}}\cdot{}e_{0}\otimes{}e_{0}=L_{\alpha{}}\cdot{}e_{0}\otimes{}L_{\beta{}}\cdot{}e_{0}=\left[\begin{array}{c}
0 \\
0 \\
0 \\
0 \\
\end{array}\right].
\end{split}\end{align}

Similarly,
\begin{align}\begin{split}
L_{\alpha{}}\otimes{}L_{\beta{}}\cdot{}\delta{}\cdot{}e_{1}=L_{\alpha{}}\otimes{}L_{\beta{}}\cdot{}e_{1}\otimes{}e_{1}=L_{\alpha{}}\cdot{}e_{1}\otimes{}L_{\beta{}}\cdot{}e_{1}=\left[\begin{array}{c}
0 \\
0 \\
0 \\
0 \\
\end{array}\right].
\end{split}\end{align}

Consequently,
\begin{align}\begin{split}
L_{\alpha{}}\otimes{}L_{\beta{}}\cdot{}\delta{}\cdot{}\left(e_{0}+e_{1}\right)=\left[\begin{array}{c}
0 \\
0 \\
0 \\
0 \\
\end{array}\right].
\end{split}\end{align}

Following similar logic,
\begin{align}\begin{split}
L_{\beta{}}\otimes{}L_{\alpha{}}\cdot{}\delta{}\cdot{}\left(e_{0}+e_{1}\right)=\left[\begin{array}{c}
0 \\
0 \\
0 \\
0 \\
\end{array}\right].
\end{split}\end{align}

Since $\delta{}$ only acts on unit vectors, we can ignore both of these terms from our expressions in a similar way to ignoring constants when performing definite integration.

Alternatively, instead of using $L_{\alpha{}}$ and $L_{\beta{}}$ as generators, we could have used the identity matrix and the matrix with off-diagonal entries being one and diagonal entries being zero. When we compare the action of the splitting operator on the rate matrix in both bases, we see why the extra terms that we have ignored appear. Having noted this freedom, we will not use it in this work.

Referring back to the expressions for the action of the splitting operator on the identity matrix and the generating matrices, we can see that for each expression the splitting operator is ``pushed back''. \citet{sumner2012algebra} showed that if we have a product of multiple transition matrices, the splitting operator can be pushed back through every transition matrix, one at a time. Eventually, the splitting operator will act on the initial probability distribution at the root.

We will now make a change of notation. Let $\mathcal{L}_{\alpha{}}^{\left[2\right]}=L_{\alpha{}}\otimes{}L_{\alpha{}}+L_{\alpha{}}\otimes{}I+I\otimes{}L_{\alpha{}}$ and $\mathcal{L}_{\beta{}}^{\left[2\right]}=L_{\beta{}}\otimes{}L_{\beta{}}+L_{\beta{}}\otimes{}I+I\otimes{}L_{\beta{}}$. Also let $R_{11}=\mathcal{L}_{\alpha{}}^{\left[2\right]}+\mathcal{L}_{\beta{}}^{\left[2\right]}$. Later, it will become apparent why we have chosen this notation.

With this notation, the action of the splitting operator on the transition matrix is
\begin{align}\begin{split}
\delta{}\cdot{}M=\left[I\otimes{}I-\frac{1}{2}\left(e^{-2\lambda{}t}-1\right)R_{11}\right]\cdot{}\delta{}.
\end{split}\end{align}

From \citet{sumner2012algebra} we can simplify this to
\begin{align}\begin{split}
\delta{}\cdot{}M&=e^{\lambda{}R_{11}t}\cdot{}\delta{} \\
&:=\mathcal{M}\cdot{}\delta{}.
\end{split}\end{align}

We now have a new transition matrix and rate matrix, with the splitting operator now acting after the transition matrix.

The rate matrix for the new transition matrix is
\begin{align}\begin{split}
\lambda{}R_{11}=\left[\begin{array}{c|cccc}
& 00 & 01 & 10 & 11 \\
\hline
00 & -\lambda{} & \lambda{} & \lambda{} & \lambda{} \\
01 & 0 & -2\lambda{} & 0 & 0 \\
10 & 0 & 0 & -2\lambda{} & 0 \\
11 & \lambda{} & \lambda{} & \lambda{} & -\lambda{} \\
\end{array}\right],
\end{split}\end{align}
where columns sum to zero, the matrix element in the $\textrm{\textbf{i}}^{th}$ row and the $\textrm{\textbf{j}}^{th}$ column, $\left[\lambda{}R_{11}\right]_{\textrm{\textbf{i}}\textrm{\textbf{j}}}\equiv{}\left[\lambda{}R_{11}\right]_{i_{1}i_{2}j_{1}j_{2}}$, is the rate of substitutions from the combination of states $j_{1}j_{2}$ to $i_{1}i_{2}$.

We now have a transition matrix that represents transitions between combinations of states for two taxa. We can see that the only substitutions that have positive rates are the substitutions that result in the two taxa having the same state. $\lambda{}R_{11}$ can be used to force two edges to remain identical. The substitutions between combinations of states can be represented as a diagram, shown below in Figure~\ref{substitutionscombinationstates}.
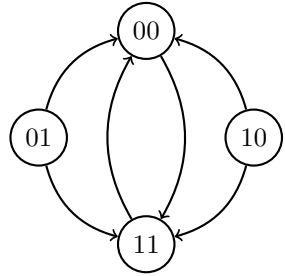
\begin{figure}[H]
	\centering
		\begin{tikzpicture}[->,node distance=2cm,thick,main node/.style={circle,draw}]
			\node[main node] (1) {00};
			\node[main node] (2) [below left of=1] {01};
			\node[main node] (3) [below right of=2] {11};
			\node[main node] (4) [below right of=1] {10};
			\path[every node/.style={}]
				(1) edge [bend left] node[above] {} (3)
				(2) edge [bend left] node[left] {} (1)
				(2) edge [bend right] node[left] {} (3)
				(3) edge [bend left] node[below] {} (1)
				(4) edge [bend right] node[right] {} (1)
				(4) edge [bend left] node[right] {} (3);
		\end{tikzpicture}
		\caption{$\lambda{}R_{11}$ on two taxa.}
		\label{substitutionscombinationstates}
\end{figure}

The direction of the arrows represents substitutions that can occur and each character state transition has the same rate, $\lambda{}$, from the binary symmetrical model.

\subsection{The Splitting Operator on Edges}

As we have seen, the splitting operator is a special function that uses the Kronecker product to increase the dimension of the phylogenetic tensor space in a way that represents the splitting of one ancestral edge into two descendant edges. Recall that the splitting operator acts directly on an individual unit vector as follows,
\begin{align}\begin{split}
\delta{}\cdot{}e_{i}=e_{i}\otimes{}e_{i},
\end{split}\end{align}
where $\cdot{}$ is the matrix product and $\otimes{}$ is the Kronecker product.

The unit column vector, $e_{i}$, corresponds to the state, $i$, in an individual character sequence representing a single taxon. It follows that $e_{i}\otimes{}e_{i}$ represents the combination of states, $ii$, at a given site in two character sequences representing two taxa. Immediately after a speciation event, when no time has elapsed, the two resultant character sequences must be identical. Hence, the splitting operator follows our intuition on speciation.

Let's now look at a basic two-taxon clock-like tree with the splitting operator acting on the tree, shown in Figure~\ref{splittingoperatortree} below.
\begin{figure}[H]
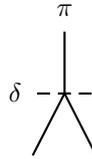

	\centering
		\psmatrix[colsep=.3cm,rowsep=.4cm,mnode=r]
		& ~ \\
		~ & ~ & ~ \\
		~ && ~
		\ncline{1,2}{2,2}^{$\pi$}
		\ncline{2,2}{3,1}
		\ncline{2,2}{3,3}
		\ncline[linestyle=dashed]{2,1}{2,3}<{$\delta$}
		\endpsmatrix
\caption{Two-taxon clock-like tree with the splitting operator, $\delta$, and the initial probability distribution, $\pi$.}
\label{splittingoperatortree}
\end{figure}

The tree starts at the top, with the initial probability distribution, $\pi$, at the root. At the root there is only a single taxon present. The initial probability distribution is the phylogenetic tensor for the single root taxon when no time has elapsed. The number of elements in the initial probability distribution will be equal to the number of states in the state space.

We will choose the initial probability distribution to be
\begin{align}\begin{split}
\pi{}=\left[\begin{array}{c} p_{0}\left(0\right) \\
p_{1}\left(0\right) \\
\end{array}\right]=\left[\begin{array}{c} \frac{1}{2}\vspace{1pt} \\
\frac{1}{2} \\
\end{array}\right],
\end{split}\end{align}
the stationary distribution. The stationary distribution is invariant under the action of any transition matrix from the chosen model on a single edge. In other words, if $\pi{}$ is the stationary distribution then $e^{Qt}\cdot{}\pi{}=\pi{}$ for any time, $t$, and for any $Q$ from the binary symmetric model. The stationary distribution is the equilibrium distribution, which cannot be left once reached. Consequently, the stationary distribution represents the long run probability distribution for a single taxon. For an $n$-taxon phylogenetic tensor, the initial probability distribution can be recovered from the marginal probability distributions for each taxon.

The vertical line down from the root is the edge representing the evolution of the single taxon over time from the root. Eventually this single taxon splits into two new descendant taxa. The dotted line represents the stage in time where the splitting operator is used to represent the splitting of the single ancestral taxon into two descendant taxa. Over time the two new taxa diverge from each other, represented by two edges which move further apart through time.

\begin{figure}[H]
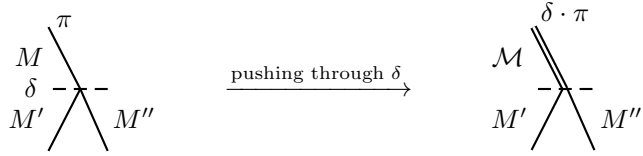

	\centering
		\begin{subfigure}[h]{0.25\textwidth}
			\centering
        \psmatrix[colsep=0.3cm,rowsep=.4cm,mnode=r]
         ~ \\
         ~ & ~ & \\
         ~ & ~ &
        \ncline{1,1}{2,2}^{$\pi$}<{$M$}
        \ncline{2,2}{3,1}<{$M'$}
        \ncline{2,2}{3,3}>{$M''$}
        \ncline[linestyle=dashed]{2,1}{2,3}<{$\delta$}
        \endpsmatrix
		\end{subfigure}
		$\xrightarrow{\text{pushing through }\delta{}}$
		\begin{subfigure}[h]{0.25\textwidth}
			\centering
				\psmatrix[colsep=0.3cm,rowsep=.4cm,mnode=r]
				~ \\
				~ & ~ & \\
				~ & ~ &
				\ncline[doubleline=true]{1,1}{2,2}^[tpos=1]{$\delta{}\cdot{}\pi{}$}<{$\mathcal{M}$}
				\ncline[offset=-\pslinewidth]{2,2}{3,1}<{$M'$}
				\ncline[offset=\pslinewidth]{2,2}{3,3}>{$M''$}
				\ncline[linestyle=dashed]{2,1}{2,3}
				\endpsmatrix
		\end{subfigure}
		\caption{Action of the splitting operator on a single taxon, transforming it into a two-taxon tree. $M$, $M'$, $M''$ and $\mathcal{M}$ refer to the Markov matrices on those edges. $\pi{}$ is the probability distribution at the root.}
		\label{splittingoperatorsingletaxon}
\end{figure}

After placing the splitting operator wherever a splitting event occurs on a tree, we can push each splitting operator through the tree. After pushing the splitting operators through edges towards the root, it is equivalent to think of a single edge as a collection of edges being forced to remain identical, with the number of edges equal to the number of descendant taxa. When one splitting operator is pushed back, two edges become identical. The number of identical edges is equal to one plus the number of splitting operators below it which have been pushed back above the edge. In Figure~\ref{splittingoperatorsingletaxon} below is an example of the action the splitting operator performs on a single taxon, transforming it into a two-taxon tree.
\vspace{12pt}

Algebraically, the above scenario is given by $\left(M'\otimes{}M''\right)\cdot{}\left(\delta{}\cdot{}M\right)\cdot{}\pi_{}\rightarrow{}\left(M'\otimes{}M''\right)\cdot{}\mathcal{M}\cdot{}\delta{}\cdot{}\pi{}$, where the edge labels are the transition matrices for the edges in question.

The consequence is that the splitting operator now acts further up on the tree or network, in this case at the root. If the root is further up the tree or network, we can repeat this process as many times as necessary until the splitting operator acts directly on the root. Likewise, we can perform this process on every splitting operator on the tree or network. Every splitting operator would then act directly on the root.

$\mathcal{M}$ plays the role of implementing correlated changes. By pushing back the splitting operator, we now have a rate matrix which models multiple identical edges being forced to remain identical. An intriguing question that arises from this result is ``what happens when this rate matrix is applied to multiple adjacent non-identical edges?''. \citet{sumner2012algebra} discussed the issue of whether the rate matrices arising from pushing back the splitting operator can be used to model convergence on parts of a network which are isolated from splitting processes.

\section{Modelling Convergence}

Recall the rate matrix that arises when we push back the splitting operator,
\begin{align}\begin{split}
\lambda{}R_{11}=\left[\begin{array}{c|cccc}
& 00 & 01 & 10 & 11 \\
\hline
00 & -\lambda{} & \lambda{} & \lambda{} & \lambda{} \\
01 & 0 & -2\lambda{} & 0 & 0 \\
10 & 0 & 0 & -2\lambda{} & 0 \\
11 & \lambda{} & \lambda{} & \lambda{} & -\lambda{} \\
\end{array}\right].
\end{split}\end{align}

We have seen that the rate matrix formed by pushing back the splitting operator can be used to model identical edges. We will argue that it can also be used to model the convergence of two edges in the same time period which have previously diverged from each other. We can simply model the convergence of these two diverged edges with a single transition matrix and the above rate matrix. For the binary symmetric model, the rates of substitutions leading to identical states will be the same as the rates of substitutions on individual edges. Rates of substitutions away from identical states will be zero.

We can generalise this concept to model the convergence of any subset of edges with each other in a given time period. We are not restricted to one group of converging edges. We can have multiple groups of converging edges in one time period. We will also see that we can have groups of more than two taxa converging with each other in a given time period.

We will represent converging edges graphically with curved lines. Below in Figure~\ref{secondtimeperiodconvergence} is a simple example of a two-taxon network with two converging edges.
\begin{figure}[H]
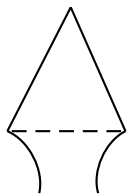

	\centering
		\psmatrix[colsep=.3cm,rowsep=.4cm,mnode=r]
		&& ~ \\
		\\
		~ && && ~ \\
		& ~ && ~
		\ncline{1,3}{3,1}
		\ncline{1,3}{3,5}
		\ncarc[arcangle=40]{3,1}{4,2}
		\ncarc[arcangle=-40]{3,5}{4,4}
		\psset{linestyle=dashed}\ncline{3,1}{3,5}
		\endpsmatrix
\caption{Two-taxon clock-like convergence-divergence network with a convergence period in the second time period.}
\label{secondtimeperiodconvergence}
\end{figure}

On our trees and networks, divergence is represented by straight lines emanating from a node, while convergence is represented by curved lines. The root is at the top of the tree or network and time progresses linearly down the page. The other dimension in the two-dimensional plane on the page has no meaning other than to create space for the diagram. By moving across the page we are not moving through time or any other dimension, we are simply moving from one edge to another. Consequently, we can reflect the edges below any node in any tree or network. In this thesis the labelling of each tree and network should be considered independently unless otherwise stated. Time parameters and leaf labels should not be equated or compared between one tree or network and another unless explicitly stated.

\section{Constructing the Trees and Networks}

Phylogenetic tensors can be expressed in many algebraically equivalent forms. It is perhaps most convenient conceptually to express the phylogenetic tensor as a series of matrix products, with each matrix being a transition matrix representing a distinct, non-overlapping period in time. We first partition the tree or network into these time periods, with time periods being separated by splitting events or the start or end of convergence periods. We can even express unrooted trees this way by choosing an arbitrary placement of the root. Each transition matrix is a matrix exponential, with its argument being the sum of the evolutionary processes happening in that time period.

We will now introduce some notation to keep track of which taxa are converging or diverging in different time periods. Let $\left[n\right]$ denote the set of all taxa and $A_{\textrm{\textbf{i}}}^{k}=i_{1}i_{2}\ldots{}i_{n}\subseteq{}\left[n\right]$ be a subset of edges for a convergence group in time period $k$, represented by an ordered binary string concatenation of length $n$, with ``$1$'' representing a taxon in the group and ``$0$'' representing a taxon not in the group.

Since they are algebraically equivalent, there is no distinction made between a group of edges undergoing convergence and a single diverging edge ``converging'' with itself. A single diverging edge can be treated as its own convergence group. We will restrict each edge to belong to exactly one convergence group. Consequently, $A_{\textrm{\textbf{i}}}^{k}\cap{}A_{\textrm{\textbf{j}}}^{k}=\emptyset{}$ for all $\textrm{\textbf{i}}\neq{}\textrm{\textbf{j}}$ and $\bigcup{}A_{\textrm{\textbf{i}}}^{k}=A_{1}^{k}\cup{}A_{2}^{k}\cup{}\ldots{}\cup{}A_{l}^{k}=\left[n\right]$. Below in Figure~\ref{notation} is an example of the notation on a three-taxon tree.
\begin{figure}[H]
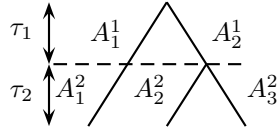

	\centering
		\psmatrix[colsep=.4cm,rowsep=.4cm,mnode=r]
		~ & && ~ \\
		~ & & ~ & & ~ & ~ \\
		~ & ~ && ~ && ~
		\ncline{1,4}{2,3}
		\tlput{$A_{1}^{1}$}
		\ncline{2,3}{3,2}
		\psset{tpos=.4}
		\tlput{$A_{1}^{2}$}
		\psset{tpos=.5}
		\ncline{1,4}{2,5}
		\trput{$A_{2}^{1}$}
		\ncline{2,5}{3,4}
		\psset{tpos=.4}
		\tlput{$A_{2}^{2}$}
		\ncline{2,5}{3,6}
		\trput{$A_{3}^{2}$}
		\ncline{<->,arrowscale=1.5}{1,1}{2,1}
		\tlput{$\tau_{1}$}
		\ncline{<->,arrowscale=1.5}{2,1}{3,1}
		\tlput{$\tau_{2}$}
		\psset{linestyle=dashed}\ncline{2,1}{2,6}
		\endpsmatrix
		\caption{Three-taxon clock-like tree. $A_{1}^{1}=A_{1}^{2}=100$, $A_{2}^{1}=011$, $A_{2}^{2}=010$ and $A_{3}^{2}=001$.}
		\label{notation}
\end{figure}

The rate matrix for convergence of edges, $A_{\textrm{\textbf{i}}}^{k}$, is expressed as
\begin{align}\begin{split}
R_{A_{\textrm{\textbf{i}}}^{k}}=\lambda{}\sum\limits_{B\subseteq{}A_{\textrm{\textbf{i}}}^{k},B\neq{}\emptyset}\left(R_{B,\alpha{}}+R_{B,\beta{}}\right),
\end{split}\end{align}
where $R_{B,x}=S_{i_{1},x}\otimes{}S_{i_{2},x}\otimes{}\ldots{}\otimes{}S_{i_{n},x}$ and $S_{i_{r},x}=\begin{cases} I & \text{if }i_{r}=0, \\
L_{x} & \text{if } i_{r}=1. \\
\end{cases}$

As a further example, suppose $n=3$. The rate matrix for the convergence of the first two taxa, with the third taxa diverging, is then
\begin{align}\begin{split}
\lambda{}R_{110}&=\lambda{}\left(L_{\alpha{}}\otimes{}L_{\alpha{}}\otimes{}I+L_{\alpha{}}\otimes{}I\otimes{}I+I\otimes{}L_{\alpha{}}\otimes{}I+L_{\beta{}}\otimes{}L_{\beta{}}\otimes{}I+L_{\beta{}}\otimes{}I\otimes{}I+I\otimes{}L_{\beta{}}\otimes{}I\right) \\
&=\left[\begin{array}{c|cccccccc}
& 000 & 001 & 010 & 011 & 100 & 101 & 110 & 111 \\
\hline
000 & -\lambda{} & 0 & \lambda{} & 0 & \lambda{} & 0 & \lambda{} & 0 \\
001 & 0 & -\lambda{} & 0 & \lambda{} & 0 & \lambda{} & 0 & \lambda{} \\
010 & 0 & 0 & -2\lambda{} & 0 & 0 & 0 & 0 & 0 \\
011 & 0 & 0 & 0 & -2\lambda{} & 0 & 0 & 0 & 0 \\
100 & 0 & 0 & 0 & 0 & -2\lambda{} & 0 & 0 & 0 \\
101 & 0 & 0 & 0 & 0 & 0 & -2\lambda{} & 0 & 0 \\
110 & \lambda{} & 0 & \lambda{} & 0 & \lambda{} & 0 & -\lambda{} & 0 \\
111 & 0 & \lambda{} & 0 & \lambda{} & 0 & \lambda{} & 0 & -\lambda{} \\
\end{array}\right].
\end{split}\end{align}

We will set $\lambda{}t_{i}=\tau_{i}$ for convenience since there is only one degree of freedom in the product $\lambda{}t_{i}$. Finally, the general phylogenetic tensor for $n$ taxa can be expressed as
\begin{align}\begin{split}
P=\left(\prod_{k=1}^{m}e^{\left(\sum\limits_{\textrm{\textbf{i}}}R_{A_{\textrm{\textbf{i}}}^{m+1-k}}\right)\tau_{m+1-k}}\right)\cdot{}\Pi{},
\end{split}\end{align}
where $m$ is the number of epochs and the summation is over all convergence groups in the time period in question and $\Pi{}=\Delta{}\cdot{}\pi{}$. We have let $\Pi{}$ denote the initial probability distribution after all splitting operators have been pushed back, with $\Delta{}$ being the splitting operator that maps the initial probability distribution on the root taxon, $\pi$, to the initial probability distribution on all $n$ taxa, $\Pi{}$. It should be noted that the product counts backwards, as $k=1$ gives the time parameter $\tau_{m}$ and $k=m$ gives the time parameter $\tau_{1}$. This is because the matrix exponentials start with the most recent time period, the $m^{th}$ time period, on the left and end with the first time period on the right.

\begin{figure}[H]
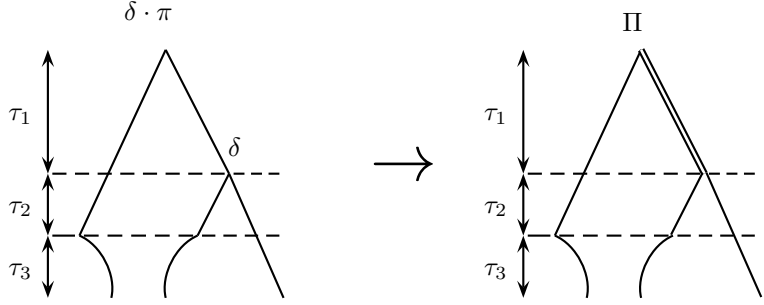

	\centering
		\begin{subfigure}[h]{0.35\textwidth}
			\centering
				\psmatrix[colsep=.3cm,rowsep=.4cm,mnode=r]
				~ && && ~ \\
				~ \\
				~ && && && ~ && ~ \\
				~ & ~ && && ~ & && ~ \\
				~ && ~ && ~ && && ~
				\ncline{1,5}{4,2}^[tpos=0.8]{$\delta{}\cdot{}\pi{}$}
				\ncline{1,5}{3,7}
				\ncline{3,7}{5,9}^[tpos=0.1]{$\delta{}$}
				\ncline{3,7}{4,6}
				\ncarc[arcangle=40]{4,2}{5,3}
				\ncarc[arcangle=-40]{4,6}{5,5}
				\ncline{<->,arrowscale=1.5}{1,1}{3,1}<{$\tau_{1}$}
				\ncline{<->,arrowscale=1.5}{3,1}{4,1}<{$\tau_{2}$}
				\ncline{<->,arrowscale=1.5}{4,1}{5,1}<{$\tau_{3}$}
				\ncline[linestyle=dashed]{3,1}{3,7}
				\ncline[linestyle=dashed]{3,7}{3,9}
				\ncline[linestyle=dashed]{4,1}{4,2}
				\ncline[linestyle=dashed]{4,2}{4,6}
				\ncline[linestyle=dashed]{4,6}{4,9}
				\endpsmatrix
		\end{subfigure}
		$\text{\Huge$\rightarrow$}$		
		\begin{subfigure}[h]{0.35\textwidth}
			\centering
				\psmatrix[colsep=.3cm,rowsep=.4cm,mnode=r]
				~ && && ~ \\
				~ \\
				~ && && && ~ && ~ \\
				~ & ~ && && ~ & && ~ \\
				~ && ~ && ~ && && ~
				\ncline{1,5}{4,2}^[tpos=0.9]{$\Pi{}$}
				\ncline[doubleline=true]{1,5}{3,7}
				\ncline[offset=\pslinewidth]{3,7}{5,9}
				\ncline[offset=-\pslinewidth]{3,7}{4,6}
				\ncarc[arcangle=40]{4,2}{5,3}
				\ncarc[arcangle=-40]{4,6}{5,5}
				\ncline{<->,arrowscale=1.5}{1,1}{3,1}<{$\tau_{1}$}
				\ncline{<->,arrowscale=1.5}{3,1}{4,1}<{$\tau_{2}$}
				\ncline{<->,arrowscale=1.5}{4,1}{5,1}<{$\tau_{3}$}
				\ncline[linestyle=dashed]{3,1}{3,7}
				\ncline[linestyle=dashed]{3,7}{3,9}
				\ncline[linestyle=dashed]{4,1}{4,2}
				\ncline[linestyle=dashed]{4,2}{4,6}
				\ncline[linestyle=dashed]{4,6}{4,9}
				\endpsmatrix
		\end{subfigure}
		\caption{A three-taxon convergence-divergence network.}
		\label{fig:convergence-divergenceexample}
\end{figure}

Now we will look at a three-taxon convergence-divergence network in Figure~\ref{fig:convergence-divergenceexample} below as an example.
\vspace{12pt}

The phylogenetic tensor for this network can be written as
\begin{align}\begin{split}
P=e^{\left(R_{110}+R_{001}\right)\tau_{3}}\cdot{}e^{\left(R_{100}+R_{010}+R_{001}\right)\tau_{2}}\cdot{}e^{\left(R_{100}+R_{011}\right)\tau_{1}}\cdot{}\Pi{}.
\end{split}\end{align}

As an example, the convergence of the first and second taxa, in conjunction with the divergence of the third taxon, in the third time interval, is described by the rate matrix,
\begin{align}\begin{split}
\lambda{}\left(R_{110}+R_{001}\right)=\left[\begin{array}{c|cccccccc}
& 000 & 001 & 010 & 011 & 100 & 101 & 110 & 111 \\
\hline
000 & -2\lambda{} & \lambda{} & \lambda{} & 0 & \lambda{} & 0 & \lambda{} & 0 \\
001 & \lambda{} & -2\lambda{} & 0 & \lambda{} & 0 & \lambda{} & 0 & \lambda{} \\
010 & 0 & 0 & -3\lambda{} & \lambda{} & 0 & 0 & 0 & 0 \\
011 & 0 & 0 & \lambda{} & -3\lambda{} & 0 & 0 & 0 & 0 \\
100 & 0 & 0 & 0 & 0 & -3\lambda{} & \lambda{} & 0 & 0 \\
101 & 0 & 0 & 0 & 0 & \lambda{} & -3\lambda{} & 0 & 0 \\
110 & \lambda{} & 0 & \lambda{} & 0 & \lambda{} & 0 & -2\lambda{} & \lambda{} \\
111 & 0 & \lambda{} & 0 & \lambda{} & 0 & \lambda{} & \lambda{} & -2\lambda{} \\
\end{array}\right].
\end{split}\end{align}

Note that this rate matrix is the sum of the rate matrix for the convergence of the first two taxa, which we gave as an example earlier, and the rate matrix for the divergence of the third taxon from the other two taxa.

The substitutions are represented diagrammatically in Figure~\ref{examplesubstitutionrates} below.
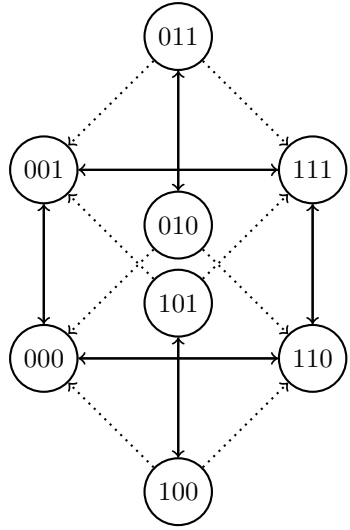
\begin{figure}[H]
	\centering
		\begin{tikzpicture}[->,node distance=2.5cm,thick,main node/.style={circle,draw}]
			\node[main node] (1) {011};
			\node[main node] (2) [below left of=1] {001};
			\node[main node] (3) [below right of=1] {111};
			\node[main node] (4) [below of=1] {010};
			\node[main node] (5) [below of=2] {000};
			\node[main node] (6) [below of=3] {110};
			\node[main node] (7) [below right of=2] {101};
			\node[main node] (8) [below right of=5] {100};
			\draw [->] (1) to (4);
			\draw [->] (4) to (1);
			\draw [->] (2) to (5);
			\draw [->] (5) to (2);
			\draw [->] (3) to (6);
			\draw [->] (6) to (3);
			\draw [->] (7) to (8);
			\draw [->] (8) to (7);
			\draw [->] (2) to (3);
			\draw [->] (3) to (2);
			\draw [->] (5) to (6);
			\draw [->] (6) to (5);
			\draw[dotted] [->] (1) to (2);
			\draw[dotted] [->] (1) to (3);
			\draw[dotted] [->] (4) to (5);
			\draw[dotted] [->] (4) to (6);
			\draw[dotted] [->] (7) to (2);
			\draw[dotted] [->] (7) to (3);
			\draw[dotted] [->] (8) to (5);
			\draw[dotted] [->] (8) to (6);		
		\end{tikzpicture}
		\caption{$\lambda{}\left(R_{110}+R_{001}\right)$ on three edges. Each transition has the same rate, $\lambda{}$, from the binary symmetrical model. Solid lines indicate ``regular'' mutations, while dotted lines indicate ``correction'' mutations, responsible for convergence.}
		\label{examplesubstitutionrates}
\end{figure}

\section{Transformed Phylogenetic Tensors in the Hadamard Basis}
\label{transformedphylo}

We transform the basis of the $n$-taxon phylogenetic tensor by multiplying it on the left by the transforming matrix, $H=h\otimes{}h\otimes{}\ldots{}\otimes{}h$, where there are $n-1$ tensor products. Matrices which are in the transformed basis will be denoted by a ``\textit{ $\widehat{}$ }'' above them. We showed in Chapter~\ref{chapter2} that the general transforming matrix, $h$, for the binary symmetric model is
\begin{align}\begin{split}
h=\left[\begin{array}{cc} k_{1} & k_{2} \\
k_{1} & -k_{2} \\
\end{array}\right].
\end{split}\end{align}

If we let $k_{1}=k_{2}=1$, the first row of $H$ multiplied by the phylogenetic tensor will give the sum of all of the probability distribution elements and must be equal to one due to probability conservation. By letting $k_{1}=k_{2}=1$, $h$ becomes the Hadamard matrix,
\begin{align}\begin{split}
h=\left[\begin{array}{cc} 1 & 1 \\
1 & -1 \\
\end{array}\right].
\end{split}\end{align}
which diagonalises $Q$ to give
\begin{align}\begin{split}
\widehat{Q}=h\cdot{}Q\cdot{}h^{-1}=\left[\begin{array}{cc} 0 & 0 \\
0 & -2\lambda{} \\
\end{array}\right].
\end{split}\end{align}

It transforms the Markov generators to
\begin{align}\begin{split}
\widehat{L}_{\alpha{}}=h\cdot{}L_{\alpha{}}\cdot{}h^{-1}=\left[\begin{array}{cc}
0 & 0 \\
-1 & -1 \\
\end{array}\right],\quad{}\widehat{L}_{\beta{}}=h\cdot{}L_{\beta{}}\cdot{}h^{-1}=\left[\begin{array}{cc}
0 & 0 \\
1 & -1 \\
\end{array}\right].
\end{split}\end{align}

The initial probability distribution becomes $\widehat{\Pi{}}=H\cdot{}\Delta{}\cdot{}\pi{}$.

Finally, the transformed basis of the general phylogenetic tensor will be
\begin{align}\begin{split}
\widehat{P}=H\cdot{}P=\left(\prod_{k=1}^{m}e^{\left(\sum\limits_{\textrm{\textbf{i}}}\widehat{R}_{A_{\textrm{\textbf{i}}}^{m+1-k}}\right)\tau_{m+1-k}}\right)\cdot{}\widehat{\Pi{}}.
\end{split}\end{align}

Applying the Hadamard transformation results in $\widehat{p}_{00\ldots{}0}=q_{00\ldots{}0}=1$, the conservation of probability. All other elements of $\widehat{P}$ will be linear combinations of the elements of $P$, with coefficients of $\pm{1}$, the elements of the Hadamard matrix.

Since there is only a single degree of freedom in $\tau_{i}=\lambda{}t_{i}$ we are free to rescale $\tau_{i}\rightarrow{}\frac{1}{2}\tau_{i}$ to simplify our expressions.

\section{Modelling Convergence on Networks}

Before we proceed to examine some two-taxon, three-taxon and four-taxon trees and networks, we will first establish some criteria for determining permissible convergence-divergence networks.

\begin{figure}[H]
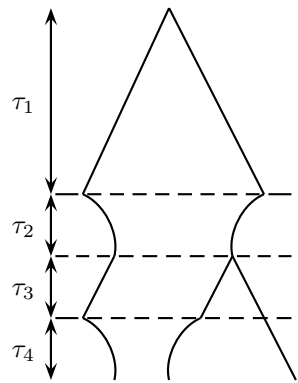

	\centering
		\psmatrix[colsep=.3cm,rowsep=.4cm,mnode=r]
		~ && && ~ \\
		~ \\
		~ \\
		~ & ~ && && && ~ & ~ \\
		~ && ~ && && ~ && ~ \\
		~ & ~ && && ~ & && ~ \\
		~ && ~ && ~ && && ~
		\ncline{1,5}{4,2}
		\ncline{1,5}{4,8}
		\ncarc[arcangle=40]{4,2}{5,3}
		\ncarc[arcangle=-40]{4,8}{5,7}
		\ncline{5,3}{6,2}
		\ncline{5,7}{6,6}
		\ncline{5,7}{7,9}
		\ncarc[arcangle=40]{6,2}{7,3}
		\ncarc[arcangle=-40]{6,6}{7,5}
		\ncline{<->,arrowscale=1.5}{1,1}{4,1}
		\tlput{$\tau_{1}$}
		\ncline{<->,arrowscale=1.5}{4,1}{5,1}
		\tlput{$\tau_{2}$}
		\ncline{<->,arrowscale=1.5}{5,1}{6,1}
		\tlput{$\tau_{3}$}
		\ncline{<->,arrowscale=1.5}{6,1}{7,1}
		\tlput{$\tau_{4}$}
		\psset{linestyle=dashed}\ncline{4,1}{4,2}
		\psset{linestyle=dashed}\ncline{4,2}{4,8}
		\psset{linestyle=dashed}\ncline{4,8}{4,9}
		\psset{linestyle=dashed}\ncline{5,1}{5,3}
		\psset{linestyle=dashed}\ncline{5,3}{5,7}
		\psset{linestyle=dashed}\ncline{5,7}{5,9}
		\psset{linestyle=dashed}\ncline{6,1}{6,2}
		\psset{linestyle=dashed}\ncline{6,2}{6,6}
		\psset{linestyle=dashed}\ncline{6,6}{6,9}
		\endpsmatrix
		\caption{An example of a three-taxon clock-like network. The two edges are converging in the second time period and the non-sister taxa are converging in the fourth time period.}
		\label{fig:threetaxondespeciationexample}
\end{figure}

Our convergence-divergence networks will be partitioned into distinct non-overlapping time periods. We will call a set of edges converging together in an time period a \emph{convergence group}. A new time period starts directly after a splitting event or when the convergence groups change. For example, there may be a splitting event giving rise to a new time period where all edges diverge from each other. A group of edges may then converge with each other in a new time period. Finally, the convergence group(s) may change, giving rise to another new time period. All time periods containing convergence groups must occur immediately after a divergence period or another convergence period. For diagrammatic examples of converge-divergence networks see Figure~\ref{fig:convergence-divergenceexample} on page~\pageref{fig:convergence-divergenceexample} or Figure~\ref{fig:threetaxondespeciationexample} on page~\pageref{fig:threetaxondespeciationexample}.

Suppose we have an $n$-taxon clock-like tree. $n$-taxon clock-like trees will have $n-1$ numerical parameters. If we wish to adopt a more parameter rich tree we can drop the clock-like assumption. Non-clock-like trees will have $2n-3$ numerical parameters. For $n\geq{}3$, $2n-3>n-1$ and non-clock-like trees will be more parameter-rich than clock-like trees. $2n-3-\left(n-1\right)=n-2$ and for large $n$ non-clock-like trees will have approximately double the number of numerical parameters that clock-like trees with the same number of taxa have. In contrast, convergence-divergence networks introduce any positive integer number of extra parameters to the clock-like tree. A clock-like tree may not be parameter rich enough to be a ``good'' fit for a set of pattern frequencies and a non-clock-like tree may overfit the pattern frequencies by providing too many parameters. In this case a convergence-divergence network could be used to introduce a number of extra parameters to the clock-like tree which is less than the number of extra parameters introduced by removing the clock assumption.

We could potentially have an unlimited number of time periods, each with different convergence groups to the time period immediately prior to it. Increasing the number of parameters raises the question of potential overfitting, however. To avoid networks that may overfit or make the phylogenetic tensor expressions too complicated, we will only look at examples with relatively few time periods.

We have seen that if a clock-like tree does not have enough parameters to fit a particular set of pattern frequencies then we can introduce more parameters by either removing the clock-like assumption or introducing time periods with convergence. We also have the option of a non-clock-like convergence-divergence network, where we both remove the clock-like assumption and introduce time periods with convergence. A non-clock-like convergence-divergence network may be preferred when no direction of time is known or there is rate heterogeneity across edges in a given time period and there is reason to believe that the assumption of independent evolution has broken down at some stage in time, leading to convergence of edges. For simplicity, we will only examine clock-like convergence-divergence networks however.

It is necessary to place some restrictions on convergence periods so that the convergence-divergence networks to examine are manageable. We will restrict the convergence periods to allow for the more biologically realistic convergence-divergence networks. For example, without any restrictions on convergence periods we could have two edges initially diverging from a node, followed by the same two edges alternating between converging and diverging indefinitely. Another potential issue is having too many allowable groups of converging edges in a time period. The more edges there are across a time period on a network, the more potential convergence groups there could be among these edges. We will place some restrictions on which groups of edges can convergence on a network.

For an $n$-taxon tree or network and a Markov model with $m$ states, there will be $m^{n}$ elements in the phylogenetic tensor. If there are greater than $m^{n}$ numerical parameters, then there is no possibility of the tree or network being identifiable. We will therefore restrict the number of convergence periods on our networks accordingly.

We will now introduce some restrictions on our convergence time periods. These restrictions will be:
\begin{enumerate}
	\item In each time period, each edge can only be included in one convergence group.
	\item There will be a maximum of one convergence period between each splitting event and after the last splitting event.
\end{enumerate}

Any subset of edges, $2$, $3$, $\ldots{}$, $w$, where $w$ is the total number of edges at that time period, after $w-1$ splitting events, can converge together. An individual edge can also be thought of as ``converging'' with itself when it is diverging from all other edges in the time period, as described earlier in Chapter~\ref{chapter3}.

Now consider a three-taxon clock-like tree. This tree has two splitting events, the first represented by the root and the second by a node which splits one of the diverged edges into two new edges. The first place where we can put a convergence period is before the second splitting event, which will represent convergence of the two diverged edges. The second place a convergence period can be located is after the second node and after divergence of the third edge has occurred. We can represent convergence of any two edges or of all three edges. Below in Figure~\ref{fig:threetaxondespeciationexample} is an example of a three-taxon network with a convergence period after the divergence of the first two edges and a convergence period involving the non-sister taxa after a third edge has split off.

\chapter{Two-Taxon and Three-Taxon Trees and Networks}
\label{chapter4}

In Chapter~\ref{chapter3} we laid out the necessary notation to express phylogenetic tensors for trees and networks in the Hadamard basis. In this chapter we will present the phylogenetic tensors for some basic two-taxon and three-taxon trees and networks. We will address the question of identifiability. That is, if the tree or network structure is fixed, is there a bijective function between the parameter space and the pattern frequencies? The constraints on each phylogenetic tensor will provide a convenient basis for determining the theoretical character sequence patterns that can arise from that tree or network. In this chapter we will compare the constraints and the probability spaces that arise from the constraints of various trees and networks under the binary symmetric model and determine which trees and networks are distinguishable from each other.

Recall from Chapter~\ref{chapter3} that rates of substitutions cannot be distinguished from time along an edge for the binary symmetric model. In other words, we can only ever know the product of the rate and the time. This product can be large from either a fast rate or a long period of time. Likewise, the product can be small from either a slow rate or a short period time. We are therefore able to rescale our rate and time parameters into a single parameter. Recall in Section~\ref{transformedphylo} that we rescaled $\tau_{i}=\lambda{}t_{i}$ by letting $\tau_{i}\rightarrow{}\frac{1}{2}\tau_{i}$ since there is only a single degree of freedom in $\tau_{i}=\lambda{}t_{i}$. This rescaling will allow our expressions for the transformed phylogenetic tensor elements to be compared more easily to distances between leaves on the tree. For example, with this rescaling one element of the transformed phylogenetic tensor for the two-taxon clock-like tree is $q_{11}=e^{-2\tau_{1}}$, where $2\tau_{1}$ is the sum of the edge lengths between the two taxa.

\section{Constraints on the Transformed Phylogenetic Tensors}
\label{twotaxondistinguish}

Before we introduce the two-taxon and three-taxon trees and networks, we will have a look at how the Hadamard transformation impacts on the phylogenetic tensors under the binary symmetric model.

The phylogenetic tensors are represented as $2^{n}$ dimensional vectors, where there are $2$ states in the Markov model and $n$ taxa on the tree or network. The tensor elements are ordered according to the base-$2$ numeral system, with each element in the form,
\begin{align}\begin{split}
p_{i_{1}i_{2}\ldots{}i_{n}},
\end{split}\end{align}
where $i_{1},i_{2},\ldots{},i_{n}\in\{0,1\}$.

In the transformed basis, the phylogenetic tensor will be expressed in vector representation as the product of the diagonalising matrix, $h^{\otimes{}n}$, and the vector representation of the phylogenetic tensor in the untransformed basis. Recall that we expressed the transformed phylogenetic tensor as
\begin{align}\begin{split}
\widehat{P}=H\cdot{}P=h^{\otimes{}n}\cdot{}P.
\end{split}\end{align}

Recall from Chapter~\ref{chapter2} that if the Markov model is the binary symmetric model then the diagonalising matrix, $h$, will be
\begin{align}\begin{split}
h=\left[\begin{array}{cc} k_{1} & k_{2} \\
k_{1} & -k_{2} \\
\end{array}\right],
\end{split}\end{align}
where $k_{1},k_{2}\in{\mathbb{C}\setminus{}\{0\}}$.
If we choose $k_{1}=k_{2}=1$, the diagonalising matrix becomes a Hadamard matrix.

For an $n$-taxon tree or network under the binary symmetric model, the diagonalising matrix will be
\begin{align}\begin{split}
H=h^{\otimes{}n}=h\otimes{}h\otimes{}\ldots{}\otimes{}h,
\end{split}\end{align}
where there are $n-1$ tensor products.

If we index rows and columns from $0$ upwards then the elements of the Hadamard matrix of order $2$ are
\begin{align}\begin{split}
h_{\textrm{\textbf{i}}\textrm{\textbf{j}}}\equiv{}h_{i_{1}i_{2},j_{1}j_{2}}=\left(-1\right)^{i_{1}j_{1}+i_{2}j_{2}}.
\end{split}\end{align}
The elements of the Hadamard matrix of order $n$ will be
\begin{align}\begin{split}
H_{\textrm{\textbf{i}}\textrm{\textbf{j}}}\equiv{}H_{i_{1}i_{2}\ldots{}i_{n},j_{1}j_{2}\ldots{}j_{n}}=\left(-1\right)^{\sum\limits_{l=1}^{n}i_{l}j_{l}},
\end{split}\end{align}
since the elements with the appropriate indices are simply multiplied together under the tensor product.

Since Hadamard matrices are orthogonal matrices and our choice of $k_{1}=k_{2}=1$ resulted in every element of the first row being equal to $1$, every other row must have exactly half of its elements being $1$ and the other half being $-1$. For the binary symmetric model, the functions for the phylogenetic tensor elements come in identical pairs in the untransformed basis. If every $0$ is replaced with a $1$ and every $1$ is replaced by a $0$ the resulting phylogenetic tensor element will be identical to the original phylogenetic tensor element. We express this as
\begin{align}\begin{split}
p_{i_{1}i_{2}\ldots{}i_{n}}=p_{\bar{i_{1}}\bar{i_{2}}\ldots{}\bar{i_{n}}},
\end{split}\end{align}
where
\begin{align}\begin{split}
\begin{cases}
\bar{i_{l}}=0 & \textrm{if } i_{l}=1 \\
\bar{i_{l}}=1 & \textrm{if } i_{l}=0
\end{cases},
\end{split}\end{align}
or $\bar{i_{l}}=\frac{1}{2}\left(1+\left(-1\right)^{i_{l}}\right)$.

The elements of the transformed phylogenetic tensor will be
\begin{align}\begin{split}
q_{i_{1}i_{2}\ldots{}i_{n}}&=\sum\limits_{j_{1},j_{2},\ldots{},j_{n}=0}^{1}H_{i_{1}i_{2}\ldots{}i_{n},j_{1}j_{2}\ldots{}j_{n}}p_{j_{1}j_{2}\ldots{}j_{n}} \\
&=\sum\limits_{j_{1},j_{2},\ldots{},j_{n}=0}^{1}\left(-1\right)^{\sum\limits_{l=1}^{n}i_{l}j_{l}}p_{j_{1}j_{2}\ldots{}j_{n}}.
\end{split}\end{align}

Note that the first element of the transformed phylogenetic tensor will represent the conservation of probability,
\begin{align}\begin{split}
q_{00\ldots{}0}=\sum\limits_{i_{1},i_{2},\ldots{}i_{n}=0}^{1}p_{i_{1}i_{2}\ldots{}i_{n}}=1.
\end{split}\end{align}

Suppose the only information about a phylogenetic tensor that is known is that each element, $p_{i_{1}i_{2}\ldots{}i_{n}}$, is a probability. Then
\begin{align}\begin{split}
0\leq{}p_{i_{1}i_{2}\ldots{}i_{n}}\leq{}1
\end{split}\end{align}
for all $i_{1},i_{2},\ldots{}i_{n}=0,1$, and
\begin{align}\begin{split}
\sum\limits_{i_{1},i_{2},\ldots{}i_{n}=0}^{1}p_{i_{1}i_{2}\ldots{}i_{n}}=1.
\end{split}\end{align}

Since every element of the Hadamard matrix is $\pm{}1$, every element of the transformed phylogenetic tensor must be equal to the sum of a subset of elements of the untransformed phylogenetic tensor minus the remaining elements of the untransformed phylogenetic tensor elements. We can express this as
\begin{align}\begin{split}
q_{i_{1}i_{2}\ldots{}i_{n}}&=\sum\limits_{A}p_{i_{1}i_{2}\ldots{}i_{n}}-\sum\limits_{A^{C}}p_{i_{1}i_{2}\ldots{}i_{n}} \\
&=\sum\limits_{A}p_{i_{1}i_{2}\ldots{}i_{n}}-\left(1-\sum\limits_{A}p_{i_{1}i_{2}\ldots{}i_{n}}\right) \\
&=2\sum\limits_{A}p_{i_{1}i_{2}\ldots{}i_{n}}-1,
\end{split}\end{align}
where $A$ is some subset of the indices of the phylogenetic tensor elements and $A^{C}$ is the complementary set and represents all of the remaining indices of the phylogenetic tensor elements.
If
\begin{align}\begin{split}
\sum\limits_{A}p_{i_{1}i_{2}\ldots{}i_{n}}=1,
\end{split}\end{align}
then
\begin{align}\begin{split}
q_{i_{1}i_{2}\ldots{}i_{n}}=1.
\end{split}\end{align}
If
\begin{align}\begin{split}
\sum\limits_{A}p_{i_{1}i_{2}\ldots{}i_{n}}=0,
\end{split}\end{align}
then
\begin{align}\begin{split}
q_{i_{1}i_{2}\ldots{}i_{n}}=-1.
\end{split}\end{align}
We can conclude that in general, with the Hadamard matrix as the diagonalising matrix, the transformed phylogenetic tensor elements will satisfy
\begin{align}\begin{split}
-1\leq{}q_{i_{1}i_{2}\ldots{}i_{n}}\leq{}1,
\end{split}\end{align}
for all $i_{1},i_{2},\ldots{}i_{n}=0,1$.

Now recall that the elements of the transformed phylogenetic tensor will be
\begin{align}\begin{split}
q_{i_{1}i_{2}\ldots{}i_{n}}&=\sum\limits_{j_{1},j_{2},\ldots{},j_{n}=0}^{1}\left(-1\right)^{\sum\limits_{l=1}^{n}i_{l}j_{l}}p_{j_{1}j_{2}\ldots{}j_{n}}.
\end{split}\end{align}
Only if $i_{l}=1$ will $j_{l}$ appear in the sum in the exponent. For example,
\begin{align}\begin{split}
q_{01101}&=\sum\limits_{j_{1},j_{2},\ldots{},j_{5}=0}^{1}\left(-1\right)^{\sum\limits_{l=1}^{5}i_{l}j_{l}}p_{j_{1}j_{2}\ldots{}j_{5}} \\
&=\sum\limits_{j_{1},j_{2},\ldots{},j_{5}=0}^{1}\left(-1\right)^{j_{2}+j_{3}+j_{5}}p_{j_{1}j_{2}\ldots{}j_{5}}.
\end{split}\end{align}

If $\sum\limits_{l=1}^{n}i_{l}$ is odd then there will be an odd number of terms in the exponent, $\sum\limits_{l=1}^{n}i_{l}j_{l}$. If the sum of the exponent terms is odd then $\left(-1\right)^{\sum\limits_{l=1}^{n}i_{l}j_{l}}$, the coefficient for $p_{j_{1}j_{2}\ldots{}j_{n}}$, will be negative. The sum of the complement combination, $\sum\limits_{l=1}^{n}i_{l}\bar{j_{l}}$, will then be even and the coefficient for $p_{\bar{j_{1}}\bar{j_{2}}\ldots{}\bar{j_{n}}}$, $\left(-1\right)^{\sum\limits_{l=1}^{n}i_{l}\bar{j_{l}}}$, will be positive. Hence, if $\sum\limits_{l=1}^{n}i_{l}$ is odd then $q_{i_{1}i_{2}\ldots{}i_{n}}=0$ for all trees and networks under the binary symmetric model.

If $\sum\limits_{l=1}^{n}i_{l}$ is even then there will be an even number of terms in the exponent, $\sum\limits_{l=1}^{n}i_{l}j_{l}$. Again, if the sum of the exponent terms is odd then $\left(-1\right)^{\sum\limits_{l=1}^{n}i_{l}j_{l}}$, the coefficient for $p_{j_{1}j_{2}\ldots{}j_{n}}$, will be negative. However, the sum of the complement combination, $\sum\limits_{l=1}^{n}i_{l}\bar{j_{l}}$, will be odd too and $\left(-1\right)^{\sum\limits_{l=1}^{n}i_{l}\bar{j_{l}}}$, the coefficient for $p_{\bar{j_{1}}\bar{j_{2}}\ldots{}\bar{j_{n}}}$, will also be negative. Likewise, if the coefficient for $p_{j_{1}j_{2}\ldots{}j_{n}}$ is positive then the coefficient for $p_{\bar{j_{1}}\bar{j_{2}}\ldots{}\bar{j_{n}}}$ will also be positive. Hence, we cannot assume under the binary symmetric model with no tree or network structure that $q_{i_{1}i_{2}\ldots{}i_{n}}=0$ if $\sum\limits_{l=1}^{n}i_{l}$ is even.

For example, we will again look at $q_{01101}$. For every combination of $j_{2}j_{3}j_{5}$ for which $j_{2}+j_{3}+j_{5}$ is even and $\left(-1\right)^{j_{2}+j_{3}+j_{5}}=1$, the complement combination, $\bar{j_{2}}\bar{j_{3}}\bar{j_{5}}$, will have $\bar{j_{2}}+\bar{j_{3}}+\bar{j_{5}}$ being odd and $\left(-1\right)^{\bar{j_{2}}+\bar{j_{3}}+\bar{j_{5}}}=-1$. This means that the coefficients for $p_{i_{1}i_{2}\ldots{}i_{5}}$ and $p_{\bar{i_{1}}\bar{i_{2}}\ldots{}\bar{i_{5}}}$ must always have the opposite sign since we are summing over all possible values of $j_{1}$ and $j_{4}$ as well. Hence, we must have $q_{01101}=0$.

As a second example, we will look at $q_{01100}$. For every combination of $j_{2}j_{3}$ for which $j_{2}+j_{3}$ is odd and $\left(-1\right)^{j_{2}+j_{3}}=-1$, the complement combination, $\bar{j_{2}}\bar{j_{3}}$, will have $\bar{j_{2}}+\bar{j_{3}}$ being odd and $\left(-1\right)^{\bar{j_{2}}+\bar{j_{3}}}=-1$. This means that the coefficients for $p_{i_{1}i_{2}\ldots{}i_{5}}$ and $p_{\bar{i_{1}}\bar{i_{2}}\ldots{}\bar{i_{5}}}$ must both be negative. Likewise, for  every combination of $j_{2}j_{3}$ for which $j_{2}+j_{3}$ is even and $\left(-1\right)^{j_{2}+j_{3}}=1$, the complement combination, $\bar{j_{2}}\bar{j_{3}}$, will have $\bar{j_{2}}+\bar{j_{3}}$ being even and $\left(-1\right)^{\bar{j_{2}}+\bar{j_{3}}}=1$. This means that the coefficients for $p_{i_{1}i_{2}\ldots{}i_{5}}$ and $p_{\bar{i_{1}}\bar{i_{2}}\ldots{}\bar{i_{5}}}$ must both be positive. This means that the coefficients for $p_{i_{1}i_{2}\ldots{}i_{5}}$ and $p_{\bar{i_{1}}\bar{i_{2}}\ldots{}\bar{i_{5}}}$ must always have the same sign since we are summing over all possible values of $j_{1}$, $j_{4}$ and $j_{5}$ as well.

\section{Two-Taxon Phylogenetic Tensors}
\label{twotaxon}

We will start by presenting expressions for the phylogenetic tensors for two-taxon trees and networks. There are two trees, the non-clock-like tree and the clock-like tree, and one network, the clock-like convergence-divergence network, that we will examine. The two-taxon clock-like convergence-divergence network is simply the two-taxon clock-like tree, with the two taxa converging with each other in the last time period.

\subsection{Non-Clock-Like Tree}

The two-taxon non-clock-like tree is shown below in Figure~\ref{twotaxonnonclock}.
\begin{figure}[H]
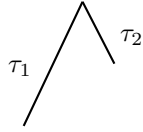

	\centering
		\psmatrix[colsep=.3cm,rowsep=.4cm,mnode=r]
		~ & && ~ \\
		~ && ~ && ~ \\
		~ & ~
		\ncline{1,4}{3,2}
		\tlput{$\tau_{1}$}
		\ncline{1,4}{2,5}
		\trput{$\tau_{2}$}
		\endpsmatrix
\caption{Two-taxon non-clock-like tree.}
\label{twotaxonnonclock}
\end{figure}
The phylogenetic tensor for this tree is
\begin{align}\begin{split}
\widehat{P}=e^{\frac{1}{2}\left(\widehat{R}_{10}\tau_{1}+\widehat{R}_{01}\tau_{2}\right)}\cdot{}\widehat{\Pi{}}=\left[\begin{array}{c} q_{00} \\
q_{01} \\
q_{10} \\
q_{11} \\
\end{array}\right]=\left[\begin{array}{c} 1 \\
0 \\
0 \\
e^{-\left(\tau_{1}+\tau_{2}\right)} \\
\end{array}\right].
\end{split}\end{align}

Clearly the numerical parameters of this tree are not recoverable and the tree is not identifiable since there are two numerical parameters and only one variable phylogenetic tensor element. For each choice of the non-negative numerical parameters there will be a single phylogenetic tensor. The converse is not true, however. For every phylogenetic tensor there will be infinitely many numerical parameter combinations that can give rise to that tensor. It is only the sum of the two numerical parameters that the phylogenetic tensor is dependent on. There are infinitely many choices of the two numerical parameters that will result in the same sum and hence the same phylogenetic tensor.

Demanding $\tau_{1},\tau_{2}\geq{}0$, the minimum length interval containing all possible values of the one variable phylogenetic tensor element is
\begin{align}\begin{split}
0<q_{11}=e^{-\left(\tau_{1}+\tau_{2}\right)}\leq{}1.
\end{split}\end{align}

In summary, the constraints on the phylogenetic tensor for the two-taxon non-clock-like tree are
\begin{align}\begin{split}
\begin{mycases}
&q_{00}=1, \\
&q_{01}=0, \\
&q_{10}=0, \\
0<&q_{11}\leq{}1.
\end{mycases}
\end{split}\end{align}

We will now look at the remaining two two-taxon trees and networks and compare the phylogenetic tensor constraints.

\subsection{Clock-Like Tree}

The next tree we will consider is the two-taxon clock-like tree, shown in Figure~\ref{twotaxonclock} below.
\begin{figure}[H]
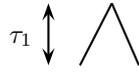

	\centering
		\psmatrix[colsep=.3cm,rowsep=.4cm,mnode=r]
		~ && ~ \\
		~ & ~ && ~
		\ncline{1,3}{2,2}
		\ncline{1,3}{2,4}
		\ncline{<->,arrowscale=1.5}{1,1}{2,1}
		\tlput{$\tau_{1}$}
		\endpsmatrix
\caption{Two-taxon clock-like tree.}
\label{twotaxonclock}
\end{figure}
The phylogenetic tensor for this tree is
\begin{align}\begin{split}
\widehat{P}=e^{\frac{1}{2}\left(\widehat{R}_{10}+\widehat{R}_{01}\right)\tau_{1}}\cdot{}\widehat{\Pi{}}=\left[\begin{array}{c} q_{00} \\
q_{01} \\
q_{10} \\
q_{11} \\
\end{array}\right]=\left[\begin{array}{c} 1 \\
0 \\
0 \\
e^{-2\tau_{1}} \\
\end{array}\right].
\end{split}\end{align}

We can see that the phylogenetic tensor for the clock-like tree can be recovered from the phylogenetic tensor for the non-clock-like tree by simply forcing the two time parameters to be equal. The phylogenetic tensor elements for the two trees are very similar, with the only difference being the argument of the exponential in the single variable element. For the clock-like tree, there is only one numerical parameter and only one variable phylogenetic tensor element. The exponential function maps the single non-positive variable to the interval $(0,1]$ and is a bijective function. Therefore, the two-taxon clock-like tree must be identifiable.

As with the non-clock-like tree, the constraints on the phylogenetic tensor for the two-taxon clock-like tree are
\begin{align}\begin{split}
\begin{mycases}
q_{00}&=1, \\
q_{01}&=0, \\
q_{10}&=0, \\
0<q_{11}&\leq{}1.
\end{mycases}
\end{split}\end{align}

\subsection{Clock-Like Convergence-Divergence Network}

The final two-taxon tree or network is the two-taxon convergence-divergence network, shwon below in Figure~\ref{twotaxoncondiv}. This is the two-taxon clock-like tree with convergence of the two taxa immediately after the divergence.
\begin{figure}[H]
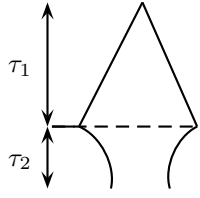

	\centering
		\psmatrix[colsep=.3cm,rowsep=.4cm,mnode=r]
		~ & && ~ \\
		~ \\
		~ & ~ && && ~ \\
		~ && ~ && ~
		\ncline{1,4}{3,2}
		\ncline{1,4}{3,6}
		\ncarc[arcangle=40]{3,2}{4,3}
		\ncarc[arcangle=-40]{3,6}{4,5}
		\ncline{<->,arrowscale=1.5}{1,1}{3,1}
		\tlput{$\tau_{1}$}
		\ncline{<->,arrowscale=1.5}{3,1}{4,1}
		\tlput{$\tau_{2}$}
		\psset{linestyle=dashed}\ncline{3,1}{3,2}
		\psset{linestyle=dashed}\ncline{3,1}{3,6}
		\endpsmatrix
\caption{Two-taxon clock-like convergence-divergence network with a convergence period in the second time period.}
\label{twotaxoncondiv}
\end{figure}
The phylogenetic tensor for this network is
\begin{align}\begin{split}
\widehat{P}=e^{\frac{1}{2}\widehat{R}_{11}\tau_{2}}\cdot{}e^{\frac{1}{2}\left(\widehat{R}_{10}+\widehat{R}_{01}\right)\tau_{1}}\cdot{}\widehat{\Pi{}}=e^{\frac{1}{2}\widehat{R}_{11}\tau_{2}}\cdot{}\widehat{P}_{cl}=\left[\begin{array}{c} q_{00} \\
q_{01} \\
q_{10} \\
q_{11} \\
\end{array}\right]=\left[\begin{array}{c} 1 \\
0 \\
0 \\
1-e^{-\tau_{2}}\left(1-e^{-2\tau_{1}}\right) \\
\end{array}\right].
\end{split}\end{align}

Clearly this network is not identifiable since there are two numerical parameters and only one variable phylogenetic tensor element and hence the numerical parameters cannot be recovered independently.

Unlike the previous two trees, it is not immediately clear what the minimum length interval that contains all of the possible phylogenetic tensors is. Demanding $\tau_{1}\geq{}0$,
\begin{align}\begin{split}
0<e^{-2\tau_{1}}&\leq{}1 \\
\Leftrightarrow{}0\leq{}1-e^{-2\tau_{1}}&<1.
\end{split}\end{align}
Also, demanding $\tau_{2}\geq{}0$,
\begin{align}\begin{split}
0<e^{-\tau_{2}}\leq{}1.
\end{split}\end{align} 
The product of these two components must also lie between $0$ and $1$, with the interval including $0$, but not including $1$.
\begin{align}\begin{split}
0\leq{}e^{-\tau_{2}}\left(1-e^{-2\tau_{1}}\right)&<1 \\
\Leftrightarrow{}0<1-e^{-\tau_{2}}\left(1-e^{-2\tau_{1}}\right)&\leq{}1.
\end{split}\end{align}

As with the other two trees, the constraints on the network are
\begin{align}\begin{split}
\begin{mycases}
q_{00}&=1, \\
q_{01}&=0, \\
q_{10}&=0, \\
0<q_{11}&\leq{}1.
\end{mycases}
\end{split}\end{align}

\subsection{Network Identifiability}

We established in the previous sections that while the clock-like tree is identifiable and the convergence-divergence network is not, both the clock-like tree and the convergence-divergence network have the same phylogenetic tensor constraints. This raises the question of whether the convergence-divergence network can be compared to the clock-like tree with less divergence. For example, suppose we have the convergence-divergence network with divergence over time $\tau'_{1}$, followed by convergence over time $\tau'_{2}$. Is this network identifiable from a clock-like tree with divergence over time $\tau_{1}\leq{}\tau'_{1}$?

Now suppose we have the phylogenetic tensor element, $q_{11}$, and wish to compare the time variables on the clock-like tree with no convergence to the clock-like convergence-divergence network. If we assume $0<q_{11}\leq{}1$ such that it could have arisen on either tree or network, we require
\begin{align}\begin{split}
q_{11}=e^{-2\tau_{1}}&=1-e^{-\tau'_{2}}\left(1-e^{-2\tau'_{1}}\right) \\
\Leftrightarrow{}e^{-2\tau_{1}}-1&=-e^{-\tau'_{2}}\left(1-e^{-2\tau'_{1}}\right) \\
\Leftrightarrow{}1-e^{-2\tau_{1}}&=e^{-\tau'_{2}}\left(1-e^{-2\tau'_{1}}\right).
\end{split}\end{align}
Since $0<1-e^{-2\tau_{1}}\leq{}1$, $0<e^{-\tau'_{2}}\leq{}1$ and $0<1-e^{-2\tau'_{1}}\leq{}1$, we can conclude that
\begin{align}\begin{split}
1-e^{-2\tau_{1}}&\leq{}1-e^{-2\tau'_{1}} \\
\Leftrightarrow{}\tau_{1}&\leq{}\tau'_{1}.
\end{split}\end{align}

Consequently, the clock-like tree, which allows only independent evolution of adjacent edges is not network identifiable from our convergence-divergence network, where the two edges evolve independently for a longer period of time, before some convergence between the two edges happens.

Comparing our three two-taxon trees and networks, each tree and network has only a single variable phylogenetic tensor element, $q_{11}$. For all trees and networks, all other elements of the phylogenetic tensors do not vary with the tree or network. For the networks to be identifiable the system must not be underdetermined. This means that since there is only one variable phylogenetic tensor element, for identifiability there must be only a single time parameter. Consequently, the clock-like tree is identifiable, while the non-clock-like tree and the convergence-divergence network are not identifiable. It has also been shown that the minimum length interval for the variable element is $0<q_{11}\leq{}1$ for all three networks. We can therefore conclude that none of the networks are distinguishable from each other. The convergence-divergence network is not network identifiable from the clock-like tree, with divergence followed by convergence equivalent to less divergence with no convergence. Given that the clock-like tree is the only identifiable network and it is the least parameter rich of the networks, the two-taxon clock-like tree should generally be preferred under the binary symmetric model.

For some networks with more than two taxa, we will see that the two-taxon convergence-divergence network is embedded as a subnetwork inside the network. An issue that arises is determining whether these networks are distinguishable from similar clock-like networks where the two-taxon convergence-divergence network embedded in the network is replaced with the two-taxon clock-like tree. For example, consider the three trees and networks in Figure~\ref{threeexamplenetworks} below.
\begin{figure}[H]
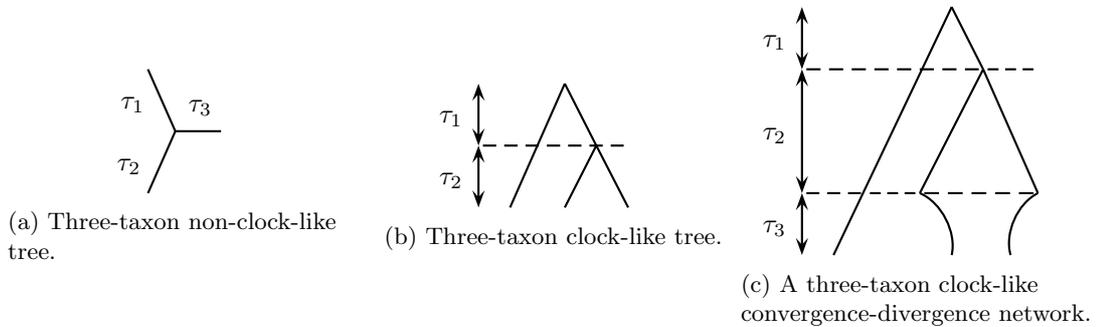

	\centering
		\begin{subfigure}[h]{0.32\textwidth}
			\centering
				\psmatrix[colsep=.3cm,rowsep=.4cm,mnode=r]
				~ \\
				& {} && ~ \\
				~
				\ncline{1,1}{2,2}
				\tlput{$\tau_{1}$}
				\ncline{2,2}{3,1}
				\tlput{$\tau_{2}$}
				\ncline{2,2}{2,4}
				\taput{$\tau_{3}$}
				\endpsmatrix
				\caption{Three-taxon non-clock-like \\*
				tree.}
		\end{subfigure}
\begin{subfigure}[h]{0.32\textwidth}
	\centering
		\psmatrix[colsep=.3cm,rowsep=.4cm,mnode=r]
		~ & && ~ \\
		~ & && & ~ & ~ \\
		~ & ~ && ~ && ~
		\ncline{1,4}{3,2}
		\ncline{1,4}{2,5}
		\ncline{2,5}{3,4}
		\ncline{2,5}{3,6}
		\ncline{<->,arrowscale=1.5}{1,1}{2,1}
		\tlput{$\tau_{1}$}
		\ncline{<->,arrowscale=1.5}{2,1}{3,1}
		\tlput{$\tau_{2}$}
		\psset{linestyle=dashed}\ncline{2,1}{2,6}
		\endpsmatrix
		\caption{Three-taxon clock-like tree.}
\end{subfigure}
\begin{subfigure}[h]{0.32\textwidth}
			\centering
				\psmatrix[colsep=.3cm,rowsep=.4cm,mnode=r]
				~ & && && ~ \\
				~ && && && ~ && ~ \\
				~ \\
				~ && ~ && ~ && && ~ \\
				~ & ~ && && ~ && ~
				\ncline{1,6}{5,2}
				\ncline{1,6}{2,7}
				\ncline{2,7}{4,5}
				\ncline{2,7}{4,9}
				\ncarc[arcangle=40]{4,5}{5,6}
				\ncarc[arcangle=-40]{4,9}{5,8}
				\ncline{<->,arrowscale=1.5}{1,1}{2,1}
				\tlput{$\tau_{1}$}
				\ncline{<->,arrowscale=1.5}{2,1}{4,1}
				\tlput{$\tau_{2}$}
				\ncline{<->,arrowscale=1.5}{4,1}{5,1}
				\tlput{$\tau_{3}$}
				\psset{linestyle=dashed}\ncline{2,1}{2,7}
				\psset{linestyle=dashed}\ncline{2,7}{2,9}
				\psset{linestyle=dashed}\ncline{4,1}{4,3}
				\psset{linestyle=dashed}\ncline{4,3}{4,5}
				\psset{linestyle=dashed}\ncline{4,5}{4,9}
				\endpsmatrix
				\caption{A three-taxon clock-like \\*
				convergence-divergence network.}
		\end{subfigure}
		\caption{Three three-taxon trees and networks.}
		\label{threeexamplenetworks}
\end{figure}

We would like to know whether any of these trees and networks are distinguishable from each other. From the results for the two-taxon case we might expect the three-taxon clock-like tree and the three-taxon clock-like convergence-divergence network shown above to be indistinguishable from each other. To see whether this is the case we must compare the two-taxon clock-like tree embedded in the three-taxon clock-like tree to the two-taxon clock-like convergence-divergence network embedded in the three-taxon clock-like convergence-divergence network. The tree and network that we will compare are shown in Figure~\ref{treenetworkcomparison} below.

\begin{figure}[H]
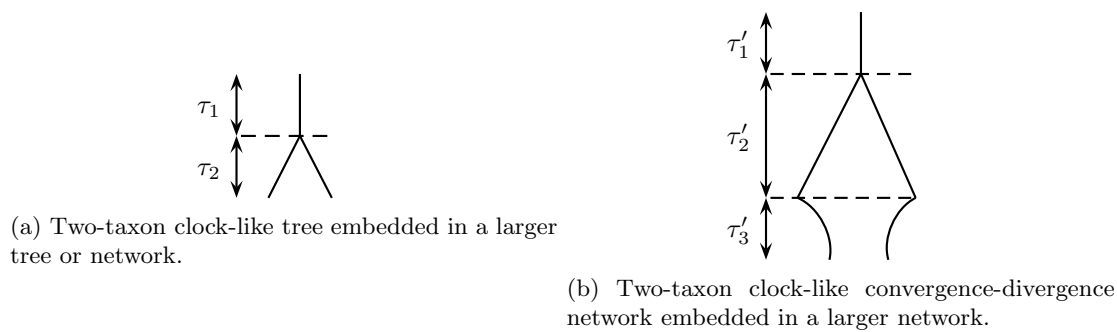

	\centering
\begin{subfigure}[h]{0.49\textwidth}
	\centering
		\psmatrix[colsep=.3cm,rowsep=.4cm,mnode=r]
		~ && ~ \\
		~ && ~ & ~ \\
		~ & ~ && ~
		\ncline{1,3}{2,3}
		\ncline{2,3}{3,2}
		\ncline{2,3}{3,4}
		\ncline{<->,arrowscale=1.5}{1,1}{2,1}
		\tlput{$\tau_{1}$}
		\ncline{<->,arrowscale=1.5}{2,1}{3,1}
		\tlput{$\tau_{2}$}
		\psset{linestyle=dashed}\ncline{2,1}{2,4}
		\endpsmatrix
		\caption{Two-taxon clock-like tree embedded in a larger tree or network.}
\end{subfigure}
\begin{subfigure}[h]{0.49\textwidth}
			\centering
				\psmatrix[colsep=.3cm,rowsep=.4cm,mnode=r]
				~ & && ~ \\
				~ & && ~ && ~ \\
				~ \\
				~ & ~ && && ~ \\
				~ && ~ && ~
				\ncline{1,4}{2,4}
				\ncline{2,4}{4,2}
				\ncline{2,4}{4,6}
				\ncarc[arcangle=40]{4,2}{5,3}
				\ncarc[arcangle=-40]{4,6}{5,5}
				\ncline{<->,arrowscale=1.5}{1,1}{2,1}
				\tlput{$\tau_{1}'$}
				\ncline{<->,arrowscale=1.5}{2,1}{4,1}
				\tlput{$\tau_{2}'$}
				\ncline{<->,arrowscale=1.5}{4,1}{5,1}
				\tlput{$\tau_{3}'$}
				\psset{linestyle=dashed}\ncline{2,1}{2,6}
				\psset{linestyle=dashed}\ncline{4,1}{4,6}
				\endpsmatrix
				\caption{Two-taxon clock-like convergence-divergence network embedded in a larger network.}
		\end{subfigure}
		\caption{Two-taxon clock-like tree and two-taxon clock-like convergence-divergence network embedded in a larger network.}
		\label{treenetworkcomparison}
\end{figure}

Recall from the two-taxon case that the tree and network are not network identifiable provided $\tau_{2}\leq{}\tau_{2}'$. Consequently,
\begin{align}\begin{split}
\tau_{2}&\leq{}\tau_{2}'+\tau_{3}'.
\end{split}\end{align}
If the tree and network are to be embedded in a larger tree or network, then we require the sum of the edge lengths to be equal,
\begin{align}\begin{split}
\tau_{1}+\tau_{2}&=\tau_{1}'+\tau_{2}'+\tau_{3}'.
\end{split}\end{align}
Rearranging in terms of $\tau_{1}$,
\begin{align}\begin{split}
\tau_{1}&=\tau_{1}'+\tau_{2}'+\tau_{3}'-\tau_{2}\geq{}\tau_{1}'.
\end{split}\end{align}
This means that we can always find a clock-like tree, with the parameters, $\{\tau_{1},\tau_{2}\}$, that is not network identifiable from the clock-like convergence-divergence network, with the parameters, $\{\tau_{1}',\tau_{2}',\tau_{3}'\}$, embedded in a larger convergence-divergence network. The clock-like tree and clock-like convergence-divergence network will not be distinguishable from each other either.

\paragraph{}\label{twotaxonresult}
We can conclude that convergence-divergence networks for which the two-taxon convergence-divergence network is embedded in will not be identifiable, nor will they be network identifiable nor distinguishable from the same convergence-divergence network with the embedded two-taxon convergence-divergence network replaced by the two-taxon clock-like tree.

\section{Three-Taxon Phylogenetic Tensors}
\label{threetaxon}

We saw that of the three two-taxon trees and networks under the binary symmetric model, only the clock-like tree was a suitable candidate tree or network. We will now compare some three-taxon trees and networks: the non-clock-like tree, the clock-like tree and the clock-like convergence-divergence network with a convergence period involving the two non-sister taxa. It should be noted that there are many more networks which we will compare. For these other networks we will simply present the results without deriving the phylogenetic tensor constraints.

\subsection{Non-Clock-Like Tree}

The first three-taxon tree that we will consider is the non-clock-like tree, shown below in Figure~\ref{threetaxonnonclock}.
\begin{figure}[H]
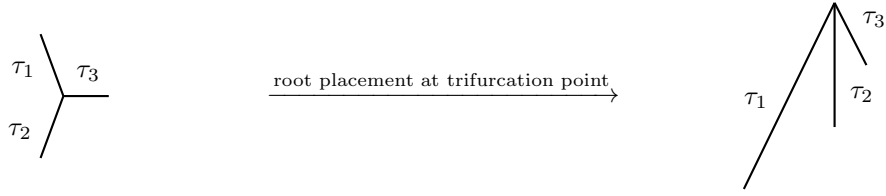

	\centering
		\begin{subfigure}[h]{0.32\textwidth}
			\centering
				\psmatrix[colsep=.3cm,rowsep=.4cm,mnode=r]
				~\\
				& {} && ~ \\
				~
				\ncline{1,1}{2,2}
				\tlput{$\tau_{1}$}
				\ncline{2,2}{3,1}
				\tlput{$\tau_{2}$}
				\ncline{2,2}{2,4}
				\taput{$\tau_{3}$}
				\endpsmatrix
		\end{subfigure}
		$\xrightarrow{\text{root placement at trifurcation point}}$
		\begin{subfigure}[h]{0.32\textwidth}
			\centering
				\psmatrix[colsep=.3cm,rowsep=.4cm,mnode=r]
				&& & ~ \\
				&& ~ & ~ & ~ \\
				& ~ && ~ \\
				~
				\ncline{1,4}{4,1}
				\tlput{$\tau_{1}$}
				\ncline{1,4}{3,4}
				\trput[tpos=.7]{$\tau_{2}$}
				\ncline{1,4}{2,5}
				\psset{tpos=0.2}
				\trput{$\tau_{3}$}
				\endpsmatrix
		\end{subfigure}
		\caption{Three-taxon non-clock-like tree.}
		\label{threetaxonnonclock}
		\end{figure}
According to the procedure from Chapter~\ref{chapter3}, the phylogenetic tensor for this tree is
\begin{align}\begin{split}
\widehat{P}=e^{\frac{1}{2}\left(\widehat{R}_{100}\tau_{1}+\widehat{R}_{010}\tau_{2}+\widehat{R}_{001}\tau_{3}\right)}\cdot{}\widehat{\Pi{}}=\left[\begin{array}{c} q_{000} \\
q_{001} \\
q_{010} \\
q_{011} \\
q_{100} \\
q_{101} \\
q_{110} \\
q_{111} \\
\end{array}\right]=\left[\begin{array}{c} 1 \\
0 \\
0 \\
e^{-\left(\tau_{2}+\tau_{3}\right)} \\
0 \\
e^{-\left(\tau_{1}+\tau_{3}\right)} \\
e^{-\left(\tau_{1}+\tau_{2}\right)} \\
0 \\
\end{array}\right].
\end{split}\end{align}

Demanding $\tau_{1},\tau_{2},\tau_{3}\geq{}0$, we can see that
\begin{align}\begin{split}
&0<e^{-\left(\tau_{2}+\tau_{3}\right)}\leq{}1,\quad{}0<e^{-\left(\tau_{1}+\tau_{3}\right)}\leq{}1,\quad{}0<e^{-\left(\tau_{1}+\tau_{2}\right)}\leq{}1.
\end{split}\end{align}
Hence $0<q_{011},q_{101},q_{110}\leq{}1$ are the tight intervals treated independently containing all possible values of the phylogenetic tensor elements.

By using simple algebraic operations, we can solve for the three numerical parameters as functions of the phylogenetic tensor elements,
\begin{align}\begin{split}
\begin{mycases}
\tau_{1}&=\ln\frac{q_{011}}{q_{101}q_{110}}, \\
\tau_{2}&=\ln\frac{q_{101}}{q_{011}q_{110}}, \\
\tau_{3}&=\ln\frac{q_{110}}{q_{011}q_{101}}.
\end{mycases}
\end{split}\end{align}
The natural logarithm is a bijective function which maps the phylogenetic tensor elements to the numerical parameters, provided that the phylogenetic tensor elements are adequately constrained. We can conclude that the non-clock-like tree is identifiable.

From the solutions for the numerical parameters, demanding that each $\tau_{i}\geq{}0$, the set of constraints on the phylogenetic tensor is
\begin{align}\begin{split}
\left\{q_{011}\geq{}q_{101}q_{110},\quad{}q_{101}\geq{}q_{011}q_{110},\quad{}q_{110}\geq{}q_{011}q_{101}\right\}.
\end{split}\end{align}

Since there is no molecular clock or sense of direction of time for the non-clock-like tree, permuting the taxa does not impact on the constraints.

\subsection{Clock-Like Tree}
\label{threetaxonclock}

The next tree to consider is the three-taxon clock-like tree, shown in Figure~\ref{threetaxonclockfour} below.
\begin{figure}[H]
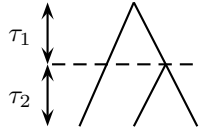

	\centering
		\psmatrix[colsep=.3cm,rowsep=.4cm,mnode=r]
		~ & && ~ \\
		~ & && & ~ & ~ \\
		~ & ~ && ~ && ~
		\ncline{1,4}{3,2}
		\ncline{1,4}{2,5}
		\ncline{2,5}{3,4}
		\ncline{2,5}{3,6}
		\ncline{<->,arrowscale=1.5}{1,1}{2,1}
		\tlput{$\tau_{1}$}
		\ncline{<->,arrowscale=1.5}{2,1}{3,1}
		\tlput{$\tau_{2}$}
		\psset{linestyle=dashed}\ncline{2,1}{2,6}
		\endpsmatrix
		\caption{Three-taxon clock-like tree.}
		\label{threetaxonclockfour}
\end{figure}
The phylogenetic tensor for this tree is
\begin{align}\begin{split}
\widehat{P}=e^{\frac{1}{2}\left(\widehat{R}_{100}+\widehat{R}_{010}+\widehat{R}_{001}\right)\tau_{2}}\cdot{}e^{\frac{1}{2}\left(\widehat{R}_{100}+\widehat{R}_{011}\right)\tau_{1}}\cdot{}\widehat{\Pi{}}=\left[\begin{array}{c} q_{000} \\
q_{001} \\
q_{010} \\
q_{011} \\
q_{100} \\
q_{101} \\
q_{110} \\
q_{111} \\
\end{array}\right]=\left[\begin{array}{c} 1 \\
0 \\
0 \\
e^{-2\tau_{2}} \\
0 \\
e^{-2\left(\tau_{1}+\tau_{2}\right)} \\
e^{-2\left(\tau_{1}+\tau_{2}\right)} \\
0 \\
\end{array}\right].
\end{split}\end{align}

Similarly to the non-clock-like tree, the clock-like tree is also identifiable.

Demanding $\tau_{1},\tau_{2}\geq{}0$,
\begin{align}\begin{split}
&0<e^{-2\tau_{2}}\leq{}1,\quad{}0<e^{-2\left(\tau_{1}+\tau_{2}\right)}\leq{}1.
\end{split}\end{align}
Hence $0<q_{011},q_{101},q_{110}\leq{}1$ are the tight intervals treated independently containing all possible values of the phylogenetic tensor elements.

We can find the constraints on the phylogenetic tensor by solving for the numerical parameters as functions of the phylogenetic tensor elements. However, we can also find the constraints simply by inspecting the phylogenetic tensor. By inspection, it can be seen that the full set of constraints on the phylogenetic tensor is
\begin{align}\begin{split}
\left\{q_{101}=q_{110},\quad{}q_{011}\geq{}q_{110}\right\}.
\end{split}\end{align}

\subsection{Clock-Like Convergence-Divergence Network}

We will look at only one example of a three-taxon convergence-divergence network here. The example we will look at is the three-taxon clock-like tree with convergence between the two non-sister taxa, shown in Figure~\ref{threetaxonnonsister} below. We will later discuss other examples of networks that fit our criteria for convergence-divergence networks detailed in Chapter~\ref{chapter3}.
\begin{figure}[H]
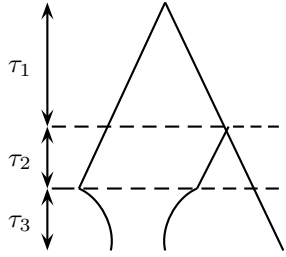

	\centering
		\psmatrix[colsep=.3cm,rowsep=.4cm,mnode=r]
		~ && && ~ \\
		~ \\
		~ && && && ~ && ~ \\
		~ & ~ && && ~ & && ~ \\
		~ && ~ && ~ && && ~
		\ncline{1,5}{4,2}
		\ncline{1,5}{5,9}
		\ncline{3,7}{4,6}
		\ncarc[arcangle=40]{4,2}{5,3}
		\ncarc[arcangle=-40]{4,6}{5,5}
		\ncline{<->,arrowscale=1.5}{1,1}{3,1}
		\tlput{$\tau_{1}$}
		\ncline{<->,arrowscale=1.5}{3,1}{4,1}
		\tlput{$\tau_{2}$}
		\ncline{<->,arrowscale=1.5}{4,1}{5,1}
		\tlput{$\tau_{3}$}
		\psset{linestyle=dashed}\ncline{3,1}{3,7}
		\psset{linestyle=dashed}\ncline{3,7}{3,9}
		\psset{linestyle=dashed}\ncline{4,1}{4,2}
		\psset{linestyle=dashed}\ncline{4,2}{4,6}
		\psset{linestyle=dashed}\ncline{4,6}{4,9}
		\endpsmatrix
		\caption{Three-taxon clock-like convergence-divergence network with convergence between non-sister taxa.}
		\label{threetaxonnonsister}
\end{figure}
The phylogenetic tensor for this network is
\begin{align}\begin{split}
\widehat{P}&=e^{\frac{1}{2}\left(\widehat{R}_{110}+\widehat{R}_{001}\right)\tau_{3}}\cdot{}e^{\frac{1}{2}\left(\widehat{R}_{100}+\widehat{R}_{010}+\widehat{R}_{001}\right)\tau_{2}}\cdot{}e^{\frac{1}{2}\left(\widehat{R}_{100}+\widehat{R}_{011}\right)\tau_{1}}\cdot{}\widehat{\Pi{}} \\
&=\left[\begin{array}{c} q_{000} \\
q_{001} \\
q_{010} \\
q_{011} \\
q_{100} \\
q_{101} \\
q_{110} \\
q_{111} \\
\end{array}\right]
=\left[\begin{array}{c} 1 \\
0 \\
0 \\
e^{-2\left(\tau_{2}+\tau_{3}\right)} \\
0 \\
e^{-2\left(\tau_{1}+\tau_{2}+\tau_{3}\right)} \\
1-e^{-\tau_{3}}\left(1-e^{-2\left(\tau_{1}+\tau_{2}\right)}\right) \\
0 \\
\end{array}\right].
\end{split}\end{align}

Using a similar argument to that used for the two-taxon convergence-divergence network and demanding $\tau_{1},\tau_{2},\tau_{3}\geq{}0$,
\begin{align}\begin{split}
&0<e^{-2\left(\tau_{2}+\tau_{3}\right)}\leq{}1,\quad{}0<e^{-2\left(\tau_{1}+\tau_{2}+\tau_{3}\right)}\leq{}1,\quad{}0<1-e^{-\tau_{3}}\left(1-e^{-2\left(\tau_{1}+\tau_{2}\right)}\right)\leq{}1.
\end{split}\end{align}
Hence $0<q_{011},q_{101},q_{110}\leq{}1$ are the tight intervals treated independently containing all possible values of the phylogenetic tensor elements.

To find the full set of constraints on the phylogenetic tensor we will express the time parameters as functions of the phylogenetic tensor elements. Since the constraints on the time parameters are known, $\tau_{1},\tau_{2},\tau_{3}\geq{}0$, we can then find the constraints on the phylogenetic tensor elements involving multiple elements. We will first make the substitutions, $e^{-\tau_{i}}=x_{i}$, for $i=1,2,3$. Solving for each $x_{i}$ and choosing solutions that allow for non-negative parameters when there is a choice, we find
\begin{align}\begin{split}
\begin{mycases}
x_{1}&=\sqrt{\frac{q_{101}}{q_{011}}}, \\
x_{2}&=\frac{\sqrt{q_{011}}\left(-\left(1-q_{110}\right)+\sqrt{\left(1-q_{110}\right)^{2}+4q_{101}}\right)}{2q_{101}}, \\
x_{3}&=\frac{1}{2}\left(1-q_{110}+\sqrt{\left(1-q_{110}\right)^{2}+4q_{101}}\right).
\end{mycases}
\end{split}\end{align}

We find that the convergence-divergence network with convergence between non-sister taxa is identifiable. We now demand that $0<x_{1},x_{2},x_{3}\leq{}1$, since $x_{1}$, $x_{2}$ and $x_{3}$ are all functions of the time parameters. We then determine what constraints $q_{011}$, $q_{101}$ and $q_{110}$ must have.

We start by demanding $x_{1}>0$. Since $q_{011},q_{101}\geq{}0$, this does not introduce any new constraints on the phylogenetic tensor elements.

Demanding $x_{1}\leq{}1$,
\begin{align}\begin{split}
q_{011}\geq{}q_{101},
\end{split}\end{align}
since $q_{011},q_{101}\geq{}0$.

Since $q_{101}>0$, demanding $x_{2}>0$,
\begin{align}\begin{split}
\sqrt{q_{011}}\left(-\left(1-q_{110}\right)+\sqrt{\left(1-q_{110}\right)^{2}+4q_{101}}\right)>&0.
\end{split}\end{align}
Since $q_{011}>0$,
\begin{align}\begin{split}
-\left(1-q_{110}\right)+\sqrt{\left(1-q_{110}\right)^{2}+4q_{101}}>&0 \\
\Leftrightarrow{}\sqrt{\left(1-q_{110}\right)^{2}+4q_{101}}>&\left(1-q_{110}\right).
\end{split}\end{align}
Since both sides of the inequality must be non-negative and the argument inside the square root must be positive,
\begin{align}\begin{split}
\left(1-q_{110}\right)^{2}+4q_{101}>&\left(1-q_{110}\right)^{2} \\
\Leftrightarrow{}4q_{101}>&0 \\
\Leftrightarrow{}q_{101}>&0.
\end{split}\end{align}
We already knew that $q_{101}>0$. Hence, demanding $x_{2}>0$ does not introduce any new constraints on the phylogenetic tensor elements. 

Demanding $x_{2}\leq{}1$,
\begin{align}\begin{split}
\sqrt{q_{011}}\left(-\left(1-q_{110}\right)+\sqrt{\left(1-q_{110}\right)^{2}+4q_{101}}\right)\leq{}&2q_{101} \\
\Leftrightarrow{}-\left(1-q_{110}\right)+\sqrt{\left(1-q_{110}\right)^{2}+4q_{101}}\leq{}&\frac{2q_{101}}{\sqrt{q_{011}}} \\
\Leftrightarrow{}\sqrt{\left(1-q_{110}\right)^{2}+4q_{101}}\leq{}&\left(1-q_{110}\right)+\frac{2q_{101}}{\sqrt{q_{011}}}.
\end{split}\end{align}
Since both sides of the inequality must be non-negative and the argument inside the square root must be positive, the inequality remains after we square both sides,
\begin{align}\begin{split}
\left(1-q_{110}\right)^{2}+4q_{101}\leq{}&\left(1-q_{110}\right)^{2}+\frac{4q_{101}^{2}}{q_{011}}+\frac{4q_{101}\left(1-q_{110}\right)}{\sqrt{q_{011}}} \\
\Leftrightarrow{}4q_{101}\leq{}&\frac{4q_{101}^{2}}{q_{011}}+\frac{4q_{101}\left(1-q_{110}\right)}{\sqrt{q_{011}}} \\
\Leftrightarrow{}1\leq{}&\frac{q_{101}}{q_{011}}+\frac{\left(1-q_{110}\right)}{\sqrt{q_{011}}} \\
\Leftrightarrow{}q_{011}\leq{}&q_{101}+\left(1-q_{110}\right)\sqrt{q_{011}} \\
\Leftrightarrow{}q_{011}-q_{101}\leq{}&\left(1-q_{110}\right)\sqrt{q_{011}}.
\end{split}\end{align}
Since $q_{011}\geq{}q_{101}$, we can square both sides and the equality will remain,
\begin{align}\begin{split}
\left(q_{011}-q_{101}\right)^{2}\leq{}&\left(1-q_{110}\right)^{2}q_{011}.
\end{split}\end{align}

We now demand $x_{3}>0$. Recall that we have already established that $q_{101}>0$ and $q_{110}\leq{}1$. Hence $1-q_{110}\geq{}0$ and
\begin{align}\begin{split}
\frac{1}{2}\left(1-q_{110}+\sqrt{\left(1-q_{110}\right)^{2}+4q_{101}}\right)>0.
\end{split}\end{align}
Hence, imposing this constraint does not introduce any new constraints to the phylogenetic tensor elements.

Finally, demanding $x_{3}\leq{}1$,
\begin{align}\begin{split}
\frac{1}{2}\left(1-q_{110}+\sqrt{\left(1-q_{110}\right)^{2}+4q_{101}}\right)&\leq{}1 \\
\Leftrightarrow{}1-q_{110}+\sqrt{\left(1-q_{110}\right)^{2}+4q_{101}}&\leq{}2 \\
\Leftrightarrow{}\sqrt{\left(1-q_{110}\right)^{2}+4q_{101}}&\leq{}1+q_{110}.
\end{split}\end{align}
Since both sides of the inequality must be positive and real, the inequality remains after we square both sides,
\begin{align}\begin{split}
\left(1-q_{110}\right)^{2}+4q_{101}&\leq{}1+2q_{110}+q_{110}^{2} \\
\Leftrightarrow{}1-2q_{110}+q_{110}^{2}+4q_{101}&\leq{}1+2q_{110}+q_{110}^{2} \\
\Leftrightarrow{}q_{101}&\leq{}q_{110}.
\end{split}\end{align}
So the constraint $q_{101}\leq{}q_{110}$ arises from demanding $x_{3}=e^{-\tau_{3}}\leq{}1$ or $\tau_{3}>0$.

Hence, the full set of constraints on the phylogenetic tensor is
\begin{align}\begin{split}
\left\{q_{011}\geq{}q_{101},\quad{}q_{110}\geq{}q_{101},\quad{}q_{011}\left(1-q_{110}\right)^{2}\geq{}\left(q_{011}-q_{101}\right)^{2}\right\}.
\end{split}\end{align}

We can conclude that if a phylogenetic tensor satisfies this set of constraints, we can find the time parameters, $\tau_{1}$, $\tau_{2}$ and $\tau_{3}$, on a convergence-divergence network with convergence between non-sister taxa. Likewise, if we start with a phylogenetic tensor and time parameters from the convergence-divergence network with convergence between non-sister taxa, this set of constraints must be satisfied.

We saw that all three of the three-taxon trees and networks that we have so far examined are identifiable.

The constant elements of the phylogenetic tensors must be the same for all three of the three-taxon trees and networks. These constraints are
\begin{align}\begin{split}
\left\{q_{000}=1,\quad{}q_{001}=q_{010}=q_{100}=q_{111}=0\right\}.
\end{split}\end{align}

The non-constant elements of the phylogenetic tensors must all satisfy the constraints,
\begin{align}\begin{split}
\left\{0<q_{011},q_{101},q_{110}\leq{}1\right\}.
\end{split}\end{align}

We can conclude that these three trees and networks are all distinguishable from each other since the sets of constraints are not identical. Recall from our definition of distinguishability that two networks are distinguishable if their probability spaces are not identical. This does not preclude a particular set of phylogenetic tensor elements satisfying the constraints for multiple distinguishable trees and networks. What it does mean is that there must be at least one particular pattern frequency that only satisfies the constraints of one of the trees or networks. The clock-like tree, however, has one less numerical parameter than the non-clock-like tree and the clock-like convergence-divergence network with convergence between non-sister taxa. Hence, from a practical model-fitting perspective it should be given preference unless there is a strong argument against the molecular clock assumption or tree behaviour or if the non-clock-like tree or the clock-like convergence-divergence network with convergence between non-sister taxa provide a significantly better fit than the clock-like tree.

\subsection{Comparisons of Other Three-Taxon Networks}

Having defined our restrictions on the convergence periods, we will now analyse the remaining three-taxon networks. There are nine trees and networks which meet our restrictions on the convergence periods. We will label these as shown below in Figure~\ref{allthreetaxonnetworks}.
\begin{figure}[H]
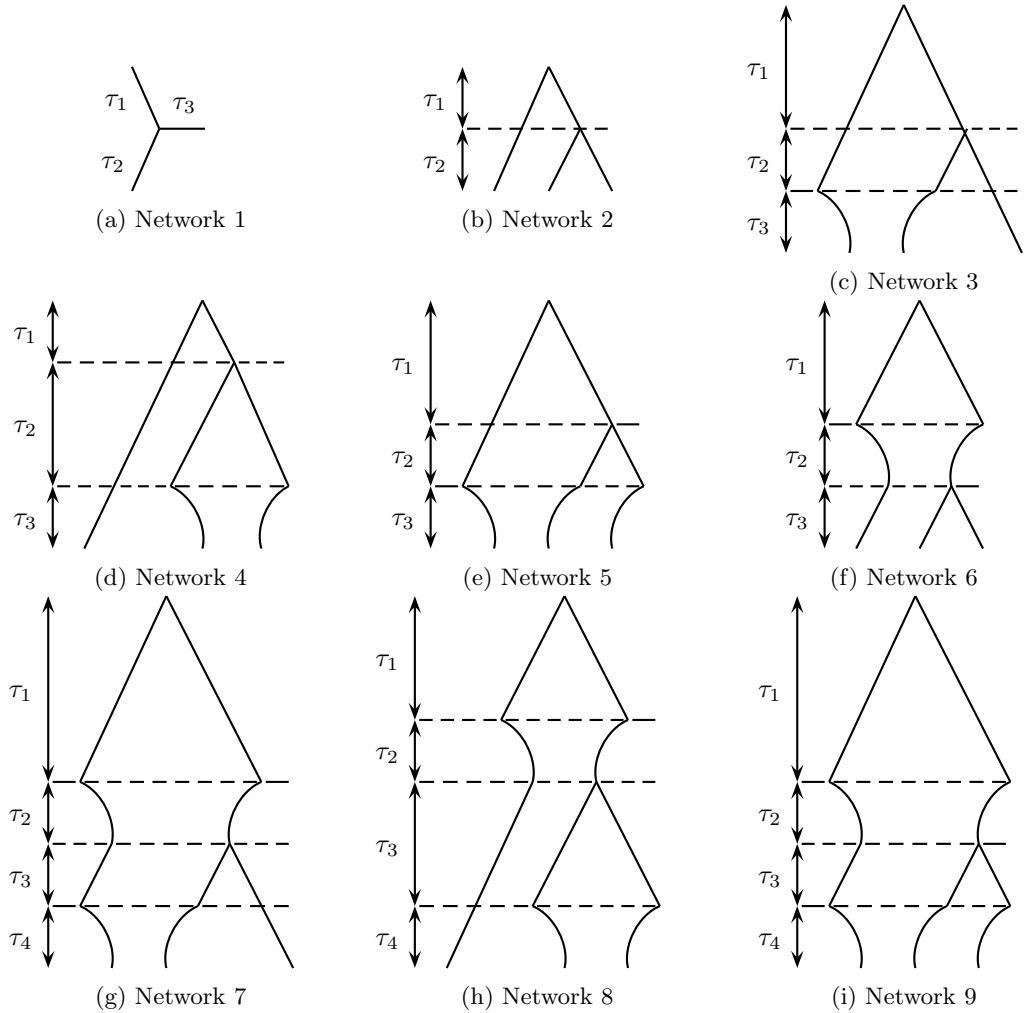

	\centering 
		\begin{subfigure}[h]{0.32\textwidth}
			\centering
				\psmatrix[colsep=.3cm,rowsep=.4cm,mnode=r]
				~ \\
				& {} && ~ \\
				~
				\ncline{1,1}{2,2}
				\tlput{$\tau_{1}$}
				\ncline{2,2}{3,1}
				\tlput{$\tau_{2}$}
				\ncline{2,2}{2,4}
				\taput{$\tau_{3}$}
				\endpsmatrix
				\caption{Network $1$}
		\end{subfigure}
		\begin{subfigure}[h]{0.32\textwidth}
			\centering
				\psmatrix[colsep=.3cm,rowsep=.4cm,mnode=r]
				~ & && ~ \\
				~ & && & ~ & ~ \\
				~ & ~ && ~ && ~
				\ncline{1,4}{3,2}
				\ncline{1,4}{2,5}
				\ncline{2,5}{3,4}
				\ncline{2,5}{3,6}
				\ncline{<->,arrowscale=1.5}{1,1}{2,1}
				\tlput{$\tau_{1}$}
				\ncline{<->,arrowscale=1.5}{2,1}{3,1}
				\tlput{$\tau_{2}$}
				\psset{linestyle=dashed}\ncline{2,1}{2,6}
				\endpsmatrix
				\caption{Network $2$}
		\end{subfigure}
		\begin{subfigure}[h]{0.32\textwidth}
			\centering
				\psmatrix[colsep=.3cm,rowsep=.4cm,mnode=r]
				~ && && ~ \\
				~ \\
				~ && && && ~ && ~ \\
				~ & ~ && && ~ & && ~ \\
				~ && ~ && ~ && && ~
				\ncline{1,5}{4,2}
				\ncline{1,5}{5,9}
				\ncline{3,7}{4,6}
				\ncarc[arcangle=40]{4,2}{5,3}
				\ncarc[arcangle=-40]{4,6}{5,5}
				\ncline{<->,arrowscale=1.5}{1,1}{3,1}
				\tlput{$\tau_{1}$}
				\ncline{<->,arrowscale=1.5}{3,1}{4,1}
				\tlput{$\tau_{2}$}
				\ncline{<->,arrowscale=1.5}{4,1}{5,1}
				\tlput{$\tau_{3}$}
				\psset{linestyle=dashed}\ncline{3,1}{3,7}
				\psset{linestyle=dashed}\ncline{3,7}{3,9}
				\psset{linestyle=dashed}\ncline{4,1}{4,2}
				\psset{linestyle=dashed}\ncline{4,2}{4,6}
				\psset{linestyle=dashed}\ncline{4,6}{4,9}
				\endpsmatrix
				\caption{Network $3$}
		\end{subfigure}
		\begin{subfigure}[h]{0.32\textwidth}
			\centering
				\psmatrix[colsep=.3cm,rowsep=.4cm,mnode=r]
				~ & && && ~ \\
				~ && && && ~ && ~ \\
				~ \\
				~ && ~ && ~ && && ~ \\
				~ & ~ && && ~ && ~
				\ncline{1,6}{5,2}
				\ncline{1,6}{2,7}
				\ncline{2,7}{4,5}
				\ncline{2,7}{4,9}
				\ncarc[arcangle=40]{4,5}{5,6}
				\ncarc[arcangle=-40]{4,9}{5,8}
				\ncline{<->,arrowscale=1.5}{1,1}{2,1}
				\tlput{$\tau_{1}$}
				\ncline{<->,arrowscale=1.5}{2,1}{4,1}
				\tlput{$\tau_{2}$}
				\ncline{<->,arrowscale=1.5}{4,1}{5,1}
				\tlput{$\tau_{3}$}
				\psset{linestyle=dashed}\ncline{2,1}{2,7}
				\psset{linestyle=dashed}\ncline{2,7}{2,9}
				\psset{linestyle=dashed}\ncline{4,1}{4,3}
				\psset{linestyle=dashed}\ncline{4,3}{4,5}
				\psset{linestyle=dashed}\ncline{4,5}{4,9}
				\endpsmatrix
				\caption{Network $4$}
		\end{subfigure}
		\begin{subfigure}[h]{0.32\textwidth}
			\centering
				\psmatrix[colsep=.3cm,rowsep=.4cm,mnode=r]
				~ && && ~ \\
				~ \\
				~ && && && ~ & ~ \\
				~ & ~ && && ~ && ~ \\
				~ && ~ && ~ && ~
				\ncline{1,5}{4,2}
				\ncline{1,5}{4,8}
				\ncline{3,7}{4,6}
				\ncarc[arcangle=40]{4,2}{5,3}
				\ncarc[arcangle=-40]{4,6}{5,5}
				\ncarc[arcangle=-40]{4,8}{5,7}
				\ncline{<->,arrowscale=1.5}{1,1}{3,1}
				\tlput{$\tau_{1}$}
				\ncline{<->,arrowscale=1.5}{3,1}{4,1}
				\tlput{$\tau_{2}$}
				\ncline{<->,arrowscale=1.5}{4,1}{5,1}
				\tlput{$\tau_{3}$}
				\psset{linestyle=dashed}\ncline{3,1}{3,7}
				\psset{linestyle=dashed}\ncline{3,7}{3,8}
				\psset{linestyle=dashed}\ncline{4,1}{4,2}
				\psset{linestyle=dashed}\ncline{4,2}{4,6}
				\psset{linestyle=dashed}\ncline{4,6}{4,8}
				\endpsmatrix
				\caption{Network $5$}
		\end{subfigure}
		\begin{subfigure}[h]{0.32\textwidth}
			\centering
				\psmatrix[colsep=.3cm,rowsep=.4cm,mnode=r]
				~ & && ~ \\
				~ \\
				~ & ~ && && ~ \\
				~ && ~ && ~ & ~ \\
				~ & ~ && ~ && ~
				\ncline{1,4}{3,2}
				\ncline{1,4}{3,6}
				\ncarc[arcangle=40]{3,2}{4,3}
				\ncarc[arcangle=-40]{3,6}{4,5}
				\ncline{4,3}{5,2}
				\ncline{4,5}{5,4}
				\ncline{4,5}{5,6}
				\ncline{<->,arrowscale=1.5}{1,1}{3,1}
				\tlput{$\tau_{1}$}
				\ncline{<->,arrowscale=1.5}{3,1}{4,1}
				\tlput{$\tau_{2}$}
				\ncline{<->,arrowscale=1.5}{4,1}{5,1}
				\tlput{$\tau_{3}$}
				\psset{linestyle=dashed}\ncline{3,1}{3,2}
				\psset{linestyle=dashed}\ncline{3,2}{3,6}
				\psset{linestyle=dashed}\ncline{4,1}{4,3}
				\psset{linestyle=dashed}\ncline{4,3}{4,5}
				\psset{linestyle=dashed}\ncline{4,5}{4,6}
				\endpsmatrix
				\caption{Network $6$}
		\end{subfigure}
		\begin{subfigure}[h]{0.32\textwidth}
			\centering
				\psmatrix[colsep=.3cm,rowsep=.4cm,mnode=r]
				~ && && ~ \\
				~ \\
				~ \\
				~ & ~ && && && ~ & ~ \\
				~ && ~ && && ~ && ~ \\
				~ & ~ && && ~ & && ~ \\
				~ && ~ && ~ && && ~
				\ncline{1,5}{4,2}
				\ncline{1,5}{4,8}
				\ncarc[arcangle=40]{4,2}{5,3}
				\ncarc[arcangle=-40]{4,8}{5,7}
				\ncline{5,3}{6,2}
				\ncline{5,7}{6,6}
				\ncline{5,7}{7,9}
				\ncarc[arcangle=40]{6,2}{7,3}
				\ncarc[arcangle=-40]{6,6}{7,5}
				\ncline{<->,arrowscale=1.5}{1,1}{4,1}
				\tlput{$\tau_{1}$}
				\ncline{<->,arrowscale=1.5}{4,1}{5,1}
				\tlput{$\tau_{2}$}
				\ncline{<->,arrowscale=1.5}{5,1}{6,1}
				\tlput{$\tau_{3}$}
				\ncline{<->,arrowscale=1.5}{6,1}{7,1}
				\tlput{$\tau_{4}$}
				\psset{linestyle=dashed}\ncline{4,1}{4,2}
				\psset{linestyle=dashed}\ncline{4,2}{4,8}
				\psset{linestyle=dashed}\ncline{4,8}{4,9}
				\psset{linestyle=dashed}\ncline{5,1}{5,3}
				\psset{linestyle=dashed}\ncline{5,3}{5,7}
				\psset{linestyle=dashed}\ncline{5,7}{5,9}
				\psset{linestyle=dashed}\ncline{6,1}{6,2}
				\psset{linestyle=dashed}\ncline{6,2}{6,6}
				\psset{linestyle=dashed}\ncline{6,6}{6,9}
				\endpsmatrix
				\caption{Network $7$}
		\end{subfigure}
		\begin{subfigure}[h]{0.32\textwidth}
			\centering
				\psmatrix[colsep=.3cm,rowsep=.4cm,mnode=r]
				~ & && && ~ \\
				~ \\
				~ & && ~ && && ~ & ~ \\
				~ && && ~ && ~ && ~ \\
				~ \\
				~ && && ~ && && ~ \\
				~ & ~ && && ~ && ~
				\ncline{1,6}{3,4}
				\ncline{1,6}{3,8}
				\ncarc[arcangle=40]{3,4}{4,5}
				\ncarc[arcangle=-40]{3,8}{4,7}
				\ncline{4,5}{7,2}
				\ncline{4,7}{6,5}
				\ncline{4,7}{6,9}
				\ncarc[arcangle=40]{6,5}{7,6}
				\ncarc[arcangle=-40]{6,9}{7,8}
				\ncline{<->,arrowscale=1.5}{1,1}{3,1}
				\tlput{$\tau_{1}$}
				\ncline{<->,arrowscale=1.5}{3,1}{4,1}
				\tlput{$\tau_{2}$}
				\ncline{<->,arrowscale=1.5}{4,1}{6,1}
				\tlput{$\tau_{3}$}
				\ncline{<->,arrowscale=1.5}{6,1}{7,1}
				\tlput{$\tau_{4}$}
				\psset{linestyle=dashed}\ncline{3,1}{3,4}
				\psset{linestyle=dashed}\ncline{3,4}{3,8}
				\psset{linestyle=dashed}\ncline{3,8}{3,9}
				\psset{linestyle=dashed}\ncline{4,1}{4,5}
				\psset{linestyle=dashed}\ncline{4,5}{4,7}
				\psset{linestyle=dashed}\ncline{4,7}{4,9}
				\psset{linestyle=dashed}\ncline{6,1}{6,5}
				\psset{linestyle=dashed}\ncline{6,5}{6,9}
				\endpsmatrix
				\caption{Network $8$}
		\end{subfigure}
		\begin{subfigure}[h]{0.32\textwidth}
			\centering
				\psmatrix[colsep=.3cm,rowsep=.4cm,mnode=r]
				~ && && ~ \\
				~ \\
				~ \\
				~ & ~ && && && ~ \\
				~ && ~ && && ~ & ~ \\
				~ & ~ && && ~ && ~ \\
				~ && ~ && ~ && ~
				\ncline{1,5}{4,2}
				\ncline{1,5}{4,8}
				\ncarc[arcangle=40]{4,2}{5,3}
				\ncarc[arcangle=-40]{4,8}{5,7}
				\ncline{5,3}{6,2}
				\ncline{5,7}{6,6}
				\ncline{5,7}{6,8}
				\ncarc[arcangle=40]{6,2}{7,3}
				\ncarc[arcangle=-40]{6,6}{7,5}
				\ncarc[arcangle=-40]{6,8}{7,7}
				\ncline{<->,arrowscale=1.5}{1,1}{4,1}
				\tlput{$\tau_{1}$}
				\ncline{<->,arrowscale=1.5}{4,1}{5,1}
				\tlput{$\tau_{2}$}
				\ncline{<->,arrowscale=1.5}{5,1}{6,1}
				\tlput{$\tau_{3}$}
				\ncline{<->,arrowscale=1.5}{6,1}{7,1}
				\tlput{$\tau_{4}$}
				\psset{linestyle=dashed}\ncline{4,1}{4,2}
				\psset{linestyle=dashed}\ncline{4,2}{4,8}
				\psset{linestyle=dashed}\ncline{5,1}{5,3}
				\psset{linestyle=dashed}\ncline{5,3}{5,7}
				\psset{linestyle=dashed}\ncline{5,7}{5,8}
				\psset{linestyle=dashed}\ncline{6,1}{6,2}
				\psset{linestyle=dashed}\ncline{6,2}{6,6}
				\psset{linestyle=dashed}\ncline{6,6}{6,8}
				\endpsmatrix
				\caption{Network $9$}
		\end{subfigure}
		\caption{Some three-taxon networks. Networks $1$, $2$ and $3$ have been introduced already. Network $1$ is the non-clock-like tree. All other trees and networks are clock-like.}
		\label{allthreetaxonnetworks}
\end{figure}

Recall from Section~\ref{twotaxondistinguish} on page~\pageref{twotaxondistinguish} that the first transformed phylogenetic tensor element is the conservation of probability, $q_{00\ldots{}0}=1$. Recall also that any transformed phylogenetic tensor element with the sum of its indices being odd must be identically zero. Hence, for any three-taxon tree or network, $q_{001}=q_{010}=q_{100}=q_{111}=0$. Recall also that any transformed phylogenetic tensor element with the sum of its indices being even will be non-zero. Hence, $q_{011},q_{101},q_{110}\neq{}0$. There are only $2^{n-1}-1=3$ variable phylogenetic tensor elements for any three-taxon tree or network. As a result, any trees or networks with more than three time parameters will not be identifiable. We can immediately rule out Network $7$, Network $8$ and Network $9$ as they cannot possibly be identifiable. From our two-taxon result, Network $2$, Network $4$, Network $6$ and Network $8$ are not distinguishable from each other, Network $3$ and Network $7$ are not distinguishable from each other and Network $5$ and Network $9$ are not distinguishable from each other. Hence, we will focus only on Network $1$, Network $2$, Network $3$ and Network $5$ as they are the least parameter rich. The four trees and networks remaining to examine are the non-clock-like tree, the clock-like tree, the clock-like convergence-divergence network with convergence between non-sister taxa and the clock-like convergence-divergence network with convergence between all three taxa.

To determine whether a tree or network is distinguishable from each other we will examine the intersection of their probability spaces, that is the intersection of the sets of constraints on the transformed phylogenetic tensor elements. If the two sets of constraints are identical then the two trees or networks will not be distinguishable from each other. If the two sets of constraints are not identical then the two trees or networks will be distinguishable from each other.

\subsubsection{Network $5$}

Before analysing the identifiability of these trees and networks and whether they are distinguishable from one another, we need to find the constraints for the clock-like convergence-divergence network with convergence between all three taxa, shown in Figure~\ref{allconvergencethreetaxon} below.
\begin{figure}[H]
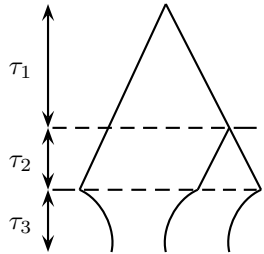

			\centering
				\psmatrix[colsep=.3cm,rowsep=.4cm,mnode=r]
				~ && && ~ \\
				~ \\
				~ && && && ~ & ~ \\
				~ & ~ && && ~ && ~ \\
				~ && ~ && ~ && ~
				\ncline{1,5}{4,2}
				\ncline{1,5}{4,8}
				\ncline{3,7}{4,6}
				\ncarc[arcangle=40]{4,2}{5,3}
				\ncarc[arcangle=-40]{4,6}{5,5}
				\ncarc[arcangle=-40]{4,8}{5,7}
				\ncline{<->,arrowscale=1.5}{1,1}{3,1}
				\tlput{$\tau_{1}$}
				\ncline{<->,arrowscale=1.5}{3,1}{4,1}
				\tlput{$\tau_{2}$}
				\ncline{<->,arrowscale=1.5}{4,1}{5,1}
				\tlput{$\tau_{3}$}
				\psset{linestyle=dashed}\ncline{3,1}{3,7}
				\psset{linestyle=dashed}\ncline{3,7}{3,8}
				\psset{linestyle=dashed}\ncline{4,1}{4,2}
				\psset{linestyle=dashed}\ncline{4,2}{4,6}
				\psset{linestyle=dashed}\ncline{4,6}{4,8}
				\endpsmatrix
				\caption{Three-taxon clock-like convergence-divergence network with convergence between all three taxa.}
				\label{allconvergencethreetaxon}
\end{figure}
The phylogenetic tensor for this network is
\begin{align}\begin{split}
\widehat{P}=e^{\frac{1}{2}\widehat{R}_{111}\tau_{3}}\cdot{}e^{\frac{1}{2}\left(\widehat{R}_{100}+\widehat{R}_{010}+\widehat{R}_{001}\right)\tau_{2}}\cdot{}e^{\frac{1}{2}\left(\widehat{R}_{100}+\widehat{R}_{011}\right)\tau_{1}}\cdot{}\widehat{\Pi{}}&=\left[\begin{array}{c} q_{000} \\
q_{001} \\
q_{010} \\
q_{011} \\
q_{100} \\
q_{101} \\
q_{110} \\
q_{111} \\
\end{array}\right] \\
&=\left[\begin{array}{c} 1 \\
0 \\
0 \\
1-e^{-\tau_{3}}\left(1-e^{-2\tau_{2}}\right) \\
0 \\
1-e^{-\tau_{3}}\left(1-e^{-2\left(\tau_{1}+\tau_{2}\right)}\right) \\
1-e^{-\tau_{3}}\left(1-e^{-2\left(\tau_{1}+\tau_{2}\right)}\right) \\
0 \\
\end{array}\right].
\end{split}\end{align}

Using a similar argument to that used previously and demanding $\tau_{1},\tau_{2},\tau_{3}\geq{}0$,
\begin{align}\begin{split}
&0<1-e^{-\tau_{3}}\left(1-e^{-2\tau_{2}}\right),\quad{}0<1-e^{-\tau_{3}}\left(1-e^{-2\left(\tau_{1}+\tau_{2}\right)}\right)\leq{}1.
\end{split}\end{align}
Hence $0<q_{011},q_{101},q_{110}\leq{}1$ are the minimum length intervals containing all possible values of the individual tensor elements.

By inspection, it can be seen that the full set of constraints on the phylogenetic tensor is
\begin{align}\begin{split}
\left\{q_{101}=q_{110},\quad{}q_{011}\geq{}q_{110}\right\}.
\end{split}\end{align}

Clearly this network is not identifiable and cannot be distinguished from the three-taxon clock-like tree. We conclude that the three-taxon clock-like tree is preferable to the three-taxon clock-like convergence-divergence network with convergence between all three taxa.

\subsection{Distinguishability of Three-Taxon Trees and Networks}

We have established that there are three remaining three-taxon trees and networks that are identifiable and distinguishable from each other: the non-clock-like tree, the clock-like tree and the clock-like convergence-divergence network with convergence between non-sister taxa. We have constructed Table~\ref{constraintsmet} below, which displays whether or not a particular constraint is met for those three trees and networks, as well as the other six networks that we originally considered. In addition to the constraints displayed in the table, there are three other constraints which must be met for the non-clock-like tree.
\begin{table}[H]
\centering
\small
\begin{tabular}{c|c|c|c} Network(s) & $q_{101}=q_{110}$ (Y/N) & $q_{011}\geq{}q_{101}$ (Y/N) & $q_{011}(1-q_{110})^{2}\geq{}(q_{011}-q_{101})^{2}$ (Y/N) \\
\hline
1 & N & N & N \\
\hline
2, 4, 5, 6, 8, 9 & Y & Y & N \\
\hline
3, 7 & N & Y & Y \\
\end{tabular}
\caption{Summary of network constraints which must be met. ``Y'' indicates that the constraint is necessary to be met for that particular network, while ``N'' indicates that the constraint is not necessary to be met for the network in question.}
\label{constraintsmet}
\end{table}

In addition, the non-clock-like tree, or Network $1$, must meet the constraints,
\begin{align}\begin{split}
\left\{q_{011}\geq{}q_{101}q_{110},\quad{}q_{101}\geq{}q_{011}q_{110},\quad{}q_{110}\geq{}q_{011}q_{101}\right\}.
\end{split}\end{align}

It should be noted that this table expresses minimum constraints that must be met for the given tree or network. If meeting a given constraint is not necessary for a particular network that constraint could still be met for the network. For example, the constraint $q_{101}=q_{110}$ being met does not discount the non-clock-like tree as a candidate, however the constraint $q_{101}=q_{110}$ not being met does discount the clock-like tree as a candidate.

The probability spaces that the networks occupy will be denoted by $\Omega_{1}$, $\Omega_{2}$ and $\Omega_{3}$, with the subscripts referring to the three networks in question as labelled earlier. We will look at the intersections of every pair of probability spaces.

\subsubsection{Network $1$ and Network $2$}

We will start with the intersection of the probability spaces of the two trees, $\Omega_{1}\cap{}\Omega_{2}$. Recall that the constraints for Network $2$ are
\begin{align}\begin{split}
\left\{q_{101}=q_{110},\quad{}q_{011}\geq{}q_{110}\right\}.
\end{split}\end{align}
$q_{110}$ is not a free element and can be replaced by $q_{101}$ for every constraint for Network $1$. Recall that the constraints for Network $1$ are
\begin{align}\begin{split}
\left\{q_{011}\geq{}q_{101}q_{110},\quad{}q_{101}\geq{}q_{011}q_{110},\quad{}q_{110}\geq{}q_{011}q_{101}\right\}.
\end{split}\end{align}
Replacing $q_{110}$ with $q_{101}$, the constraints become
\begin{align}\begin{split}
\left\{q_{011}\geq{}q_{101}^{2},\quad{}1\geq{}q_{011},\quad{}1\geq{}q_{011}\right\}.
\end{split}\end{align}
From the second constraint for Network $2$ and since $1\geq{}q_{101}\geq{}q_{101}^{2}$, we can conclude that $q_{011}\geq{}q_{101}^{2}$. The intersection of the two probability spaces must then be the probability space for Network $2$,
\begin{align}\begin{split}
\Omega_{1}\cap{}\Omega_{2}&=\Omega_{2}, \\
\Omega_{2}&\subset{}\Omega_{1}.
\end{split}\end{align}
$\Omega_{2}$ is a proper subset of $\Omega_{1}$ since it does not have the same set of constraints as this tree and therefore does not have the same probability space.

\subsubsection{Network $2$ and Network $3$}

We will now examine the intersection of the probability spaces of Network $2$ and Network $3$, $\Omega_{2}\cap{}\Omega_{3}$. Recall the constraints for Network $3$ are
\begin{align}\begin{split}
\left\{q_{011}\geq{}q_{101},\quad{}q_{110}\geq{}q_{101},\quad{}q_{011}\left(1-q_{110}\right)^{2}\geq{}\left(q_{011}-q_{101}\right)^{2}\right\}.
\end{split}\end{align}
Clearly the two networks have the constraint, $q_{011}\geq{}q_{101}$, in common. Recall that $q_{110}$ is not a free element for Network $2$ and can be replaced by $q_{101}$ for every constraint for Network $3$. Replacing $q_{110}$ with $q_{101}$, the constraints become
\begin{align}\begin{split}
\left\{q_{011}\geq{}q_{101},\quad{}q_{101}\geq{}q_{101},\quad{}q_{011}\left(1-q_{101}\right)^{2}\geq{}\left(q_{011}-q_{101}\right)^{2}\right\}.
\end{split}\end{align}
Clearly the second constraint is trivial. The only constraint left to examine is the last constraint, which when expanded out becomes
\begin{align}\begin{split}
q_{011}+q_{011}q_{101}^{2}-2q_{011}q_{101}&\geq{}q_{011}^{2}+q_{101}^{2}-2q_{011}q_{101} \\
\Leftrightarrow{}q_{011}+q_{011}q_{101}^{2}&\geq{}q_{011}^{2}+q_{101}^{2} \\
\Leftrightarrow{}q_{011}\left(1-q_{011}\right)&\geq{}q_{101}^{2}\left(1-q_{011}\right).
\end{split}\end{align}
Since $q_{011}\neq{}1$ generally,
\begin{align}\begin{split}
q_{011}&\geq{}q_{101}^{2}.
\end{split}\end{align}
From the second constraint for Network $2$ and since $1\geq{}q_{101}\geq{}q_{101}^{2}$, we can conclude that $q_{011}\geq{}q_{101}^{2}$. The intersection of the two probability spaces must then be the probability space for Network $2$,
\begin{align}\begin{split}
\Omega_{2}\cap{}\Omega_{3}&=\Omega_{2}, \\
\Omega_{2}&\subset{}\Omega_{3}.
\end{split}\end{align}
$\Omega_{2}$ is a proper subset of $\Omega_{3}$ since it does not have the same set of constraints as this tree and therefore does not have the same probability space.

We have shown that if the constraints for the clock-like tree are met, the constraints for the non-clock-like tree and the constraints for the clock-like convergence-divergence network must also be met. This is to be expected since the clock-like tree can be recovered from both the non-clock-like tree and the clock-like convergence-divergence network with the appropriate time parameters. Hence,
\begin{align}\begin{split}
\Omega_{2}\subset{}\Omega_{1}, \\
\Omega_{2}\subset{}\Omega_{3}.
\end{split}\end{align}
$\Omega_{2}$ is a proper subset of both $\Omega_{1}$ and $\Omega_{3}$ since it does not have the same set of constraints as these two trees and networks and therefore does not have the same probability space.

\subsubsection{Network $1$ and Network $3$}

The last intersection of probability spaces is the intersection of the probability spaces of Network $1$ and Network $3$, $\Omega_{1}\cap{}\Omega_{3}$. The set of constraints in the region will be
\begin{align}\begin{split}
\Omega_{1}\cap{}\Omega_{3}=\{&q_{011}\geq{}q_{101}q_{110},\quad{}q_{101}\geq{}q_{011}q_{110},\quad{}q_{110}\geq{}q_{011}q_{101},\quad{}q_{011}\geq{}q_{101},\quad{}q_{110}\geq{}q_{101}, \\
&q_{011}(1-q_{110})^{2}\geq{}(q_{011}-q_{101})^{2}\}.
\end{split}\end{align}
Some of these constraints are implied by others. We will now find out which of these constraints are implied by the other constraints. Since $0<q_{110}\leq{}1$,
\begin{align}\begin{split}
q_{011}\geq{}q_{101}\Rightarrow{}q_{011}\geq{}q_{101}q_{110}.
\end{split}\end{align}
Since $0<q_{011}\leq{}1$,
\begin{align}\begin{split}
q_{110}\geq{}q_{101}\Rightarrow{}q_{110}\geq{}q_{011}q_{101}.
\end{split}\end{align}
Hence, the constraints $q_{011}\geq{}q_{101}q_{110}$ and $q_{110}\geq{}q_{011}q_{101}$ are implied by the other constraints. Since $0<q_{011},q_{101},q_{110}\leq{}1$,
\begin{align}\begin{split}
q_{101}\geq{}q_{011}q_{110}\Leftrightarrow{}\frac{q_{101}}{q_{011}}&\geq{}q_{110} \\
\Leftrightarrow{}\frac{q_{101}^{2}}{q_{011}^{2}}&\geq{}q_{110}^{2} \\
\Leftrightarrow{}-q_{110}^{2}&\geq{}-\frac{q_{101}^{2}}{q_{011}^{2}} \\
\Leftrightarrow{}1-q_{110}^{2}&\geq{}1-\frac{q_{101}^{2}}{q_{011}^{2}}.
\end{split}\end{align}
Since $q_{101}\geq{}q_{011}q_{110}$,
\begin{align}\begin{split}
\frac{q_{101}}{q_{011}}&\geq{}q_{110} \\
\Leftrightarrow{}-2q_{110}&\geq{}-2\frac{q_{101}}{q_{011}}.
\end{split}\end{align}
Combining the two inequalities, $1-q_{110}^{2}\geq{}1-\frac{q_{101}^{2}}{q_{011}^{2}}$ and $-2q_{110}\geq{}-2\frac{q_{101}}{q_{011}}$,
\begin{align}\begin{split}
1-2q_{110}-q_{110}^{2}&\geq{}1-2\frac{q_{101}}{q_{011}}-\frac{q_{101}^{2}}{q_{011}^{2}} \\
\Leftrightarrow{}\left(1-q_{110}\right)^{2}&\geq{}\left(1-\frac{q_{101}}{q_{011}}\right)^{2} \\
\Rightarrow{}\left(1-q_{110}\right)^{2}&\geq{}q_{011}\left(1-\frac{q_{101}}{q_{011}}\right)^{2} \\
\Leftrightarrow{}\left(1-q_{110}\right)^{2}&\geq{}q_{011}\left(\frac{q_{011}-q_{101}}{q_{011}}\right)^{2} \\
\Leftrightarrow{}q_{011}\left(1-q_{110}\right)^{2}&\geq{}\left(q_{011}-q_{101}\right)^{2}.
\end{split}\end{align}
This is one of the constraints in the probability space for Network $3$. Finally, the set can be simplified to
\begin{align}\begin{split}
\Omega_{1}\cap{}\Omega_{3}=\{q_{011}\geq{}q_{101},\quad{}q_{110}\geq{}q_{101},\quad{}q_{101}\geq{}q_{011}q_{110}\}.
\end{split}\end{align}

Consequently, there must be three distinct regions involving the non-clock-like tree and the clock-like convergence-divergence network. There must be a region where the constraints of both networks are met and regions where the constraints of only one network are met. Below in Figure~\ref{probabilityspacesthreetaxon} is a diagram representing the four regions in the probability space involving the three trees and networks. The diagram is not to scale, nor do the shapes of the regions have any meaning other than the reflection that some regions border each other in the probability space.
\begin{figure}[H]
	\centering
		\begin{tikzpicture}
			\draw[ultra thick, fill=red!60] (0cm,0cm) -- (4cm,0cm) -- (4cm,-1cm) -- (0cm,-1cm) -- cycle;
			\draw[ultra thick, fill=yellow] (0cm,-1cm) -- (2cm,-1cm) -- (2cm,-2cm) -- (0cm,-2cm) -- cycle;
			\draw[ultra thick, fill=green] (2cm,-1cm) -- (4cm,-1cm) -- (4cm,-2cm) -- (2cm,-2cm) -- cycle;
			\draw[ultra thick, fill=blue!30] (0cm,-2cm) -- (4cm,-2cm) -- (4cm,-3cm) -- (0cm,-3cm) -- cycle;
			\draw (2cm,-0.5cm) node {$\Omega_{1}\cap{}\Omega_{3}^{C}$};
			\draw (1cm,-1.5cm) node {\scriptsize{$\left(\Omega_{1}\cap{}\Omega_{3}\right)\cap{}\Omega_{2}^{C}$}};
			\draw (3cm,-1.5cm) node {$\Omega_{2}$};
			\draw (2cm,-2.5cm) node {$\Omega_{1}^{C}\cap{}\Omega_{3}$};
		\end{tikzpicture}
	\caption{Probability spaces of the networks. $C$ in the superscripts refers to the complement of the set. In other words, the constraints for that network are not met in that region. In the region outside of the probability space formed by the union of the probability spaces for the three trees and networks, none of the sets of constraints of the trees or networks examined are met. In part of this region the constraints for more complex models than the binary symmetric model will be met. For example, for the binary symmetric model, $p_{\textbf{i}}=p_{\bar{\textbf{i}}}$, where $\textbf{i}$ is an element of the phylogenetic tensor and $\bar{\textbf{i}}$ is the term in the phylogenetic tensor where every `$0$' has been replaced by a `$1$' and every `$1$' has been replaced by a `$0$'. If these constraints are not met then the constraints on a more complex model could still be met.}
	\label{probabilityspacesthreetaxon}
\end{figure}
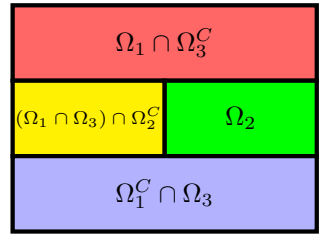

We can write a pseudocode algorithm to determine which tree or network a set of constraints could have arisen on. Below is the pseudocode algorithm.

\begin{algorithmic}
\If {Network $1$ constraints $==$ TRUE}
	\If {Network $3$ constraints $==$ TRUE}
		\If {Network $2$ constraints $==$ TRUE}
			\State $\textrm{tree or network}\gets \textrm{Network $1$, Network $2$ or Network $3$}$
		\Else
			\State $\textrm{tree or network}\gets \textrm{Network $1$ or Network $3$}$
		\EndIf
	\Else
		\State $\textrm{tree or network}\gets \textrm{Network $1$}$
	\EndIf
\Else
	\State $\textrm{tree or network}\gets \textrm{Network $3$}$
\EndIf
\end{algorithmic}

In summary, there are four interesting regions in the probability space of the networks. A phylogenetic tensor either belongs to the non-clock-like tree exclusively, the clock-like convergence-divergence network exclusively, either of the non-clock-like tree or the clock-like convergence-divergence network, or all three trees and networks.

\chapter{Network Identifiability of Taxon Label Permutations}
\label{chapter5}

In this chapter we will introduce the concept of permuting taxon labels and examine whether different taxon label permutations are network identifiable from each other on a given tree or network. Recall from Chapter~\ref{chapter1} that a pair of trees or networks are network identifiable from each other if there is a pattern frequency that could have only arisen on one of the trees or networks. If we have non-negative time parameters on one of the trees or networks, can we find a set of non-negative time parameters on the other tree or network that gives rise to an identical phylogenetic tensor? If we can find a set of non-negative time parameters then the pair of trees or networks are not network identifiable from each other. If some of the time parameters can be negative then the pair of trees or networks are network identifiable from each other.

For the two-taxon clock-like tree and the two-taxon convergence-divergence network the choice of taxon labelling will be of no consequence since all two-taxon trees and networks are symmetrical below every node. We will start by examining the three-taxon clock-like tree, before examining the three-taxon convergence-divergence network with convergence between non-sister taxa.

\section{Permuting the Taxon Labels}

We now wish to see how taxon label permutations may affect the transformed phylogenetic tensor elements on a three-taxon tree or network. There are $3!=6$ taxon label permutations on three-taxon networks, including the original tree or network with no taxon label permutation. These permutations are described by the symmetric group,
\begin{align}\begin{split}
S_{3}=\left\{\left(\right),\left(12\right),\left(13\right),\left(23\right),\left(123\right),\left(132\right)\right\}.
\end{split}\end{align}

Suppose the element, $\left(\right)$, corresponds to the three-taxon clock-like tree shown in Figure~\ref{threetaxonclocknopermutation} below.
\begin{figure}[H]
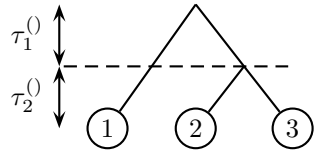

		\centering
				\psmatrix[colsep=.3cm,rowsep=.4cm,mnode=r]
				~ & && ~ \\
				~ & && & ~ & ~ \\
				~ & [mnode=circle] 1 && [mnode=circle] 2 && [mnode=circle] 3
				\ncline{1,4}{3,2}
				\ncline{1,4}{2,5}
				\ncline{2,5}{3,4}
				\ncline{2,5}{3,6}
				\ncline{<->,arrowscale=1.5}{1,1}{2,1}
				\tlput{$\tau_{1}^{\left(\right)}$}
				\ncline{<->,arrowscale=1.5}{2,1}{3,1}
				\tlput{$\tau_{2}^{\left(\right)}$}
				\psset{linestyle=dashed}\ncline{2,1}{2,6}
				\endpsmatrix
				\caption{The $\left(\right)$ taxon label permutation, where the taxon labels have not been permuted.}
				\label{threetaxonclocknopermutation}
\end{figure}

Suppose it also corresponds to the three-taxon clock-like network shown in Figure~\ref{threetaxonnetworknopermutation} below.
\begin{figure}[H]
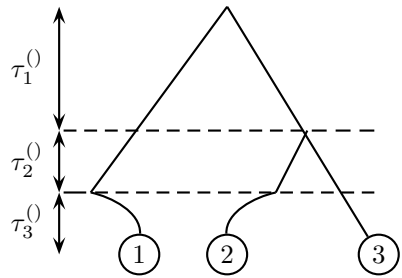

		\centering
				\psmatrix[colsep=.3cm,rowsep=.4cm,mnode=r]
				~ && && ~ \\
				~ \\
				~ && && && ~ && ~ \\
				~ & ~ && && ~ & && ~ \\
				~ && [mnode=circle] 1 && [mnode=circle] 2 && && [mnode=circle] 3
				\ncline{1,5}{4,2}
				\ncline{1,5}{5,9}
				\ncline{3,7}{4,6}
				\ncarc[arcangle=40]{4,2}{5,3}
				\ncarc[arcangle=-40]{4,6}{5,5}
				\ncline{<->,arrowscale=1.5}{1,1}{3,1}
				\tlput{$\tau_{1}^{\left(\right)}$}
				\ncline{<->,arrowscale=1.5}{3,1}{4,1}
				\tlput{$\tau_{2}^{\left(\right)}$}
				\ncline{<->,arrowscale=1.5}{4,1}{5,1}
				\tlput{$\tau_{3}^{\left(\right)}$}
				\psset{linestyle=dashed}\ncline{3,1}{3,7}
				\psset{linestyle=dashed}\ncline{3,7}{3,9}
				\psset{linestyle=dashed}\ncline{4,1}{4,2}
				\psset{linestyle=dashed}\ncline{4,2}{4,6}
				\psset{linestyle=dashed}\ncline{4,6}{4,9}
				\endpsmatrix
				\caption{The $\left(\right)$ taxon label permutation, where the taxon labels have not been permuted.}
				\label{threetaxonnetworknopermutation}
\end{figure}

The other elements of the symmetric group will permute this taxon labelling accordingly. Referring to Figure~\ref{threetaxonclockpermutations}~on~page~\pageref{threetaxonclockpermutations}, it should be noted that the $\left(23\right)$ taxon label permutation on the three-taxon clock-like tree will not change the structure of the tree and is not particularly interesting to us. Similarly, since $\left(23\right)\left(12\right)=\left(132\right)$, the $\left(12\right)$ and $\left(132\right)$ taxon label permutations will result in the same structure on the three-taxon clock-like tree. Since $\left(23\right)\left(13\right)=\left(123\right)$, the $\left(13\right)$ and $\left(123\right)$ taxon label permutations will also result in the same structure on the three-taxon clock-like tree.

There are two issues that arise when we permute the taxon labels.
\begin{enumerate}
	\item Suppose we permute the taxon labels. The structure of the tree or network remains unchanged. The only change that has been made is the labelling of the leaves. The tree or network must therefore have the same set of time parameters. However, the elements of the phylogenetic tensor must be permuted according to the taxon label permutation in question. For example, suppose we wish to permute the taxon labels on the first two leaves of the tree or network, counting from the left. The transformed phylogenetic tensor elements that must be equal functions are
\begin{align}\begin{split}
q_{i_{1}i_{2}i_{3}}^{\left(12\right)}\left(u,v,w\right)=q_{i_{2}i_{1}i_{3}}^{\left(\right)}\left(u,v,w\right),
\end{split}\end{align}
where $\left(\right)$ refers to the original taxon labelling, $\left(12\right)$ refers to the permuting of the first and second taxon labels, and $u$, $v$ and $w$ are any arbitrary time parameters for the convergence-divergence network.
	\item Suppose we start with the site pattern frequency, $i_{1}i_{2}i_{3}$. We now wish to equate the transformed phylogenetic tensor elements with the same site patterns for different taxon label permutations. If we equate each element of the transformed phylogenetic tensor for the tree or network with no taxon label permutation to the tree or network with the first and second taxon labels permuted, we must have
\begin{align}\begin{split}
q_{i_{1}i_{2}i_{3}}^{\left(\right)}\left(\tau_{1}^{\left(\right)},\tau_{2}^{\left(\right)},\tau_{3}^{\left(\right)}\right)=q_{i_{1}i_{2}i_{3}}^{\left(12\right)}\left(\tau_{1}^{\left(12\right)},\tau_{2}^{\left(12\right)},\tau_{3}^{\left(12\right)}\right).
\end{split}\end{align}
The two transformed phylogenetic tensor elements will not have identical functions. We don't know a priori whether there will be any set of non-negative time parameters on one taxon label permutation that will correspond to a set of non-negative time parameters on the other taxon label permutation. The transformed phylogenetic tensor elements will only be equal for certain sets of time parameters, if any. We want to determine whether the probability distributions overlap and to find the time parameters if they do. This second issue is a question of network identifiability.
\end{enumerate}

\section{Determining How Often Time Parameters are Non-Negative}

Suppose we have found the time parameters for a second taxon label permutation in terms of the time parameters for a first taxon label permutation on a given tree or network. Suppose that one of the time parameters on the second taxon label permutation, when expressed in terms of the non-negative time parameters on the first taxon label permutation, can take any real number. We can simulate data to determine how often the time parameter on the second taxon label permutation will be non-negative. We can generate $x_{i}=e^{-\tau_{i}}\in{}U\left(0,1\right)$ for the time parameters on the first taxon label permutation using the ``runif'' function in \emph{R}. We can then evaluate the time parameter on the second taxon label permutation and determine whether it is non-negative or not. By performing many simulations we can get an idea of how often the time parameter on the second taxon label permutation will be non-negative. Below is a sample code for determining how often the time parameter on the second taxon label permutation will be non-negative.

\begin{lstlisting}[breaklines=true]
# Set the number of iterations, n'.
n <- n'

# Vector for the xi=exp(-taui) in U(0,1), the i time parameters on the tree or network for the first taxon label permutation.
x <- vector()

# Define the number of time parameters on the tree or network for the first taxon label permutation, m'.
m <- m'

# Counter for the number of times that the constraint is met.
c <- 0

# Perform n iterations.
for (i in 1:n) {

# Randomly generate xi=exp(-taui) in U(0,1) for the i time parameters on the tree or network for the first taxon label permutation.
for (j in 1:m) {
x[j] <- runif(1)
}

# Find one of the time parameters for the second taxon label permutation, where f(x[1], x[2], ..., x[m]) is a function of the m time parameters on the tree or network for the first taxon label permutation.
t2 <- f(x[1], x[2], ..., x[m])

# If the time parameter is non-negative, increase the total count of non-negative time parameters by one, otherwise keep the count of non-negative time parameters the same.
c <- ifelse(t2>=0,c+1,c)
}

# Output the number of times the time parameter is non-negative.
c
\end{lstlisting}

It should be noted that these simulations assume that each $x_{i}=e^{-\tau_{i}}$ is uniformly distributed, $x_{i}=e^{-\tau_{i}}\in{}U\left(0,1\right)$.

\section{The Three-Taxon Clock-Like Tree}

We can compare the sets of constraints on the transformed phylogenetic tensor elements for all of the taxon label permutations. Below in Figure~\ref{threetaxonclockpermutations} are the six taxon label permutations on the three-taxon clock-like tree.
\begin{figure}[H]
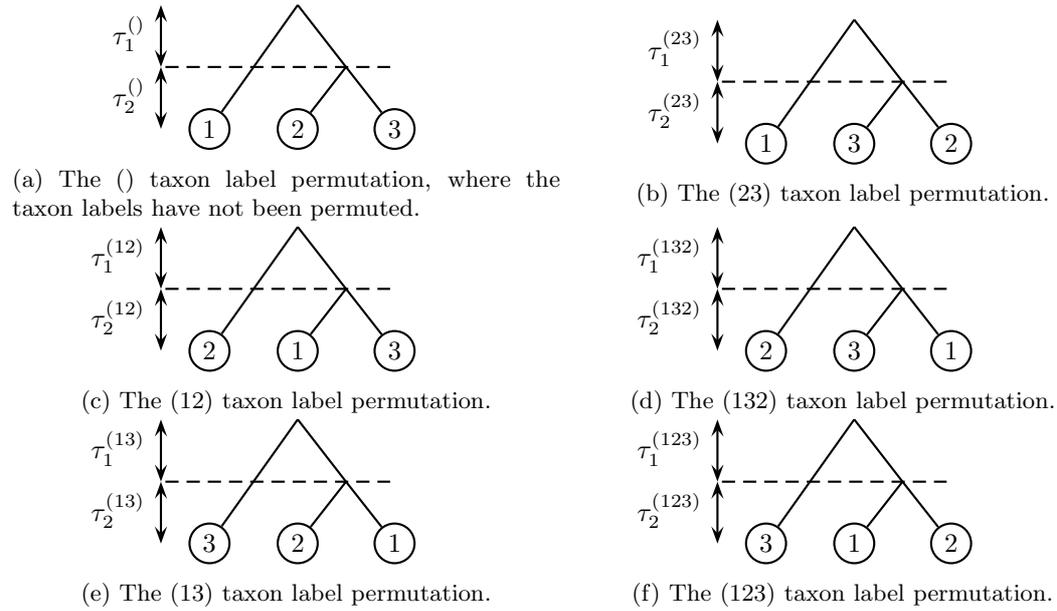

	\centering 
		\begin{subfigure}[h]{0.49\textwidth}
		\centering
				\psmatrix[colsep=.3cm,rowsep=.4cm,mnode=r]
				~ & && ~ \\
				~ & && & ~ & ~ \\
				~ & [mnode=circle] 1 && [mnode=circle] 2 && [mnode=circle] 3
				\ncline{1,4}{3,2}
				\ncline{1,4}{2,5}
				\ncline{2,5}{3,4}
				\ncline{2,5}{3,6}
				\ncline{<->,arrowscale=1.5}{1,1}{2,1}
				\tlput{$\tau_{1}^{\left(\right)}$}
				\ncline{<->,arrowscale=1.5}{2,1}{3,1}
				\tlput{$\tau_{2}^{\left(\right)}$}
				\psset{linestyle=dashed}\ncline{2,1}{2,6}
				\endpsmatrix
				\caption{The $\left(\right)$ taxon label permutation, where the taxon labels have not been permuted.}
		\end{subfigure}
		\begin{subfigure}[h]{0.49\textwidth}
		\centering
				\psmatrix[colsep=.3cm,rowsep=.4cm,mnode=r]
				~ & && ~ \\
				~ & && & ~ & ~ \\
				~ & [mnode=circle] 1 && [mnode=circle] 3 && [mnode=circle] 2
				\ncline{1,4}{3,2}
				\ncline{1,4}{2,5}
				\ncline{2,5}{3,4}
				\ncline{2,5}{3,6}
				\ncline{<->,arrowscale=1.5}{1,1}{2,1}
				\tlput{$\tau_{1}^{\left(23\right)}$}
				\ncline{<->,arrowscale=1.5}{2,1}{3,1}
				\tlput{$\tau_{2}^{\left(23\right)}$}
				\psset{linestyle=dashed}\ncline{2,1}{2,6}
				\endpsmatrix
				\caption{The $\left(23\right)$ taxon label permutation.}
		\end{subfigure}
				\begin{subfigure}[h]{0.49\textwidth}
		\centering
				\psmatrix[colsep=.3cm,rowsep=.4cm,mnode=r]
				~ & && ~ \\
				~ & && & ~ & ~ \\
				~ & [mnode=circle] 2 && [mnode=circle] 1 && [mnode=circle] 3
				\ncline{1,4}{3,2}
				\ncline{1,4}{2,5}
				\ncline{2,5}{3,4}
				\ncline{2,5}{3,6}
				\ncline{<->,arrowscale=1.5}{1,1}{2,1}
				\tlput{$\tau_{1}^{\left(12\right)}$}
				\ncline{<->,arrowscale=1.5}{2,1}{3,1}
				\tlput{$\tau_{2}^{\left(12\right)}$}
				\psset{linestyle=dashed}\ncline{2,1}{2,6}
				\endpsmatrix
				\caption{The $\left(12\right)$ taxon label permutation.}
		\end{subfigure}
				\begin{subfigure}[h]{0.49\textwidth}
		\centering
				\psmatrix[colsep=.3cm,rowsep=.4cm,mnode=r]
				~ & && ~ \\
				~ & && & ~ & ~ \\
				~ & [mnode=circle] 2 && [mnode=circle] 3 && [mnode=circle] 1
				\ncline{1,4}{3,2}
				\ncline{1,4}{2,5}
				\ncline{2,5}{3,4}
				\ncline{2,5}{3,6}
				\ncline{<->,arrowscale=1.5}{1,1}{2,1}
				\tlput{$\tau_{1}^{\left(132\right)}$}
				\ncline{<->,arrowscale=1.5}{2,1}{3,1}
				\tlput{$\tau_{2}^{\left(132\right)}$}
				\psset{linestyle=dashed}\ncline{2,1}{2,6}
				\endpsmatrix
				\caption{The $\left(132\right)$ taxon label permutation.}
		\end{subfigure}
				\begin{subfigure}[h]{0.49\textwidth}
		\centering
				\psmatrix[colsep=.3cm,rowsep=.4cm,mnode=r]
				~ & && ~ \\
				~ & && & ~ & ~ \\
				~ & [mnode=circle] 3 && [mnode=circle] 2 && [mnode=circle] 1
				\ncline{1,4}{3,2}
				\ncline{1,4}{2,5}
				\ncline{2,5}{3,4}
				\ncline{2,5}{3,6}
				\ncline{<->,arrowscale=1.5}{1,1}{2,1}
				\tlput{$\tau_{1}^{\left(13\right)}$}
				\ncline{<->,arrowscale=1.5}{2,1}{3,1}
				\tlput{$\tau_{2}^{\left(13\right)}$}
				\psset{linestyle=dashed}\ncline{2,1}{2,6}
				\endpsmatrix
				\caption{The $\left(13\right)$ taxon label permutation.}
		\end{subfigure}
				\begin{subfigure}[h]{0.49\textwidth}
		\centering
				\psmatrix[colsep=.3cm,rowsep=.4cm,mnode=r]
				~ & && ~ \\
				~ & && & ~ & ~ \\
				~ & [mnode=circle] 3 && [mnode=circle] 1 && [mnode=circle] 2
				\ncline{1,4}{3,2}
				\ncline{1,4}{2,5}
				\ncline{2,5}{3,4}
				\ncline{2,5}{3,6}
				\ncline{<->,arrowscale=1.5}{1,1}{2,1}
				\tlput{$\tau_{1}^{\left(123\right)}$}
				\ncline{<->,arrowscale=1.5}{2,1}{3,1}
				\tlput{$\tau_{2}^{\left(123\right)}$}
				\psset{linestyle=dashed}\ncline{2,1}{2,6}
				\endpsmatrix
				\caption{The $\left(123\right)$ taxon label permutation.}
		\end{subfigure}
		\caption{The six taxon label permutations on the three-taxon clock-like tree.}
		\label{threetaxonclockpermutations}
\end{figure}

Since the order of the two taxa sharing the most recent common ancestor is inconsequential, we can see that there will be three pairs of taxon label permutations with the same set of constraints, shown in Table~\ref{threetaxonclockpairs} below.
\begin{table}[H]
\centering
\begin{tabular}{c|c} Taxon Label Permutations & Constraints \\
\hline
$\left(\right)$, $\left(23\right)$ & $\left\{q_{101}^{\left(\right)}=q_{110}^{\left(\right)},\quad{}q_{011}^{\left(\right)}\geq{}q_{110}^{\left(\right)}\right\}$ \\
\hline
$\left(12\right)$, $\left(132\right)$ & $\left\{q_{011}^{\left(\right)}=q_{110}^{\left(\right)},\quad{}q_{101}^{\left(\right)}\geq{}q_{110}^{\left(\right)}\right\}$ \\
\hline
$\left(13\right)$, $\left(123\right)$ & $\left\{q_{101}^{\left(\right)}=q_{011}^{\left(\right)},\quad{}q_{110}^{\left(\right)}\geq{}q_{011}^{\left(\right)}\right\}$
\end{tabular}
\caption{Summary of constraints for each taxon label permutation.}
\label{threetaxonclockpairs}
\end{table}

When choosing a taxon label permutation on the clock-like tree, there are three choices for the taxon label for the non-sister taxa. Each of these three choices of taxon labellings will have two choices for the labellings of the sister taxa, which must give rise to identical phylogenetic tensors. We can see from the summary of constraints that the three choices for the taxon label for the non-sister taxa must give rise to different phylogenetic tensors. The only circumstance where two different taxon labelling choices for the non-sister taxa could have identical phylogenetic tensors is when the first time parameter is zero, $\tau_{1}=0$. In this circumstance the three-taxon clock-like tree becomes the tripod tree with all edges being of equal length.

Comparing taxon label permutations on the clock-like tree is straightforward, however for more complicated networks, such as the convergence-divergence network with convergence between non-sister taxa, the permuting of taxon labels is much more complicated.

\section{The Convergence-Divergence Network}

We will now examine the convergence-divergence network with convergence between non-sister taxa, shown in Figure~\ref{threetaxonnonsisterconvergence} below.
\begin{figure}[H]
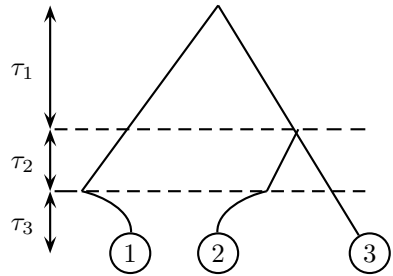

	\centering
		\psmatrix[colsep=.3cm,rowsep=.4cm,mnode=r]
		~ && && ~ \\
		~ \\
		~ && && && ~ && ~ \\
		~ & ~ && && ~ & && ~ \\
		~ && [mnode=circle] 1 && [mnode=circle] 2 && && [mnode=circle] 3
		\ncline{1,5}{4,2}
		\ncline{1,5}{5,9}
		\ncline{3,7}{4,6}
		\ncarc[arcangle=40]{4,2}{5,3}
		\ncarc[arcangle=-40]{4,6}{5,5}
		\ncline{<->,arrowscale=1.5}{1,1}{3,1}
		\tlput{$\tau_{1}$}
		\ncline{<->,arrowscale=1.5}{3,1}{4,1}
		\tlput{$\tau_{2}$}
		\ncline{<->,arrowscale=1.5}{4,1}{5,1}
		\tlput{$\tau_{3}$}
		\psset{linestyle=dashed}\ncline{3,1}{3,7}
		\psset{linestyle=dashed}\ncline{3,7}{3,9}
		\psset{linestyle=dashed}\ncline{4,1}{4,2}
		\psset{linestyle=dashed}\ncline{4,2}{4,6}
		\psset{linestyle=dashed}\ncline{4,6}{4,9}
		\endpsmatrix
		\caption{The three-taxon clock-like convergence-divergence network with convergence between non-sister taxa.}
		\label{threetaxonnonsisterconvergence}
\end{figure}

We wish to determine whether various taxon label permutations are network identifiable from each other. Recall that there are six taxon label permutations for three-taxon trees and networks. Below in Figure~\ref{taxonlabellingscondiv} are the six taxon labellings for the convergence-divergence network.
\begin{figure}[H]
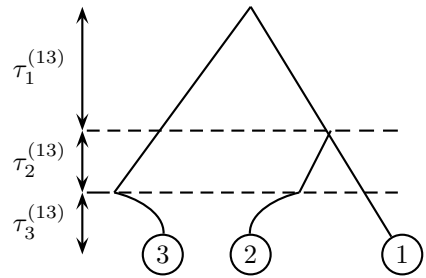
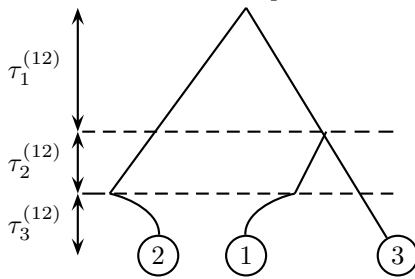
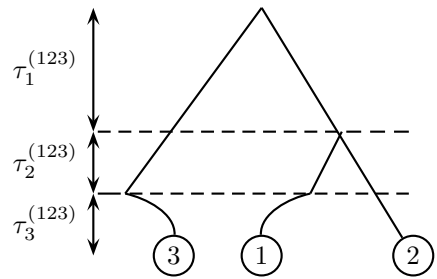
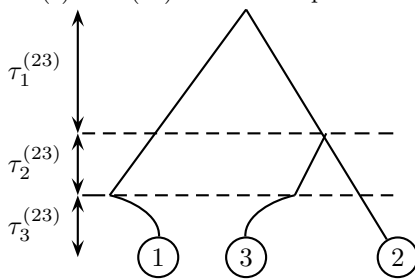
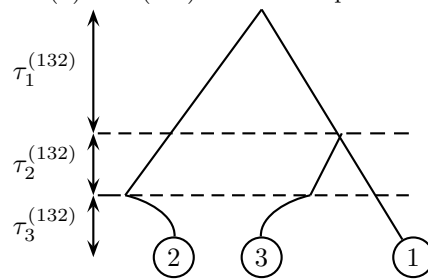

	\centering 
		\begin{subfigure}[h]{0.49\textwidth}
		\centering
				\psmatrix[colsep=.3cm,rowsep=.4cm,mnode=r]
				~ && && ~ \\
				~ \\
				~ && && && ~ && ~ \\
				~ & ~ && && ~ & && ~ \\
				~ && [mnode=circle] 1 && [mnode=circle] 2 && && [mnode=circle] 3
				\ncline{1,5}{4,2}
				\ncline{1,5}{5,9}
				\ncline{3,7}{4,6}
				\ncarc[arcangle=40]{4,2}{5,3}
				\ncarc[arcangle=-40]{4,6}{5,5}
				\ncline{<->,arrowscale=1.5}{1,1}{3,1}
				\tlput{$\tau_{1}^{\left(\right)}$}
				\ncline{<->,arrowscale=1.5}{3,1}{4,1}
				\tlput{$\tau_{2}^{\left(\right)}$}
				\ncline{<->,arrowscale=1.5}{4,1}{5,1}
				\tlput{$\tau_{3}^{\left(\right)}$}
				\psset{linestyle=dashed}\ncline{3,1}{3,7}
				\psset{linestyle=dashed}\ncline{3,7}{3,9}
				\psset{linestyle=dashed}\ncline{4,1}{4,2}
				\psset{linestyle=dashed}\ncline{4,2}{4,6}
				\psset{linestyle=dashed}\ncline{4,6}{4,9}
				\endpsmatrix
				\caption{The $\left(\right)$ taxon label permutation, where the taxon labels have not been permuted.}
		\end{subfigure}
		\begin{subfigure}[h]{0.49\textwidth}
		\centering
				\psmatrix[colsep=.3cm,rowsep=.4cm,mnode=r]
				~ && && ~ \\
				~ \\
				~ && && && ~ && ~ \\
				~ & ~ && && ~ & && ~ \\
				~ && [mnode=circle] 3 && [mnode=circle] 2 && && [mnode=circle] 1
				\ncline{1,5}{4,2}
				\ncline{1,5}{5,9}
				\ncline{3,7}{4,6}
				\ncarc[arcangle=40]{4,2}{5,3}
				\ncarc[arcangle=-40]{4,6}{5,5}
				\ncline{<->,arrowscale=1.5}{1,1}{3,1}
				\tlput{$\tau_{1}^{\left(13\right)}$}
				\ncline{<->,arrowscale=1.5}{3,1}{4,1}
				\tlput{$\tau_{2}^{\left(13\right)}$}
				\ncline{<->,arrowscale=1.5}{4,1}{5,1}
				\tlput{$\tau_{3}^{\left(13\right)}$}
				\psset{linestyle=dashed}\ncline{3,1}{3,7}
				\psset{linestyle=dashed}\ncline{3,7}{3,9}
				\psset{linestyle=dashed}\ncline{4,1}{4,2}
				\psset{linestyle=dashed}\ncline{4,2}{4,6}
				\psset{linestyle=dashed}\ncline{4,6}{4,9}
				\endpsmatrix
				\caption{The $\left(13\right)$ taxon label permutation.}
		\end{subfigure}
		\begin{subfigure}[h]{0.49\textwidth}
		\centering
				\psmatrix[colsep=.3cm,rowsep=.4cm,mnode=r]
				~ && && ~ \\
				~ \\
				~ && && && ~ && ~ \\
				~ & ~ && && ~ & && ~ \\
				~ && [mnode=circle] 2 && [mnode=circle] 1 && && [mnode=circle] 3
				\ncline{1,5}{4,2}
				\ncline{1,5}{5,9}
				\ncline{3,7}{4,6}
				\ncarc[arcangle=40]{4,2}{5,3}
				\ncarc[arcangle=-40]{4,6}{5,5}
				\ncline{<->,arrowscale=1.5}{1,1}{3,1}
				\tlput{$\tau_{1}^{\left(12\right)}$}
				\ncline{<->,arrowscale=1.5}{3,1}{4,1}
				\tlput{$\tau_{2}^{\left(12\right)}$}
				\ncline{<->,arrowscale=1.5}{4,1}{5,1}
				\tlput{$\tau_{3}^{\left(12\right)}$}
				\psset{linestyle=dashed}\ncline{3,1}{3,7}
				\psset{linestyle=dashed}\ncline{3,7}{3,9}
				\psset{linestyle=dashed}\ncline{4,1}{4,2}
				\psset{linestyle=dashed}\ncline{4,2}{4,6}
				\psset{linestyle=dashed}\ncline{4,6}{4,9}
				\endpsmatrix
				\caption{The $\left(12\right)$ taxon label permutation.}
		\end{subfigure}
		\begin{subfigure}[h]{0.49\textwidth}
		\centering
				\psmatrix[colsep=.3cm,rowsep=.4cm,mnode=r]
				~ && && ~ \\
				~ \\
				~ && && && ~ && ~ \\
				~ & ~ && && ~ & && ~ \\
				~ && [mnode=circle] 3 && [mnode=circle] 1 && && [mnode=circle] 2
				\ncline{1,5}{4,2}
				\ncline{1,5}{5,9}
				\ncline{3,7}{4,6}
				\ncarc[arcangle=40]{4,2}{5,3}
				\ncarc[arcangle=-40]{4,6}{5,5}
				\ncline{<->,arrowscale=1.5}{1,1}{3,1}
				\tlput{$\tau_{1}^{\left(123\right)}$}
				\ncline{<->,arrowscale=1.5}{3,1}{4,1}
				\tlput{$\tau_{2}^{\left(123\right)}$}
				\ncline{<->,arrowscale=1.5}{4,1}{5,1}
				\tlput{$\tau_{3}^{\left(123\right)}$}
				\psset{linestyle=dashed}\ncline{3,1}{3,7}
				\psset{linestyle=dashed}\ncline{3,7}{3,9}
				\psset{linestyle=dashed}\ncline{4,1}{4,2}
				\psset{linestyle=dashed}\ncline{4,2}{4,6}
				\psset{linestyle=dashed}\ncline{4,6}{4,9}
				\endpsmatrix
				\caption{The $\left(123\right)$ taxon label permutation.}
		\end{subfigure}
		\begin{subfigure}[h]{0.49\textwidth}
		\centering
				\psmatrix[colsep=.3cm,rowsep=.4cm,mnode=r]
				~ && && ~ \\
				~ \\
				~ && && && ~ && ~ \\
				~ & ~ && && ~ & && ~ \\
				~ && [mnode=circle] 1 && [mnode=circle] 3 && && [mnode=circle] 2
				\ncline{1,5}{4,2}
				\ncline{1,5}{5,9}
				\ncline{3,7}{4,6}
				\ncarc[arcangle=40]{4,2}{5,3}
				\ncarc[arcangle=-40]{4,6}{5,5}
				\ncline{<->,arrowscale=1.5}{1,1}{3,1}
				\tlput{$\tau_{1}^{\left(23\right)}$}
				\ncline{<->,arrowscale=1.5}{3,1}{4,1}
				\tlput{$\tau_{2}^{\left(23\right)}$}
				\ncline{<->,arrowscale=1.5}{4,1}{5,1}
				\tlput{$\tau_{3}^{\left(23\right)}$}
				\psset{linestyle=dashed}\ncline{3,1}{3,7}
				\psset{linestyle=dashed}\ncline{3,7}{3,9}
				\psset{linestyle=dashed}\ncline{4,1}{4,2}
				\psset{linestyle=dashed}\ncline{4,2}{4,6}
				\psset{linestyle=dashed}\ncline{4,6}{4,9}
				\endpsmatrix
				\caption{The $\left(23\right)$ taxon label permutation.}
		\end{subfigure}
		\begin{subfigure}[h]{0.49\textwidth}
		\centering
				\psmatrix[colsep=.3cm,rowsep=.4cm,mnode=r]
				~ && && ~ \\
				~ \\
				~ && && && ~ && ~ \\
				~ & ~ && && ~ & && ~ \\
				~ && [mnode=circle] 2 && [mnode=circle] 3 && && [mnode=circle] 1
				\ncline{1,5}{4,2}
				\ncline{1,5}{5,9}
				\ncline{3,7}{4,6}
				\ncarc[arcangle=40]{4,2}{5,3}
				\ncarc[arcangle=-40]{4,6}{5,5}
				\ncline{<->,arrowscale=1.5}{1,1}{3,1}
				\tlput{$\tau_{1}^{\left(132\right)}$}
				\ncline{<->,arrowscale=1.5}{3,1}{4,1}
				\tlput{$\tau_{2}^{\left(132\right)}$}
				\ncline{<->,arrowscale=1.5}{4,1}{5,1}
				\tlput{$\tau_{3}^{\left(132\right)}$}
				\psset{linestyle=dashed}\ncline{3,1}{3,7}
				\psset{linestyle=dashed}\ncline{3,7}{3,9}
				\psset{linestyle=dashed}\ncline{4,1}{4,2}
				\psset{linestyle=dashed}\ncline{4,2}{4,6}
				\psset{linestyle=dashed}\ncline{4,6}{4,9}
				\endpsmatrix
				\caption{The $\left(132\right)$ taxon label permutation.}
		\end{subfigure}
		\caption{The six taxon label permutations for the three-taxon convergence-divergence network.}
		\label{taxonlabellingscondiv}
\end{figure}

Unlike with the three-taxon clock-like tree, it is not immediately obvious whether the six different taxon labellings will be network identifiable from each other. To see whether the various taxon labellings are network identifiable we will solve for the time parameters for various taxon labellings in terms of the time parameters of a different taxon labelling.

\section{Pairwise Distances}

Before looking at a taxon label permutation, we wish to determine which pairs of taxa are ``closer'' to each other than other pairs on the network. We will do this by introducing the notions of \emph{tree distance} and \emph{pairwise distance}. \citet{hendy1993spectral} introduced the notion of \emph{tree distance} as a measure of how close two nodes are to each other on a tree. We will only use the term tree distance when referring to leaves on a tree.

\begin{definition}[\textbf{Tree Distance}]
The \textbf{tree distance}, $d\left(a,b\right)$, between two leaves, $a$ and $b$, on a tree is the sum of the edge lengths connecting the two leaves.
\end{definition}
\label{treedistance}

\citet{hendy1993spectral} noted that the tree distance could be found by using the Hadamard transformation. To see how this is done, let's look at the transformed phylogenetic tensor for the two-taxon clock-like tree as an example. This tree has only one variable element in the transformed phylogenetic tensor,
\begin{align}\begin{split}
q_{11}=e^{-2\tau_{1}}.
\end{split}\end{align}
Clearly the tree distance between the two leaves is
\begin{align}\begin{split}
d\left(1,2\right)=2\tau_{1}=-\ln{\left(q_{11}\right)}.
\end{split}\end{align}

We will see from the expressions for the phylogenetic tensor elements that on a tree the tree distance will simply be
\begin{align}\begin{split}
d\left(a,b\right)=-\ln\left(q_{A}\right),
\end{split}\end{align}
where $A=i_{1}i_{2}\ldots{}i_{n}$, $a,b\in{}\left\{1,2,\ldots{},n\right\}$, $i_{a}=i_{b}=1$ and $i_{k}=0$ for all $k\in{}\left\{1,2,\ldots{},n\right\}\setminus{}\left\{a,b\right\}$.

Without considering any taxon labelling, we will now examine the distances on the three-taxon convergence-divergence network. It is immediately clear that the distance between taxon two and taxon three, $d\left(2,3\right)$, is simply
\begin{align}\begin{split}
d\left(2,3\right)=2\left(\tau_{2}+\tau_{3}\right)=-\ln{\left(q_{011}\right)}.
\end{split}\end{align}

Although taxon two is converging with taxon one, it is always diverging from taxon three. It is therefore clear that the distance between taxon one and taxon three is
\begin{align}\begin{split}
d\left(1,3\right)=2\left(\tau_{1}+\tau_{2}+\tau_{3}\right)=-\ln{\left(q_{101}\right)}.
\end{split}\end{align}

Unlike $d\left(2,3\right)$ and $d\left(1,3\right)$, it is not immediately clear what $d\left(1,2\right)$ is intuitively. The convergence between taxon one and taxon two should decrease the distance between the two taxa. It is not immediately clear whether our definition of tree distance will be appropriate when we generalise our trees to convergence-divergence networks. We want to know whether our distance function is appropriate when convergence has occurred.

If we are to assume that the distance function is an appropriate choice, the distance between the two converging taxa on the three-taxon convergence-divergence network will be
\begin{align}\begin{split}
d\left(1,2\right)=-\ln{\left(q_{110}\right)}=-\ln{\left(1-e^{-\tau_{3}}\left(1-e^{-2\left(\tau_{1}+\tau_{2}\right)}\right)\right)}.
\end{split}\end{align}
We will see whether this is in fact an appropriate choice for the distance between the two converging taxa.

We will now take a moment to determine whether our distances satisfy some key criteria of a distance on a tree or network. Clearly
\begin{align}\begin{split}
d\left(2,3\right)=2\left(\tau_{2}+\tau_{3}\right)
\end{split}\end{align}
and
\begin{align}\begin{split}
d\left(1,3\right)=2\left(\tau_{1}+\tau_{2}+\tau_{3}\right)
\end{split}\end{align}
are simply the sums of the lengths of the edges along the paths from one leaf to the other leaf. These distances are intuitively what we would expect them to be. However, it is not clear intuitively whether the third distance,
\begin{align}\begin{split}
d\left(1,2\right)=-\ln{\left(1-e^{-\tau_{3}}\left(1-e^{-2\left(\tau_{1}+\tau_{2}\right)}\right)\right)},
\end{split}\end{align}
is an appropriate choice. To determine whether it is appropriate choice, we will look at its behaviour for various choices of $\tau_{3}$. Firstly, we will see what happens when $\tau_{3}=0$. When $\tau_{3}=0$, we simply have the three-taxon clock-like tree. The distance on the convergence-divergence network becomes
\begin{align}\begin{split}
d\left(1,2\right)&=-\ln{\left(1-e^{-0}\left(1-e^{-2\left(\tau_{1}+\tau_{2}\right)}\right)\right)} \\
&=-\ln{\left(1-1\left(1-e^{-2\left(\tau_{1}+\tau_{2}\right)}\right)\right)} \\
&=-\ln{\left(1-1+e^{-2\left(\tau_{1}+\tau_{2}\right)}\right)} \\
&=-\ln{\left(e^{-2\left(\tau_{1}+\tau_{2}\right)}\right)} \\
&=2\left(\tau_{1}+\tau_{2}\right),
\end{split}\end{align}
as is reasonable, since it is equal to the distance between the two equivalent leaves on the three-taxon clock-like tree.

Next, we will take the limit as $\tau_{3}\rightarrow{}\infty{}$ so that $e^{-\tau_{3}}\rightarrow{}0$. The distance on the convergence-divergence satisfies
\begin{align}\begin{split}
d\left(1,2\right)&\rightarrow{}-\ln{\left(1-0\left(1-e^{-2\left(\tau_{1}+\tau_{2}\right)}\right)\right)} \\
&=-\ln{\left(1\right)} \\
&=0.
\end{split}\end{align}
If taxon one and taxon two have converged for an infinitely long period of time the distance between the two taxa will be zero. This is consistent with what we expect with convergence.

For a ``small'' choice of $\tau_{3}$, the distance should be ``slightly'' less than when $\tau_{3}=0$. We will let $\tau_{3}=\epsilon{}$, a ``small'' positive number. When $\tau_{3}=\epsilon{}$,
\begin{align}\begin{split}
d\left(1,2\right)&=-\ln{\left(1-e^{-\epsilon{}}\left(1-e^{-2\left(\tau_{1}+\tau_{2}\right)}\right)\right)} \\
&\approx{}-\ln{\left(1-\left(1-\epsilon{}\right)\left(1-e^{-2\left(\tau_{1}+\tau_{2}\right)}\right)\right)} \\
&=-\ln{\left(1-1+\epsilon{}+e^{-2\left(\tau_{1}+\tau_{2}\right)}-\epsilon{}e^{-2\left(\tau_{1}+\tau_{2}\right)}\right)} \\
&=-\ln{\left(e^{-2\left(\tau_{1}+\tau_{2}\right)}+\epsilon{}\left(1-e^{-2\left(\tau_{1}+\tau_{2}\right)}\right)\right)} \\
&=-\ln{\left(e^{-2\left(\tau_{1}+\tau_{2}\right)}\left(1+\epsilon{}\frac{\left(1-e^{-2\left(\tau_{1}+\tau_{2}\right)}\right)}{e^{-2\left(\tau_{1}+\tau_{2}\right)}}\right)\right)} \\
&=-\ln{e^{-2\left(\tau_{1}+\tau_{2}\right)}}-\ln{\left(1+\epsilon{}\frac{\left(1-e^{-2\left(\tau_{1}+\tau_{2}\right)}\right)}{e^{-2\left(\tau_{1}+\tau_{2}\right)}}\right)} \\
&=2\left(\tau_{1}+\tau_{2}\right)-\ln{\left(1+\epsilon{}\frac{\left(1-e^{-2\left(\tau_{1}+\tau_{2}\right)}\right)}{e^{-2\left(\tau_{1}+\tau_{2}\right)}}\right)}
\end{split}\end{align}

Before we expand the second term into its appropriate Taylor series, we need to know that the argument for the natural logarithm is in the interval $\left(0,2\right)$,
\begin{align}\begin{split}
0<1+\epsilon{}\frac{\left(1-e^{-2\left(\tau_{1}+\tau_{2}\right)}\right)}{e^{-2\left(\tau_{1}+\tau_{2}\right)}}<2.
\end{split}\end{align}

Since $e^{-2\left(\tau_{1}+\tau_{2}\right)}\in{}(0,1]$ and $1-e^{-2\left(\tau_{1}+\tau_{2}\right)}\in{}[0,1)$, it is clear that
\begin{align}\begin{split}
1+\epsilon{}\frac{\left(1-e^{-2\left(\tau_{1}+\tau_{2}\right)}\right)}{e^{-2\left(\tau_{1}+\tau_{2}\right)}}>0.
\end{split}\end{align}

For the other inequality,
\begin{align}\begin{split}
1+\epsilon{}\frac{\left(1-e^{-2\left(\tau_{1}+\tau_{2}\right)}\right)}{e^{-2\left(\tau_{1}+\tau_{2}\right)}}&<2 \\
\Leftrightarrow{}\epsilon{}\frac{\left(1-e^{-2\left(\tau_{1}+\tau_{2}\right)}\right)}{e^{-2\left(\tau_{1}+\tau_{2}\right)}}&<1 \\
\Leftrightarrow{}\epsilon{}\left(1-e^{-2\left(\tau_{1}+\tau_{2}\right)}\right)&<e^{-2\left(\tau_{1}+\tau_{2}\right)} \\
\Leftrightarrow{}\epsilon{}-\epsilon{}e^{-2\left(\tau_{1}+\tau_{2}\right)}&<e^{-2\left(\tau_{1}+\tau_{2}\right)} \\
\Leftrightarrow{}\epsilon{}&<\left(1+\epsilon{}\right)e^{-2\left(\tau_{1}+\tau_{2}\right)} \\
\Leftrightarrow{}\frac{\epsilon{}}{1+\epsilon{}}&<e^{-2\left(\tau_{1}+\tau_{2}\right)}.
\end{split}\end{align}

Since we are taking $\tau_{1}$ and $\tau_{2}$ to be finite, $e^{-2\left(\tau_{1}+\tau_{2}\right)}$ will also be finite. As $\epsilon{}\rightarrow{}0$, $\frac{\epsilon{}}{1+\epsilon{}}\rightarrow{}0$. Therefore, there will always be some ``small'' $\epsilon{}$ such that
\begin{align}\begin{split}
\frac{\epsilon{}}{1+\epsilon{}}&<e^{-2\left(\tau_{1}+\tau_{2}\right)}.
\end{split}\end{align}

We will now go ahead and take the Taylor series expansion of the distance function,
\begin{align}\begin{split}
d\left(1,2\right)&\approx{}2\left(\tau_{1}+\tau_{2}\right)-\ln{\left(1+\epsilon{}\frac{\left(1-e^{-2\left(\tau_{1}+\tau_{2}\right)}\right)}{e^{-2\left(\tau_{1}+\tau_{2}\right)}}\right)} \\
&=2\left(\tau_{1}+\tau_{2}\right)-\epsilon{}\frac{\left(1-e^{-2\left(\tau_{1}+\tau_{2}\right)}\right)}{e^{-2\left(\tau_{1}+\tau_{2}\right)}}+O\left(\epsilon{}^{2}\right),
\end{split}\end{align}
where $O\left(\epsilon{}^{2}\right)$ is all the terms of order $\epsilon{}^{2}$ or higher.

If $\epsilon{}$ is ``small'' then the lower order terms will dominate,
\begin{align}\begin{split}
d\left(1,2\right)\approx{}2\left(\tau_{1}+\tau_{2}\right)-\epsilon{}\frac{\left(1-e^{-2\left(\tau_{1}+\tau_{2}\right)}\right)}{e^{-2\left(\tau_{1}+\tau_{2}\right)}}\leq{}2\left(\tau_{1}+\tau_{2}\right),
\end{split}\end{align}
as we require.

Hence, when $\tau_{3}$ is ``small'' the two taxa will be closer in distance than when $\tau_{3}=0$ by a ``small'' amount.

The last criterion that we want to check is that $d\left(1,2\right)$ decreases monotonically as $\tau_{3}$ increases from $\tau_{3}=0$. To check this, we will look at the partial derivative of $d\left(1,2\right)$ with respect to $\tau_{3}$. The partial derivative is
\begin{align}\begin{split}
\frac{\partial{}d\left(1,2\right)}{\partial{}\tau_{3}}&=\frac{\partial{}}{\partial{}\tau_{3}}\left(-\ln{\left(1-e^{-\tau_{3}}\left(1-e^{-2\left(\tau_{1}+\tau_{2}\right)}\right)\right)}\right) \\
&=-\frac{e^{-\tau_{3}}\left(1-e^{-2\left(\tau_{1}+\tau_{2}\right)}\right)}{1-e^{-\tau_{3}}\left(1-e^{-2\left(\tau_{1}+\tau_{2}\right)}\right)}\leq{}0,
\end{split}\end{align}
as we require.

It is important to note that our function, $d\left(1,2\right)$, is not the only function that could behave in the appropriate manner for a distance. For $d\left(1,2\right)$ to be an appropriate distance function, it is necessary, but potentially not sufficient, for it to meet the criteria above.

We will expand the definition of tree distance to include distances between two leaves on a convergence-divergence network. We will call this distance the \emph{pairwise distance}.

\begin{definition}[\textbf{Pairwise Distance}]
The \textbf{pairwise distance}, $d\left(a,b\right)$, between two leaves, $a$ and $b$, on a convergence-divergence network is expressed as
\begin{align*}\begin{split}
d\left(a,b\right)=-\ln{\left(q_{A}\right)},
\end{split}\end{align*}
where $A=i_{1}i_{2}\ldots{}i_{n}$, $a,b\in{}\left\{1,2,\ldots{},n\right\}$, $a\neq{}b$, $i_{a}=i_{b}=1$ and $i_{k}=0$ for all $k\in{}\left\{1,2,\ldots{},n\right\}\setminus{}\left\{a,b\right\}$.
\end{definition}
\label{pairwisedistance}

If the convergence-divergence network has no convergence and is a tree, then the pairwise distance will be equivalent to the tree distance.

We will now go ahead and compare the distances on different taxon label permutations of the three-taxon convergence-divergence network.

\section{The $\left(12\right)$ Taxon Label Permutation}

\begin{figure}[H]
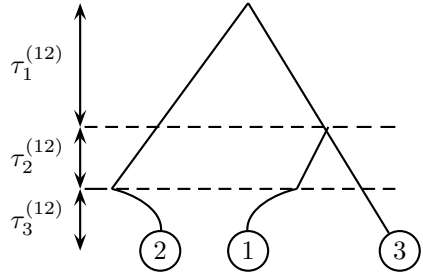

	\centering
		\psmatrix[colsep=.3cm,rowsep=.4cm,mnode=r]
		~ && && ~ \\
		~ \\
		~ && && && ~ && ~ \\
		~ & ~ && && ~ & && ~ \\
		~ && [mnode=circle] 2 && [mnode=circle] 1 && && [mnode=circle] 3
		\ncline{1,5}{4,2}
		\ncline{1,5}{5,9}
		\ncline{3,7}{4,6}
		\ncarc[arcangle=40]{4,2}{5,3}
		\ncarc[arcangle=-40]{4,6}{5,5}
		\ncline{<->,arrowscale=1.5}{1,1}{3,1}
		\tlput{$\tau_{1}^{(12)}$}
		\ncline{<->,arrowscale=1.5}{3,1}{4,1}
		\tlput{$\tau_{2}^{(12)}$}
		\ncline{<->,arrowscale=1.5}{4,1}{5,1}
		\tlput{$\tau_{3}^{(12)}$}
		\psset{linestyle=dashed}\ncline{3,1}{3,7}
		\psset{linestyle=dashed}\ncline{3,7}{3,9}
		\psset{linestyle=dashed}\ncline{4,1}{4,2}
		\psset{linestyle=dashed}\ncline{4,2}{4,6}
		\psset{linestyle=dashed}\ncline{4,6}{4,9}
		\endpsmatrix
		\caption{The $\left(12\right)$ taxon label permutation on the three-taxon clock-like convergence-divergence network.}
		\label{12permutationnonsisterconvergence}
\end{figure}

We will now examine the convergence-divergence network with convergence between non-sister taxa, but with the first and second taxon labels permuted, the $\left(12\right)$ taxon label permutation, shown in Figure~\ref{12permutationnonsisterconvergence}.

Recall from Chapter~\ref{chapter4} that the constraints on the transformed phylogenetic tensor elements for the three-taxon clock-like convergence-divergence network were
\begin{align}\begin{split}
\left\{q_{011}\geq{}q_{101},\quad{}q_{110}\geq{}q_{101},\quad{}q_{011}\left(1-q_{110}\right)^{2}\geq{}\left(q_{011}-q_{101}\right)^{2}\right\}.
\end{split}\end{align}
On the $\left(12\right)$ taxon label permutation the set of constraints becomes
\begin{align}\begin{split}
\left\{q_{101}^{\left(12\right)}\geq{}q_{011}^{\left(12\right)},\quad{}q_{110}^{\left(12\right)}\geq{}q_{011}^{\left(12\right)},\quad{}q_{101}^{\left(12\right)}\left(1-q_{110}^{\left(12\right)}\right)^{2}\geq{}\left(q_{101}^{\left(12\right)}-q_{011}^{\left(12\right)}\right)^{2}\right\}.
\end{split}\end{align}
If we compare this set of constraints to the set of constraints with no taxon label permutation, we find that the intersection of the two sets of constraints is empty, except for when $q_{011}=q_{101}$, where the two sets of constraints reduce to
\begin{align}\begin{split}
\left\{q_{011}=q_{101},\quad{}q_{110}\geq{}q_{101}\right\}.
\end{split}\end{align}
This is the same set of constraints as the clock-like tree with the $\left(13\right)$ and $\left(123\right)$ taxon label permutations.

We can conclude that the clock-like convergence-divergence network with the $\left(12\right)$ taxon label permutation is distinguishable from the clock-like convergence-divergence network with no taxon label permutation. We will not look at the distinguishability of other taxon label permutations from each other. Instead, we will examine whether they are network identifiable from each other.

Suppose now that we have a set of non-negative time parameters, $\tau_{1}^{\left(\right)}$, $\tau_{2}^{\left(\right)}$ and $\tau_{3}^{\left(\right)}$, for the three-taxon clock-like convergence-divergence network with no taxon label permutation. Now suppose that we wish to find the time parameters on the $\left(12\right)$ taxon label permutation that give rise to the same transformed phylogenetic tensor elements.

The expressions for the transformed phylogenetic tensor elements on the network with the $\left(12\right)$ taxon label permutation in terms of the phylogenetic tensor elements on the network with no taxon label permutation are
\begin{align}\begin{split}
\begin{mycases}
q_{101}^{\left(12\right)}\left(\tau_{1}^{\left(12\right)},\tau_{2}^{\left(12\right)},\tau_{3}^{\left(12\right)}\right)&=q_{011}^{\left(\right)}\left(\tau_{1}^{\left(\right)},\tau_{2}^{\left(\right)},\tau_{3}^{\left(\right)}\right), \\
q_{011}^{\left(12\right)}\left(\tau_{1}^{\left(12\right)},\tau_{2}^{\left(12\right)},\tau_{3}^{\left(12\right)}\right)&=q_{101}^{\left(\right)}\left(\tau_{1}^{\left(\right)},\tau_{2}^{\left(\right)},\tau_{3}^{\left(\right)}\right), \\
q_{110}^{\left(12\right)}\left(\tau_{1}^{\left(12\right)},\tau_{2}^{\left(12\right)},\tau_{3}^{\left(12\right)}\right)&=q_{110}^{\left(\right)}\left(\tau_{1}^{\left(\right)},\tau_{2}^{\left(\right)},\tau_{3}^{\left(\right)}\right).
\end{mycases}
\end{split}\end{align}

We will now find expressions for the time parameters for the network with the $\left(12\right)$ taxon label permutation in terms of the time parameters for the network with no taxon label permutation. The three pairwise distances for the network with no taxon label permutation are
\begin{align}\begin{split}
\begin{mycases}
d\left(2,3\right)&=-\ln{\left(q_{011}^{\left(\right)}\left(\tau_{1}^{\left(\right)},\tau_{2}^{\left(\right)},\tau_{3}^{\left(\right)}\right)\right)}, \\
d\left(1,3\right)&=-\ln{\left(q_{101}^{\left(\right)}\left(\tau_{1}^{\left(\right)},\tau_{2}^{\left(\right)},\tau_{3}^{\left(\right)}\right)\right)}, \\
d\left(1,2\right)&=-\ln{\left(q_{110}^{\left(\right)}\left(\tau_{1}^{\left(\right)},\tau_{2}^{\left(\right)},\tau_{3}^{\left(\right)}\right)\right)}.
\end{mycases}
\end{split}\end{align}

The three pairwise distances in terms of the transformed phylogenetic tensor elements for the $\left(12\right)$ taxon label permutation are then
\begin{align}\begin{split}
\begin{mycases}
d\left(2,3\right)&=-\ln{\left(q_{101}^{\left(12\right)}\left(\tau_{1}^{\left(12\right)},\tau_{2}^{\left(12\right)},\tau_{3}^{\left(12\right)}\right)\right)}, \\
d\left(1,3\right)&=-\ln{\left(q_{011}^{\left(12\right)}\left(\tau_{1}^{\left(12\right)},\tau_{2}^{\left(12\right)},\tau_{3}^{\left(12\right)}\right)\right)}, \\
d\left(1,2\right)&=-\ln{\left(q_{110}^{\left(12\right)}\left(\tau_{1}^{\left(12\right)},\tau_{2}^{\left(12\right)},\tau_{3}^{\left(12\right)}\right)\right)}.
\end{mycases}
\end{split}\end{align}

We can now solve for the time parameters for the $\left(12\right)$ taxon label permutation in terms of the time parameters for the network with no taxon label permutation by equating the two sets of expressions for the pairwise distances. For the derivation of the solutions to these equations see Appendix~\ref{derivation1}. The time parameters for the network with the $\left(12\right)$ taxon label permutation in terms of the time parameters for the network with no taxon label permutation are
\begin{align}\begin{split}
\begin{mycases}
\tau_{1}^{\left(12\right)}&=-\tau_{1}^{\left(\right)}, \\
\tau_{2}^{\left(12\right)}&=\ln{\left(\frac{\left(1-e^{-2\left(\tau_{1}^{\left(\right)}+\tau_{2}^{\left(\right)}\right)}\right)+\sqrt{\left(1-e^{-2\left(\tau_{1}^{\left(\right)}+\tau_{2}^{\left(\right)}\right)}\right)^{2}+4e^{-2\tau_{2}^{\left(\right)}}}}{2e^{-\left(\tau_{1}^{\left(\right)}+\tau_{2}^{\left(\right)}\right)}}\right)}, \\
\tau_{3}^{\left(12\right)}&=\ln{\left(\frac{-\left(1-e^{-2\left(\tau_{1}^{\left(\right)}+\tau_{2}^{\left(\right)}\right)}\right)+\sqrt{\left(1-e^{-2\left(\tau_{1}^{\left(\right)}+\tau_{2}^{\left(\right)}\right)}\right)^{2}+4e^{-2\tau_{2}^{\left(\right)}}}}{2e^{-\left(2\tau_{2}^{\left(\right)}+\tau_{3}^{\left(\right)}\right)}}\right)}.
\end{mycases}
\end{split}\end{align}

Clearly $\tau_{1}^{\left(12\right)}\leq{}0$. We can conclude that the $\left(12\right)$ taxon label permutation is network identifiable from no taxon label permutation. Although it is not necessary to look at $\tau_{2}^{\left(12\right)}$ or $\tau_{3}^{\left(12\right)}$, we will for curiosity anyway. From Appendix~\ref{derivation1}, we can conclude that $\tau_{2}^{\left(12\right)}$ will always be non-negative, while $\tau_{3}^{\left(12\right)}$ can be non-negative or negative. From simulations for $x_{1},x_{2},x_{3}\in{}U\left(0,1\right)$, we found that $\tau_{3}^{\left(12\right)}$ was non-negative $434$ times out of $500$. We can conclude that the two taxon label permutations, the $\left(12\right)$ taxon label permutation and no taxon label permutation, are network identifiable from each other.

The only case where both sets of time parameters are non-negative is when $\tau_{1}^{\left(\right)}=\tau_{1}^{\left(12\right)}=0$, where there is a trifurcation at the root of both of the convergence-divergence networks. When this happens all other time parameters for both taxon label permutations are non-negative, with $\tau_{2}^{\left(\right)}=\tau_{2}^{\left(12\right)}$ and $\tau_{3}^{\left(\right)}=\tau_{3}^{\left(12\right)}$. If $\tau_{1}^{\left(\right)}>0$ then it will not be possible to find a set of non-negative time parameters on the $\left(12\right)$ taxon label permutation which gives rise to equivalent pairwise distances or transformed phylogenetic tensor elements on the convergence-divergence network with no taxon label permutation.

If we were given a pattern frequency and we wished to find the maximum likelihood for both of the taxon label permutations on the three-taxon convergence-divergence network then we would find that the maximum likelihoods would not be equal unless $\tau_{1}^{\left(\right)}=\tau_{1}^{\left(12\right)}=0$. Suppose $\tau_{1}^{\left(\right)}$, $\tau_{2}^{\left(\right)}$ and $\tau_{3}^{\left(\right)}$ are all non-negative and are the maximum likelihood time parameters. For the maximum likelihood to be equal under the $\left(12\right)$ taxon label permutation $\tau_{1}^{\left(12\right)}$ cannot be positive and $\tau_{3}^{\left(12\right)}$ may also be negative. If we constrain $\tau_{1}^{\left(12\right)}$, $\tau_{2}^{\left(12\right)}$ and $\tau_{3}^{\left(12\right)}$ to all be non-negative then the maximum likelihoods for both taxon label permutations will not be equal.

\section{The $\left(13\right)$ Taxon Label Permutation}

We will now address the question of network identifiability of the $\left(13\right)$ taxon label permutation from no taxon label permutation on the three-taxon clock-like convergence-divergence network. The $\left(13\right)$ taxon label permutation on the three-taxon clock-like convergence-divergence network is shown below in Figure~\ref{13permutationnonsisterconvergence}.
\begin{figure}[H]
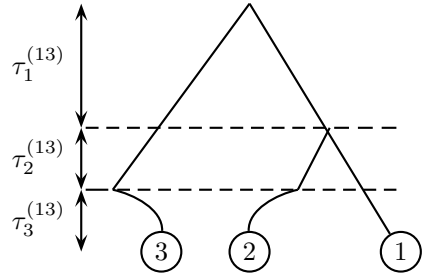

	\centering
		\psmatrix[colsep=.3cm,rowsep=.4cm,mnode=r]
		~ && && ~ \\
		~ \\
		~ && && && ~ && ~ \\
		~ & ~ && && ~ & && ~ \\
		~ && [mnode=circle] 3 && [mnode=circle] 2 && && [mnode=circle] 1
		\ncline{1,5}{4,2}
		\ncline{1,5}{5,9}
		\ncline{3,7}{4,6}
		\ncarc[arcangle=40]{4,2}{5,3}
		\ncarc[arcangle=-40]{4,6}{5,5}
		\ncline{<->,arrowscale=1.5}{1,1}{3,1}
		\tlput{$\tau_{1}^{(13)}$}
		\ncline{<->,arrowscale=1.5}{3,1}{4,1}
		\tlput{$\tau_{2}^{(13)}$}
		\ncline{<->,arrowscale=1.5}{4,1}{5,1}
		\tlput{$\tau_{3}^{(13)}$}
		\psset{linestyle=dashed}\ncline{3,1}{3,7}
		\psset{linestyle=dashed}\ncline{3,7}{3,9}
		\psset{linestyle=dashed}\ncline{4,1}{4,2}
		\psset{linestyle=dashed}\ncline{4,2}{4,6}
		\psset{linestyle=dashed}\ncline{4,6}{4,9}
		\endpsmatrix
		\caption{The $\left(13\right)$ taxon label permutation on the three-taxon clock-like convergence-divergence network.}
		\label{13permutationnonsisterconvergence}
\end{figure}

Equating the transformed phylogenetic tensor elements for this taxon label permutation to no taxon label permutation,
\begin{align}\begin{split}
\begin{mycases}
q_{110}^{\left(13\right)}\left(\tau_{1}^{\left(13\right)},\tau_{2}^{\left(13\right)},\tau_{3}^{\left(13\right)}\right)&=q_{011}^{\left(\right)}\left(\tau_{1}^{\left(\right)},\tau_{2}^{\left(\right)},\tau_{3}^{\left(\right)}\right), \\
q_{101}^{\left(13\right)}\left(\tau_{1}^{\left(13\right)},\tau_{2}^{\left(13\right)},\tau_{3}^{\left(13\right)}\right)&=q_{101}^{\left(\right)}\left(\tau_{1}^{\left(\right)},\tau_{2}^{\left(\right)},\tau_{3}^{\left(\right)}\right), \\
q_{011}^{\left(13\right)}\left(\tau_{1}^{\left(13\right)},\tau_{2}^{\left(13\right)},\tau_{3}^{\left(13\right)}\right)&=q_{110}^{\left(\right)}\left(\tau_{1}^{\left(\right)},\tau_{2}^{\left(\right)},\tau_{3}^{\left(\right)}\right).
\end{mycases}
\end{split}\end{align}
The three pairwise distances in terms of the transformed phylogenetic tensor elements for the $\left(13\right)$ taxon label permutation are then
\begin{align}\begin{split}
\begin{mycases}
d\left(2,3\right)&=-\ln{\left(q_{110}^{\left(13\right)}\left(\tau_{1}^{\left(13\right)},\tau_{2}^{\left(13\right)},\tau_{3}^{\left(13\right)}\right)\right)}, \\
d\left(1,3\right)&=-\ln{\left(q_{101}^{\left(13\right)}\left(\tau_{1}^{\left(13\right)},\tau_{2}^{\left(13\right)},\tau_{3}^{\left(13\right)}\right)\right)}, \\
d\left(1,2\right)&=-\ln{\left(q_{011}^{\left(13\right)}\left(\tau_{1}^{\left(13\right)},\tau_{2}^{\left(13\right)},\tau_{3}^{\left(13\right)}\right)\right)}.
\end{mycases}
\end{split}\end{align}
In Appendix~\ref{derivation2} we solve for the time parameters for the $\left(13\right)$ taxon label permutation in terms of no taxon label permutation. Those time parameters are
\begin{align}\begin{split}
\begin{mycases}
\tau_{1}^{\left(13\right)}=&\ln{\sqrt{\frac{1-e^{-\tau_{3}^{\left(\right)}}\left(1-e^{-2\left(\tau_{1}^{\left(\right)}+\tau_{2}^{\left(\right)}\right)}\right)}{e^{-2\left(\tau_{1}^{\left(\right)}+\tau_{2}^{\left(\right)}+\tau_{3}^{\left(\right)}\right)}}}}, \\
\tau_{2}^{\left(13\right)}=&\ln{\left(\frac{\left(1-e^{-2\left(\tau_{2}^{\left(\right)}+\tau_{3}^{\left(\right)}\right)}\right)+\sqrt{\left(1-e^{-2\left(\tau_{2}^{\left(\right)}+\tau_{3}^{\left(\right)}\right)}\right)^{2}+4e^{-2\left(\tau_{1}^{\left(\right)}+\tau_{2}^{\left(\right)}+\tau_{3}^{\left(\right)}\right)}}}{2\sqrt{1-e^{-\tau_{3}^{\left(\right)}}\left(1-e^{-2\left(\tau_{1}^{\left(\right)}+\tau_{2}^{\left(\right)}\right)}\right)}}\right)}, \\
\tau_{3}^{\left(13\right)}=&\ln{\left(\frac{-\left(1-e^{-2\left(\tau_{2}^{\left(\right)}+\tau_{3}^{\left(\right)}\right)}\right)+\sqrt{\left(1-e^{-2\left(\tau_{2}^{\left(\right)}+\tau_{3}^{\left(\right)}\right)}\right)^{2}+4e^{-2\left(\tau_{1}^{\left(\right)}+\tau_{2}^{\left(\right)}+\tau_{3}^{\left(\right)}\right)}}}{2e^{-2\left(\tau_{1}^{\left(\right)}+\tau_{2}^{\left(\right)}+\tau_{3}^{\left(\right)}\right)}}\right)}.
\end{mycases}
\end{split}\end{align}

In Appendix~\ref{derivation2} we have shown that two of the time parameters on the $\left(13\right)$ taxon label permutation, $\tau_{1}^{\left(13\right)}$ and $\tau_{3}^{\left(13\right)}$, are non-negative, however we have also shown that $\tau_{2}^{\left(13\right)}$ is almost always non-negative but can occasionally be negative. We can conclude that the $\left(13\right)$ taxon label permutation is network identifiable from no taxon label permutation. However, for $497$ out of $500$ simulations for $x_{1},x_{2},x_{3}\in{}U\left(0,1\right)$ all three of the time parameters were non-negative. In these cases the maximum likelihoods for the two taxon label permutations will be identical and we will have a choice between no taxon label permutation or the $\left(13\right)$ taxon label permutation. We will discuss at the end of the chapter how we could choose between the two taxon label permutations.

\section{The $\left(23\right)$ Taxon Label Permutation}

We will now address the question of network identifiability of the $\left(23\right)$ taxon label permutation from no taxon label permutation on the three-taxon clock-like convergence-divergence network. The $\left(23\right)$ taxon label permutation on the three-taxon clock-like convergence-divergence network is shown below in Figure~\ref{23permutationnonsisterconvergence}.
\begin{figure}[H]
	\centering
		\psmatrix[colsep=.3cm,rowsep=.4cm,mnode=r]
		~ && && ~ \\
		~ \\
		~ && && && ~ && ~ \\
		~ & ~ && && ~ & && ~ \\
		~ && [mnode=circle] 1 && [mnode=circle] 3 && && [mnode=circle] 2
		\ncline{1,5}{4,2}
		\ncline{1,5}{5,9}
		\ncline{3,7}{4,6}
		\ncarc[arcangle=40]{4,2}{5,3}
		\ncarc[arcangle=-40]{4,6}{5,5}
		\ncline{<->,arrowscale=1.5}{1,1}{3,1}
		\tlput{$\tau_{1}^{(23)}$}
		\ncline{<->,arrowscale=1.5}{3,1}{4,1}
		\tlput{$\tau_{2}^{(23)}$}
		\ncline{<->,arrowscale=1.5}{4,1}{5,1}
		\tlput{$\tau_{3}^{(23)}$}
		\psset{linestyle=dashed}\ncline{3,1}{3,7}
		\psset{linestyle=dashed}\ncline{3,7}{3,9}
		\psset{linestyle=dashed}\ncline{4,1}{4,2}
		\psset{linestyle=dashed}\ncline{4,2}{4,6}
		\psset{linestyle=dashed}\ncline{4,6}{4,9}
		\endpsmatrix
		\caption{The $\left(23\right)$ taxon label permutation on the three-taxon clock-like convergence-divergence network.}
		\label{23permutationnonsisterconvergence}
\end{figure}

Equating the transformed phylogenetic tensor elements for this taxon label permutation to no taxon label permutation,
\begin{align}\begin{split}
\begin{mycases}
q_{011}^{\left(23\right)}\left(\tau_{1}^{\left(23\right)},\tau_{2}^{\left(23\right)},\tau_{3}^{\left(23\right)}\right)&=q_{011}^{\left(\right)}\left(\tau_{1}^{\left(\right)},\tau_{2}^{\left(\right)},\tau_{3}^{\left(\right)}\right), \\
q_{110}^{\left(23\right)}\left(\tau_{1}^{\left(23\right)},\tau_{2}^{\left(23\right)},\tau_{3}^{\left(23\right)}\right)&=q_{101}^{\left(\right)}\left(\tau_{1}^{\left(\right)},\tau_{2}^{\left(\right)},\tau_{3}^{\left(\right)}\right), \\
q_{101}^{\left(23\right)}\left(\tau_{1}^{\left(23\right)},\tau_{2}^{\left(23\right)},\tau_{3}^{\left(23\right)}\right)&=q_{110}^{\left(\right)}\left(\tau_{1}^{\left(\right)},\tau_{2}^{\left(\right)},\tau_{3}^{\left(\right)}\right).
\end{mycases}
\end{split}\end{align}
The three pairwise distances in terms of the transformed phylogenetic tensor elements for the $\left(23\right)$ taxon label permutation are then
\begin{align}\begin{split}
\begin{mycases}
d\left(2,3\right)&=-\ln{\left(q_{011}^{\left(23\right)}\left(\tau_{1}^{\left(23\right)},\tau_{2}^{\left(23\right)},\tau_{3}^{\left(23\right)}\right)\right)}, \\
d\left(1,3\right)&=-\ln{\left(q_{110}^{\left(23\right)}\left(\tau_{1}^{\left(23\right)},\tau_{2}^{\left(23\right)},\tau_{3}^{\left(23\right)}\right)\right)}, \\
d\left(1,2\right)&=-\ln{\left(q_{101}^{\left(23\right)}\left(\tau_{1}^{\left(23\right)},\tau_{2}^{\left(23\right)},\tau_{3}^{\left(23\right)}\right)\right)}.
\end{mycases}
\end{split}\end{align}
In Appendix~\ref{derivation3} we solve for the time parameters for the $\left(23\right)$ taxon label permutation in terms of no taxon label permutation. Those time parameters are
\begin{align}\begin{split}
\begin{mycases}
\tau_{1}^{\left(23\right)}&=\ln{\sqrt{\frac{e^{-2\left(\tau_{2}^{\left(\right)}+\tau_{3}^{\left(\right)}\right)}}{1-e^{-\tau_{3}^{\left(\right)}}\left(1-e^{-2\left(\tau_{1}^{\left(\right)}+\tau_{2}^{\left(\right)}\right)}\right)}}}, \\
\tau_{2}^{\left(23\right)}&=\scriptstyle{\ln{\left(\frac{\left(1-e^{-2\left(\tau_{1}^{\left(\right)}+\tau_{2}^{\left(\right)}+\tau_{3}^{\left(\right)}\right)}\right)+\sqrt{\left(1-e^{-2\left(\tau_{1}^{\left(\right)}+\tau_{2}^{\left(\right)}+\tau_{3}^{\left(\right)}\right)}\right)^{2}+4\left(1-e^{-\tau_{3}^{\left(\right)}}\left(1-e^{-2\left(\tau_{1}^{\left(\right)}+\tau_{2}^{\left(\right)}\right)}\right)\right)}}{2e^{-\left(\tau_{2}^{\left(\right)}+\tau_{3}^{\left(\right)}\right)}}\right)}}, \\
\tau_{3}^{\left(23\right)}&=\scriptstyle{\ln{\left(\frac{-\left(1-e^{-2\left(\tau_{1}^{\left(\right)}+\tau_{2}^{\left(\right)}+\tau_{3}^{\left(\right)}\right)}\right)+\sqrt{\left(1-e^{-2\left(\tau_{1}^{\left(\right)}+\tau_{2}^{\left(\right)}+\tau_{3}^{\left(\right)}\right)}\right)^{2}+4\left(1-e^{-\tau_{3}^{\left(\right)}}\left(1-e^{-2\left(\tau_{1}^{\left(\right)}+\tau_{2}^{\left(\right)}\right)}\right)\right)}}{2\left(1-e^{-\tau_{3}^{\left(\right)}}\left(1-e^{-2\left(\tau_{1}^{\left(\right)}+\tau_{2}^{\left(\right)}\right)}\right)\right)}\right)}}. \\
\end{mycases}
\end{split}\end{align}

From Appendix~\ref{derivation3}, we have shown that $\tau_{1}^{\left(23\right)}$ can be either non-negative or negative, $\tau_{2}^{\left(23\right)}$ must be non-negative and $\tau_{3}^{\left(23\right)}$ must be non-positive. When simulating samples for $x_{1},x_{2},x_{3}\in{}U\left(0,1\right)$, we found that $\tau_{1}^{\left(23\right)}$ was non-negative $52$ times out of $500$. We can conclude that the $\left(23\right)$ taxon label permutation is network identifiable from no taxon label permutation.

\section{Comparing the $\left(12\right)$ Permutation to the $\left(23\right)$ Permutation}

Having compared the $\left(12\right)$ taxon label permutation to no taxon label permutation and the $\left(23\right)$ taxon label permutation to no taxon label permutation, we will now compare the $\left(12\right)$ taxon label permutation to the $\left(23\right)$ taxon label permutation. We will address the question of network identifiability of the $\left(23\right)$ taxon label permutation from the $\left(12\right)$ taxon label permutation on the three-taxon clock-like convergence-divergence network. The $\left(23\right)$ taxon label permutation on the three-taxon clock-like convergence-divergence network is shown in Figure~\ref{23permutationnonsisterconvergence}~on~page~\pageref{23permutationnonsisterconvergence}. The $\left(12\right)$ taxon label permutation on the three-taxon clock-like convergence-divergence network is shown in Figure~\ref{12permutationnonsisterconvergence}~on~page~\pageref{12permutationnonsisterconvergence}.

Equating the transformed phylogenetic tensor elements for the two taxon label permutations,
\begin{align}\begin{split}
\begin{mycases}
q_{011}^{\left(23\right)}\left(\tau_{1}^{\left(23\right)},\tau_{2}^{\left(23\right)},\tau_{3}^{\left(23\right)}\right)&=q_{101}^{\left(12\right)}\left(\tau_{1}^{\left(12\right)},\tau_{2}^{\left(12\right)},\tau_{3}^{\left(12\right)}\right), \\
q_{110}^{\left(23\right)}\left(\tau_{1}^{\left(23\right)},\tau_{2}^{\left(23\right)},\tau_{3}^{\left(23\right)}\right)&=q_{011}^{\left(12\right)}\left(\tau_{1}^{\left(12\right)},\tau_{2}^{\left(12\right)},\tau_{3}^{\left(12\right)}\right), \\
q_{101}^{\left(23\right)}\left(\tau_{1}^{\left(23\right)},\tau_{2}^{\left(23\right)},\tau_{3}^{\left(23\right)}\right)&=q_{110}^{\left(12\right)}\left(\tau_{1}^{\left(12\right)},\tau_{2}^{\left(12\right)},\tau_{3}^{\left(12\right)}\right).
\end{mycases}
\end{split}\end{align}
The three pairwise distances in terms of the transformed phylogenetic tensor elements for the two taxon label permutations are then
\begin{align}\begin{split}
\begin{mycases}
d\left(2,3\right)=-\ln{\left(q_{011}^{\left(23\right)}\left(\tau_{1}^{\left(23\right)},\tau_{2}^{\left(23\right)},\tau_{3}^{\left(23\right)}\right)\right)}&=-\ln{\left(q_{101}^{\left(12\right)}\left(\tau_{1}^{\left(12\right)},\tau_{2}^{\left(12\right)},\tau_{3}^{\left(12\right)}\right)\right)}, \\
d\left(1,3\right)=-\ln{\left(q_{110}^{\left(23\right)}\left(\tau_{1}^{\left(23\right)},\tau_{2}^{\left(23\right)},\tau_{3}^{\left(23\right)}\right)\right)}&=-\ln{\left(q_{011}^{\left(12\right)}\left(\tau_{1}^{\left(12\right)},\tau_{2}^{\left(12\right)},\tau_{3}^{\left(23\right)}\right)\right)}, \\
d\left(1,2\right)=-\ln{\left(q_{101}^{\left(23\right)}\left(\tau_{1}^{\left(23\right)},\tau_{2}^{\left(23\right)},\tau_{3}^{\left(23\right)}\right)\right)}&=-\ln{\left(q_{110}^{\left(12\right)}\left(\tau_{1}^{\left(12\right)},\tau_{2}^{\left(12\right)},\tau_{3}^{\left(23\right)}\right)\right)}.
\end{mycases}
\end{split}\end{align}
In Appendix~\ref{derivation4} we solve for the time parameters for the $\left(23\right)$ taxon label permutation in terms of the $\left(12\right)$ label permutation. Those time parameters are
\begin{align}\begin{split}
\begin{mycases}
\tau_{1}^{\left(23\right)}&=\ln{\sqrt{\frac{e^{-2\left(\tau_{1}^{\left(12\right)}+\tau_{2}^{\left(12\right)}+\tau_{3}^{\left(12\right)}\right)}}{1-e^{-\tau_{3}^{\left(12\right)}}\left(1-e^{-2\left(\tau_{1}^{\left(12\right)}+\tau_{2}^{\left(12\right)}\right)}\right)}}}, \\
\tau_{2}^{\left(23\right)}&=\scriptstyle{\ln{\left(\frac{\left(1-e^{-2\left(\tau_{2}^{\left(12\right)}+\tau_{3}^{\left(12\right)}\right)}\right)+\sqrt{\left(1-e^{-2\left(\tau_{2}^{\left(12\right)}+\tau_{3}^{\left(12\right)}\right)}\right)^{2}+4\left(1-e^{-\tau_{3}^{\left(12\right)}}\left(1-e^{-2\left(\tau_{1}^{\left(12\right)}+\tau_{2}^{\left(12\right)}\right)}\right)\right)}}{2e^{-\left(\tau_{1}^{\left(12\right)}+\tau_{2}^{\left(12\right)}+\tau_{3}^{\left(12\right)}\right)}}\right)}}, \\
\tau_{3}^{\left(23\right)}&=\scriptstyle{\ln{\left(\frac{-\left(1-e^{-2\left(\tau_{2}^{\left(12\right)}+\tau_{3}^{\left(12\right)}\right)}\right)+\sqrt{\left(1-e^{-2\left(\tau_{2}^{\left(12\right)}+\tau_{3}^{\left(12\right)}\right)}\right)^{2}+4\left(1-e^{-\tau_{3}^{\left(12\right)}}\left(1-e^{-2\left(\tau_{1}^{\left(12\right)}+\tau_{2}^{\left(12\right)}\right)}\right)\right)}}{2\left(1-e^{-\tau_{3}^{\left(12\right)}}\left(1-e^{-2\left(\tau_{1}^{\left(12\right)}+\tau_{2}^{\left(12\right)}\right)}\right)\right)}\right)}}. \\
\end{mycases}
\end{split}\end{align}

From Appendix~\ref{derivation4}, we have shown that $\tau_{1}^{\left(23\right)}$ must be non-positive, $\tau_{2}^{\left(23\right)}$ must be non-negative and $\tau_{3}^{\left(23\right)}$ can be non-negative or negative. When simulating samples for $x_{1},x_{2},x_{3}\in{}U\left(0,1\right)$, we found that $\tau_{3}^{\left(23\right)}$ was non-negative $74$ times out of $500$. We can conclude that the $\left(23\right)$ taxon label permutation is network identifiable from the $\left(12\right)$ taxon label permutation.

\section{The Six Taxon Label Permutations}

Recall that there are six taxon label permutations, described by the symmetric group,
\begin{align}\begin{split}
S_{3}=\left\{\left(\right),\left(12\right),\left(13\right),\left(23\right),\left(123\right),\left(132\right)\right\}.
\end{split}\end{align}
We can express these group elements as products of other group elements. For example, we can express every group element as a product of another group element and $\left(13\right)$,
\begin{align}\begin{split}
\begin{mycases}
\left(\right)&=\left(13\right)\cdot{}\left(13\right), \\
\left(12\right)&=\left(13\right)\cdot{}\left(123\right), \\
\left(13\right)&=\left(13\right)\cdot{}\left(\right), \\
\left(23\right)&=\left(13\right)\cdot{}\left(132\right), \\
\left(132\right)&=\left(13\right)\cdot{}\left(23\right), \\
\left(123\right)&=\left(13\right)\cdot{}\left(12\right).
\end{mycases}
\end{split}\end{align}
Recall that when the time parameters on no taxon label permutation are non-negative, the time parameters on the $\left(13\right)$ taxon label permutation are almost always non-negative as well. When we compare the two taxon label permutations we see that they have the same pairwise distance, $d\left(1,3\right)$. This pairwise distance is twice the sum of the time parameters on the network. However, for the pairwise distances to be identical on both taxon label permutations the second time parameter on the $\left(13\right)$ taxon label permutation, $\tau_{2}^{\left(12\right)}$, must occasionally be negative.

From a symmetry argument, we can conclude that when the time parameters for the $\left(13\right)$ taxon label permutation in terms of no taxon label permutation are all non-negative, the time parameters for the $\left(123\right)$ taxon label permutation in terms of the $\left(12\right)$ taxon label permutation must all be non-negative and the time parameters for the $\left(132\right)$ taxon label permutation in terms of the $\left(23\right)$ taxon label permutation must all be non-negative also. In these cases there will be three pairs of identical maximum likelihoods: no taxon label permutation and the $\left(13\right)$ taxon label permutation, the $\left(12\right)$ taxon label permutation and the $\left(123\right)$ taxon label permutation, the $\left(23\right)$ taxon label permutation and the $\left(132\right)$ taxon label permutation.

In these circumstances we will have a choice between two taxon label permutations that both maximise the likelihood. The time parameters, however, will be different. If it is known or suspected that two taxa have converged then we may wish to choose the taxon label permutation that allows for this convergence. We may also choose the taxon label permutation that allows for the shortest convergence time or we may choose the taxon label permutation that converges the two leaves that have the smallest pairwise distance before convergence.

\chapter[Gr\"{o}bner Bases]{Distinguishability of Trees and Networks using Gr\"{o}bner Bases}
\label{chapter6}

In Chapter~\ref{chapter4} we looked at the identifiability and distinguishability of trees and networks for two and three taxa. Determining the distinguishability of two-taxon and three-taxon trees and networks was fairly straightforward. To address the issue of distinguishability we compared the constraints on the phylogenetic tensors for various trees and networks, as well as taxon label permutations. For trees and networks with more than three taxa or with a large number of convergence periods we will need a more sophisticated method for determining distinguishability.

Recall that the basis of the phylogenetic tensors was first transformed. The phylogenetic tensor elements, both in the original and transformed bases, are expressions in terms of the time parameters on the tree or network. The issue of identifiability was addressed by finding expressions for the time parameters on the trees and networks in terms of the transformed phylogenetic tensor elements. If expressions for every time parameter were able to be found independently of every other time parameter then the tree or network was identifiable. For both two and three taxa, the equality and inequality constraints on the elements of the transformed phylogenetic tensor were then found and compared between various trees and networks to address the issue of distinguishability between the trees and networks.

For trees and networks with large numbers of taxa the phylogenetic tensors are of greater dimension with more time parameters, corresponding to a system with more equations and more time parameters to solve for. For the binary symmetric model, the phylogenetic tensors are of dimension $2^{n}$, where $n$ is the number of taxa. It is easy to see that for large $n$ the system will contain a large number of equations and solving the system will become progressively more challenging as $n$ becomes larger. To compare trees and networks with a large number of taxa it will be necessary to employ a more efficient technique for dealing with our large systems of equations.

Recall that the systems of equations for the transformed phylogenetic tensor elements from our two-taxon and three-taxon examples are non-linear. Since the systems are non-linear we cannot simply use Gaussian elimination to solve them. Likewise, the systems of equations for the transformed phylogenetic tensor elements for four-taxon trees and networks, and more generally $n$-taxon trees and networks, will be non-linear. Our equations involve exponential quantities, with the arguments of these exponential quantities being the negative of a sum of time parameters. The arguments of the exponential quantities are therefore non-positive. We made the simple substitutions, $y_{i}=e^{-\tau_{i}}$, where $\tau_{i}$ is a dimensionless time parameter dependent on the product of two scalars, the positive rate parameter from the binary symmetric model and a non-negative time parameter from the tree or network. Since the units for the rate parameters are the inverse of time, $\tau_{i}$ will be a non-negative real number with no units. We saw in Chapter~\ref{chapter4} and Chapter~\ref{chapter5} that the systems of equations become systems of polynomial equations in the variables $y_{i}$ after these substitutions are made. From here, we can use techniques from algebraic geometry to solve the systems and find their constraints. We will see how this is done by looking at a three-taxon example that we have already examined in Section~\ref{threetaxonclock} before addressing several four-taxon examples. We will show that the technique is useful for networks with many taxa or many convergence periods.

Before we solve the systems of polynomials which define the phylogenetic tensors, we must first introduce the basic concepts we require from algebraic geometry. An excellent source to introduce readers to the algebraic geometric concepts discussed here, including the proofs of all of the theorems we require, is \citet{little1992ideals}. The algebraic geometric concepts that we will discuss, including the notation, will be similar to that used by \citet{little1992ideals}.

The example we will use to illustrate the processes involved will be the three-taxon clock-like tree, shown below in Figure~\ref{threetaxonclockfive}.
\begin{figure}[H]
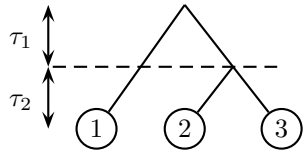

	\centering
		\psmatrix[colsep=.3cm,rowsep=.4cm,mnode=r]
		~ & && ~ \\
		~ & && & ~ & ~ \\
		~ & [mnode=circle] 1 && [mnode=circle] 2 && [mnode=circle] 3
		\ncline{1,4}{3,2}
		\ncline{1,4}{2,5}
		\ncline{2,5}{3,4}
		\ncline{2,5}{3,6}
		\ncline{<->,arrowscale=1.5}{1,1}{2,1}
		\tlput{$\tau_{1}$}
		\ncline{<->,arrowscale=1.5}{2,1}{3,1}
		\tlput{$\tau_{2}$}
		\psset{linestyle=dashed}\ncline{2,1}{2,6}
		\endpsmatrix
		\caption{Three-taxon clock-like tree.}
		\label{threetaxonclockfive}
\end{figure}

The transformed phylogenetic tensor for the three-taxon clock-like tree is
\begin{align}\begin{split}
\widehat{P}=\left[\begin{array}{c} q_{000} \\
q_{001} \\
q_{010} \\
q_{011} \\
q_{100} \\
q_{101} \\
q_{110} \\
q_{111} \\
\end{array}\right]=\left[\begin{array}{c} 1 \\
0 \\
0 \\
e^{-2\tau_{2}} \\
0 \\
e^{-2\left(\tau_{1}+\tau_{2}\right)} \\
e^{-2\left(\tau_{1}+\tau_{2}\right)} \\
0 \\
\end{array}\right].
\end{split}\end{align}

We have already found the constraints for this tree, however this is a simple example to introduce to the reader the techniques in algebraic geometry required to find the equality and inequality constraints on the transformed phylogenetic tensor.

\section{Algebraic Geometric Techniques for Finding the Constraints}

Suppose we are given a set of linear equations. Provided the system is not underdetermined then it can be solved through Gaussian elimination. However, if we are given a set of non-linear equations then there is no general algorithm to solve them unless they are polynomial equations. We have seen that for every two-taxon and three-taxon tree or network the equations for the transformed phylogenetic tensor elements could be converted into polynomial equations. We will see that the equations for the transformed phylogenetic tensor elements for every four-taxon tree or network can also be converted into polynomial equations. We will discuss the relevant techniques in algebraic geometry required to solve polynomial equations and apply these techniques to the example we have introduced before examining some four-taxon trees and networks.

We will begin by formally defining monomials and polynomials. We will follow this with a discussion on how the terms of a polynomial in one variable can be ordered by their degrees. Next, we will show how the ordering of polynomials is used in the division algorithm for polynomial division. We will then discuss how multi-variable monomials can be ordered when there is no immediately obvious generalisation from the single variable ordering by degrees.

The next step will be to define ideals, subrings that can be used to solve systems of polynomial equations. Given a system of polynomial equations, $f_{1},f_{2},\ldots{},f_{s}=0$, we can make the polynomials, $f_{1},f_{2},\ldots{},f_{s}$, a generating set of polynomials for an ideal. The set of polynomial generators of a polynomial ideal is called the basis and is not unique.

We will then introduce two specific types of monomial orderings: lex ordering and product ordering. Lex ordering is analogous to the ordering of words in the dictionary and will be applied to each of the ``types'' of variable: the time variables and the transformed phylogenetic tensor elements. The product ordering allows us to partition the two different types of variables.

We will mention Buchberger's algorithm, which uses a multi-variable generalisation of the division algorithm to find the reduced Gr\"{o}bner basis of the ideal. By choosing the reduced Gr\"{o}bner basis, the remainder from the division algorithm will always be unique, a property which is not generally true of the division algorithm in multiple variables. The set of polynomials in the reduced Gr\"{o}bner basis, or any basis for that matter, generates the same ideal that the original set of polynomials generated.

We can find the solutions to the system of polynomial equations from the set of polynomials in the reduced Gr\"{o}bner basis of the ideal. By using a lex monomial ordering we can eliminate the variables successively in a fashion analogous to back-substitution in Gaussian elimination. By using the product order we can separate the two types of variables, as mentioned earlier. The generators in the reduced Gr\"{o}bner basis will then be a mix of two ``types'' of polynomials: the first containing only the transformed phylogenetic tensor elements and the second containing both the transformed phylogenetic tensor elements and the time variables.

The equality constraints on the transformed phylogenetic tensor elements can be found from the polynomials of the first type. From the second type of polynomials we can find solutions for the time variables in terms of the transformed phylogenetic tensor elements. Recall that the time variables involved exponential quantities and were of the form $y_{i}=e^{-\tau_{i}}$, where $\tau_{i}$ is a dimensionless time parameter. We will demand the time variables be constrained to $0<y_{i}\leq{}1$. Demanding these constraints on the time variables will then give us a set of inequality constraints on the transformed phylogenetic tensor elements.

To avoid confusion, we let $x_{1}$, $x_{2}$, $\ldots{}$, $x_{m}$ be the variables in our definitions and let $y_{1}$, $y_{2}$, $\ldots{}$, $y_{m}$ be the variables for our examples.

We will start by defining a \emph{monomial}.

\begin{definition}[\textbf{Monomial}]
In the scalar variables, $x_{1}$, $x_{2}$, $\ldots{}$, $x_{m}$, a \textbf{monomial} is a single term,
\begin{align*}\begin{split}
x_{1}^{d_{1}}\cdot{}x_{2}^{d_{2}}\cdot{}\ldots{}\cdot{}x_{m}^{d_{m}},
\end{split}\end{align*}
where $d_{i}\in{}\mathbb{N}_{\geq{}0}$ for all $1\leq{}i\leq{}m$ and $\mathbb{N}_{\geq{}0}=\{0,1,\ldots{}\}$.
\end{definition}

We can express the monomial as
\begin{align}\begin{split}
x^{d}=x_{1}^{d_{1}}\cdot{}x_{2}^{d_{2}}\cdot{}\ldots{}\cdot{}x_{m}^{d_{m}},
\end{split}\end{align}
where $d\in{}\mathbb{N}_{\geq{}0}^{m}$.

\begin{definition}[\textbf{Total degree}]
The \textbf{total degree} of the monomial, $\left|d\right|$, is
\begin{align*}\begin{split}
\left|d\right|=\sum\limits_{i=1}^{m}d_{i}.
\end{split}\end{align*}
\end{definition}

For example, we make the substitution, $y_{2}=e^{-\tau_{2}}$. Thus $e^{-2\tau_{2}}\mapsto{}y_{2}^{2}$ and is now a monomial. Having defined a monomial, we are now in a position to define a \emph{polynomial}.

\begin{definition}[\textbf{Polynomial}]
A \textbf{polynomial}, $f$, is a finite linear sum of monomials, with the coefficients of the monomials belonging to some arbitrary field, $\mathbb{F}$. In terms of monomials,
\begin{align*}\begin{split}
f=\sum\limits_{d}a_{d}x^{d},
\end{split}\end{align*}
where $a_{d}\in{}\mathbb{F}$.
\end{definition}

The polynomial ring $\mathbb{F}\left[x_{1},x_{2},\ldots{},x_{m}\right]$ is formed from the set of all polynomials in $x_{1}$, $x_{2}$, $\ldots{}$, $x_{m}$, with the coefficients belonging to $\mathbb{F}$.

Referring to our example, we can now see that if $q_{011}\in{}\mathbb{F}$ then $y_{2}^{2}-q_{011}=0$ is an equation with $y_{2}^{2}-q_{011}$ being a polynomial in the ring, $\mathbb{F}\left[y_{2}\right]$. We will now discuss some definitions relating to polynomials before introducing the notion of an \emph{ordering} of monomials, which is required for our process.

\begin{definition}
Given the polynomial, $f=\sum\limits_{d}a_{d}x^{d}$, in $\mathbb{F}\left[x_{1},x_{2},\ldots{},x_{m}\right]$,
\begin{align*}\begin{split}
(i)&\text{	$a_{d}$ is the \textbf{coefficient} of $x^{d}$,} \\
(ii)&\text{	$a_{d}x^{d}$ is a \textbf{term} of $f$, provided $a_{d}\neq{}0$,} \\
(iii)&\text{	The \textbf{total degree} of $f$ is $deg\left(f\right)=max\left(\left|d\right|\right)$, where $a_{d}\neq{}0$.}
\end{split}\end{align*}
\end{definition}

We can order the polynomial terms based on the degrees of those terms. For example, we can order an arbitrary polynomial in one variable, $x$, (or any number of variables, for that matter) in decreasing order, from the term with the highest degree first to the term with the lowest degree last. An arbitrary polynomial in one variable can be written in the form
\begin{align}\begin{split}
f=a_{1}x^{p}+a_{2}x^{p-1}+\ldots{}+a_{p}x+a_{p+1},
\end{split}\end{align}
where $a_{i}\in{}\mathbb{F}$, $a_{1}\neq{}0$, $p\in{}\mathbb{N}_{\geq{}0}$ and some coefficients may be zero.

\begin{definition}[\textbf{Leading Term}]
The term with the highest degree, or the total degree of the polynomial, $LT\left(f\right)=a_{1}x^{p}$, is called the \textbf{leading term}.
\end{definition}

We can now define polynomial division, which is necessary for finding the Gr\"{o}bner basis of the ideal. By comparing the leading terms of two polynomials we can determine whether the leading term of one of those polynomials \emph{divides} the leading term of the other polynomial. Suppose we have two polynomials, $f_{1}$ and $f_{2}$, in one variable, $x$.

\begin{definition}[\textbf{Polynomial Division}]
For two polynomials, $f_{1},f_{2}\in{}\mathbb{F}\left[x\right]$, $LT\left(f_{2}\right)$ \emph{divides} $LT\left(f_{1}\right)$ if and only if $deg\left(f_{2}\right)\leq{}deg\left(f_{1}\right)$.
\end{definition}

If $LT\left(f_{2}\right)$ \emph{divides} $LT\left(f_{1}\right)$, then $f_{1}$ can be expressed as
\begin{align}\begin{split}
f_{1}=sf_{2}+r,
\end{split}\end{align}
where $r,s\in{}\mathbb{F}\left[x\right]$, $deg\left(r\right)<deg\left(f_{2}\right)$ and possibly $r=0$. If $r=0$ then not only does $LT\left(f_{2}\right)$ divide $LT\left(f_{1}\right)$, but $f_{2}$ divides $f_{1}$.

Given two polynomials, $f_{1}$ and $f_{2}$, in one variable, $x$, $r$ and $s$ will be unique. The \emph{division algorithm} can be used to find $r$ and $s$.

\begin{proposition}[\textbf{Division Algorithm}]
Start with $s=0$ and $r=f_{1}$.
\begin{algorithmic}
\While {$r\neq{}0$ and $LT\left(f_{2}\right)$ divides $LT\left(r\right)$}
    \State $s\gets s+\frac{LT\left(r\right)}{LT\left(f_{2}\right)}$
		\State $r\gets r-\left(\frac{LT\left(r\right)}{LT\left(f_{2}\right)}\right)f_{2}$
\EndWhile
\end{algorithmic}
\end{proposition}

We will now look at a simple example to illustrate how one polynomial can divide another. Suppose we wanted to divide $f_{1}=x^{2}-1$ by $f_{2}=x+1$ and express $f_{1}$ as $f_{1}=sf_{2}+r$. We start by setting $s=0$ and $r=f_{1}=x^{2}-1$. We now check that $r=x^{2}-1\neq{}0$ and $deg\left(LT\left(f_{2}\right)\right)=deg\left(x\right)=1\leq{}deg\left(LT\left(r\right)\right)=deg\left(x^{2}\right)=2$. Since both are true, $s$ and $r$ become
\begin{align}\begin{split}
s&\gets 0+\frac{x^{2}}{x}=x, \\
r&\gets x^{2}-1-\frac{x^{2}}{x}\left(x+1\right)=-x-1.
\end{split}\end{align}

We again check that $r=-x-1\neq{}0$ and $deg\left(LT\left(f_{2}\right)\right)=deg\left(x\right)=1\leq{}deg\left(LT\left(r\right)\right)=deg\left(-x\right)=1$. Since the coefficient of the leading term has no influence on its degree, both statements are again true. $s$ and $r$ then become
\begin{align}\begin{split}
s&\gets x+\frac{-x}{x}=x-1, \\
r&\gets -x-1-\frac{-x}{x}\left(x+1\right)=0.
\end{split}\end{align}

Since $r=0$ the algorithm terminates and we have successfully divided $f_{1}=x^{2}-1$ by $f_{2}=x+1$. $f_{1}$ can now be expressed as,
\begin{align}\begin{split}
f_{1}=sf_{2}+r,
\end{split}\end{align}
where $r=0$ and $s=x-1$.

It can be seen that the division algorithm in $\mathbb{F}\left[x\right]$ and row-reduction algorithm for linear systems (Gaussian elimination with matrices) depend on an \emph{ordering of terms} of polynomials. The polynomials can be ordered by total degree. For our multi-variable scenario it is necessary to introduce the notion of \emph{monomial ordering}.

\begin{definition}[\textbf{Monomial Ordering}]
Suppose we have the monomials, $x^{d}$, $x^{e}$ and $x^{g}$, where $d,e,g\in{}\mathbb{N}_{\geq{}0}^{m}$. A \textbf{monomial ordering} is an ordering, $>$, on the monomials, which meets the following conditions:
\begin{align*}\begin{split}
(i)&\text{	$x^{d}\leq{}x^{e}$ and $x^{e}\leq{}x^{d}\Rightarrow{}x^{d}=x^{e}$. This is the antisymmetry condition,} \\
(ii)&\text{	$x^{d}\leq{}x^{e}$ and $x^{e}\leq{}x^{g}\Rightarrow{}x^{d}\leq{}x^{g}$. This is the transitivity condition,} \\
(iii)&\text{	$x^{d}\leq{}x^{e}$ or $x^{e}\leq{}x^{d}$. This is the totality condition,} \\
(iv)&\text{	$x^{d}>x^{e}$ and $g\in{}\mathbb{N}_{\geq{}0}^{m}\Rightarrow{}x^{d}x^{g}>x^{e}x^{g}$,} \\
(v)&\text{	There is a minimum monomial for every non-empty subset of $\mathbb{N}_{\geq{}0}^{m}$ under $>$,} \\
&\text{	with the ordering $0<1<\ldots{}$.}
\end{split}\end{align*}
\end{definition}

Conditions $(i)$ to $(iii)$ are the restriction that $>$ must be a total (or linear) ordering on $\mathbb{N}_{\geq{}0}^{m}$. Condition $(iv)$ says that the ordering of two monomials is only dependent on the variables that differ in their degrees. Condition $(v)$ says that $>$ must be a well-ordering on $\mathbb{N}_{\geq{}0}^{m}$, a technical point not needed to understand our general procedure.

Suppose we have three monomials, $x^{a}$, $x^{b}$ and $x^{c}$, where $a,b,c\in{}\mathbb{N}_{\geq{}0}^{3}$, where the three monomials are
\begin{align}\begin{split}
\begin{mycases}
x^{a}=x_{1}^{a_{1}}x_{2}^{a_{2}}x_{3}^{a_{3}}&=x_{1}^{2}x_{2}^{5}x_{3}, \\
x^{b}=x_{1}^{b_{1}}x_{2}^{b_{2}}x_{3}^{b_{3}}&=x_{1}^{3}x_{3}, \\
x^{c}=x_{1}^{c_{1}}x_{2}^{c_{2}}x_{3}^{c_{3}}&=x_{1}^{3}x_{2}^{2}x_{3}.
\end{mycases}
\end{split}\end{align}

An example of a monomial ordering is lex ordering, which will be discussed in more detail later. Under lex ordering of monomials, the degrees of each variable in each monomial are compared successively. Comparing two monomials, the degrees of the first variable are compared first. If these degrees are equal, then the degrees of the second variable are then compared. The process is continued until the two monomials differ in their degrees for a given variable. If the two monomials never differ in their degrees then they are identical and equal under any monomial ordering. Comparing $x^{a}$, $x^{b}$ and $x^{c}$, $b_{1}=c_{1}=3>a_{1}=2$. We can conclude that $x^{b},x^{c}>x^{a}$. To determine whether $x^{b}>x^{c}$, $x^{b}=x^{c}$ or $x^{b}<x^{c}$, we must look at the second variable. Clearly, $c_{2}=2>b_{2}=0$ and $x^{c}>x^{b}$. In conclusion, $x^{c}=x_{1}^{3}x_{2}^{2}x_{3}>x^{b}=x_{1}^{3}x_{3}>x^{a}=x_{1}^{2}x_{2}^{5}x_{3}$ under lex ordering.

\begin{definition}
Suppose we have a nonzero polynomial in $\mathbb{F}\left[x_{1},x_{2},\ldots{},x_{m}\right]$, $f=\sum\limits_{d}a_{d}x^{d}$, with the monomial order, $>$.
\begin{align*}\begin{split}
(i)&\text{	$multideg\left(f\right)=max\left(d\in{}\mathbb{N}_{\geq{}0}^{m}:a_{d}\neq{}0\right)$ is the \textbf{multidegree} of f}, \\
(ii)&\text{	$LC\left(f\right)=a_{multideg\left(f\right)}\in{}\mathbb{F}$ is the \textbf{leading coefficient} of f}, \\
(iii)&\text{	$LM\left(f\right)=x^{multideg\left(f\right)}$ is the \textbf{leading monomial} of f}, \\
(iv)&\text{	$LT\left(f\right)=LC\left(f\right)\cdot{}LM\left(f\right)$ is the \textbf{leading term} of f.}
\end{split}\end{align*}
\end{definition}

In order to solve a system of polynomial equations, $f_{1},f_{2},\ldots{},f_{s}=0$, we will introduce an ideal, which can be expressed in terms of a system of polynomial equations.

\begin{definition}[\textbf{Ideal}]
An \textbf{ideal} is a set, $I\subset{}\mathbb{F}\left[x_{1},x_{2},\ldots{},x_{m}\right]$, that satisfies the conditions:
\begin{align*}\begin{split}
(i)&\text{	}0\in{}I, \\
(ii)&\text{	If $f_{1},f_{2}\in{}I$ then $f_{1}+f_{2}\in{}I$,} \\
(iii)&\text{	If $f_{1}\in{}I$ and $s\in{}\mathbb{F}\left[x_{1},x_{2},\ldots{},x_{m}\right]$ then $sf_{1}\in{}I$.}
\end{split}\end{align*}
\end{definition}

An example of an ideal can be expressed in terms of a set of \emph{generators}.

\begin{definition}
For any set of polynomials, $f_{1},f_{2},\ldots{},f_{s}\in{}\mathbb{F}\left[x_{1},x_{2},\ldots{},x_{m}\right]$, we define the set $I=\left<f_{1},f_{2},\ldots{},f_{s}\right>$, as follows:
\begin{align*}\begin{split}
I=\left<f_{1},f_{2},\ldots{},f_{s}\right>=\left\{\sum\limits_{1}^{s}h_{i}f_{i}:\quad{}h_{1},h_{2},\ldots{},h_{s}\in{}\mathbb{F}\left[x_{1},x_{2},\ldots{},x_{m}\right]\right\},
\end{split}\end{align*}
where $f_{1}$, $f_{2}$, $\ldots{}$, $f_{s}$ are called the \textbf{generators} of the ideal. Different sets of generators are referred to as different \textbf{bases} for the ideal.
\end{definition}

It is easy to check that $I=\left<f_{1},f_{2},\ldots{},f_{s}\right>$ is an ideal for \emph{any} polynomials $f_{1}$, $f_{2}$, $\ldots{}$, $f_{s}$ $\in{}\mathbb{F}\left[x_{1},x_{2},\ldots{},x_{m}\right]$. The set of generators of an ideal is not unique. If we have multiple sets of generators for an ideal then the sets of polynomial equations that we form from the generators must all have the same solutions. Some sets of generators may, however, allow for the polynomial equations to be solved more easily.​

After making the appropriate substitutions, $y_{i}=e^{-\tau_{i}}$, our equations for the transformed phylogenetic tensor elements are all of the form,
\begin{align}\begin{split}
q_{i_{1}i_{2}\ldots{}i_{n}}&=f\left(y_{1},y_{2},\ldots{},y_{m}\right),
\end{split}\end{align}
where $q_{i_{1}i_{2}\ldots{}i_{n}}\in{}\mathbb{F}$ and $f\left(y_{1},y_{2},\ldots{},y_{m}\right)\in{}\mathbb{F}\left[y_{1},y_{2},\ldots{},y_{m}\right]$.

The polynomial equations are then rearranged to be
\begin{align}\begin{split}
f\left(y_{1},y_{2},\ldots{},y_{m}\right)-q_{i_{1}i_{2}\ldots{}i_{n}}&=0.
\end{split}\end{align}

We then take each $f\left(y_{1},y_{2},\ldots{},y_{m}\right)-q_{i_{1}i_{2}\ldots{}i_{n}}$ to be a generator in the generating set of polynomials for an ideal.

For the three-taxon clock-like tree, our set of generating polynomials is $\{y_{2}^{2}-q_{011},y_{1}^{2}y_{2}^{2}-q_{101},y_{1}^{2}y_{2}^{2}-q_{110}\}$, which generates the ideal $\left<y_{2}^{2}-q_{011},y_{1}^{2}y_{2}^{2}-q_{101},y_{1}^{2}y_{2}^{2}-q_{110}\right>$.

We would now like to transform the basis of the ideal. By transforming the basis of the ideal, the new generators of the ideal will form an equivalent set of equations which may be able to be solved more easily. Likewise, any equality constraints involving the phylogenetic tensor elements can be found from the generators of the ideal if they are in the appropriate basis. We will now specify the monomial ordering necessary for our basis transformation process.

\begin{definition}[\textbf{Lexicographic (Lex) Order}]
$x^{d}>_{\text{lex}}x^{e}$, where $d,e\in{}\mathbb{N}_{\geq{}0}^{m}$, if the first non-zero element in $d-e=\left(d_{1}-e_{1},d_{2}-e_{2},\ldots{},d_{m}-e_{m}\right)\in{}\mathbb{N}_{\geq{}0}^{m}$, is greater than zero.
\end{definition}

We will apply the lex order $y_{1}>y_{2}>\ldots{}>y_{m}$ to our $m$ time variables, where $y_{i}=e^{-\tau_{i}}$ for all $1\leq{}i\leq{}m$. Lex order derives it's name from an analogy to the ordering of words in a dictionary. For example, if we take the words ``algorithm'' and ``arbitrary'' and assign them the strings ``$x_{1}x_{2}\ldots{}x_{9}$'' and ``$x_{1}'x_{2}'\ldots{}x_{9}'$'' respectively, we see that $x_{1}=a=_{\text{lex}}x_{1}'=a$ and $x_{2}=l>_{\text{lex}}x_{2}'=r$. Hence, $algorithm$ $>_{\text{lex}}$ $arbitrary$. If words are ordered in descending order according to this choice of lex ordering then it is equivalent to the reverse alphabetical order. Two more examples of lex order in use are $x_{1}x_{2}x_{3}>_{\text{lex}}x_{2}x_{3}$ and $x_{1}^{2}x_{2}x_{3}>_{\text{lex}}x_{1}x_{2}x_{3}^{2}$.

\begin{definition}[\textbf{Block (or Elimination) Order}]
A \textbf{block order} is a monomial order on $\mathbb{F}\left[x_{1},x_{2},\ldots{},x_{m},x_{1}',x_{2}',\ldots{}x_{p}'\right]$ for $x_{1},x_{2},\ldots{},x_{m}$ if every polynomial which has its leading monomial in $\mathbb{F}\left[x_{1}',x_{2}',\ldots{}x_{p}'\right]$ is inside $\mathbb{F}\left[x_{1}',x_{2}',\ldots{}x_{p}'\right]$. In other words,
\begin{align*}\begin{split}
LM\left(g\right)\in{}\mathbb{F}\left[x_{1}',x_{2}',\ldots{}x_{p}'\right]\Rightarrow{}g\in{}\mathbb{F}\left[x_{1}',x_{2}',\ldots{}x_{p}'\right].
\end{split}\end{align*}
\end{definition}
For more on block ordering and its relevance to the Elimination Theorem, a theorem necessary to our process that will not be discussed further, see \citep{hassett2007introduction}. For our variables, we will set $y_{1}$, $y_{2}$, $\ldots{}$, $y_{m}$ as the $m$ time variables, as before. The remaining variables will be the phylogenetic tensor elements, $q_{i_{1}i_{2}\ldots{}i_{n}}$. We will use lex order for the two sets of variables, the $m$ time variables and the variables for the phylogenetic tensor elements.

\begin{definition}[\textbf{Product Order}]
The product order, $>_{\text{prod}}$, is defined as
\begin{align*}\begin{split}
x^{d}x'^{e}>_{\text{prod}}x^{g}x'^{h}\text{ if }\begin{cases} x^{d}>_{\text{lex}}x^{g}, \\
\text{or }x^{d}=_{\text{lex}}x^{g}\text{ and }x'^{e}>_{\text{lex}}x'^{h}.
\end{cases}
\end{split}\end{align*}
\end{definition}

A product order is a type of block order. The product ordering will allow us to solve the system of equations for the time variables, $y_{1}$, $y_{2}$, $\ldots{}$, $y_{m}$, in terms of the transformed phylogenetic tensor variables, $q_{i_{1}i_{2}\ldots{}i_{n}}$, while retaining any equality constraints involving the transformed phylogenetic tensor variables.

Having chosen a monomial order for our variables, we will restrict our coefficients to belong to the rationals, $\mathbb{Q}$. We can now apply the multivariable division algorithm to attempt to find an ideal equivalent to our existing ideal, but in a different basis with different generators. This will allow our polynomial equations to be solved more easily and will give any equality constraints on the transformed phylogenetic tensor variables. Recall that the remainder for the division algorithm is unique in the one variable case. Unfortunately, in multiple variables the division algorithm is not sufficient to uniquely characterise the remainder, $r$, upon division by a generating set of polynomials. By choosing the \emph{Gr\"{o}bner basis} as the basis for our ideals, $r$ is uniquely determined and $r=0$ is equivalent to the polynomial belonging to the ideal.

\begin{definition}[\textbf{Gr\"{o}bner (or Standard) Basis}]
Suppose we are given an ideal, $I=\left<f_{1},f_{2},\ldots{},f_{s}\right>$, and a monomial ordering. A finite subset, $G=\{g_{1},g_{1},\ldots{},g_{t}\}$, of the ideal, is a Gr\"{o}bner (or standard) basis if
\begin{align*}\begin{split}
\left<LT\left(g_{1}\right),LT\left(g_{2}\right),\ldots{},LT\left(g_{t}\right)\right>=\left<LT\left(I\right)\right>=\left<LT\left(f_{1}\right),LT\left(f_{2}\right),\ldots{},LT\left(f_{s}\right)\right>.
\end{split}\end{align*}
\end{definition}

\begin{corollary}
For a given monomial order, a Gr\"{o}bner basis exists for every ideal, $I\subset{}\mathbb{F}\left[x_{1},x_{2},\ldots{},x_{m}\right]$, $I\neq{}\left\{0\right\}$.
\begin{proof}
Proof can be found in \citet{little1992ideals}.
\end{proof}
\end{corollary}

If we are given a basis for an ideal, we can determine whether the basis is a Gr\"{o}bner basis for the ideal. Following from the definition of an ideal, we can determine if a monomial lies in a monomial ideal.

\begin{lemma}
Suppose $I=\left<x^{d}:\quad{}d\in{}D\right>$ is a monomial ideal, where $D\subset{}\mathbb{N}_{\geq{}0}^{m}$. Then a given monomial, $x^{e}$, is in $I$ if and only if there is a $d\in{}D$ for which $x^{e}$ is divisible by $x^{d}$.
\begin{proof}
Proof can be found in \citet{little1992ideals}.
\end{proof}
\label{monomialideal}
\end{lemma}

\begin{corollary}
For a given monomial order, every ideal, $I\subset{}\mathbb{F}\left[x_{1},x_{2},\ldots{},x_{m}\right]$, $I\neq{}\{0\}$, has a Gr\"{o}bner basis and any Gr\"{o}bner basis for an ideal, $I$, is a basis of $I$.
\begin{proof}
Proof can be found in \citet{little1992ideals}.
\end{proof}
\end{corollary}

\begin{definition}[\textbf{Reduced Gr\"{o}bner Basis}]
A reduced Gr\"{o}bner basis is a Gr\"{o}bner basis, $G=\left\{g_{1},g_{2},\ldots{}g_{t}\right\}$, for an ideal, $I$, which meets the requirements:
\begin{align*}\begin{split}
(i)&\text{	$LC\left(p\right)=1$ for all $p\in{}G$}, \\
(ii)&\text{	No monomial of $p$ is in $\left<LT\left(G-\{p\}\right)\right>$ for all $p\in{}G$.}
\end{split}\end{align*}
\end{definition}

\begin{proposition}
Suppose $I\neq{}\{0\}$ is an ideal. For a fixed monomial order, $I$ will have a unique reduced Gr\"{o}bner basis.
\begin{proof}
Proof can be found in \citet{little1992ideals}.
\end{proof}
\end{proposition}

A Gr\"{o}bner basis for a polynomial ideal, $I\neq{}\left\{0\right\}$, can be found using \emph{Buchberger's algorithm}. Buchberger's algorithm uses the multivariable division algorithm to transform the basis of an ideal into the Gr\"{o}bner basis. For more on Buchberger's algorithm and the proof that it produces a Gr\"{o}bner basis, see \citet{little1992ideals}. We will use the computer algebra system $Macaulay2$ to compute our Gr\"{o}bner bases. \citet{sturmfels2002ideals} explains how to use $Macaulay2$ to compute a reduced Gr\"{o}bner basis for a given monomial order.

We will choose the field to be the rationals, $\mathbb{Q}$, the variables to be $y_{1}$, $y_{2}$, $\ldots{}$, $y_{m}$ and $q_{00\ldots{}00}$, $q_{00\ldots{}01}$, $q_{00\ldots{}10}$, $\ldots{}$, $q_{11\ldots{}11}$, the ring to be the set of all polynomials in our stated variables, with coefficients belonging to our field. We will choose the lex ordering of $y_{1}>y_{2}>\ldots{}>y_{m}$, the lex ordering of the transformed phylogenetic tensor variables of $q_{00\ldots{}00}>q_{00\ldots{}01}>q_{00\ldots{}10}>\ldots{}>q_{11\ldots{}11}$ and product ordering for $y_{1}$, $y_{2}$, $\ldots{}$, $y_{m}$. The lex ordering on the transformed phylogenetic tensor variables is the ordering of the tensor elements determined by the Kronecker product.

Once we have found the generators of the Gr\"{o}bner basis, the transformed phylogenetic tensor variables, $q_{i_{1}i_{2}\ldots{}i_{n}}$, will then be interpreted as constants. This allows the polynomial equations to be solved for the variables, $y_{1}$, $y_{2}$, $\ldots{}$, $y_{m}$, as functions of the transformed phylogenetic tensor variables, $q_{i_{1}i_{2}\ldots{}i_{n}}$. Additionally, any constraints involving only the transformed phylogenetic tensor variables, $q_{i_{1}i_{2}\ldots{}i_{n}}$, and not $y_{1}$, $y_{2}$, $\ldots{}$, $y_{m}$ will have also been found explicitly.

We will now refer back to our ideal for the three-taxon clock-like tree,
\begin{align}\begin{split}
I=\left<y_{2}^{2}-q_{011},y_{1}^{2}y_{2}^{2}-q_{101},y_{1}^{2}y_{2}^{2}-q_{110}\right>.
\end{split}\end{align}

For any of our trees or networks, we can use the following \emph{Macaulay2} code to find the Gr\"{o}bner basis of the ideal.
\begin{lstlisting}[breaklines=true, mathescape]
--Field of rationals, $\mathbb{Q}$.
S = QQ

--A polynomial ring with lex ordering for $y_{1}$, $y_{2}$, $\ldots{}$, $y_{m}$, lex ordering for the variable transformed phylogenetic tensor elements, $q_{1}$, $q_{2}$, $\ldots{}$, $q_{p}$, which are indexed according to the ordering determined by the Kronecker product, and product ordering for $y_{1}$, $y_{2}$, $\ldots{}$, $y_{m}$.
R = S[y_1..y_m, q_1..q_p, MonomialOrder => {Lex => m, Lex => p}]

--The generators of the ideal.
I = ideal(f1, f2, ..., fs)

--The generators of the ideal in the Gr$\textrm{\"{o}}$bner basis, which have been transposed so they print down the page instead of across the page.
transpose gens gb I
\end{lstlisting}
\label{macaulay2}

For the three-taxon clock-like tree we have two time variables, $m=2$, and three variable transformed phylogenetic tensor elements, $p=3$. We will change the labelling of the transformed phylogenetic tensor elements from $\left\{q_{011},q_{101},q_{110}\right\}$ to $\left\{q_1,q_2,q_3\right\}$. The \emph{Macaulay2} code for the three-taxon clock-like tree is shown below.
\begin{lstlisting}
S = QQ
R = S[y_1..y_2, q_1..q_3, MonomialOrder => {Lex => 2, Lex => 3}]
I = ideal(y_2^2 - q_1, y_1^2*y_2^2 - q_2, y_1^2*y_2^2 - q_3)
transpose gens gb I
\end{lstlisting}

The ideal in the Gr\"{o}bner basis is found to be
\begin{align}\begin{split}
I=\left<q_{101}-q_{110},y_{2}^{2}-q_{011},y_{1}^{2}q_{011}-q_{110}\right>.
\end{split}\end{align}

We will now check that this is indeed a reduced Gr\"{o}bner basis for the ideal. It follows from Lemma \ref{monomialideal} that to show that this is a Gr\"{o}bner basis we need to show that the leading terms of all of the elements of $I$ will be divisible by at least one of the leading monomials of the generators of the Gr\"{o}bner basis, $G$. The ideal in the original basis can be expressed as
\begin{align}\begin{split}
I=&\left\{h_{1}\left(y_{2}^{2}-q_{011}\right)+h_{2}\left(y_{1}^{2}y_{2}^{2}-q_{101}\right)+h_{3}\left(y_{1}^{2}y_{2}^{2}-q_{110}\right):\right. \\
&\quad{}\left.h_{1},h_{2},h_{3}\in{}\mathbb{F}\left[y_{1},y_{2},q_{011},q_{101},q_{110}\right]\right\} \\
=&\left\{\left(h_{2}+h_{3}\right)y_{1}^{2}y_{2}^{2}+h_{1}y_{2}^{2}-h_{1}q_{011}-h_{2}q_{101}-h_{3}q_{110}:\right. \\
&\quad{}\left.h_{1},h_{2},h_{3}\in{}\mathbb{F}\left[y_{1},y_{2},q_{011},q_{101},q_{110}\right]\right\}.
\end{split}\end{align}

By restricting $h_{1}$, $h_{2}$, $h_{3}$ to $h_{1},h_{2},h_{3}\in\{-1,0,1\}$ and taking the leading monomials we can find the set of all of the monomials that the leading terms of the elements of $I$ will be divisible by. This set of monomials will be $S=\{y_{1}^{2}y_{2}^{2},y_{2}^{2},q_{101}\}$. The leading terms of every element of $I$ must be divisible by at least one of the monomials in $S$. For $G$ to be a Gr\"{o}bner basis, these three monomials in $S$ must be divisible by at least one of the leading monomials of the generators of $G$. Clearly $y_{1}^{2}y_{2}^{2}$ will be divisible by $y_{2}^{2}$, while $y_{2}^{2}$ will also be divisible by $y_{2}^{2}$ and $q_{101}$ will be divisible by $q_{101}$. Since the monomials of $S$ will all be divisible by at least one of the leading monomials of the generators of $G$ and every leading monomial of an element of $I$ must be divisible by a monomial of $S$ we can conclude that $G$ is indeed a Gr\"{o}bner basis for $I$.

We will now check that the Gr\"{o}bner basis is a reduced Gr\"{o}bner basis. The set of polynomial generators for the Gr\"{o}bner basis is $\left\{q_{101}-q_{110},y_{2}^{2}-q_{011},y_{1}^{2}q_{011}-q_{110}\right\}$. We can see that $LC\left(p\right)=1$ for all of the polynomial generators, the first requirement for a reduced Gr\"{o}bner basis. The first polynomial generator is $p_{1}=q_{101}-q_{110}$. It follows that
\begin{align}\begin{split}
\left<LT\left(G-\left\{p_{1}\right\}\right)\right>=\left<y_{2}^{2},y_{1}^{2}q_{011}\right>.
\end{split}\end{align}
The two monomials of $p_{1}$ are $q_{101}$ and $q_{110}$. $q_{101}\notin{}\left\{y_{2}^{2},y_{1}^{2}q_{011}\right\}$ and $q_{110}\notin{}\left\{y_{2}^{2},y_{1}^{2}q_{011}\right\}$. The second polynomial generator is $p_{2}=y_{2}^{2}-q_{011}$. It follows that
\begin{align}\begin{split}
\left<LT\left(G-\left\{p_{2}\right\}\right)\right>=\left<q_{101},y_{1}^{2}q_{011}\right>.
\end{split}\end{align}
The two monomials of $p_{2}$ are $y_{2}^{2}$ and $q_{011}$. $y_{2}^{2}\notin{}\left\{q_{101},y_{1}^{2}q_{011}\right\}$ and $q_{011}\notin{}\left\{q_{101},y_{1}^{2}q_{011}\right\}$. The third polynomial generator is $p_{3}=y_{1}^{2}q_{011}-q_{110}$. It follows that
\begin{align}\begin{split}
\left<LT\left(G-\left\{p_{3}\right\}\right)\right>=\left<q_{101},y_{2}^{2}\right>.
\end{split}\end{align}
The two monomials of $p_{3}$ are $y_{1}^{2}q_{011}$ and $q_{110}$. $y_{1}^{2}q_{011}\notin{}\left\{q_{101},y_{2}^{2}\right\}$ and $q_{110}\notin{}\left\{q_{101},y_{2}^{2}\right\}$. We can conclude that $G$ is a reduced Gr\"{o}bner basis for $I$.

Now solving this system of polynomial equations defined by the generators in the Gr\"{o}bner basis gives
\begin{align}\begin{split}
\left\{q_{101}=q_{110},\quad{}y_{1}=\sqrt{\frac{q_{110}}{q_{011}}},\quad{}y_{2}=\sqrt{q_{011}}\right\}.
\end{split}\end{align}

Recalling that $0<y_{1}=e^{-\tau_{1}},y_{2}=e^{-\tau_{2}}\leq{}1$, the constraints on the non-constant elements of the transformed phylogenetic tensor for the non-clock-like tree are then
\begin{align}\begin{split}
\left\{q_{101}=q_{110},\quad{}q_{101}\leq{}q_{011}\right\}.
\end{split}\end{align}

By using the same techniques, we will address the issue of identifiability for four-taxon trees and networks. We will see that this method will be much more efficient than finding the constraints manually when dealing with trees and networks with a large number of taxa or more convergence periods. We will then address the issue of distinguishability by finding the intersections of the constraints for pairs of trees or networks.

\section{Four-Taxon Convergence-Divergence Networks}

We now consider some four-taxon trees and networks as examples. There are two ``structures'' for four-taxon clock-like trees, defined by different sets of clusters. Below in Figure~\ref{twofourtaxonclockstructures} are the two structures for four-taxon clock-like trees.
\begin{figure}[H]
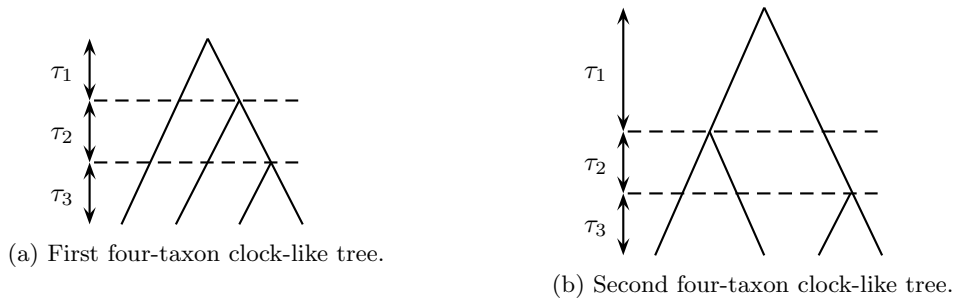

	\centering 
		\begin{subfigure}[h]{0.49\textwidth}
			\centering
				\psmatrix[colsep=.3cm,rowsep=.4cm,mnode=r]
				~ && && ~ \\
				~ & && && ~ && ~ \\
				~ && && && ~ & ~ \\
				~ & ~ && ~ && ~ && ~
				\ncline{1,5}{4,2}
				\ncline{1,5}{4,8}
				\ncline{2,6}{4,4}
				\ncline{3,7}{4,6}
				\ncline{<->,arrowscale=1.5}{1,1}{2,1}
				\tlput{$\tau_{1}$}
				\ncline{<->,arrowscale=1.5}{2,1}{3,1}
				\tlput{$\tau_{2}$}
				\ncline{<->,arrowscale=1.5}{3,1}{4,1}
				\tlput{$\tau_{3}$}
				\psset{linestyle=dashed}\ncline{2,1}{2,8}
				\psset{linestyle=dashed}\ncline{3,1}{3,8}
				\endpsmatrix
				\caption{First four-taxon clock-like tree.}
		\end{subfigure}
		\begin{subfigure}[h]{0.49\textwidth}
			\centering
				\psmatrix[colsep=.3cm,rowsep=.4cm,mnode=r]
				~ & && && ~ \\
				\\
				~ & && ~ & && && & ~ \\
				~ && && && && ~ & ~ \\
				~ & ~ && && ~ && ~ && ~
				\ncline{1,6}{5,2}
				\ncline{3,4}{5,6}
				\ncline{1,6}{5,10}
				\ncline{4,9}{5,8}
				\ncline{<->,arrowscale=1.5}{1,1}{3,1}
				\tlput{$\tau_{1}$}
				\ncline{<->,arrowscale=1.5}{3,1}{4,1}
				\tlput{$\tau_{2}$}
				\ncline{<->,arrowscale=1.5}{4,1}{5,1}
				\tlput{$\tau_{3}$}
				\psset{linestyle=dashed}\ncline{3,1}{3,10}
				\psset{linestyle=dashed}\ncline{4,1}{4,10}
				\endpsmatrix
				\caption{Second four-taxon clock-like tree.}
		\end{subfigure}
				\caption{The two structures for four-taxon clock-like trees.}
				\label{twofourtaxonclockstructures}
\end{figure}

We will look at both of the four-taxon trees, as well as all of the convergence-divergence networks involving two leaves converging with each other and the non-clock-like tree. We give each tree or network an arbitrary name. For the first structure of the four-taxon clock-like tree there are four choices for pairs of converging leaves. When counting from the left, the inequivalent pairs of converging taxa are taxa $1$ and $2$, taxa $1$ and $3$, taxa $2$ and $3$ and taxa $3$ and $4$. Since the positioning of an edge below a node is arbitrary, the taxon pair $1$ and $4$ is equivalent to $1$ and $3$, while the taxon pair $2$ and $4$ is equivalent to $2$ and $3$. The first structure of the four-taxon clock-like tree and the four convergence-divergence networks with pairs of converging leaves are shown in Figure~\ref{firststructurefourtaxon} below.
\begin{figure}[H]
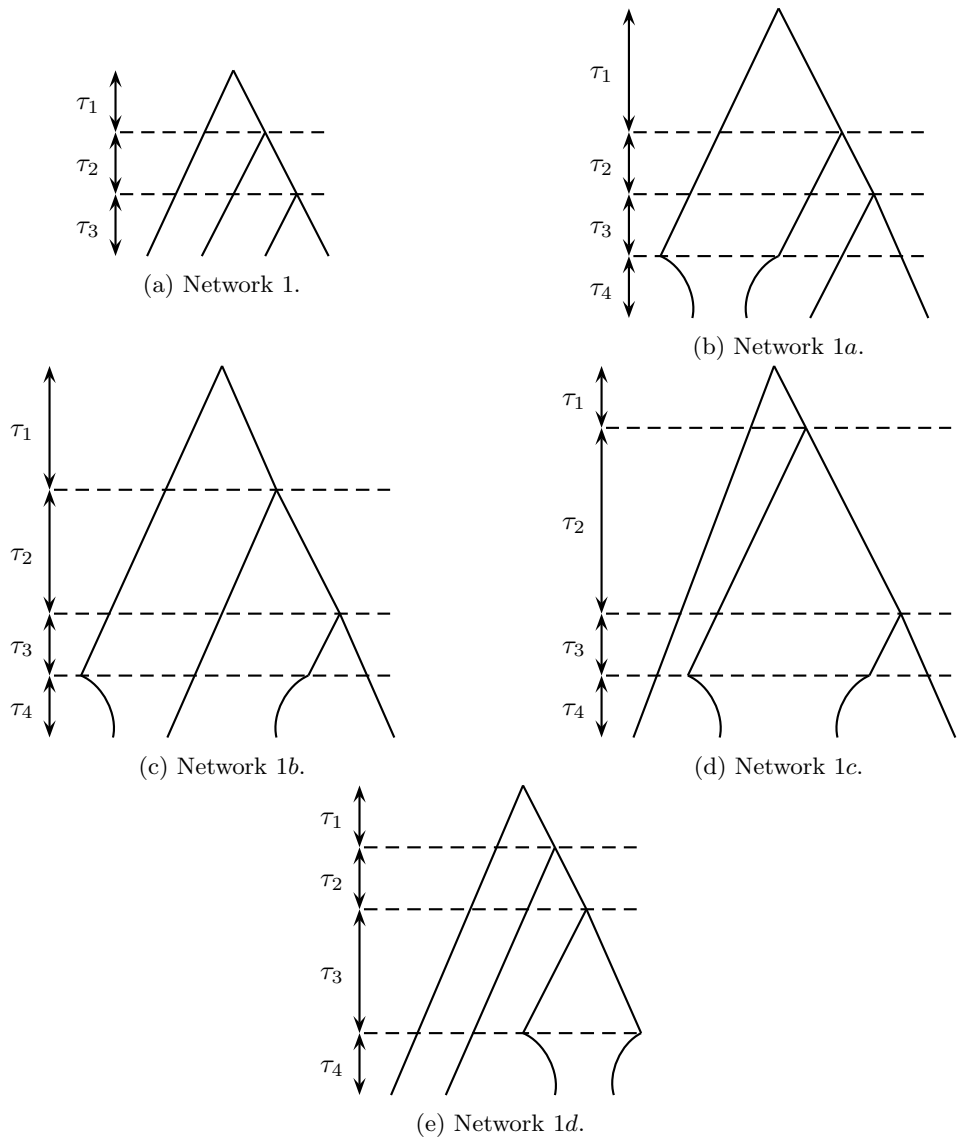

	\centering
		\begin{subfigure}[h]{0.49\textwidth}
			\centering
				\psmatrix[colsep=.3cm,rowsep=.4cm,mnode=r]
				~ && && ~ \\
				~ & && && ~ && ~ \\
				~ && && && ~ & ~ \\
				~ & ~ && ~ && ~ && ~
				\ncline{1,5}{4,2}
				\ncline{1,5}{4,8}
				\ncline{2,6}{4,4}
				\ncline{3,7}{4,6}
				\ncline{<->,arrowscale=1.5}{1,1}{2,1}
				\tlput{$\tau_{1}$}
				\ncline{<->,arrowscale=1.5}{2,1}{3,1}
				\tlput{$\tau_{2}$}
				\ncline{<->,arrowscale=1.5}{3,1}{4,1}
				\tlput{$\tau_{3}$}
				\psset{linestyle=dashed}\ncline{2,1}{2,8}
				\psset{linestyle=dashed}\ncline{3,1}{3,8}
				\endpsmatrix
				\caption{Network $1$.}
		\end{subfigure}
		\begin{subfigure}[h]{0.49\textwidth}
			\centering
				\psmatrix[colsep=.3cm,rowsep=.4cm,mnode=r]
				~ & && && ~ \\
				\\
				~ & && && && ~ & && ~ \\
				~ && && && && ~ && ~ \\
				~ & ~ && && ~ & && && ~ \\
				~ && ~ && ~ && ~ && && ~
				\ncline{1,6}{5,2}
				\ncline{1,6}{3,8}
				\ncline{3,8}{4,9}
				\ncline{4,9}{6,11}
				\ncline{3,8}{5,6}
				\ncline{4,9}{6,7}
				\ncarc[arcangle=40]{5,2}{6,3}
				\ncarc[arcangle=-40]{5,6}{6,5}
				\ncline{<->,arrowscale=1.5}{1,1}{3,1}
				\tlput{$\tau_{1}$}
				\ncline{<->,arrowscale=1.5}{3,1}{4,1}
				\tlput{$\tau_{2}$}
				\ncline{<->,arrowscale=1.5}{4,1}{5,1}
				\tlput{$\tau_{3}$}
				\ncline{<->,arrowscale=1.5}{5,1}{6,1}
				\tlput{$\tau_{4}$}
				\psset{linestyle=dashed}\ncline{3,1}{3,11}
				\psset{linestyle=dashed}\ncline{4,1}{4,11}
				\psset{linestyle=dashed}\ncline{5,1}{5,11}
				\endpsmatrix
				\caption{Network $1a$.}
		\end{subfigure}
		\begin{subfigure}[h]{0.49\textwidth}
			\centering
				\psmatrix[colsep=.3cm,rowsep=.4cm,mnode=r]
				~ && && && ~ \\
				\\
				~ && && && && ~ && && ~ \\
				\\
				~ && && && && && ~ && ~ \\
				~ & ~ && && && && ~ & && ~ \\
				~ && ~ && ~ && && ~ && && ~
				\ncline{1,7}{6,2}
				\ncline{1,7}{3,9}
				\ncline{3,9}{5,11}
				\ncline{5,11}{7,13}
				\ncline{3,9}{7,5}
				\ncline{5,11}{6,10}
				\ncarc[arcangle=40]{6,2}{7,3}
				\ncarc[arcangle=-40]{6,10}{7,9}
				\ncline{<->,arrowscale=1.5}{1,1}{3,1}
				\tlput{$\tau_{1}$}
				\ncline{<->,arrowscale=1.5}{3,1}{5,1}
				\tlput{$\tau_{2}$}
				\ncline{<->,arrowscale=1.5}{5,1}{6,1}
				\tlput{$\tau_{3}$}
				\ncline{<->,arrowscale=1.5}{6,1}{7,1}
				\tlput{$\tau_{4}$}
				\psset{linestyle=dashed}\ncline{3,1}{3,13}
				\psset{linestyle=dashed}\ncline{5,1}{5,13}
				\psset{linestyle=dashed}\ncline{6,1}{6,13}
				\endpsmatrix
				\caption{Network $1b$.}
		\end{subfigure}
		\begin{subfigure}[h]{0.49\textwidth}
			\centering
				\psmatrix[colsep=.3cm,rowsep=.4cm,mnode=r]
				~ && && && ~ \\
				~ & && && && ~ & && && ~ \\
				\\
				\\
				~ && && && && && ~ && ~ \\
				~ && & ~ && && && ~ & && ~ \\
				~ & ~ & && ~ && && ~ && && ~
				\ncline{1,7}{7,2}
				\ncline{1,7}{2,8}
				\ncline{2,8}{5,11}
				\ncline{5,11}{7,13}
				\ncline{2,8}{6,4}
				\ncline{5,11}{6,10}
				\ncarc[arcangle=40]{6,4}{7,5}
				\ncarc[arcangle=-40]{6,10}{7,9}
				\ncline{<->,arrowscale=1.5}{1,1}{2,1}
				\tlput{$\tau_{1}$}
				\ncline{<->,arrowscale=1.5}{2,1}{5,1}
				\tlput{$\tau_{2}$}
				\ncline{<->,arrowscale=1.5}{5,1}{6,1}
				\tlput{$\tau_{3}$}
				\ncline{<->,arrowscale=1.5}{6,1}{7,1}
				\tlput{$\tau_{4}$}
				\psset{linestyle=dashed}\ncline{2,1}{2,13}
				\psset{linestyle=dashed}\ncline{5,1}{5,13}
				\psset{linestyle=dashed}\ncline{6,1}{6,13}
				\endpsmatrix
				\caption{Network $1c$.}
		\end{subfigure}
				\begin{subfigure}[h]{0.49\textwidth}
			\centering
				\psmatrix[colsep=.3cm,rowsep=.4cm,mnode=r]
				~ && && && ~ \\
				~ & && && && ~ & && ~ \\
				~ && && && && ~ && ~ \\
				\\
				~ && && && ~ && && ~ \\
				~ & ~ && ~ && && ~ && ~
				\ncline{1,7}{6,2}
				\ncline{1,7}{2,8}
				\ncline{2,8}{3,9}
				\ncline{3,9}{5,11}
				\ncline{2,8}{6,4}
				\ncline{3,9}{5,7}
				\ncarc[arcangle=40]{5,7}{6,8}
				\ncarc[arcangle=-40]{5,11}{6,10}
				\ncline{<->,arrowscale=1.5}{1,1}{2,1}
				\tlput{$\tau_{1}$}
				\ncline{<->,arrowscale=1.5}{2,1}{3,1}
				\tlput{$\tau_{2}$}
				\ncline{<->,arrowscale=1.5}{3,1}{5,1}
				\tlput{$\tau_{3}$}
				\ncline{<->,arrowscale=1.5}{5,1}{6,1}
				\tlput{$\tau_{4}$}
				\psset{linestyle=dashed}\ncline{2,1}{2,11}
				\psset{linestyle=dashed}\ncline{3,1}{3,11}
				\psset{linestyle=dashed}\ncline{5,1}{5,11}
				\endpsmatrix
				\caption{Network $1d$.}
		\end{subfigure}
				\caption{The first clock-like tree and the corresponding convergence-divergence networks.}
				\label{firststructurefourtaxon}
\end{figure}

There are three convergence-divergence networks with inequivalent pairs of leaves converging for the second structure of the four-taxon clock-like tree. The three inequivalent pairs of converging leaves are the two pairs of sister taxa convergence and the single non-sister taxa convergence pair. The four examples, including the second structure of the four-taxon clock-like tree, are shown below in Figure~\ref{secondstructurefourtaxon}.
\begin{figure}[H]
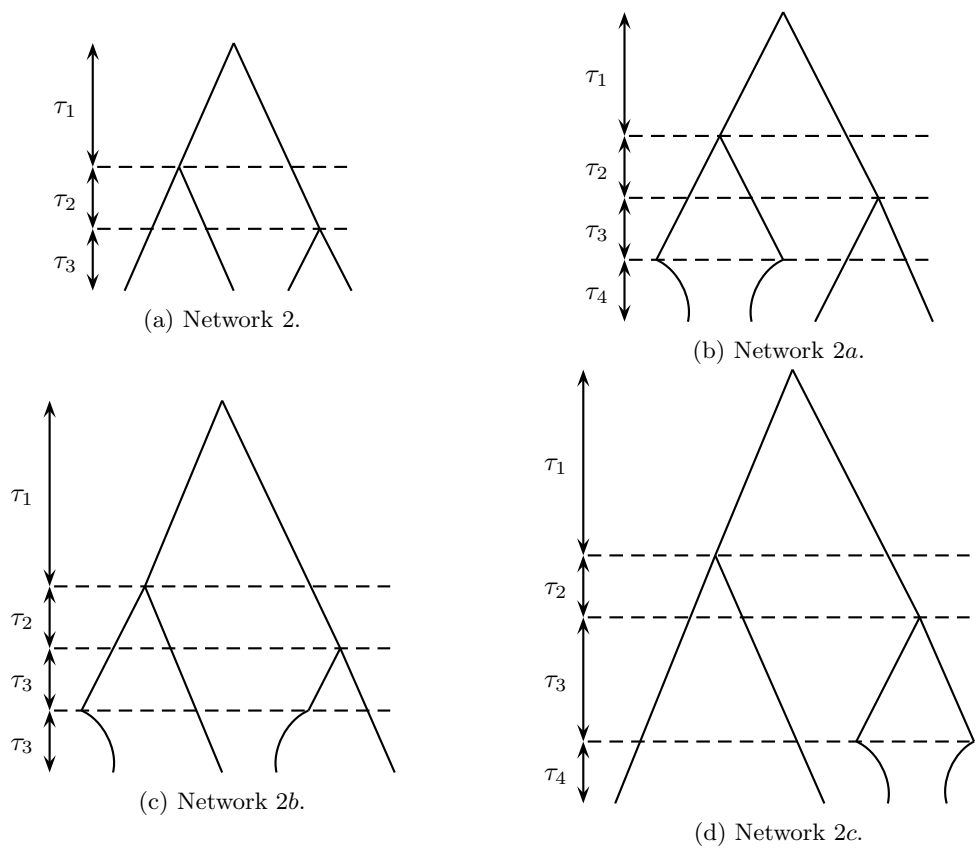

	\centering
		\begin{subfigure}[h]{0.49\textwidth}
			\centering
				\psmatrix[colsep=.3cm,rowsep=.4cm,mnode=r]
				~ & && && ~ \\
				\\
				~ & && ~ & && && & ~ \\
				~ && && && && ~ & ~ \\
				~ & ~ && && ~ && ~ && ~
				\ncline{1,6}{3,4}
				\ncline{3,4}{5,2}
				\ncline{3,4}{5,6}
				\ncline{1,6}{4,9}
				\ncline{4,9}{5,10}
				\ncline{4,9}{5,8}
				\ncline{<->,arrowscale=1.5}{1,1}{3,1}
				\tlput{$\tau_{1}$}
				\ncline{<->,arrowscale=1.5}{3,1}{4,1}
				\tlput{$\tau_{2}$}
				\ncline{<->,arrowscale=1.5}{4,1}{5,1}
				\tlput{$\tau_{3}$}
				\psset{linestyle=dashed}\ncline{3,1}{3,10}
				\psset{linestyle=dashed}\ncline{4,1}{4,10}
				\endpsmatrix
				\caption{Network $2$.}
		\end{subfigure}
		\begin{subfigure}[h]{0.49\textwidth}
			\centering
				\psmatrix[colsep=.3cm,rowsep=.4cm,mnode=r]
				~ & && && ~ \\
				\\
				~ & && ~ & && && && ~ \\
				~ && && && && ~ && ~ \\
				~ & ~ && && ~ && ~ & && ~ \\
				~ && ~ && ~ && ~ && && ~
				\ncline{1,6}{3,4}
				\ncline{3,4}{5,2}
				\ncline{3,4}{5,6}
				\ncline{1,6}{4,9}
				\ncline{4,9}{6,11}
				\ncline{4,9}{6,7}
				\ncarc[arcangle=40]{5,2}{6,3}
				\ncarc[arcangle=-40]{5,6}{6,5}
				\ncline{<->,arrowscale=1.5}{1,1}{3,1}
				\tlput{$\tau_{1}$}
				\ncline{<->,arrowscale=1.5}{3,1}{4,1}
				\tlput{$\tau_{2}$}
				\ncline{<->,arrowscale=1.5}{4,1}{5,1}
				\tlput{$\tau_{3}$}
				\ncline{<->,arrowscale=1.5}{5,1}{6,1}
				\tlput{$\tau_{4}$}
				\psset{linestyle=dashed}\ncline{3,1}{3,11}
				\psset{linestyle=dashed}\ncline{4,1}{4,11}
				\psset{linestyle=dashed}\ncline{5,1}{5,11}
				\endpsmatrix
				\caption{Network $2a$.}
		\end{subfigure}
		\begin{subfigure}[h]{0.49\textwidth}
			\centering
				\psmatrix[colsep=.3cm,rowsep=.4cm,mnode=r]
				~ && && && ~ \\
				\\
				\\
				~ & && ~ & && && && && ~ \\
				~ && && && && && ~ && ~ \\
				~ & ~ && && && && ~ & && ~ \\
				~ && ~ && && ~ && ~ && && ~
				\ncline{1,7}{4,4}
				\ncline{4,4}{6,2}
				\ncline{4,4}{7,7}
				\ncline{1,7}{5,11}
				\ncline{5,11}{6,10}
				\ncline{5,11}{7,13}
				\ncarc[arcangle=40]{6,2}{7,3}
				\ncarc[arcangle=-40]{6,10}{7,9}
				\ncline{<->,arrowscale=1.5}{1,1}{4,1}
				\tlput{$\tau_{1}$}
				\ncline{<->,arrowscale=1.5}{4,1}{5,1}
				\tlput{$\tau_{2}$}
				\ncline{<->,arrowscale=1.5}{5,1}{6,1}
				\tlput{$\tau_{3}$}
				\ncline{<->,arrowscale=1.5}{6,1}{7,1}
				\tlput{$\tau_{3}$}
				\psset{linestyle=dashed}\ncline{4,1}{4,13}
				\psset{linestyle=dashed}\ncline{5,1}{5,13}
				\psset{linestyle=dashed}\ncline{6,1}{6,13}
				\endpsmatrix
				\caption{Network $2b$.}
		\end{subfigure}
		\begin{subfigure}[h]{0.49\textwidth}
			\centering
				\psmatrix[colsep=.3cm,rowsep=.4cm,mnode=r]
				~ && && && && ~ \\
				\\
				\\
				~ & && && ~ & && && && && ~ \\
				~ && && && && && && ~ && ~ \\
				\\
				~ && && && && && ~ && && ~ \\
				~ & ~ && && && && ~ && ~ && ~
				\ncline{1,9}{4,6}
				\ncline{4,6}{8,2}
				\ncline{4,6}{8,10}
				\ncline{1,9}{5,13}
				\ncline{5,13}{7,11}
				\ncline{5,13}{7,15}
				\ncarc[arcangle=40]{7,11}{8,12}
				\ncarc[arcangle=-40]{7,15}{8,14}
				\ncline{<->,arrowscale=1.5}{1,1}{4,1}
				\tlput{$\tau_{1}$}
				\ncline{<->,arrowscale=1.5}{4,1}{5,1}
				\tlput{$\tau_{2}$}
				\ncline{<->,arrowscale=1.5}{5,1}{7,1}
				\tlput{$\tau_{3}$}
				\ncline{<->,arrowscale=1.5}{7,1}{8,1}
				\tlput{$\tau_{4}$}
				\psset{linestyle=dashed}\ncline{4,1}{4,15}
				\psset{linestyle=dashed}\ncline{5,1}{5,15}
				\psset{linestyle=dashed}\ncline{7,1}{7,15}
				\endpsmatrix
				\caption{Network $2c$.}
		\end{subfigure}
				\caption{The second clock-like tree and the corresponding convergence-divergence networks.}
				\label{secondstructurefourtaxon}
\end{figure}

Recall from Result~\ref{twotaxonresult}~on~page~\pageref{twotaxonresult} that clock-like networks with convergence periods between sister taxa cannot be distinguished from the same networks with the convergence period replaced with a divergence period. From this result we can immediately rule that Network $1$ and Network $1d$ are indistinguishable and disregard Network $1d$. Similarly, Network $2a$ and Network $2c$ will both be indistinguishable from Network $2$ and can be disregarded. This leaves six networks to examine: $1$, $1a$, $1b$, $1c$, $2$ and $2b$.

We will begin by stating the phylogenetic tensors in the Hadamard basis for each tree and network. Recall from Section~\ref{twotaxondistinguish}~on~page~\pageref{twotaxondistinguish} that $q_{0000}=1$ and $q_{0001}=q_{0010}=q_{0100}=q_{0111}=q_{1000}=q_{1011}=q_{1101}=q_{1110}=0$. Similarly to the two-taxon and three-taxon cases, for the binary symmetric model the transformed phylogenetic tensor elements for each tree and network that we will examine will satisfy the constraints,
\begin{align}\begin{split}
\begin{mycases}
q_{0000}&=1, \\
q_{0001}=q_{0010}=q_{0100}=q_{0111}=q_{1000}=q_{1011}=q_{1101}=q_{1110}&=0, \\
0<q_{0011},q_{0101},q_{0110},q_{1001},q_{1010},q_{1100},q_{1111}&\leq{}1.
\end{mycases}
\end{split}\end{align}

Consequently, we will not consider the transformed phylogenetic tensor elements, $q_{0000}$, $q_{0001}$, $q_{0010}$, $q_{0100}$, $q_{0111}$, $q_{1000}$, $q_{1011}$, $q_{1101}$ and $q_{1110}$ since they are uninformative.

We will then find the remaining constraints on the transformed phylogenetic tensor elements for each tree and network from their Gr\"{o}bner bases before comparing these constraints between trees and networks. We will only write down the Gr\"{o}bner bases for some of the trees and networks, since for the non-clock-like tree and some of the convergence-divergence networks there are many generators with many monomials in the Gr\"{o}bner basis. The Gr\"{o}bner bases that are not shown in Chapter~\ref{chapter6} will be shown in Appendices~\ref{cha.Appendix1},\ref{cha.Appendix2},\ref{cha.Appendix3}~and~\ref{cha.Appendix4}.

\subsection{Network $1$}

The first tree or network that we will examine is Network $1$, shown in Figure~\ref{Network1} below.
\begin{figure}[H]
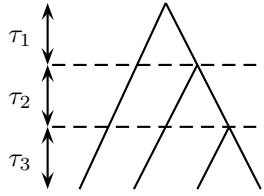

\centering
				\psmatrix[colsep=.3cm,rowsep=.4cm,mnode=r]
				~ && && ~ \\
				~ & && && ~ && ~ \\
				~ && && && ~ & ~ \\
				~ & ~ && ~ && ~ && ~
				\ncline{1,5}{4,2}
				\ncline{1,5}{4,8}
				\ncline{2,6}{4,4}
				\ncline{3,7}{4,6}
				\ncline{<->,arrowscale=1.5}{1,1}{2,1}
				\tlput{$\tau_{1}$}
				\ncline{<->,arrowscale=1.5}{2,1}{3,1}
				\tlput{$\tau_{2}$}
				\ncline{<->,arrowscale=1.5}{3,1}{4,1}
				\tlput{$\tau_{3}$}
				\psset{linestyle=dashed}\ncline{2,1}{2,8}
				\psset{linestyle=dashed}\ncline{3,1}{3,8}
				\endpsmatrix
				\caption{Network $1$.}
				\label{Network1}
\end{figure}
The variable transformed phylogenetic tensor elements are
\begin{align}\begin{split}
\left[\begin{array}{c} q_{0011} \\
q_{0101} \\
q_{0110} \\
q_{1001} \\
q_{1010} \\
q_{1100} \\
q_{1111} \\
\end{array}\right]=\left[\begin{array}{c} e^{-2\tau_{3}} \\
e^{-2\left(\tau_{2}+\tau_{3}\right)} \\
e^{-2\left(\tau_{2}+\tau_{3}\right)} \\
e^{-2\left(\tau_{1}+\tau_{2}+\tau_{3}\right)} \\
e^{-2\left(\tau_{1}+\tau_{2}+\tau_{3}\right)} \\
e^{-2\left(\tau_{1}+\tau_{2}+\tau_{3}\right)} \\
e^{-2\left(\tau_{1}+\tau_{2}+2\tau_{3}\right)} \\
\end{array}\right]&=\left[\begin{array}{c} y_{3}^{2} \\
y_{2}^{2}y_{3}^{2} \\
y_{2}^{2}y_{3}^{2} \\
y_{1}^{2}y_{2}^{2}y_{3}^{2} \\
y_{1}^{2}y_{2}^{2}y_{3}^{2} \\
y_{1}^{2}y_{2}^{2}y_{3}^{2} \\
y_{1}^{2}y_{2}^{2}y_{3}^{4} \\
\end{array}\right].
\end{split}\end{align}
We form the ideal,
\begin{align}\begin{split}
I=&\left<y_{3}^{2}-q_{0011},y_{2}^{2}y_{3}^{2}-q_{0101},y_{2}^{2}y_{3}^{2}-q_{0110},y_{1}^{2}y_{2}^{2}y_{3}^{2}-q_{1001},y_{1}^{2}y_{2}^{2}y_{3}^{2}-q_{1010},y_{1}^{2}y_{2}^{2}y_{3}^{2}-q_{1100},\right. \\
&\qquad{}\left.y_{1}^{2}y_{2}^{2}y_{3}^{4}-q_{1111}\right>.
\end{split}\end{align}
Using the \emph{Macaulay2} code from Section~\ref{macaulay2}~on~page~\pageref{macaulay2}, the Gr\"{o}bner basis of the ideal for Network $1$ is
\begin{align}\begin{split}
I=&\left<q_{1010}-q_{1100},q_{1001}-q_{1100},q_{0101}-q_{0110},q_{0011}q_{1100}-q_{1111},y_{3}^{2}-q_{0011},\right. \\
&\qquad{}\left.y_{2}^{2}q_{1111}-q_{0110}q_{1100},y_{2}^{2}q_{0011}-q_{0110},y_{1}^{2}q_{0110}-q_{1100}\right>.
\end{split}\end{align}
After finding the Gr\"{o}bner basis of the ideal, solving for the time parameters,
\begin{align}\begin{split}
\begin{cases}
y_{1}&=\sqrt{\frac{q_{1001}}{q_{0101}}}, \\
y_{2}&=\sqrt{\frac{q_{0101}}{q_{0011}}}, \\
y_{3}&=\sqrt{q_{0011}}.
\end{cases}
\end{split}\end{align}
Demanding $0<y_{1},y_{2},y_{3}\leq{}1$ from $y_{i}=e^{-\tau_{i}}$, where $\tau_{i}\geq{}0$ is a time parameter, the constraints on Network $1$ are then
\begin{align}\begin{split}
\begin{mycases}
q_{0101}&=q_{0110}, \\
q_{1001}&=q_{1010}, \\
q_{1001}&=q_{1100}, \\
q_{0011}q_{1001}&=q_{1111}, \\
q_{0101}&\leq{}q_{0011}, \\
q_{1001}&\leq{}q_{0101}.
\end{mycases}
\end{split}\end{align}

Recall from Definition~\ref{pairwisedistance}~on~page~\pageref{pairwisedistance} that the distance between any two leaves on a network is called the pairwise distance. The pairwise distance between two leaves, $a$ and $b$, is defined as
\begin{align}\begin{split}
d\left(a,b\right)=-\ln{\left(q_{A}\right)},
\end{split}\end{align}
where $A=i_{1}i_{2}\ldots{}i_{n}$, $a,b\in{}\left\{1,2,\ldots{},n\right\}$, $a\neq{}b$, $i_{a}=i_{b}=1$ and $i_{k}=0$ for all $k\in{}\left\{1,2,\ldots{},n\right\}\setminus{}\left\{a,b\right\}$.

The first equality, $q_{0101}=q_{0110}$, implies that the pairwise distance between taxa $2$ and $4$ must be equal to the pairwise distance between taxa $2$ and $3$, $d\left(2,4\right)=d\left(2,3\right)$. The second equality, $q_{1001}=q_{1010}$, implies that the pairwise distance between taxa $1$ and $4$ must be equal to the pairwise distance between taxa $1$ and $3$, $d\left(1,4\right)=d\left(1,3\right)$. These pairs of equal distances are obvious since taxa $3$ and $4$ are sister taxa. The third equality, $q_{1001}=q_{1100}$, implies that the pairwise distance between taxa $1$ and taxa $4$ must be equal to the pairwise distance between taxa $1$ and $2$, $d\left(1,4\right)=d\left(1,2\right)$. The pairwise distances are the sums of the edge lengths connecting the two leaves. We can see that these two pairwise distances must be equal by looking at Figure~\ref{Network1}.

We will now look at the fourth equality, $q_{0011}q_{1001}=q_{1111}$. We have defined pairwise distances in terms of the transformed phylogenetic tensor elements with two elements being one and the rest being zero. It is not immediately clear how to interpret this equality since we have not defined any of the pairwise distances in terms of $q_{1111}$. Focussing on the left hand side of the equality,
\begin{align}\begin{split}
q_{0011}q_{1001}=e^{-d\left(3,4\right)}e^{-d\left(1,4\right)}=e^{-\left(d\left(3,4\right)+d\left(1,4\right)\right)}=e^{-2\left(\tau_{1}+\tau_{2}+2\tau_{3}\right)}=q_{1111}.
\end{split}\end{align}

We will now define a \emph{total network distance} in a similar way to how we defined tree distances and pairwise distances.
\begin{definition}[\textbf{Total Network Distance}]
The \textbf{total network distance}, $d\left(1,2,3,4\right)$, on a four-taxon convergence network is expressed as
\begin{align*}\begin{split}
d\left(1,2,3,4\right)=-\ln{\left(q_{1111}\right)}.
\end{split}\end{align*}
\end{definition}
\label{totalnetworkdistance}
Notice that on Network $1$, the total network distance is $d\left(1,2,3,4\right)=d\left(3,4\right)+d\left(1,4\right)=d\left(3,4\right)+d\left(1,2\right)$, which is the sum of the pairwise distances either side of the cluster, $12|34$.

The first inequality, $q_{0101}\leq{}q_{0011}$, implies that the pairwise distance between taxa $2$ and $4$ must be greater than the pairwise distance between taxa $3$ and $4$, which can again be seen by looking at Figure~\ref{Network1}. The second inequality, $q_{1001}\leq{}q_{0101}$, implies that the pairwise distance between taxa $1$ and $4$ must be greater than the pairwise distance between taxa $2$ and $4$, which can also be seen by looking at Figure~\ref{Network1}. In summary,
\begin{align}\begin{split}
\begin{mycases}
d\left(3,4\right)\leq{}d\left(2,4\right)=d\left(2,3\right)\leq{}d\left(1,4\right)=d\left(1,3\right)&=d\left(1,2\right), \\
d\left(3,4\right)+d\left(1,2\right)&=d\left(1,2,3,4\right).
\end{mycases}
\end{split}\end{align}

\citet{semple2003phylogenetics} defined the four-point condition in terms of dissimilarity maps on a tree. We will define it in terms of the sums of pairwise distance.
\begin{theorem}[\textbf{Four-Point Condition}]
Given three different pairwise distances, $d_{A}$, $d_{B}$ and $d_{C}$, where $d_{A},d_{B},d_{C}\in{}\left\{d\left(3,4\right)+d\left(1,2\right),d\left(2,4\right)+d\left(1,3\right),d\left(2,3\right)+d\left(1,4\right)\right\}$,
\begin{align*}\begin{split}
d_{A}\leq{}d_{B}=d_{C}.
\end{split}\end{align*}
The four-point condition is satisfied if we have a phylogenetic tree.
\begin{proof}
Proof can be found in \citet{semple2003phylogenetics}.
\end{proof}
\end{theorem}

Notice that on Network $1$,
\begin{align}\begin{split}
d\left(3,4\right)+d\left(1,2\right)&=2\left(\tau_{1}+\tau_{2}+2\tau_{3}\right) \\
&\leq{}d\left(2,4\right)+d\left(1,3\right)=d\left(2,3\right)+d\left(1,4\right) \\
&=2\left(\tau_{1}+2\left(\tau_{2}+\tau_{3}\right)\right) \\
&=2\left(\tau_{1}+\tau_{2}+2\tau_{3}\right)+2\tau_{2} \\
&=d\left(3,4\right)+d\left(1,2\right)+2\tau_{2}.
\end{split}\end{align}
We can conclude that Network $1$ satisfies the four-point condition, which is to be expected since it is a tree.

We can use the four-point condition as a relatively simple check for distinguishability. If a given network does not satisfy the four-point condition then it will be distinguishable from trees. The reverse is not generally true, however. A tree or network can be distinguishable from another tree or network despite both satisfying the four-point condition.

\subsection{Network $1a$}

The next tree or network that we will examine is Network $1a$, shown in Figure~\ref{Network1a} below.
\begin{figure}[H]
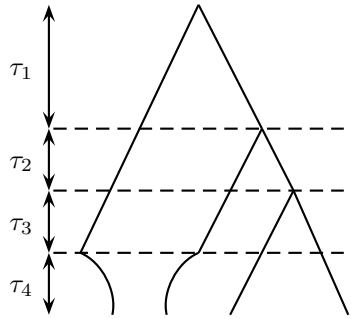

\centering
				\psmatrix[colsep=.3cm,rowsep=.4cm,mnode=r]
				~ & && && ~ \\
				\\
				~ & && && && ~ & && ~ \\
				~ && && && && ~ && ~ \\
				~ & ~ && && ~ & && && ~ \\
				~ && ~ && ~ && ~ && && ~
				\ncline{1,6}{5,2}
				\ncline{1,6}{3,8}
				\ncline{3,8}{4,9}
				\ncline{4,9}{6,11}
				\ncline{3,8}{5,6}
				\ncline{4,9}{6,7}
				\ncarc[arcangle=40]{5,2}{6,3}
				\ncarc[arcangle=-40]{5,6}{6,5}
				\ncline{<->,arrowscale=1.5}{1,1}{3,1}
				\tlput{$\tau_{1}$}
				\ncline{<->,arrowscale=1.5}{3,1}{4,1}
				\tlput{$\tau_{2}$}
				\ncline{<->,arrowscale=1.5}{4,1}{5,1}
				\tlput{$\tau_{3}$}
				\ncline{<->,arrowscale=1.5}{5,1}{6,1}
				\tlput{$\tau_{4}$}
				\psset{linestyle=dashed}\ncline{3,1}{3,11}
				\psset{linestyle=dashed}\ncline{4,1}{4,11}
				\psset{linestyle=dashed}\ncline{5,1}{5,11}
				\endpsmatrix
				\caption{Network $1a$.}
				\label{Network1a}
\end{figure}
The variable transformed phylogenetic tensor elements are
\begin{align}\begin{split}
\left[\begin{array}{c} q_{0011} \\
q_{0101} \\
q_{0110} \\
q_{1001} \\
q_{1010} \\
q_{1100} \\
q_{1111} \\
\end{array}\right]&=\left[\begin{array}{c} e^{-2\left(\tau_{3}+\tau_{4}\right)} \\
e^{-2\left(\tau_{2}+\tau_{3}+\tau_{4}\right)} \\
e^{-2\left(\tau_{2}+\tau_{3}+\tau_{4}\right)} \\
e^{-2\left(\tau_{1}+\tau_{2}+\tau_{3}+\tau_{4}\right)} \\
e^{-2\left(\tau_{1}+\tau_{2}+\tau_{3}+\tau_{4}\right)} \\
1-e^{-\tau_{4}}\left(1-e^{-2\left(\tau_{1}+\tau_{2}+\tau_{3}\right)}\right) \\
e^{-2\left(\tau_{1}+\tau_{2}+\tau_{3}+\tau_{4}\right)}\left(1-e^{-\tau_{4}}\left(1-e^{-2\left(\tau_{1}+\tau_{2}+\tau_{3}\right)}\right)\right) \\
\end{array}\right] \\
&=\left[\begin{array}{c} y_{3}^{2}y_{4}^{2} \\
y_{2}^{2}y_{3}^{2}y_{4}^{2} \\
y_{2}^{2}y_{3}^{2}y_{4}^{2} \\
y_{1}^{2}y_{2}^{2}y_{3}^{2}y_{4}^{2} \\
y_{1}^{2}y_{2}^{2}y_{3}^{2}y_{4}^{2} \\
1-y_{4}\left(1-y_{1}^{2}y_{2}^{2}y_{3}^{2}\right) \\
y_{3}^{2}y_{4}^{2}\left(1-y_{4}\left(1-y_{1}^{2}y_{2}^{2}y_{3}^{2}\right)\right) \\
\end{array}\right].
\end{split}\end{align}
We form the ideal,
\begin{align}\begin{split}
I=&\left<y_{3}^{2}y_{4}^{2}-q_{0011},y_{2}^{2}y_{3}^{2}y_{4}^{2}-q_{0101},y_{2}^{2}y_{3}^{2}y_{4}^{2}-q_{0110},y_{1}^{2}y_{2}^{2}y_{3}^{2}y_{4}^{2}-q_{1001},y_{1}^{2}y_{2}^{2}y_{3}^{2}y_{4}^{2}-q_{1010},\right. \\
&\qquad{}\left.1-y_{4}\left(1-y_{1}^{2}y_{2}^{2}y_{3}^{2}\right)-q_{1100},y_{3}^{2}y_{4}^{2}\left(1-y_{4}\left(1-y_{1}^{2}y_{2}^{2}y_{3}^{2}\right)\right)-q_{1111}\right>.
\end{split}\end{align}
Using the \emph{Macaulay2} code from Section~\ref{macaulay2}~on~page~\pageref{macaulay2}, the Gr\"{o}bner basis of the ideal for Network $1a$ is
\begin{align}\begin{split}
I=&\left<q_{1001}-q_{1010},q_{0101}-q_{0110},q_{0011}q_{1100}-q_{1111},y_{4}^{2}+y_{4}q_{1100}-y_{4}-q_{1010},\right. \\
&\qquad{}\left.y_{3}^{2}q_{1010}^{2}+y_{4}q_{0011}-y_{4}q_{1111}-q_{0011}q_{1010}-q_{0011}-q_{1100}q_{1111}+2q_{1111},\right. \\
&\qquad{}\left.y_{3}^{2}y_{4}q_{1100}-y_{3}^{2}y_{4}-y_{3}^{2}q_{1010}+q_{0011},y_{3}^{2}y_{4}q_{1010}-y_{4}q_{0011}+q_{0011}-q_{1111},\right. \\
&\qquad{}\left.y_{3}^{2}y_{4}q_{0011}-y_{3}^{2}y_{4}q_{1111}+y_{3}^{2}q_{0011}q_{1010}-q_{0011}^{2},y_{2}^{2}q_{1111}-q_{0110}q_{1100},y_{2}^{2}q_{0011}-q_{0110},\right. \\
&\qquad{}\left.y_{1}^{2}q_{0110}-q_{1010},y_{1}^{2}y_{2}^{2}y_{3}^{2}q_{1010}-y_{4}q_{1100}+y_{4}-q_{1010}-q_{1100}^{2}+2q_{1100}-1,\right. \\
&\qquad{}\left.y_{1}^{2}y_{2}^{2}y_{3}^{2}y_{4}-y_{4}-q_{1100}+1\right>.
\end{split}\end{align}
After finding the Gr\"{o}bner basis of the ideal, solving for the time parameters,
\begin{align}\begin{split}
\begin{cases}
y_{1}&=\sqrt{\frac{q_{1001}}{q_{0101}}}, \\
y_{2}&=\sqrt{\frac{q_{0101}}{q_{0011}}}, \\
y_{3}&=\frac{\sqrt{q_{0011}}\left(-\left(1-q_{1100}\right)+\sqrt{\left(1-q_{1100}\right)^{2}+4q_{1001}}\right)}{2q_{1001}}, \\
y_{4}&=\frac{1}{2}\left(\left(1-q_{1100}\right)+\sqrt{\left(1-q_{1100}\right)^{2}+4q_{1001}}\right).
\end{cases}
\end{split}\end{align}

We demand $0<y_{1},y_{2},y_{3},y_{4}\leq{}1$ from $y_{i}=e^{-\tau_{i}}$.
Demanding $y_{1}\leq{}1$ implies
\begin{align}
\begin{split}
q_{1001}&\leq{}q_{0101}.
\end{split}
\label{q1}
\end{align}
Demanding $y_{2}\leq{}1$ implies
\begin{align}
\begin{split}
q_{0101}&\leq{}q_{0011}.
\end{split}
\label{q2}
\end{align}

Now suppose we have some expression similar to the expression for $y_{4}$ in the form
\begin{align}\begin{split}
y_{i}&=\frac{1}{2}\left(\left(1-q_{a}\right)+\sqrt{\left(1-q_{a}\right)^{2}+4q_{b}}\right),
\end{split}\end{align}
where $y_{i}$ is a time parameter and $q_{a}$ and $q_{b}$ are transformed phylogenetic tensor elements.
Demanding $y_{i}\leq{}1$,
\begin{align}\begin{split}
\frac{1}{2}\left(\left(1-q_{a}\right)+\sqrt{\left(1-q_{a}\right)^{2}+4q_{b}}\right)&\leq{}1 \\
\Leftrightarrow{}\left(1-q_{a}\right)+\sqrt{\left(1-q_{a}\right)^{2}+4q_{b}}&\leq{}2 \\
\Leftrightarrow{}\sqrt{\left(1-q_{a}\right)^{2}+4q_{b}}&\leq{}1+q_{a}.
\end{split}\end{align}
Since $q_{a},q_{b}>0$,
\begin{align}
\begin{split}
\Leftrightarrow{}\left(1-q_{a}\right)^{2}+4q_{b}&\leq{}\left(1+q_{a}\right)^{2} \\
\Leftrightarrow{}1+q_{a}^{2}-2q_{a}+4q_{b}&\leq{}1+q_{a}^{2}+2q_{a} \\
\Leftrightarrow{}4q_{b}&\leq{}4q_{a} \\
\Leftrightarrow{}q_{b}&\leq{}q_{a}.
\end{split}
\label{yresult1}
\end{align}
From (\ref{yresult1}) we can conclude that
\begin{align}\begin{split}
q_{1001}\leq{}q_{1100}.
\end{split}\end{align}

Now suppose we have an expression similar to the expression for $y_{3}$ in the form
\begin{align}\begin{split}
y_{j}&=\frac{\sqrt{q_{e}}\left(-\left(1-q_{c}\right)+\sqrt{\left(1-q_{c}\right)^{2}+4q_{d}}\right)}{2q_{d}}.
\end{split}\end{align}
Now, demanding $y_{b}\leq{}1$,
\begin{align}\begin{split}
\frac{\sqrt{q_{e}}\left(-\left(1-q_{c}\right)+\sqrt{\left(1-q_{c}\right)^{2}+4q_{d}}\right)}{2q_{d}}&\leq{}1.
\end{split}\end{align}
Since $q_{d}>0$,
\begin{align}\begin{split}
\sqrt{q_{e}}\left(-\left(1-q_{c}\right)+\sqrt{\left(1-q_{c}\right)^{2}+4q_{d}}\right)&\leq{}2q_{d}.
\end{split}\end{align}
Since $q_{e}>0$,
\begin{align}\begin{split}
-\left(1-q_{c}\right)+\sqrt{\left(1-q_{c}\right)^{2}+4q_{d}}&\leq{}\frac{2q_{d}}{\sqrt{q_{e}}} \\
\Leftrightarrow{}\sqrt{\left(1-q_{c}\right)^{2}+4q_{d}}&\leq{}\frac{2q_{d}}{\sqrt{q_{e}}}+\left(1-q_{c}\right).
\end{split}\end{align}
Since $q_{c}\leq{}1$,
\begin{align}
\begin{split}
\left(1-q_{c}\right)^{2}+4q_{d}&\leq{}\frac{4q_{d}^{2}}{q_{e}}+\left(1-q_{c}\right)^{2}+\frac{4q_{d}\left(1-q_{c}\right)}{\sqrt{q_{e}}} \\
\Leftrightarrow{}4q_{d}&\leq{}\frac{4q_{d}^{2}}{q_{e}}+\frac{4q_{d}\left(1-q_{c}\right)}{\sqrt{q_{e}}} \\
\Leftrightarrow{}1&\leq{}\frac{q_{d}}{q_{e}}+\frac{\left(1-q_{c}\right)}{\sqrt{q_{e}}} \\
\Leftrightarrow{}q_{e}&\leq{}q_{d}+\sqrt{q_{e}}\left(1-q_{c}\right) \\
\Leftrightarrow{}\left(q_{e}-q_{d}\right)&\leq{}\sqrt{q_{e}}\left(1-q_{c}\right).
\end{split}
\label{yresult2}
\end{align}
If $q_{d}\leq{}q_{e}$ then
\begin{align}
\begin{split}
\Leftrightarrow{}\left(q_{e}-q_{d}\right)^{2}&\leq{}q_{e}\left(1-q_{c}\right)^{2}.
\end{split}
\label{yresult3}
\end{align}
Since from (\ref{q1}), $q_{1001}\leq{}q_{0101}$, and from (\ref{q2}), $q_{0101}\leq{}q_{0011}$, then
\begin{align}\begin{split}
q_{1001}\leq{}q_{0011}.
\end{split}\end{align}
From (\ref{yresult2}) and (\ref{yresult3}) we can conclude that
\begin{align}\begin{split}
\left(q_{0011}-q_{1001}\right)^{2}&\leq{}q_{0011}\left(1-q_{1100}\right)^{2}.
\end{split}\end{align}

The constraints on Network $1a$ are then
\begin{align}\begin{split}
\begin{mycases}
q_{0101}&=q_{0110}, \\
q_{1001}&=q_{1010}, \\
q_{0011}q_{1100}&=q_{1111}, \\
q_{0101}&\leq{}q_{0011}, \\
q_{1001}&\leq{}q_{0101}, \\
q_{1001}&\leq{}q_{1100}, \\
\left(q_{0011}-q_{1001}\right)^{2}&\leq{}q_{0011}\left(1-q_{1100}\right)^{2}.
\end{mycases}
\end{split}\end{align}

The constraints on the distances for Network $1a$ that are equivalent to the constraints on the distances for Network $1$ are
\begin{align}\begin{split}
\begin{mycases}
d\left(3,4\right)\leq{}d\left(2,4\right)=d\left(2,3\right)\leq{}d\left(1,4\right)&=d\left(1,3\right), \\
d\left(3,4\right)+d\left(1,2\right)&=d\left(1,2,3,4\right).
\end{mycases}
\end{split}\end{align}
Network $1a$ also has the distance constraint, $d\left(1,2\right)\leq{}d\left(1,4\right)$, while on Network $1$, $d\left(1,2\right)=d\left(1,4\right)$. We will see later how the last inequality constraint compares to the constraints for the other trees and networks.

From $d\left(1,3\right)=d\left(1,4\right)$ and $d\left(2,4\right)=d\left(2,3\right)$ we can conclude that
\begin{align}\begin{split}
d\left(2,4\right)+d\left(1,3\right)=d\left(2,3\right)+d\left(1,4\right).
\end{split}\end{align}
From $d\left(3,4\right)\leq{}d\left(2,3\right)$ and $d\left(1,2\right)\leq{}d\left(1,4\right)$ we can conclude that
\begin{align}\begin{split}
d\left(3,4\right)+d\left(1,2\right)\leq{}d\left(2,3\right)+d\left(1,4\right).
\end{split}\end{align}
Despite being a convergence-divergence network and not a tree, we can conclude that Network $1a$ must satisfy the four-point condition,
\begin{align}\begin{split}
d\left(3,4\right)+d\left(1,2\right)\leq{}d\left(2,4\right)+d\left(1,3\right)=d\left(2,3\right)+d\left(1,4\right).
\end{split}\end{align}

\subsection{Network $1b$}

The next tree or network that we will examine is Network $1b$, shown in Figure~\ref{Network1b} below.
\begin{figure}[H]
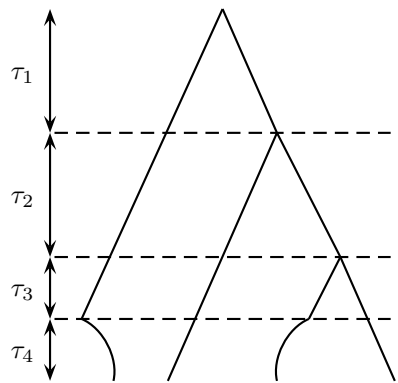

\centering
				\psmatrix[colsep=.3cm,rowsep=.4cm,mnode=r]
				~ && && && ~ \\
				\\
				~ && && && && ~ && && ~ \\
				\\
				~ && && && && && ~ && ~ \\
				~ & ~ && && && && ~ & && ~ \\
				~ && ~ && ~ && && ~ && && ~
				\ncline{1,7}{6,2}
				\ncline{1,7}{3,9}
				\ncline{3,9}{5,11}
				\ncline{5,11}{7,13}
				\ncline{3,9}{7,5}
				\ncline{5,11}{6,10}
				\ncarc[arcangle=40]{6,2}{7,3}
				\ncarc[arcangle=-40]{6,10}{7,9}
				\ncline{<->,arrowscale=1.5}{1,1}{3,1}
				\tlput{$\tau_{1}$}
				\ncline{<->,arrowscale=1.5}{3,1}{5,1}
				\tlput{$\tau_{2}$}
				\ncline{<->,arrowscale=1.5}{5,1}{6,1}
				\tlput{$\tau_{3}$}
				\ncline{<->,arrowscale=1.5}{6,1}{7,1}
				\tlput{$\tau_{4}$}
				\psset{linestyle=dashed}\ncline{3,1}{3,13}
				\psset{linestyle=dashed}\ncline{5,1}{5,13}
				\psset{linestyle=dashed}\ncline{6,1}{6,13}
				\endpsmatrix
				\caption{Network $1b$.}
				\label{Network1b}
\end{figure}
The variable transformed phylogenetic tensor elements are
\begin{align}\begin{split}
\left[\begin{array}{c} q_{0011} \\
q_{0101} \\
q_{0110} \\
q_{1001} \\
q_{1010} \\
q_{1100} \\
q_{1111} \\
\end{array}\right]&=\left[\begin{array}{c} e^{-2\left(\tau_{3}+\tau_{4}\right)} \\
e^{-2\left(\tau_{2}+\tau_{3}+\tau_{4}\right)} \\
e^{-2\left(\tau_{2}+\tau_{3}+\tau_{4}\right)} \\
1-e^{-\tau_{4}}\left(1-e^{-2\left(\tau_{1}+\tau_{2}+\tau_{3}\right)}\right) \\
e^{-2\left(\tau_{1}+\tau_{2}+\tau_{3}+\tau_{4}\right)} \\
e^{-2\left(\tau_{1}+\tau_{2}+\tau_{3}+\tau_{4}\right)} \\
e^{-2\left(\tau_{2}+\tau_{3}+\tau_{4}\right)}\left(1-e^{-\tau_{4}}\left(1-e^{-2\left(\tau_{1}+\tau_{3}\right)}\right)\right) \\
\end{array}\right] \\
&=\left[\begin{array}{c} y_{3}^{2}y_{4}^{2} \\
y_{2}^{2}y_{3}^{2}y_{4}^{2} \\
y_{2}^{2}y_{3}^{2}y_{4}^{2} \\
1-y_{4}\left(1-y_{1}^{2}y_{2}^{2}y_{3}^{2}\right) \\
y_{1}^{2}y_{2}^{2}y_{3}^{2}y_{4}^{2} \\
y_{1}^{2}y_{2}^{2}y_{3}^{2}y_{4}^{2} \\
y_{2}^{2}y_{3}^{2}y_{4}^{2}\left(1-y_{4}\left(1-y_{1}^{2}y_{3}^{2}\right)\right) \\
\end{array}\right].
\end{split}\end{align}
We form the ideal,
\begin{align}\begin{split}
I=&\left<y_{3}^{2}y_{4}^{2}-q_{0011},y_{2}^{2}y_{3}^{2}y_{4}^{2}-q_{0101},y_{2}^{2}y_{3}^{2}y_{4}^{2}-q_{0110},1-y_{4}\left(1-y_{1}^{2}y_{2}^{2}y_{3}^{2}\right)-q_{1001},y_{1}^{2}y_{2}^{2}y_{3}^{2}y_{4}^{2}-q_{1010},\right. \\
&\qquad{}\left.y_{1}^{2}y_{2}^{2}y_{3}^{2}y_{4}^{2}-q_{1100},y_{2}^{2}y_{3}^{2}y_{4}^{2}\left(1-y_{4}\left(1-y_{1}^{2}y_{3}^{2}\right)\right)-q_{1111}\right>.
\end{split}\end{align}
The Gr\"{o}bner basis of the ideal is given in Appendix~\ref{cha.Appendix1}, it has $24$ generators. After finding the Gr\"{o}bner basis of the ideal, solving for the time parameters,
\begin{align}\begin{split}
\begin{cases}
y_{1}&=\sqrt{\frac{q_{1010}}{q_{0101}}}, \\
y_{2}&=\sqrt{\frac{q_{0101}}{q_{0011}}}, \\
y_{3}&=\frac{\sqrt{q_{0011}}\left(-\left(1-q_{1001}\right)+\sqrt{\left(1-q_{1001}\right)^{2}+4q_{1010}}\right)}{2q_{1010}}, \\
y_{4}&=\frac{1}{2}\left(\left(1-q_{1001}\right)+\sqrt{\left(1-q_{1001}\right)^{2}+4q_{1010}}\right).
\end{cases}
\end{split}\end{align}

Demanding $y_{1},y_{2}\leq{}1$ and from (\ref{yresult1}), (\ref{yresult2}) and (\ref{yresult3}), the constraints on Network $1b$ are then
\begin{align}\begin{split}
\begin{mycases}
q_{0101}&=q_{0110}, \\
q_{1010}&=q_{1100}, \\
q_{0011}^2q_{1010}-q_{0011}q_{0101}q_{1001}^2+q_{0011}q_{0101}q_{1001}-2q_{0011}q_{0101}q_{1010}& \\
+q_{0011}q_{1001}q_{1111}-q_{0011}q_{1111}-q_{0101}^2q_{1001}+q_{0101}^2q_{1010}& \\
+q_{0101}q_{1001}q_{1111}+q_{0101}q_{1111}-q_{1111}^2&=0, \\
q_{0101}&\leq{}q_{0011}, \\
q_{1010}&\leq{}q_{0101}, \\
q_{1010}&\leq{}q_{1001}, \\
\left(q_{0011}-q_{1010}\right)^{2}&\leq{}q_{0011}\left(1-q_{1001}\right)^{2}.
\end{mycases}
\end{split}\end{align}

The constraints on the distances for Network $1b$ that are equivalent to the constraints on the distances for Network $1$ are
\begin{align}\begin{split}
d\left(3,4\right)\leq{}d\left(2,4\right)=d\left(2,3\right)\leq{}d\left(1,3\right)=d\left(1,2\right).
\end{split}\end{align}
Network $1b$ also has the distance constraint, $d\left(1,4\right)\leq{}d\left(1,3\right)$, while on Network $1$, $d\left(1,4\right)=d\left(1,3\right)$. While on Network $1$ the total network distance satisfies $d\left(1,2,3,4\right)=d\left(3,4\right)+d\left(1,2\right)$, on Network $1a$, $d\left(1,2,3,4\right)\neq{}d\left(3,4\right)+d\left(1,2\right)$.

From $d\left(3,4\right)\leq{}d\left(2,4\right)$ and $d\left(1,2\right)=d\left(1,3\right)$ we can conclude that
\begin{align}\begin{split}
d\left(3,4\right)+d\left(1,2\right)\leq{}d\left(2,4\right)+d\left(1,3\right).
\end{split}\end{align}
From $d\left(2,4\right)=d\left(2,3\right)$ and $d\left(1,4\right)\leq{}d\left(1,3\right)$ we can conclude that
\begin{align}\begin{split}
d\left(2,3\right)+d\left(1,4\right)\leq{}d\left(2,4\right)+d\left(1,3\right).
\end{split}\end{align}
We can conclude that the sum of pairwise distances, $d\left(2,4\right)+d\left(1,3\right)$, must be greater than or equal to each of the other two sums of pairwise distances. For the four-point condition to be satisfied the two greatest sums of pairwise distances must be equal. We can conclude that Network $1b$ does not generally satisfy the four-point condition.

\subsection{Network $1c$}

The next tree or network that we will examine is Network $1c$, shown in Figure~\ref{Network1c} below.
\begin{figure}[H]
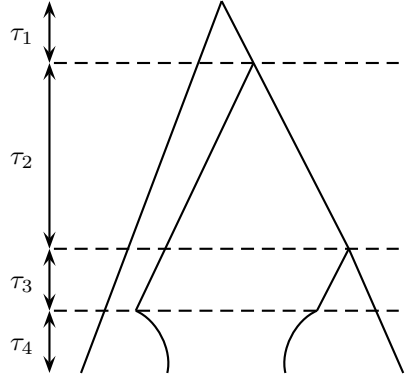

\centering
				\psmatrix[colsep=.3cm,rowsep=.4cm,mnode=r]
				~ && && && ~ \\
				~ & && && && ~ & && && ~ \\
				\\
				\\
				~ && && && && && ~ && ~ \\
				~ && & ~ && && && ~ & && ~ \\
				~ & ~ & && ~ && && ~ && && ~
				\ncline{1,7}{7,2}
				\ncline{1,7}{2,8}
				\ncline{2,8}{5,11}
				\ncline{5,11}{7,13}
				\ncline{2,8}{6,4}
				\ncline{5,11}{6,10}
				\ncarc[arcangle=40]{6,4}{7,5}
				\ncarc[arcangle=-40]{6,10}{7,9}
				\ncline{<->,arrowscale=1.5}{1,1}{2,1}
				\tlput{$\tau_{1}$}
				\ncline{<->,arrowscale=1.5}{2,1}{5,1}
				\tlput{$\tau_{2}$}
				\ncline{<->,arrowscale=1.5}{5,1}{6,1}
				\tlput{$\tau_{3}$}
				\ncline{<->,arrowscale=1.5}{6,1}{7,1}
				\tlput{$\tau_{4}$}
				\psset{linestyle=dashed}\ncline{2,1}{2,13}
				\psset{linestyle=dashed}\ncline{5,1}{5,13}
				\psset{linestyle=dashed}\ncline{6,1}{6,13}
				\endpsmatrix
				\caption{Network $1c$.}
				\label{Network1c}
\end{figure}
The variable transformed phylogenetic tensor elements are
\begin{align}\begin{split}
\left[\begin{array}{c} q_{0011} \\
q_{0101} \\
q_{0110} \\
q_{1001} \\
q_{1010} \\
q_{1100} \\
q_{1111} \\
\end{array}\right]&=\left[\begin{array}{c} e^{-2\left(\tau_{3}+\tau_{4}\right)} \\
1-e^{-\tau_{4}}\left(1-e^{-2\left(\tau_{2}+\tau_{3}\right)}\right) \\
e^{-2\left(\tau_{2}+\tau_{3}+\tau_{4}\right)} \\
e^{-2\left(\tau_{1}+\tau_{2}+\tau_{3}+\tau_{4}\right)} \\
e^{-2\left(\tau_{1}+\tau_{2}+\tau_{3}+\tau_{4}\right)} \\
e^{-2\left(\tau_{1}+\tau_{2}+\tau_{3}+\tau_{4}\right)} \\
e^{-2\left(\tau_{1}+\tau_{2}+\tau_{3}+\tau_{4}\right)}\left(1-e^{-\tau_{4}}\left(1-e^{-2\tau_{3}}\right)\right) \\
\end{array}\right] \\
&=\left[\begin{array}{c} y_{3}^{2}y_{4}^{2} \\
1-y_{4}\left(1-y_{2}^{2}y_{3}^{2}\right) \\
y_{2}^{2}y_{3}^{2}y_{4}^{2} \\
y_{1}^{2}y_{2}^{2}y_{3}^{2}y_{4}^{2} \\
y_{1}^{2}y_{2}^{2}y_{3}^{2}y_{4}^{2} \\
y_{1}^{2}y_{2}^{2}y_{3}^{2}y_{4}^{2} \\
y_{1}^{2}y_{2}^{2}y_{3}^{2}y_{4}^{2}\left(1-y_{4}\left(1-y_{3}^{2}\right)\right) \\
\end{array}\right].
\end{split}\end{align}
We form the ideal,
\begin{align}\begin{split}
I=&\left<y_{3}^{2}y_{4}^{2}-q_{0011},1-y_{4}\left(1-y_{2}^{2}y_{3}^{2}\right)-q_{0101},y_{2}^{2}y_{3}^{2}y_{4}^{2}-q_{0110},y_{1}^{2}y_{2}^{2}y_{3}^{2}y_{4}^{2}-q_{1001},y_{1}^{2}y_{2}^{2}y_{3}^{2}y_{4}^{2}-q_{1010},\right. \\
&\qquad{}\left.y_{1}^{2}y_{2}^{2}y_{3}^{2}y_{4}^{2}-q_{1100},y_{1}^{2}y_{2}^{2}y_{3}^{2}y_{4}^{2}\left(1-y_{4}\left(1-y_{3}^{2}\right)\right)-q_{1111}\right>.
\end{split}\end{align}
The Gr\"{o}bner basis of the ideal is given in Appendix~\ref{cha.Appendix2}, it has $30$ generators. After finding the Gr\"{o}bner basis of the ideal, solving for the time parameters,
\begin{align}\begin{split}
\begin{cases}
y_{1}&=\sqrt{\frac{q_{1001}}{q_{0110}}}, \\
y_{2}&=\sqrt{\frac{q_{0110}}{q_{0011}}}, \\
y_{3}&=\frac{\sqrt{q_{0011}}\left(-\left(1-q_{0101}\right)+\sqrt{\left(1-q_{0101}\right)^{2}+4q_{0110}}\right)}{2q_{0110}}, \\
y_{4}&=\frac{1}{2}\left(\left(1-q_{0101}\right)+\sqrt{\left(1-q_{0101}\right)^{2}+4q_{0110}}\right).
\end{cases}
\end{split}\end{align}

Demanding $y_{1},y_{2}\leq{}1$ and from (\ref{yresult1}), (\ref{yresult2}) and (\ref{yresult3}), the constraints on Network $1c$ are then
\begin{align}\begin{split}
\begin{mycases}
q_{1001}&=q_{1010}, \\
q_{1001}&=q_{1100}, \\
q_{0011}^2q_{1001}^2-q_{0011}q_{0101}^2q_{1001}^2+q_{0011}q_{0101}q_{1001}^2+q_{0011}q_{0101}q_{1001}q_{1111}& \\
-2q_{0011}q_{0110}q_{1001}^2-q_{0011}q_{1001}q_{1111}-q_{0101}q_{0110}q_{1001}^2& \\
+q_{0101}q_{0110}q_{1001}q_{1111}+q_{0110}^2q_{1001}^2+q_{0110}q_{1001}q_{1111}-q_{0110}q_{1111}^2&=0, \\
q_{0110}&\leq{}q_{0011}, \\
q_{0110}&\leq{}q_{0101}, \\
q_{1001}&\leq{}q_{0110}, \\
\left(q_{0011}-q_{0110}\right)^{2}&\leq{}q_{0011}\left(1-q_{0101}\right)^{2}.
\end{mycases}
\end{split}\end{align}

The constraints on the distances for Network $1c$ that are equivalent to the constraints on the distances for Network $1$ are
\begin{align}\begin{split}
d\left(3,4\right)\leq{}d\left(2,3\right)\leq{}d\left(1,4\right)=d\left(1,3\right)=d\left(1,2\right).
\end{split}\end{align}
Network $1c$ also has the distance constraint, $d\left(2,4\right)\leq{}d\left(2,3\right)$, while on Network $1$, $d\left(2,4\right)=d\left(2,3\right)$. While on Network $1$ the total network distance satisfies $d\left(1,2,3,4\right)=d\left(3,4\right)+d\left(1,2\right)$, on Network $1c$, $d\left(1,2,3,4\right)\neq{}d\left(3,4\right)+d\left(1,2\right)$. We will see later how the quintic equality constraint with $11$ terms compares to the constraints for the other trees and networks.

From $d\left(3,4\right)\leq{}d\left(2,3\right)$ and $d\left(1,2\right)=d\left(1,4\right)$ we can conclude that
\begin{align}\begin{split}
d\left(3,4\right)+d\left(1,2\right)\leq{}d\left(2,3\right)+d\left(1,4\right).
\end{split}\end{align}
From $d\left(2,4\right)\leq{}d\left(2,3\right)$ and $d\left(1,3\right)=d\left(1,4\right)$ we can conclude that
\begin{align}\begin{split}
d\left(2,4\right)+d\left(1,3\right)\leq{}d\left(2,3\right)+d\left(1,4\right).
\end{split}\end{align}
We can conclude that the sum of pairwise distances, $d\left(2,3\right)+d\left(1,4\right)$, must be greater than or equal to each of the other two sums of pairwise distances. For the four-point condition to be satisfied the two greatest sums of pairwise distances must be equal. We can conclude that Network $1c$ does not generally satisfy the four-point condition.

\subsection{Network $2$}

The next tree or network that we will examine is Network $2$, shown in Figure~\ref{Network2} below.
\begin{figure}[H]
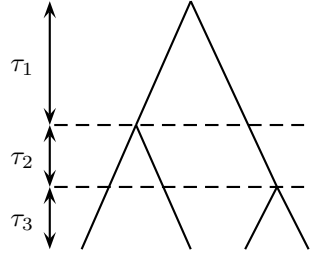

\centering
				\psmatrix[colsep=.3cm,rowsep=.4cm,mnode=r]
				~ & && && ~ \\
				\\
				~ & && ~ & && && & ~ \\
				~ && && && && ~ & ~ \\
				~ & ~ && && ~ && ~ && ~
				\ncline{1,6}{3,4}
				\ncline{3,4}{5,2}
				\ncline{3,4}{5,6}
				\ncline{1,6}{4,9}
				\ncline{4,9}{5,10}
				\ncline{4,9}{5,8}
				\ncline{<->,arrowscale=1.5}{1,1}{3,1}
				\tlput{$\tau_{1}$}
				\ncline{<->,arrowscale=1.5}{3,1}{4,1}
				\tlput{$\tau_{2}$}
				\ncline{<->,arrowscale=1.5}{4,1}{5,1}
				\tlput{$\tau_{3}$}
				\psset{linestyle=dashed}\ncline{3,1}{3,10}
				\psset{linestyle=dashed}\ncline{4,1}{4,10}
				\endpsmatrix
				\caption{Network $2$.}
				\label{Network2}
\end{figure}
The variable transformed phylogenetic tensor elements are
\begin{align}\begin{split}
\left[\begin{array}{c} q_{0011} \\
q_{0101} \\
q_{0110} \\
q_{1001} \\
q_{1010} \\
q_{1100} \\
q_{1111} \\
\end{array}\right]=\left[\begin{array}{c} e^{-2\tau_{3}} \\
e^{-2\left(\tau_{1}+\tau_{2}+\tau_{3}\right)} \\
e^{-2\left(\tau_{1}+\tau_{2}+\tau_{3}\right)} \\
e^{-2\left(\tau_{1}+\tau_{2}+\tau_{3}\right)} \\
e^{-2\left(\tau_{1}+\tau_{2}+\tau_{3}\right)} \\
e^{-2\left(\tau_{2}+\tau_{3}\right)} \\
e^{-2\left(\tau_{2}+2\tau_{3}\right)} \\
\end{array}\right]
&=\left[\begin{array}{c} y_{3}^{2} \\
y_{1}^{2}y_{2}^{2}y_{3}^{2} \\
y_{1}^{2}y_{2}^{2}y_{3}^{2} \\
y_{1}^{2}y_{2}^{2}y_{3}^{2} \\
y_{1}^{2}y_{2}^{2}y_{3}^{2} \\
y_{2}^{2}y_{3}^{2} \\
y_{2}^{2}y_{3}^{4} \\
\end{array}\right].
\end{split}\end{align}
We form the ideal,
\begin{align}\begin{split}
I=&\left<y_{3}^{2}-q_{0011},y_{1}^{2}y_{2}^{2}y_{3}^{2}-q_{0101},y_{1}^{2}y_{2}^{2}y_{3}^{2}-q_{0110},y_{1}^{2}y_{2}^{2}y_{3}^{2}-q_{1001},y_{1}^{2}y_{2}^{2}y_{3}^{2}-q_{1010},y_{2}^{2}y_{3}^{2}-q_{1100},\right. \\
&\qquad{}\left.y_{2}^{2}y_{3}^{4}-q_{1111}\right>.
\end{split}\end{align}
Using the \emph{Macaulay2} code from Section~\ref{macaulay2}~on~page~\pageref{macaulay2}, the Gr\"{o}bner basis of the ideal for Network $2$ is
\begin{align}\begin{split}
I=&\left<q_{1001}-q_{1010},q_{0110}-q_{1010},q_{0101}-q_{1010},q_{0011}q_{1100}-q_{1111},y_3^2-q_{0011},\right. \\
&\qquad{}\left.y_2^2q_{1111}-q_{1100}^2,y_2^2q_{0011}-q_{1100},y_1^2q_{1111}-q_{0011}q_{1010},y_1^2q_{1100}-q_{1010}\right>.
\end{split}\end{align}
After finding the Gr\"{o}bner basis of the ideal, solving for the time parameters,
\begin{align}\begin{split}
\begin{cases}
y_{1}&=\sqrt{\frac{q_{0101}}{q_{1100}}}, \\
y_{2}&=\sqrt{\frac{q_{1100}}{q_{0011}}}, \\
y_{3}&=\sqrt{q_{0011}}.
\end{cases}
\end{split}\end{align}

Demanding $y_{1},y_{2}\leq{}1$, the constraints on Network $2$ are then
\begin{align}\begin{split}
\begin{mycases}
q_{0101}&=q_{0110}, \\
q_{0101}&=q_{1001}, \\
q_{0101}&=q_{1010}, \\
q_{0011}q_{1100}&=q_{1111}, \\
q_{0101}&\leq{}q_{1100}, \\
q_{1100}&\leq{}q_{0011}. \\
\end{mycases}
\end{split}\end{align}

The constraints on the distances for Network $2$ that are equivalent to the constraints on the distances for Network $1$ are
\begin{align}\begin{split}
\begin{mycases}
d\left(3,4\right)\leq{}d\left(2,4\right)&=d\left(2,3\right), \\
d\left(3,4\right)&\leq{}d\left(1,2\right), \\
d\left(1,4\right)&=d\left(1,3\right), \\
d\left(3,4\right)+d\left(1,2\right)&=d\left(1,2,3,4\right).
\end{mycases}
\end{split}\end{align}
Network $2$ also has the distance constraints, $d\left(1,2\right)\leq{}d\left(2,4\right)=d\left(1,4\right)$, while on Network $1$, $d\left(2,4\right)\leq{}d\left(1,4\right)=d\left(1,2\right)$.

Notice that on Network $2$,
\begin{align}\begin{split}
d\left(3,4\right)+d\left(1,2\right)&=2\left(\tau_{2}+2\tau_{3}\right) \\
&\leq{}d\left(2,4\right)+d\left(1,3\right)=d\left(2,3\right)+d\left(1,4\right) \\
&=4\left(\tau_{1}+\tau_{2}+\tau_{3}\right) \\
&=2\left(\tau_{2}+2\tau_{3}\right)+2\left(2\tau_{1}+\tau_{2}\right) \\
&=d\left(3,4\right)+d\left(1,2\right)+2\left(2\tau_{1}+\tau_{2}\right).
\end{split}\end{align}
We can conclude that Network $2$ satisfies the four-point condition, which is to be expected since it is a tree.

\subsection{Network $2b$}

The next tree or network that we will examine is Network $2b$, shown in Figure~\ref{Network2b} below.
\begin{figure}[H]
\centering
				\psmatrix[colsep=.3cm,rowsep=.4cm,mnode=r]
				~ && && && ~ \\
				\\
				\\
				~ & && ~ & && && && && ~ \\
				~ && && && && && ~ && ~ \\
				~ & ~ && && && && ~ & && ~ \\
				~ && ~ && && ~ && ~ && && ~
				\ncline{1,7}{4,4}
				\ncline{4,4}{6,2}
				\ncline{4,4}{7,7}
				\ncline{1,7}{5,11}
				\ncline{5,11}{6,10}
				\ncline{5,11}{7,13}
				\ncarc[arcangle=40]{6,2}{7,3}
				\ncarc[arcangle=-40]{6,10}{7,9}
				\ncline{<->,arrowscale=1.5}{1,1}{4,1}
				\tlput{$\tau_{1}$}
				\ncline{<->,arrowscale=1.5}{4,1}{5,1}
				\tlput{$\tau_{2}$}
				\ncline{<->,arrowscale=1.5}{5,1}{6,1}
				\tlput{$\tau_{3}$}
				\ncline{<->,arrowscale=1.5}{6,1}{7,1}
				\tlput{$\tau_{3}$}
				\psset{linestyle=dashed}\ncline{4,1}{4,13}
				\psset{linestyle=dashed}\ncline{5,1}{5,13}
				\psset{linestyle=dashed}\ncline{6,1}{6,13}
				\endpsmatrix
				\caption{Network $2b$.}
				\label{Network2b}
\end{figure}
The variable transformed phylogenetic tensor elements are
\begin{align}\begin{split}
\left[\begin{array}{c} q_{0011} \\
q_{0101} \\
q_{0110} \\
q_{1001} \\
q_{1010} \\
q_{1100} \\
q_{1111} \\
\end{array}\right]&=\left[\begin{array}{c} e^{-2\left(\tau_{3}+\tau_{4}\right)} \\
e^{-2\left(\tau_{1}+\tau_{2}+\tau_{3}+\tau_{4}\right)} \\
e^{-2\left(\tau_{1}+\tau_{2}+\tau_{3}+\tau_{4}\right)} \\
e^{-2\left(\tau_{1}+\tau_{2}+\tau_{3}+\tau_{4}\right)} \\
1-e^{-\tau_{4}}\left(1-e^{-2\left(\tau_{1}+\tau_{2}+\tau_{3}\right)}\right) \\
e^{-2\left(\tau_{2}+\tau_{3}+\tau_{4}\right)} \\
e^{-2\left(\tau_{2}+\tau_{3}+\tau_{4}\right)}\left(e^{-2\tau_{3}+\tau_{4}}+e^{-2\tau_{1}}\left(1-e^{-\tau_{4}}\right)\right) \\
\end{array}\right] \\
&=\left[\begin{array}{c} y_{3}^{2}y_{4}^{2} \\
y_{1}^{2}y_{2}^{2}y_{3}^{2}y_{4}^{2} \\
y_{1}^{2}y_{2}^{2}y_{3}^{2}y_{4}^{2} \\
y_{1}^{2}y_{2}^{2}y_{3}^{2}y_{4}^{2} \\
1-y_{4}\left(1-y_{1}^{2}y_{2}^{2}y_{3}^{2}\right) \\
y_{2}^{2}y_{3}^{2}y_{4}^{2} \\
y_{2}^{2}y_{3}^{2}y_{4}^{2}\left(y_{3}^{2}y_{4}+y_{1}^{2}\left(1-y_{4}\right)\right) \\
\end{array}\right].
\end{split}\end{align}
We form the ideal,
\begin{align}\begin{split}
I=&\left<y_{3}^{2}y_{4}^{2}-q_{0011},y_{1}^{2}y_{2}^{2}y_{3}^{2}y_{4}^{2}-q_{0101},y_{1}^{2}y_{2}^{2}y_{3}^{2}y_{4}^{2}-q_{0110},y_{1}^{2}y_{2}^{2}y_{3}^{2}y_{4}^{2}-q_{1001},\right. \\
&\qquad{}\left.1-y_{4}\left(1-y_{1}^{2}y_{2}^{2}y_{3}^{2}\right)-q_{1010},y_{2}^{2}y_{3}^{2}y_{4}^{2}-q_{1100},y_{2}^{2}y_{3}^{2}y_{4}^{2}\left(y_{3}^{2}y_{4}+y_{1}^{2}\left(1-y_{4}\right)\right)-q_{1111}\right>.
\end{split}\end{align}
The Gr\"{o}bner basis of the ideal is given in Appendix~\ref{cha.Appendix3}, it has $29$ generators. After finding the Gr\"{o}bner basis of the ideal, solving for the time parameters,
\begin{align}\begin{split}
\begin{cases}
y_{1}&=\sqrt{\frac{q_{0101}}{q_{1100}}}, \\
y_{2}&=\sqrt{\frac{q_{1100}}{q_{0011}}}, \\
y_{3}&=\frac{\sqrt{q_{0011}}\left(-\left(1-q_{1010}\right)+\sqrt{\left(1-q_{1010}\right)^{2}+4q_{0101}}\right)}{2q_{0101}}, \\
y_{4}&=\frac{1}{2}\left(\left(1-q_{1010}\right)+\sqrt{\left(1-q_{1010}\right)^{2}+4q_{0101}}\right).
\end{cases}
\end{split}\end{align}

Demanding $y_{1},y_{2}\leq{}1$ and from (\ref{yresult1}), (\ref{yresult2}) and (\ref{yresult3}), the constraints on Network $2b$ are then
\begin{align}\begin{split}
\begin{mycases}
q_{0101}&=q_{0110}, \\
q_{0101}&=q_{1001}, \\
q_{0011}^2q_{1100}^2-2q_{0011}q_{0101}^2q_{1100}-q_{0011}q_{0101}q_{1010}^2q_{1100}& \\
+q_{0011}q_{0101}q_{1010}q_{1100}+q_{0011}q_{1010}q_{1100}q_{1111}-q_{0011}q_{1100}q_{1111}+q_{0101}^4& \\
-q_{0101}^3q_{1010}+q_{0101}^2q_{1010}q_{1111}+q_{0101}^2q_{1111}-q_{0101}q_{1111}^2&=0, \\
q_{0101}&\leq{}q_{1010}, \\
q_{0101}&\leq{}q_{1100}, \\
q_{1100}&\leq{}q_{0011}, \\
\left(q_{0011}-q_{0101}\right)^{2}&\leq{}q_{0011}\left(1-q_{1010}\right)^{2}.
\end{mycases}
\end{split}\end{align}

Networks $1b$, $1c$ and $2b$ all have an equality constraint with $11$ terms. These constraints look similar but are not equivalent. A constraint from one network cannot be transformed into a constraint from another network from a permutation on a transformed phylogenetic tensor element. For Network $1b$ the constraint is quartic, while for Networks $1c$ and $2b$ the constraint is quintic.

The constraints on the distances for Network $2b$ that are equivalent to the constraints on the distances for Network $2$ are
\begin{align}\begin{split}
d\left(3,4\right)\leq{}d\left(1,2\right)\leq{}d\left(2,4\right)=d\left(2,3\right)=d\left(1,4\right).
\end{split}\end{align}
Network $2b$ also has the distance constraint, $d\left(1,3\right)\leq{}d\left(2,4\right)$, while on Network $2$, $d\left(2,4\right)=d\left(1,3\right)$. While on Network $2$ the total network distance satisfies $d\left(1,2,3,4\right)=d\left(3,4\right)+d\left(1,2\right)$, on Network $2b$, $d\left(1,2,3,4\right)\neq{}d\left(3,4\right)+d\left(1,2\right)$.

From $d\left(3,4\right)\leq{}d\left(2,3\right)$ and $d\left(1,2\right)\leq{}d\left(1,4\right)$ we can conclude that
\begin{align}\begin{split}
d\left(3,4\right)+d\left(1,2\right)\leq{}d\left(2,3\right)+d\left(1,4\right).
\end{split}\end{align}
From $d\left(2,4\right)=d\left(2,3\right)$ and $d\left(1,3\right)\leq{}d\left(1,4\right)$ we can conclude that
\begin{align}\begin{split}
d\left(2,4\right)+d\left(1,3\right)\leq{}d\left(2,3\right)+d\left(1,4\right).
\end{split}\end{align}
We can conclude that the sum of pairwise distances, $d\left(2,3\right)+d\left(1,4\right)$, must be greater than or equal to each of the other two sums of pairwise distances. For the four-point condition to be satisfied the two greatest sums of pairwise distances must be equal. We can conclude that Network $2b$ does not generally satisfy the four-point condition.

\subsection{Network $3$}

The last tree or network that we will examine is the non-clock-like tree, which we will call Network $3$, shown in Figure~\ref{Network3}  below.
\begin{figure}[H]
  \centering
  \psmatrix[colsep=3mm,rowsep=.5cm,mnode=r]
    ~ &    &&    & ~ \\
                     & {} && {} & \\
    ~ &    &&    & ~
    \ncline{1,1}{2,2}\tlput{$\tau_{1}$}
    \ncline{2,2}{3,1}\tlput{$\tau_{2}$}
    \ncline{2,2}{2,4}\taput{$\tau_{1234}$}
    \ncline{2,4}{1,5}\trput{$\tau_{3}$}
    \ncline{2,4}{3,5}\trput{$\tau_{4}$}
  \endpsmatrix
	\caption{Network $3$.}
	\label{Network3}
\end{figure}
The variable transformed phylogenetic tensor elements are
\begin{align}\begin{split}
\left[\begin{array}{c} q_{0011} \\
q_{0101} \\
q_{0110} \\
q_{1001} \\
q_{1010} \\
q_{1100} \\
q_{1111} \\
\end{array}\right]=\left[\begin{array}{c} e^{-\left(\tau_{3}+\tau_{4}\right)} \\
e^{-\left(\tau_{2}+\tau_{4}+\tau_{1234}\right)} \\
e^{-\left(\tau_{2}+\tau_{3}+\tau_{1234}\right)} \\
e^{-\left(\tau_{1}+\tau_{4}+\tau_{1234}\right)} \\
e^{-\left(\tau_{1}+\tau_{3}+\tau_{1234}\right)} \\
e^{-\left(\tau_{1}+\tau_{2}\right)} \\
e^{-\left(\tau_{1}+\tau_{2}+\tau_{3}+\tau_{4}\right)} \\
\end{array}\right]&=\left[\begin{array}{c} y_{3}y_{4} \\
y_{2}y_{4}y_{1234} \\
y_{2}y_{3}y_{1234} \\
y_{1}y_{4}y_{1234} \\
y_{1}y_{3}y_{1234} \\
y_{1}y_{2} \\
y_{1}y_{2}y_{3}y_{4} \\
\end{array}\right].
\end{split}\end{align}
We form the ideal,
\begin{align}\begin{split}
I=&\left<y_{3}y_{4}-q_{0011},y_{2}y_{4}y_{1234}-q_{0101},y_{2}y_{3}y_{1234}-q_{0110},y_{1}y_{4}y_{1234}-q_{1001},\right. \\
&\qquad{}\left.y_{1}y_{3}y_{1234}-q_{1010},y_{1}y_{2}-q_{1100},y_{1}y_{2}y_{3}y_{4}-q_{1111}\right>.
\end{split}\end{align}
The Gr\"{o}bner basis of the ideal is given in Appendix~\ref{cha.Appendix4}, it has $23$ generators. After finding the Gr\"{o}bner basis of the ideal, solving for the time parameters,
\begin{align}\begin{split}
\begin{cases}
y_{1}&=\sqrt{\frac{q_{1001}q_{1100}}{q_{0101}}}, \\
y_{2}&=\sqrt{\frac{q_{0101}q_{1100}}{q_{1001}}}, \\
y_{3}&=\sqrt{\frac{q_{0011}q_{0110}}{q_{0101}}}, \\
y_{4}&=\sqrt{\frac{q_{0011}q_{0101}}{q_{0110}}}, \\
y_{1234}&=\sqrt{\frac{q_{0101}q_{1010}}{q_{0011}q_{1100}}}.
\end{cases}
\end{split}\end{align}

Demanding $y_{1},y_{2}\leq{}1$, the constraints on the network are then
\begin{align}\begin{split}
\begin{mycases}
q_{0101}q_{1010}&=q_{0110}q_{1001}, \\
q_{0011}q_{1100}&=q_{1111}, \\
q_{1001}q_{1100}&\leq{}q_{0101}, \\
q_{0101}q_{1100}&\leq{}q_{1001}, \\
q_{0011}q_{0110}&\leq{}q_{0101}, \\
q_{0011}q_{0101}&\leq{}q_{0110}, \\
q_{0101}q_{1010}&\leq{}q_{0011}q_{1100}.
\end{mycases}
\end{split}\end{align}

The constraints on the distances for Network $3$ are
\begin{align}\begin{split}
\begin{mycases}
d\left(2,4\right)+d\left(1,3\right)&=d\left(2,3\right)+d\left(1,4\right), \\
d\left(3,4\right)+d\left(1,2\right)&=d\left(1,2,3,4\right), \\
d\left(2,4\right)&\leq{}d\left(1,4\right)+d\left(1,2\right), \\
d\left(1,4\right)&\leq{}d\left(2,4\right)+d\left(1,2\right), \\
d\left(2,4\right)&\leq{}d\left(3,4\right)+d\left(2,3\right), \\
d\left(2,3\right)&\leq{}d\left(3,4\right)+d\left(2,4\right), \\
d\left(3,4\right)+d\left(1,2\right)&\leq{}d\left(2,4\right)+d\left(1,3\right).
\end{mycases}
\end{split}\end{align}

It can be shown that if the distance constraints for Network $1$ or Network $2$ are met then the distance constraints for Network $3$ will also be met. This is to be expected since the clock-like trees can be recovered from the non-clock-like tree with the appropriate time parameter choices.

The second equality constraint is the constraint for the total network distance, $d\left(3,4\right)+d\left(1,2\right)=d\left(1,2,3,4\right)$. This is the same as for Networks $1$, $1a$ and $2$. From the last inequality constraint and the first equality constraint,
\begin{align}\begin{split}
d\left(3,4\right)+d\left(1,2\right)\leq{}d\left(2,4\right)+d\left(1,3\right)&=d\left(2,3\right)+d\left(1,4\right).
\end{split}\end{align}
We can conclude that Network $3$ satisfies the four-point condition, which is to be expected since it is a tree.

\section{Distinguishability of the Networks}
\label{fourtaxondistinguish}

To determine if one network is distinguishable from another we must compare the constraints on their transformed phylogenetic tensors. Suppose we have two sets, $\Omega{}_{1}$ and $\Omega{}_{2}$, representing the probability spaces from the constraints for two networks. The intersection of two sets can have four options:
\begin{align}\begin{split}
(i)&\text{	$\Omega{}_{1}=\Omega{}_{2}$. The sets are equal,} \\
(ii)&\text{	$\Omega{}_{1}\neq{}\Omega{}_{2}$ and $\Omega{}_{1}\cap{}\Omega{}_{2}=\Omega{}_{1}\Leftrightarrow{}\Omega{}_{1}\subset{}\Omega{}_{2}$. $\Omega{}_{1}$ is a proper subset of $\Omega{}_{2}$,} \\
(iii)&\text{	$\Omega{}_{1}\cap{}\Omega{}_{2}=\emptyset{}$. The sets are disjoint,} \\
(iv)&\text{	$\Omega{}_{1}\neq{}\Omega{}_{2}$, $\Omega{}_{1}\cap{}\Omega{}_{2}\subset{}\Omega{}_{1}$ and $\Omega{}_{1}\cap{}\Omega{}_{2}\subset{}\Omega{}_{2}$. The intersection of the sets is non-empty, but} \\
&\text{	neither set is a proper subset of the other.}
\end{split}\end{align}
In scenario $(i)$ the two networks are indistinguishable from each other. In scenarios $(ii)$, $(iii)$ and $(iv)$ the two networks are distinguishable from each other.

We will let $\Omega{}_{i}$ denote the space of the constraints for Network $i$. We will start by examining the intersection of the constraints for Network $1$ and Network $1a$, $\Omega{}_{1}\cap{}\Omega{}_{1a}$. Below in Table~\ref{constraintscomparison11a} is a comparison of the constraints for the two networks.
\begin{table}[H]
\centering
\begin{tabular}{c|c|c}
 & \multicolumn{2}{c}{Network} \\
\hline
 & $1$ & $1a$ \\
\hline
\multirow{7}{*}{Constraints} & $q_{0101}=q_{0110}$ & $q_{0101}=q_{0110}$ \\
\cline{2-3}
& $q_{1001}=q_{1010}$ & $q_{1001}=q_{1010}$ \\
\cline{2-3}
& $q_{1001}=q_{1100}$ & $q_{1001}\leq{}q_{1100}$ \\
\cline{2-3}
& $q_{0011}q_{1001}=q_{1111}$ & $q_{0011}q_{1100}=q_{1111}$ \\
\cline{2-3}
& $q_{0101}\leq{}q_{0011}$ & $q_{0101}\leq{}q_{0011}$ \\
\cline{2-3}
& $q_{1001}\leq{}q_{0101}$ & $q_{1001}\leq{}q_{0101}$ \\
\cline{2-3}
&  & \scriptsize{\textrm{$\left(q_{0011}-q_{1001}\right)^{2}\leq{}q_{0011}\left(1-q_{1100}\right)^{2}$}} \\
\hline
\end{tabular}
\caption{Summary of constraints for Network $1$ and Network $1a$.}
\label{constraintscomparison11a}
\end{table}
Clearly the first two rows of constraints show identical equalities for both networks. We will consider the transformed phylogenetic tensor elements to be determined by others when there is an equality constraint. Consider $q_{0110}$ to be determined by $q_{0101}$ and $q_{1010}$ to be determined by $q_{1001}$. $q_{0110}$ and $q_{1010}$ are not free elements and can be replaced by $q_{0101}$ and $q_{1001}$, respectively. Likewise, for Network $1$, but not for Network $1a$, we consider $q_{1100}$ to be determined by $q_{1001}$. In the intersection of the two sets of constraints $q_{1100}$ will be determined by $q_{1001}$. For the constraints in the intersection of the two sets $q_{1100}$ is not a free element and can be replaced by $q_{1001}$. The equalities in the fourth row of constraints must therefore be identical. We can see that the inequalities in the fifth and sixth rows are identical. Hence, the constraints in Network $1a$ are the same constraints as in Network $1$, with one extra inequality constraint. We will now look at the final constraint for Network $1a$. Recognising that $q_{1100}$ is not a free element, the last constraint on Network $1a$ becomes
\begin{align}\begin{split}
\left(q_{0011}-q_{1001}\right)^{2}&\leq{}q_{0011}\left(1-q_{1001}\right)^{2}.
\end{split}\end{align}
Suppose we had an inequality in a similar form,
\begin{align}\begin{split}
\left(q_{a}-q_{b}\right)^{2}&\leq{}q_{a}\left(1-q_{b}\right)^{2}.
\end{split}\end{align}
Then
\begin{align}\begin{split}
\left(q_{a}-q_{b}\right)^{2}&\leq{}q_{a}\left(1-q_{b}\right)^{2} \\
\Leftrightarrow{}q_{a}^{2}+q_{b}^{2}-2q_{a}q_{b}&\leq{}q_{a}+q_{a}q_{b}^{2}-2q_{a}q_{b} \\
\Leftrightarrow{}q_{a}^{2}+q_{b}^{2}&\leq{}q_{a}+q_{a}q_{b}^{2} \\
\Leftrightarrow{}q_{b}^{2}-q_{a}q_{b}^{2}&\leq{}q_{a}-q_{a}^{2} \\
\Leftrightarrow{}q_{b}^{2}\left(1-q_{a}\right)&\leq{}q_{a}\left(1-q_{a}\right).
\end{split}\end{align}
Since $\left(1-q_{a}\right)\neq{}0$ in general,
\begin{align}
\begin{split}
q_{b}^{2}\left(1-q_{a}\right)\leq{}q_{a}\left(1-q_{a}\right)\Leftrightarrow{}q_{b}^{2}\leq{}q_{a}.
\end{split}
\label{yresult4}
\end{align}
Since $0<q_{b}\leq{}1$,
\begin{align}\begin{split}
q_{b}^{2}\leq{}q_{b}&\leq{}q_{a}.
\end{split}\end{align}
From the two inequality constraints, $q_{1001}\leq{}q_{0101}$ and $q_{0101}\leq{}q_{0011}$, we can conclude that $q_{1001}\leq{}q_{0011}$.
From (\ref{yresult4}) we can conclude that
\begin{align}\begin{split}
\left(q_{0011}-q_{1001}\right)^{2}&\leq{}q_{0011}\left(1-q_{1001}\right)^{2}.
\end{split}\end{align}
Consequently, the last constraint on Network $1a$ can be ignored and the constraints in the intersection of the probability spaces of Network $1$ and Network $1a$ will be
\begin{align}\begin{split}
\begin{mycases}
q_{0101}&=q_{0110}, \\
q_{1001}&=q_{1010}, \\
q_{1001}&=q_{1100}, \\
q_{0011}q_{1001}&=q_{1111}, \\
q_{0101}&\leq{}q_{0011}, \\
q_{1001}&\leq{}q_{0101}.
\end{mycases}
\end{split}\end{align}
This is the same set of constraints as Network $1$. Hence,
\begin{align}\begin{split}
\Omega{}_{1}\cap{}\Omega{}_{1a}&=\Omega{}_{1}.
\end{split}\end{align}

It is interesting to note that the distances on Network $1a$ met the four-point condition, yet Network $1$, the clock-like tree, and Network $1a$ are still distinguishable from each other. For the other intersections of the probability spaces for the other pairs of networks we used \emph{Mathematica}. Below is as example of the \emph{Mathematica} code we used.
\begin{lstlisting}[breaklines=true, mathescape]
(* All of the constraints on both Network $i$ and Network $j$. Each transformed phylogenetic tensor element must be in the interval, $(0,1]$. ``Reduce'' removes any constraints that are implied by the other constraints. *)
FullSimplify[Reduce[{Network $i$ constraints, Network $j$ constraints}, {q0011, q0101, q0110, q1001, q1010, q1100, q1111}, Reals, Backsubstitution -> True], 0 < q0011 <= 1 && 0 < q0101 <= 1 && 0 < q0110 <= 1 && 0 < q1001 <= 1 && 0 < q1010 <= 1 && 0 < q1100 <= 1 && 0 < q1111 <= 1]
\end{lstlisting}

For example, for the intersection of the probability spaces for Network $1$ and Network $1a$, we used the code below.
\begin{lstlisting}[breaklines=true]
FullSimplify[Reduce[{q0101 == q0110, q1001 == q1010, q1001 == q1100, q0011*q1001 == q1111, q0101 <= q0011, q1001 <= q0101, q0101 == q0110, q1001 == q1010, q0011*q1100 == q1111, q0101 <= q0011, q1001 <= q0101, q1001 <= q1100, (q0011 - q1001)^2 <= q0011*(1 - q1100)^2}, {q0011, q0101, q0110, q1001, q1010, q1100, q1111}, Reals, Backsubstitution -> True], 0 < q0011 <= 1 && 0 < q0101 <= 1 && 0 < q0110 <= 1 && 0 < q1001 <= 1 && 0 < q1010 <= 1 && 0 < q1100 <= 1 && 0 < q1111 <= 1]
\end{lstlisting}

Using \emph{Mathematica} to find all of the intersections of the probability spaces for each pair of networks, the intersections of the sets of constraints, denoted $R$ with arbitrary subscripts, are
\begin{align}
\begin{mycases}
\begin{split}
R_{1}=\Omega{}_{1}\cap{}\Omega{}_{1a}=\Omega{}_{1}\cap{}\Omega{}_{1b}=\Omega{}_{1}\cap{}\Omega{}_{1c}=\Omega{}_{1}\cap{}\Omega{}_{3}=\Omega{}_{1a}\cap{}\Omega{}_{1b}&=\Omega{}_{1a}\cap{}\Omega{}_{1c} \\
=\Omega{}_{1b}\cap{}\Omega{}_{1c}=\Omega{}_{1b}\cap{}\Omega{}_{3}=\Omega{}_{1c}\cap{}\Omega{}_{3}&=\Omega{}_{1}, \\
R_{2}=\Omega{}_{1}\cap{}\Omega{}_{2}=\Omega{}_{1}\cap{}\Omega{}_{2b}=\Omega{}_{1b}\cap{}\Omega{}_{2}=\Omega{}_{1b}\cap{}\Omega{}_{2b}&=\Omega{}_{1c}\cap{}\Omega{}_{2} \\
=\Omega{}_{1c}\cap{}\Omega{}_{2b}&\subset{}\Omega{}_{1}\qquad{}\left(\textrm{or }\subset{}\Omega{}_{2}\right), \\
R_{3}=\Omega{}_{1a}\cap{}\Omega{}_{3}&\subset{}\Omega{}_{1a}\qquad{}\left(\textrm{or }\subset{}\Omega{}_{3}\right), \\
R_{4}=\Omega{}_{1a}\cap{}\Omega{}_{2}=\Omega{}_{1a}\cap{}\Omega{}_{2b}&\subset{}\Omega{}_{2}, \\
R_{5}=\Omega{}_{2}\cap{}\Omega{}_{2b}=\Omega{}_{2}\cap{}\Omega{}_{3}=\Omega{}_{2b}\cap{}\Omega{}_{3}&=\Omega{}_{2}.
\end{split}
\label{regions}
\end{mycases}
\end{align}
From the first region we can conclude that $\Omega{}_{1}$ is a proper subset of each of $\Omega{}_{1a}$, $\Omega{}_{1b}$, $\Omega{}_{1c}$ and $\Omega{}_{3}$. $\Omega{}_{1}$ must be a proper subset of each of $\Omega{}_{1a}$, $\Omega{}_{1b}$, $\Omega{}_{1c}$ and $\Omega{}_{3}$ and not equal to any of these probability spaces since each of the probability spaces has constraints that are not generally met on Network $1$. This follows our intuition since the first structure of the clock-like tree, Network $1$, can be reached from the convergence-divergence networks, Network $1a$, Network $1b$ and Network $1c$, or the non-clock-like tree, Network $3$, by setting the appropriate time parameters to zero. The intersection of these five probability spaces will be $\Omega{}_{1}$. Summarising,
\begin{align}
\begin{mycases}
\begin{split}
R_{1}=\Omega{}_{1}\cap{}\Omega{}_{1a}\cap{}\Omega{}_{1b}\cap{}\Omega{}_{1c}\cap{}\Omega{}_{3}&=\Omega{}_{1}, \\
R_{1}=\Omega{}_{1}&\subset{}\Omega{}_{1a}, \\
R_{1}=\Omega{}_{1}&\subset{}\Omega{}_{1b}, \\
R_{1}=\Omega{}_{1}&\subset{}\Omega{}_{1c}, \\
R_{1}=\Omega{}_{1}&\subset{}\Omega{}_{3}.
\end{split}
\label{regions2}
\end{mycases}
\end{align}
Likewise, from the fifth region we can conclude that $\Omega{}_{2}$ is a proper subset of each of $\Omega{}_{2b}$ and $\Omega{}_{3}$. This again follows our intuition since the second structure of the clock-like tree, Network $2$, can be recovered from the convergence-divergence network, Network $2b$, and the non-clock-like tree, Network $3$, by setting the appropriate time parameters to zero. The intersection of the three probability spaces is $\Omega{}_{2}$. Summarising,
\begin{align}
\begin{mycases}
\begin{split}
R_{5}=\Omega{}_{2}\cap{}\Omega{}_{2b}\cap{}\Omega{}_{3}&=\Omega{}_{2}, \\
R_{5}=\Omega{}_{2}&\subset{}\Omega{}_{2b}, \\
R_{5}=\Omega{}_{2}&\subset{}\Omega{}_{3}.
\end{split}
\label{regions3}
\end{mycases}
\end{align}
The intersections of the pairs of regions will be
\begin{align}\begin{split}
\begin{mycases}
R_{1}\cap{}R_{2}=\Omega{}_{1}\cap{}\left(\Omega{}_{1}\cap{}\Omega{}_{2}\right)=\Omega{}_{1}\cap{}\Omega{}_{2}&=R_{2}, \\
R_{1}\cap{}R_{3}=\Omega{}_{1}\cap{}\left(\Omega{}_{1a}\cap{}\Omega{}_{3}\right)=\left(\Omega{}_{1}\cap{}\Omega{}_{1a}\right)\cap{}\Omega{}_{3}=\Omega{}_{1}\cap{}\Omega{}_{3}=\Omega{}_{1}&=R_{1}, \\
R_{1}\cap{}R_{4}=\Omega{}_{1}\cap{}\left(\Omega{}_{1a}\cap{}\Omega{}_{2}\right)=\left(\Omega{}_{1}\cap{}\Omega{}_{1a}\right)\cap{}\Omega{}_{2}=\Omega{}_{1}\cap{}\Omega{}_{2}&=R_{2}, \\
R_{1}\cap{}R_{5}=\Omega{}_{1}\cap\Omega{}_{2}&=R_{2}, \\
R_{2}\cap{}R_{3}=\left(\Omega{}_{1}\cap{}\Omega{}_{2}\right)\cap{}\left(\Omega{}_{1a}\cap{}\Omega{}_{3}\right)=\left(\Omega{}_{1}\cap{}\Omega{}_{1a}\right)\cap{}\left(\Omega{}_{2}\cap{}\Omega{}_{3}\right)&=\Omega{}_{1}\cap{}\left(\Omega{}_{2}\cap{}\Omega{}_{3}\right) \\
=\left(\Omega{}_{1}\cap{}\Omega{}_{3}\right)\cap{}\Omega{}_{2}=\Omega{}_{1}\cap{}\Omega{}_{2}&=R_{2}, \\
R_{2}\cap{}R_{4}=\left(\Omega{}_{1}\cap{}\Omega{}_{2}\right)\cap{}\left(\Omega{}_{1a}\cap{}\Omega{}_{2}\right)=\left(\Omega{}_{1}\cap{}\Omega{}_{2}\right)\cap{}\Omega{}_{2}=\Omega{}_{1}\cap{}\Omega{}_{2}&=R_{2}, \\
R_{2}\cap{}R_{5}=\left(\Omega{}_{1}\cap{}\Omega{}_{2}\right)\cap{}\Omega{}_{2}=\Omega{}_{1}\cap{}\Omega{}_{2}&=R_{2}, \\
R_{3}\cap{}R_{4}=\left(\Omega{}_{1a}\cap{}\Omega{}_{3}\right)\cap{}\left(\Omega{}_{1a}\cap{}\Omega{}_{2}\right)=\left(\Omega{}_{1a}\cap{}\Omega{}_{3}\right)\cap{}\Omega{}_{2}&=\Omega{}_{1a}\cap{}\left(\Omega{}_{2}\cap{}\Omega{}_{3}\right) \\
=\Omega{}_{1a}\cap{}\Omega{}_{2}&=R_{4}, \\
R_{3}\cap{}R_{5}=\left(\Omega{}_{1a}\cap{}\Omega{}_{3}\right)\cap{}\Omega{}_{2}=\Omega{}_{1a}\cap{}\left(\Omega{}_{2}\cap{}\Omega{}_{3}\right)=\Omega{}_{1a}\cap{}\Omega{}_{2}&=R_{4}, \\
R_{4}\cap{}R_{5}=\left(\Omega{}_{1a}\cap{}\Omega{}_{2}\right)\cap{}\Omega{}_{2}=\Omega{}_{1a}\cap{}\Omega{}_{2}&=R_{4}.
\end{mycases}
\end{split}\end{align}
In summary,
\begin{align}\begin{split}
\begin{mycases}
R_{1}\cap{}R_{2}&=R_{2}, \\
R_{1}\cap{}R_{3}&=R_{1}, \\
R_{1}\cap{}R_{4}&=R_{2}, \\
R_{1}\cap{}R_{5}&=R_{2}, \\
R_{2}\cap{}R_{3}&=R_{2}, \\
R_{2}\cap{}R_{4}&=R_{2}, \\
R_{2}\cap{}R_{5}&=R_{2}, \\
R_{3}\cap{}R_{4}&=R_{4}, \\
R_{3}\cap{}R_{5}&=R_{4}, \\
R_{4}\cap{}R_{5}&=R_{4}.
\end{mycases}
\end{split}\end{align}
From the first, fifth, sixth and seventh equations,
\begin{align}\begin{split}
R_{2}\subset{}R_{1},R_{2}\subset{}R_{3},R_{2}\subset{}R_{4},R_{2}\subset{}R_{5}.
\end{split}\end{align}
We can conclude that in any region that $R_{2}$ appears, $R_{1}$, $R_{3}$, $R_{4}$ and $R_{5}$ will also appear. This will be the region $R_{1}\cap{}R_{2}\cap{}R_{3}\cap{}R_{4}\cap{}R_{5}$.
From the eighth, ninth and tenth equations,
\begin{align}\begin{split}
R_{4}\subset{}R_{3},R_{4}\subset{}R_{5},R_{3}\cap{}R_{5}&=R_{4}.
\end{split}\end{align}
We can conclude that in any region that $R_{4}$ appears, $R_{3}$ and $R_{5}$ will also appear.
From the second, third and fourth equations,
\begin{align}\begin{split}
R_{1}\subset{}R_{3},R_{1}\cap{}R_{4}=R_{1}\cap{}R_{5}&=R_{2}.
\end{split}\end{align}
We can conclude that in any region that $R_{1}$ appears, $R_{3}$ will also appear and the only region that will contain $R_{1}$ and $R_{4}$ or $R_{1}$ and $R_{5}$ will be the entire region that $R_{2}$ appears in, $R_{1}\cap{}R_{2}\cap{}R_{3}\cap{}R_{4}\cap{}R_{5}$. There will also be the region where none of $R_{1}$, $R_{2}$, $R_{3}$, $R_{4}$ or $R_{5}$ will appear. There must therefore be six non-overlapping regions, each denoted $S$, with an arbitrary subscript. The six regions are
\begin{align}\begin{split}
\begin{mycases}
S_{1}&=R_{1}\cap{}R_{2}\cap{}R_{3}\cap{}R_{4}\cap{}R_{5}, \\
S_{2}&=R_{1}\cap{}R_{2}^{C}\cap{}R_{3}\cap{}R_{4}^{C}\cap{}R_{5}^{C}, \\
S_{3}&=R_{1}^{C}\cap{}R_{2}^{C}\cap{}R_{3}\cap{}R_{4}\cap{}R_{5}, \\
S_{4}&=R_{1}^{C}\cap{}R_{2}^{C}\cap{}R_{3}\cap{}R_{4}^{C}\cap{}R_{5}^{C}, \\
S_{5}&=R_{1}^{C}\cap{}R_{2}^{C}\cap{}R_{3}^{C}\cap{}R_{4}^{C}\cap{}R_{5}, \\
S_{6}&=R_{1}^{C}\cap{}R_{2}^{C}\cap{}R_{3}^{C}\cap{}R_{4}^{C}\cap{}R_{5}^{C},
\end{mycases}
\end{split}\end{align}
where $C$ refers to the complement of the region.
We will relabel the regions with the subscripts referring to the trees and networks that can be found in the region. $S_{6}$ will be the region,
\begin{align}\begin{split}
S_{6}&=S_{1a}\cup{}S_{1b}\cup{}S_{1c}\cup{}S_{2b}\cup{}S_{3}.
\end{split}\end{align}
In other words, it is the region where only one tree or network can be found.
There will be ten non-overlapping regions in total since $S_{6}$ will have five non-overlapping regions, one for each tree or network in the region. See Appendix~\ref{cha.Appendix5} for the working. These ten regions will be
\begin{align}\begin{split}
\begin{mycases}
S_{1,1a,1b,1c,2,2b,3}&=\Omega{}_{1}\cap{}\Omega{}_{1a}\cap{}\Omega{}_{1b}\cap{}\Omega{}_{1c}\cap{}\Omega{}_{2}\cap{}\Omega{}_{2b}\cap{}\Omega{}_{3}, \\
S_{1,1a,1b,1c,3}&=\Omega{}_{1}\cap{}\Omega{}_{1a}\cap{}\Omega{}_{1b}\cap{}\Omega{}_{1c}\cap{}\Omega{}_{2}^{C}\cap{}\Omega{}_{2b}^{2b}\cap{}\Omega{}_{3}, \\
S_{1a,2,2b,3}&=\Omega{}_{1}^{C}\cap{}\Omega{}_{1a}\cap{}\Omega{}_{1b}^{C}\cap{}\Omega{}_{1c}^{C}\cap{}\Omega{}_{2}\cap{}\Omega{}_{2b}\cap{}\Omega{}_{3}, \\
S_{1a,3}&=\Omega{}_{1}^{C}\cap{}\Omega{}_{1a}\cap{}\Omega{}_{1b}^{C}\cap{}\Omega{}_{1c}^{C}\cap{}\Omega{}_{2}^{C}\cap{}\Omega{}_{2b}^{C}\cap{}\Omega{}_{3}, \\
S_{2,2b,3}&=\Omega{}_{1}^{C}\cap{}\Omega{}_{1a}^{C}\cap{}\Omega{}_{1b}^{C}\cap{}\Omega{}_{1c}^{C}\cap{}\Omega{}_{2}\cap{}\Omega{}_{2b}\cap{}\Omega{}_{3}, \\
S_{1a}&=\Omega{}_{1}^{C}\cap{}\Omega{}_{1a}\cap{}\Omega{}_{1b}^{C}\cap{}\Omega{}_{1c}^{C}\cap{}\Omega{}_{2}^{C}\cap{}\Omega{}_{2b}^{C}\cap{}\Omega{}_{3}^{C}, \\
S_{1b}&=\Omega{}_{1}^{C}\cap{}\Omega{}_{1a}^{C}\cap{}\Omega{}_{1b}\cap{}\Omega{}_{1c}^{C}\cap{}\Omega{}_{2}^{C}\cap{}\Omega{}_{2b}^{C}\cap{}\Omega{}_{3}^{C}, \\
S_{1c}&=\Omega{}_{1}^{C}\cap{}\Omega{}_{1a}^{C}\cap{}\Omega{}_{1b}^{C}\cap{}\Omega{}_{1c}\cap{}\Omega{}_{2}^{C}\cap{}\Omega{}_{2b}^{C}\cap{}\Omega{}_{3}^{C}, \\
S_{2b}&=\Omega{}_{1}^{C}\cap{}\Omega{}_{1a}^{C}\cap{}\Omega{}_{1b}^{C}\cap{}\Omega{}_{1c}^{C}\cap{}\Omega{}_{2}^{C}\cap{}\Omega{}_{2b}\cap{}\Omega{}_{3}^{C}, \\
S_{3}&=\Omega{}_{1}^{C}\cap{}\Omega{}_{1a}^{C}\cap{}\Omega{}_{1b}^{C}\cap{}\Omega{}_{1c}^{C}\cap{}\Omega{}_{2}^{C}\cap{}\Omega{}_{2b}^{C}\cap{}\Omega{}_{3}.
\end{mycases}
\end{split}\end{align}
Below in Figure~\ref{probabilityspace} is a diagram representing the regions in the probability space. The region is in a two-dimensional space for convenience. The areas and shapes of each region are also inconsequential other than the reflection that some regions border each other in the probability space.
\begin{figure}[H]
	\centering
		\begin{tikzpicture}
			\draw[ultra thick, fill=red!60] (0.4122cm,3.8867cm) -- (-1.2058cm,2.7111cm) -- (-1.7936cm,3.5201cm) -- (-0.1756cm,4.6957cm) -- cycle;
			\draw[ultra thick, fill=orange] (2cm,0cm) -- (2.6180cm,1.9021cm) -- (3.5691cm,1.5931cm) -- (2.9511cm,-0.3090cm) -- cycle;
			\draw[ultra thick, fill=yellow] (2.6180cm,1.9021cm) -- (1cm,3.0777cm) -- (1.5878cm,3.88677cm) -- (3.2058cm,2.7111cm) -- cycle;
			\draw[ultra thick, fill=GreenYellow] (-0.6180cm,1.9021cm) -- (1cm,3.0777cm) -- (0.4122cm,3.8867cm) -- (-1.2058cm,2.7111cm) -- cycle;
			\draw[ultra thick, fill=green] (0cm,0cm) -- (-0.6180cm,1.9021cm) -- (-1.5691cm,1.5931cm) -- (-0.9511cm,-0.3090cm) -- cycle;
			\draw[ultra thick, fill=BlueGreen] (0cm,0cm) -- (2cm,0cm) -- (2.6180cm,1.9021cm) -- (1cm,3.0777cm) -- (-0.6180cm,1.9021cm) -- cycle;
			\draw[ultra thick, fill=cyan] (0cm,0cm) -- (2cm,0cm) -- (2cm,-1cm) -- (0cm,-1cm) -- cycle;
			\draw[ultra thick, fill=NavyBlue!30] (0cm,-1cm) -- (2cm,-1cm) -- (2cm,-2cm) -- (0cm,-2cm) -- cycle;
			\draw[ultra thick, fill=blue!30] (0cm,-2cm) -- (2cm,-2cm) -- (2cm,-3cm) -- (0cm,-3cm) -- cycle;
			\draw[ultra thick, fill=violet!30] (0cm,-3cm) -- (2cm,-3cm) -- (2cm,-4cm) -- (0cm,-4cm) -- cycle;
			\draw (-0.7846cm,0.9511cm) node {$S_{3}$};
			\draw (2.7846cm,0.9511cm) node {$S_{1c}$};
			\draw (-0.1029cm,2.8944cm) node {$S_{1a,3}$};
			\draw (-0.6907cm,3.7034cm) node {$S_{1a}$};
			\draw (1cm,1.5389cm) node {$S_{1,1a,1b,1c,3}$};
			\draw (2.1029cm,2.8944cm) node {$S_{1b}$};
			\draw (1cm,-0.5cm) node {$S_{\ast}$};
			\draw (1cm,-1.5cm) node {$S_{1a,2,2b,3}$};
			\draw (1cm,-2.5cm) node {$S_{2,2b,3}$};
			\draw (1cm,-3.5cm) node {$S_{2b}$};
		\end{tikzpicture}
	\caption{The probability space for every network. The subscripts refer to the networks that can be found in that region of the probability space. $S_{\ast}=S_{1,1a,1b,1c,2,2b,3}$.}
	\label{probabilityspace}
\end{figure}
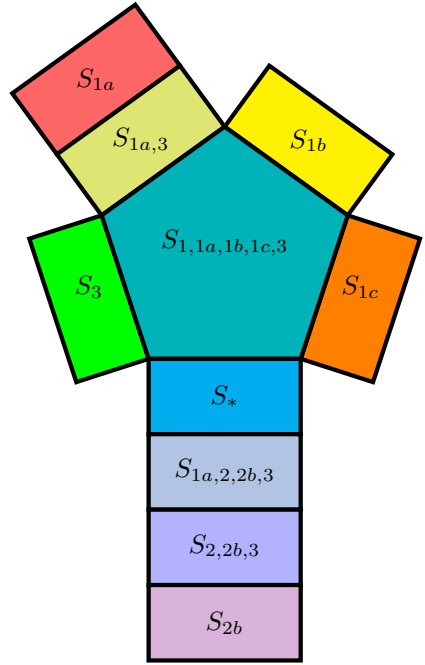

Clearly each network is distinguishable from each other network since there are no two networks that share the same probability space despite there being some overlap in the probability spaces.

If a set of constraints could have arisen on a clock-like tree structure then it could have also arisen on any convergence-divergence network that is the clock-like tree structure with convergence at the leaves or the non-clock-like tree. This is to be expected since by setting the time parameter for convergence to zero we will recover the clock-like tree. Likewise, we can recover a clock-like tree from a non-clock-like tree by making the appropriate time parameters equal.

An interesting region is the region, $S_{1a,3}$, where the constraints could have arisen on a non-clock-like tree or a convergence-divergence network from the first clock-like tree structure, but no other trees or networks.

An important result is that we have regions where we can state that a set of constraints could only have arisen on a single convergence-divergence network or non-clock-like tree and no other trees or networks that we have examined. Interestingly, there is a region where the constraints could have arisen on a convergence-divergence network from either tree structure, or the second clock-like tree structure or the non-clock-like tree. This is the region, $S_{1a,2,2b,3}$. The interesting result is that there is a region where the constraints could have arisen on the clock-like tree with one structure or on a convergence-divergence network with a different structure. Let's look at two of the networks in question, Network $1a$ and Network $2$, as well as the other clock-like tree, Network $1$. The three networks are shown in Figure~\ref{threefourtaxonnetworks} below.
\begin{figure}[H]
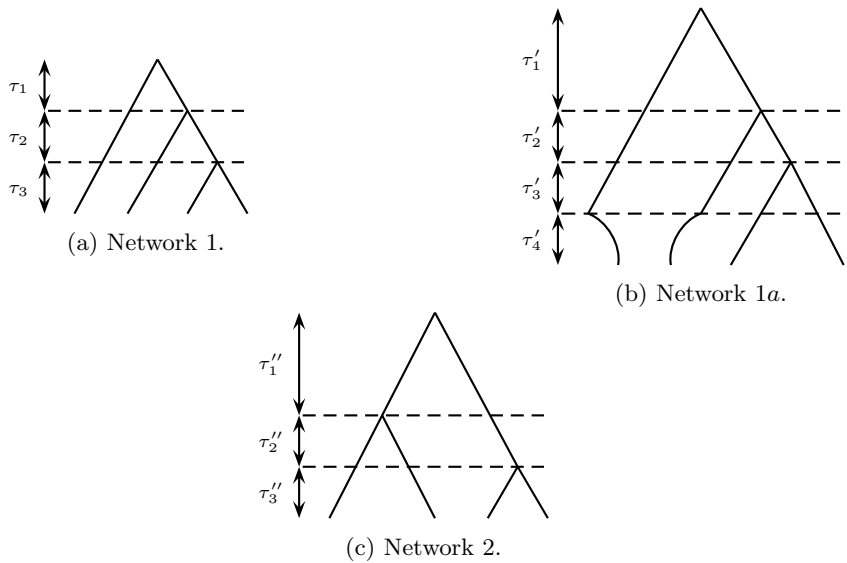

	\centering
	\scriptsize
		\begin{subfigure}[h]{0.49\textwidth}
			\centering
				\psmatrix[colsep=.3cm,rowsep=.4cm,mnode=r]
				~ && && ~ \\
				~ & && && ~ && ~ \\
				~ && && && ~ & ~ \\
				~ & ~ && ~ && ~ && ~
				\ncline{1,5}{4,2}
				\ncline{1,5}{4,8}
				\ncline{2,6}{4,4}
				\ncline{3,7}{4,6}
				\ncline{<->,arrowscale=1.5}{1,1}{2,1}
				\tlput{$\tau_{1}$}
				\ncline{<->,arrowscale=1.5}{2,1}{3,1}
				\tlput{$\tau_{2}$}
				\ncline{<->,arrowscale=1.5}{3,1}{4,1}
				\tlput{$\tau_{3}$}
				\psset{linestyle=dashed}\ncline{2,1}{2,8}
				\psset{linestyle=dashed}\ncline{3,1}{3,8}
				\endpsmatrix
				\caption{Network $1$.}
		\end{subfigure}
		\begin{subfigure}[h]{0.49\textwidth}
			\centering
				\psmatrix[colsep=.3cm,rowsep=.4cm,mnode=r]
				~ & && && ~ \\
				\\
				~ & && && && ~ & && ~ \\
				~ && && && && ~ && ~ \\
				~ & ~ && && ~ & && && ~ \\
				~ && ~ && ~ && ~ && && ~
				\ncline{1,6}{5,2}
				\ncline{1,6}{3,8}
				\ncline{3,8}{4,9}
				\ncline{4,9}{6,11}
				\ncline{3,8}{5,6}
				\ncline{4,9}{6,7}
				\ncarc[arcangle=40]{5,2}{6,3}
				\ncarc[arcangle=-40]{5,6}{6,5}
				\ncline{<->,arrowscale=1.5}{1,1}{3,1}
				\tlput{$\tau_{1}'$}
				\ncline{<->,arrowscale=1.5}{3,1}{4,1}
				\tlput{$\tau_{2}'$}
				\ncline{<->,arrowscale=1.5}{4,1}{5,1}
				\tlput{$\tau_{3}'$}
				\ncline{<->,arrowscale=1.5}{5,1}{6,1}
				\tlput{$\tau_{4}'$}
				\psset{linestyle=dashed}\ncline{3,1}{3,11}
				\psset{linestyle=dashed}\ncline{4,1}{4,11}
				\psset{linestyle=dashed}\ncline{5,1}{5,11}
				\endpsmatrix
				\caption{Network $1a$.}
		\end{subfigure}
		\begin{subfigure}[h]{0.49\textwidth}
			\centering
				\psmatrix[colsep=.3cm,rowsep=.4cm,mnode=r]
				~ & && && ~ \\
				\\
				~ & && ~ & && && & ~ \\
				~ && && && && ~ & ~ \\
				~ & ~ && && ~ && ~ && ~
				\ncline{1,6}{3,4}
				\ncline{3,4}{5,2}
				\ncline{3,4}{5,6}
				\ncline{1,6}{4,9}
				\ncline{4,9}{5,10}
				\ncline{4,9}{5,8}
				\ncline{<->,arrowscale=1.5}{1,1}{3,1}
				\tlput{$\tau_{1}''$}
				\ncline{<->,arrowscale=1.5}{3,1}{4,1}
				\tlput{$\tau_{2}''$}
				\ncline{<->,arrowscale=1.5}{4,1}{5,1}
				\tlput{$\tau_{3}''$}
				\psset{linestyle=dashed}\ncline{3,1}{3,10}
				\psset{linestyle=dashed}\ncline{4,1}{4,10}
				\endpsmatrix
				\caption{Network $2$.}
		\end{subfigure}
				\caption{The three networks in question.}
				\label{threefourtaxonnetworks}
\end{figure}

Recall from Definition \ref{treedistance} on page \pageref{treedistance} that the \emph{tree distance}, $d\left(a,b\right)$, is the sum of the edge lengths connecting two leaves, $a$ and $b$. The tree distances on Network $1$ will satisfy
\begin{align}\begin{split}
d\left(3,4\right)\leq{}d\left(2,3\right)=d\left(2,4\right)\leq{}d\left(1,2\right)=d\left(1,3\right)=d\left(1,4\right).
\end{split}\end{align}
The tree distances on Network $2$ will satisfy
\begin{align}\begin{split}
d\left(3,4\right)''\leq{}d\left(1,2\right)''\leq{}d\left(1,3\right)''=d\left(1,4\right)''=d\left(2,3\right)''=d\left(2,4\right)''.
\end{split}\end{align}
Recall from Definition \ref{pairwisedistance} on page \pageref{pairwisedistance} that the \emph{pairwise distance} was used to define the distance between two leaves on a convergence-divergence network. Ignoring the distance between taxon $1$ and taxon $2$ on Network $1a$ for the time being, the distances on Network $1a$ will satisfy
\begin{align}\begin{split}
d\left(3,4\right)'\leq{}d\left(2,3\right)'=d\left(2,4\right)'\leq{}d\left(1,3\right)'=d\left(1,4\right)'.
\end{split}\end{align}
Clearly if $\tau_{4}'=0$ then $d\left(1,2\right)'=d\left(1,3\right)'=d\left(1,4\right)'$ and the distances on Network $1a$ will fit on Network $1$. Now suppose that we let $\tau_{4}'\rightarrow{}\infty$. The distance $d\left(1,2\right)'$ will become the smallest and the distances will satisfy
\begin{align}\begin{split}
d\left(1,2\right)'\leq{}d\left(3,4\right)'\leq{}d\left(2,3\right)'=d\left(2,4\right)'\leq{}d\left(1,3\right)'=d\left(1,4\right)'.
\end{split}\end{align}
If $\tau_{1}'=0$ then the distances become
\begin{align}\begin{split}
d\left(1,2\right)'\leq{}d\left(3,4\right)'\leq{}d\left(2,3\right)'=d\left(2,4\right)'=d\left(1,3\right)'=d\left(1,4\right)'.
\end{split}\end{align}
These distances will now fit on Network $2$ if we permute taxa $1$ and $3$ and permute taxa $2$ and $4$.

Recall our discussion of clusters in Chapter~\ref{chapter1}. Labelling the leaves $a$, $b$, $c$, $d$, from left to right, the clusters on Network $1$ will be
\begin{align}\begin{split}
\begin{mycases}
a,b,c,d,& \\
b,c,d,& \\
c,d,& \\
a,& \\
b,& \\
c,& \\
d.&
\end{mycases}
\end{split}\end{align}

The clusters on Network $2$ will be
\begin{align}\begin{split}
\begin{mycases}
a,b,c,d,& \\
a,b,& \\
c,d,& \\
a,& \\
b,& \\
c,& \\
d.&
\end{mycases}
\end{split}\end{align}

The clusters on Network $1a$ are not immediately obvious. If $\tau_{4}'=0$ then clearly Network $1a$ will be identical to Network $1$ and they will have the same clusters. If $\tau_{4}'$ is ``large'' then $b$ will no longer cluster with $c$ and $d$, but will instead cluster with $a$. The clusters would then be the same as those on Network $2$. It therefore follows our intuition that there is a region where the constraints for both Network $1a$ and Network $2$ are met.

\subsection{Pseudocode Algorithm to Distinguish Trees and Networks}

We can write an algorithm to determine which trees or networks a pattern frequency could have arisen on. We start by determining whether the constraints for an individual tree or network are met or not. In this algorithm we start with Network $1a$, but we could have started with any network. We then determine which regions of the probability space the pattern frequency could lie in and which trees and networks are in that space. We then determine whether the constraints for a second tree or network are met or not and repeat the process. The algorithm terminates when we have reached one of the ten regions of the probability space in Figure \ref{probabilityspace} on page \pageref{probabilityspace}. Below is a pseudocode algorithm to find which trees or networks the pattern frequency could have arisen on.

Given some pattern frequency input, $P$,
\begin{algorithmic}
\If {Network $1a$ constraints $==$ TRUE}
	\If {Network $1$ constraints $==$ TRUE}
		\If {Network $2$ constraints $==$ TRUE}
			\State {$\textrm{tree or network}\gets \textrm{Networks $1$, $1a$, $1b$, $1c$, $2$, $2b$ or $3$}$}
		\Else
			\State {$\textrm{tree or network}\gets \textrm{Networks $1$, $1a$, $1b$, $1c$ or $3$}$}
		\EndIf
	\Else
		\If {Network $2$ constraints $==$ TRUE}
			\State {$\textrm{tree or network}\gets \textrm{Networks $1a$, $2$, $2b$ or $3$}$}
		\Else
			\If {Network $3$ constraints $==$ TRUE}
				\State {$\textrm{tree or network}\gets \textrm{Networks $1a$ or $3$}$}
			\Else
				\State {$\textrm{tree or network}\gets \textrm{Network $1a$}$}
			\EndIf
		\EndIf
	\EndIf
\Else
	\If {Network $3$ constraints $==$ TRUE}
		\If {Network $2$ constraints $==$ TRUE}
			\State {$\textrm{tree or network}\gets \textrm{Networks $2$, $2b$ or $3$}$}
		\Else
			\State {$\textrm{tree or network}\gets \textrm{Network $3$}$}
		\EndIf
	\Else
		\If {Network $1b$ constraints $==$ TRUE}
			\State {$\textrm{tree or network}\gets \textrm{Network $1b$}$}
		\ElsIf {Network $1c$ constraints $==$ TRUE}
			\State {$\textrm{tree or network}\gets \textrm{Network $1c$}$}
		\Else
			\State {$\textrm{tree or network}\gets \textrm{Network $2b$}$}
		\EndIf
	\EndIf
\EndIf
\end{algorithmic}

We can determine which trees and networks a pattern frequency could have arisen on in either three or four steps. Given the restrictive nature of the binary symmetric model, it is possible that a pattern frequency could not have arisen on any of these trees or networks.

\section{An Example with Multiple Convergence Periods}

We have examined all of the four-taxon convergence-divergence trees and networks with at most a single convergence period, with the convergence period involving two leaves. We will now consider an example of a four-taxon convergence-divergence network with many convergence periods. This network is shown in Figure~\ref{multipleconvergencenetwork} below.
\begin{figure}[H]
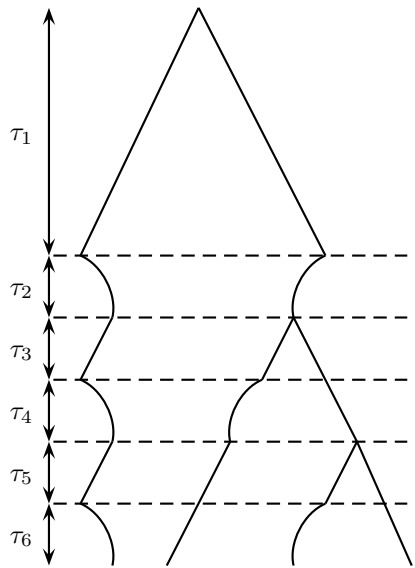

	\centering
		\psmatrix[colsep=.3cm,rowsep=.4cm,mnode=r]
		~ && && & ~ \\
		\\
		\\
		\\
		~ & ~ && && && && ~ & && ~ \\
		~ && ~ && && && ~ && && ~ \\
		~ & ~ && && && ~ & && && ~ \\
		~ && ~ && && ~ && && ~ && ~ \\
		~ & ~ && && && && ~ & && ~ \\
		~ && ~ && ~ && && ~ && && ~
		\ncline{1,6}{5,2}
		\ncline{1,6}{5,10}
		\ncarc[arcangle=40]{5,2}{6,3}
		\ncarc[arcangle=-40]{5,10}{6,9}
		\ncline{6,3}{7,2}
		\ncline{6,9}{7,8}
		\ncline{6,9}{8,11}
		\ncarc[arcangle=40]{7,2}{8,3}
		\ncarc[arcangle=-40]{7,8}{8,7}
		\ncline{8,3}{9,2}
		\ncline{8,7}{10,5}
		\ncline{8,11}{9,10}
		\ncline{8,11}{10,13}
		\ncarc[arcangle=40]{9,2}{10,3}
		\ncarc[arcangle=-40]{9,10}{10,9}
		\ncline{<->,arrowscale=1.5}{1,1}{5,1}
		\tlput{$\tau_{1}$}
		\ncline{<->,arrowscale=1.5}{5,1}{6,1}
		\tlput{$\tau_{2}$}
		\ncline{<->,arrowscale=1.5}{6,1}{7,1}
		\tlput{$\tau_{3}$}
		\ncline{<->,arrowscale=1.5}{7,1}{8,1}
		\tlput{$\tau_{4}$}
		\ncline{<->,arrowscale=1.5}{8,1}{9,1}
		\tlput{$\tau_{5}$}
		\ncline{<->,arrowscale=1.5}{9,1}{10,1}
		\tlput{$\tau_{6}$}
		\psset{linestyle=dashed}\ncline{5,1}{5,13}
		\psset{linestyle=dashed}\ncline{6,1}{6,13}
		\psset{linestyle=dashed}\ncline{7,1}{7,13}
		\psset{linestyle=dashed}\ncline{8,1}{8,13}
		\psset{linestyle=dashed}\ncline{9,1}{9,13}
		\endpsmatrix
		\caption{A four-taxon clock-like convergence-divergence network with multiple convergence periods.}
		\label{multipleconvergencenetwork}
\end{figure}
It may be unlikely that this network would be biologically relevant. We know already that this convergence-divergence network will not be identifiable since it has a convergence period in one of the time periods when there are only two edges, nor will it be network identifiable or distinguishable from the convergence-divergence network that results when $\tau_{2}=0$. However, it will highlight the power of our technique for analysing complicated networks. The variable transformed phylogenetic tensor elements are
\begin{align}\begin{split}
\left[\begin{array}{c} q_{0011} \\
q_{0101} \\
q_{0110} \\
q_{1001} \\
q_{1010} \\
q_{1100} \\
q_{1111} \\
\end{array}\right]
&=\left[\begin{array}{c} e^{-2\left(\tau_{5}+\tau_{6}\right)} \\
e^{-2\left(\tau_{3}+\tau_{4}+\tau_{5}+\tau_{6}\right)} \\
e^{-2\left(\tau_{3}+\tau_{4}+\tau_{5}+\tau_{6}\right)} \\
e^{-2\left(\tau_{3}+\tau_{4}+\tau_{5}+\tau_{6}\right)}\left(1-e^{-\tau_{2}}\left(1-e^{-2\tau_{1}}\right)\right) \\
1-e^{-\tau_{6}}\left(1-e^{-2\left(\tau_{3}+\tau_{4}+\tau_{5}\right)}\left(1-e^{-\tau_{2}}\left(1-e^{-2\tau_{1}}\right)\right)\right) \\
e^{-2\left(\tau_{5}+\tau_{6}\right)}\left(1-e^{-\tau_{4}}\left(1-e^{-2\tau_{3}}\left(1-e^{-\tau_{2}}\left(1-e^{-2\tau_{1}}\right)\right)\right)\right) \\
\scriptstyle{e^{-2\left(\tau_{5}+\tau_{6}\right)}\left(e^{-2\left(\tau_{3}+\tau_{4}\right)}\left(1-e^{-\tau_{6}}\right)+e^{-\left(2\tau_{5}+\tau_{6}\right)}\left(1-e^{-\tau_{4}}\left(1-e^{-2\tau_{3}}\left(1-e^{-\tau_{2}}\left(1-e^{-2\tau_{1}}\right)\right)\right)\right)\right)} \\
\end{array}\right] \\
&=\left[\begin{array}{c} y_{5}^{2}y_{6}^{2} \\
y_{3}^{2}y_{4}^{2}y_{5}^{2}y_{6}^{2} \\
y_{3}^{2}y_{4}^{2}y_{5}^{2}y_{6}^{2} \\
y_{3}^{2}y_{4}^{2}y_{5}^{2}y_{6}^{2}\left(1-y_{2}\left(1-y_{1}^{2}\right)\right) \\
1-y_{6}\left(1-y_{3}^{2}y_{4}^{2}y_{5}^{2}\left(1-y_{2}\left(1-y_{1}^{2}\right)\right)\right) \\
y_{5}^{2}y_{6}^{2}\left(1-y_{4}\left(1-y_{3}^{2}\left(1-y_{2}\left(1-y_{1}^{2}\right)\right)\right)\right) \\
y_{5}^{2}y_{6}^{2}\left(y_{3}^{2}y_{4}^{2}\left(1-y_{6}\right)+y_{5}^{2}y_{6}\left(1-y_{4}\left(1-y_{3}^{2}\left(1-y_{2}\left(1-y_{1}^{2}\right)\right)\right)\right)\right) \\
\end{array}\right].
\end{split}\end{align}
We form the ideal,
\begin{align}\begin{split}
I=&\left<y_{5}^{2}y_{6}^{2}-q_{0011},y_{3}^{2}y_{4}^{2}y_{5}^{2}y_{6}^{2}-q_{0101},y_{3}^{2}y_{4}^{2}y_{5}^{2}y_{6}^{2}-q_{0110},y_{3}^{2}y_{4}^{2}y_{5}^{2}y_{6}^{2}\left(1-y_{2}\left(1-y_{1}^{2}\right)\right)-q_{1001},\right. \\
&\qquad{}\left.1-y_{6}\left(1-y_{3}^{2}y_{4}^{2}y_{5}^{2}\left(1-y_{2}\left(1-y_{1}^{2}\right)\right)\right)-q_{1010},\right. \\
&\qquad{}\left.y_{5}^{2}y_{6}^{2}\left(1-y_{4}\left(1-y_{3}^{2}\left(1-y_{2}\left(1-y_{1}^{2}\right)\right)\right)\right)-q_{1100},\right. \\
&\qquad{}\left.y_{5}^{2}y_{6}^{2}\left(y_{3}^{2}y_{4}^{2}\left(1-y_{6}\right)+y_{5}^{2}y_{6}\left(1-y_{4}\left(1-y_{3}^{2}\left(1-y_{2}\left(1-y_{1}^{2}\right)\right)\right)\right)\right)-q_{1111}\right>.
\end{split}\end{align}
When we transform the basis of the ideal into the Gr\"{o}bner basis we find that the ideal has $98$ generators. Of these polynomial generators, the highest degree of the polynomials is $17$ and the polynomial with the most monomials has $266$ monomials. Clearly it is not practical to write down the Gr\"{o}bner basis of the ideal here. We will not include the Gr\"{o}bner basis of the ideal in the Appendix either. Instead, we will include the $Macaulay2$ code for finding the Gr\"{o}bner basis of the ideal, shown below.
\begin{lstlisting}
S = QQ
R = S[y_1..y_6, q_1..q_7, MonomialOrder => {Lex => 6, Lex => 7}]
I = ideal(y_5^2*y_6^2 - q_1, y_3^2*y_4^2*y_5^2*y_6^2 - q_2,
y_3^2*y_4^2*y_5^2*y_6^2 - q_3,
y_3^2*y_4^2*y_5^2*y_6^2*(1 - y_2*(1 - y_1^2)) - q_4,
1 - y_6*(1 - y_3^2*y_4^2*y_5^2*(1 - y_2*(1 - y_1^2))) - q_5,
y_5^2*y_6^2*(1 - y_4*(1 - y_3^2*(1 - y_2*(1 - y_1^2)))) - q_6,
y_5^2*y_6^2*(y_3^2*y_4^2*(1 - y_6)
+ y_5^2*y_6*(1 - y_4*(1 - y_3^2*(1 - y_2*(1 - y_1^2))))) - q_7)
timing transpose gens gb I
\end{lstlisting}
The last line of code prints the Gr\"{o}bner basis of the ideal and the CPU time. For this example, the CPU time was $0.249602$ seconds.

Only two of the generators in the Gr\"{o}bner basis do not contain any of the time parameters, $y_{i}$. Hence, there are only two equality constraints,
\begin{align}\begin{split}
\begin{mycases}
q_{0101}&=q_{0110}, \\
q_{0011}^{2}q_{1100}^{2}-2q_{0011}q_{0101}q_{1001}q_{1100}-q_{0011}q_{0101}q_{1010}^{2}q_{1100}+q_{0011}q_{0101}q_{1010}q_{1100}& \\
+q_{0011}q_{1010}q_{1100}q_{1111}-q_{0011}q_{1100}q_{1111}+q_{0101}^{2}q_{1010}^{2}-q_{0101}^{2}q_{1001}q_{1010}& \\
+q_{0101}q_{1001}q_{1010}q_{1111}+q_{0101}q_{1001}q_{1111}-q_{1001}q_{1111}^{2}&=0.
\end{mycases}
\end{split}\end{align}
All other constraints will be inequalities that arise from the expressions for the time parameters and the constraints on these time parameters. Fortunately we do not need to know every generator of the Gr\"{o}bner basis of the ideal to solve for the six time parameters. We only need six generators that we can form independent equations from that will allow us to solve for the six time parameters as functions of the transformed phylogenetic tensor elements.

If we look at the expressions for the transformed phylogenetic tensor elements in terms of the time parameters we can see that the time parameters, $y_{1}$ and $y_{2}$, always appear in the same form. Every transformed phylogenetic tensor element expression containing these time parameters is of the form
\begin{align}\begin{split}
q_{i}&=f_{1}\left(y_{3},y_{4},y_{5},y_{6}\right)+f_{2}\left(y_{3},y_{4},y_{5},y_{6}\right)\left(1-y_{2}\left(1-y_{1}^{2}\right)\right),
\end{split}\end{align}
where $f_{1}\left(y_{3},y_{4},y_{5},y_{6}\right)$ and $f_{2}\left(y_{3},y_{4},y_{5},y_{6}\right)$ are expressions of the time parameters which do not include $y_{1}$ or $y_{2}$.

Consequently, $y_{1}$ and $y_{2}$ will not be able to be recovered independently of each other and the network will not be identifiable. This is not particularly surprising since there is an obvious similarity in the structure of this network to the structure of the two-taxon convergence-divergence network introduced in Chapter~\ref{chapter4}. Both networks have two edges diverging from the root, which then converge back together. Recall that the two-taxon convergence-divergence network was not identifiable since the two time parameters, representing the divergence and convergence respectively, could not be recovered independently from the expressions for the transformed phylogenetic tensor elements.

We will use a set of five independent equations formed from five generators, some from the Gr\"{o}bner basis and some from the original basis, to find expressions for the time parameters in terms of the transformed phylogenetic tensor elements. There will be one expression involving the first two time parameters, $y_{1}$ and $y_{2}$, since they cannot be recovered independently. One of these sets of five equations formed from five of the generators after some rearranging is
\begin{align}\begin{split}
\begin{mycases}
y_{6}^2+q_{1010}y_{6}-y_{6}-q_{1001}&=0, \\
y_{5}^2y_{6}^2-q_{0011}&=0, \\
q_{0011}y_{4}^2-q_{0011}y_{4}+q_{1100}y_{4}-q_{1001}&=0, \\
y_{3}^2y_{4}^2y_{5}^2y_{6}^2-q_{0101}&=0, \\
q_{0110}y_{1}^{2}y_{2}-q_{0110}y_{2}+q_{0110}-q_{1001}&=0.
\end{mycases}
\end{split}\end{align}
This set of equations was chosen because the polynomials in them had fewer terms than most of the other polynomials and because they contained fewer time parameters. For example, from the first equation it can easily be seen that an expression for $y_{6}$ can be found since it contains none of the other time parameters.

A set of expressions for these time parameters is
\begin{align}\begin{split}
\begin{mycases}
1-y_{2}\left(1-y_{1}^{2}\right)&=\frac{q_{1001}}{q_{0101}}, \\
y_{3}&=\frac{\sqrt{q_{0101}}\left(-\left(q_{0011}-q_{1100}\right)+\sqrt{\left(q_{0011}-q_{1100}\right)^{2}+4q_{0011}q_{1001}}\right)}{2\sqrt{q_{0011}}q_{1001}}, \\
y_{4}&=\frac{\left(q_{0011}-q_{1100}\right)+\sqrt{\left(q_{0011}-q_{1100}\right)^{2}+4q_{0011}q_{1001}}}{2q_{0011}}, \\
y_{5}&=\frac{\sqrt{q_{0011}}\left(-\left(1-q_{1010}\right)+\sqrt{\left(1-q_{1010}\right)^{2}+4q_{1001}}\right)}{2q_{1001}}, \\
y_{6}&=\frac{1}{2}\left(\left(1-q_{1010}\right)+\sqrt{\left(1-q_{1010}\right)^{2}+4q_{1001}}\right).
\end{mycases}
\end{split}\end{align}

The inequality constraints on the transformed phylogenetic tensor elements can be found by demanding, $0<y_{j}\leq{}1$. Earlier in the thesis we established that if $0<y_{1},y_{2}\leq{}1$ then $0<1-y_{2}\left(1-y_{1}^{2}\right)\leq{}1$. Demanding $0<y_{3},y_{4},y_{5},y_{6}\leq{}1$ and $0<1-y_{2}\left(1-y_{1}^{2}\right)\leq{}1$, gives the constraints on the transformed phylogenetic tensor elements of
\begin{align}\begin{split}
\begin{mycases}
q_{1001}&\leq{}q_{0101}, \\
q_{1001}&\leq{}q_{1010}, \\
q_{1001}&\leq{}q_{1100}, \\
\sqrt{q_{0011}}\left(q_{0101}-q_{1001}\right)&\leq{}\sqrt{q_{0101}}\left(q_{0011}-q_{1100}\right), \\
\left(q_{0011}-q_{1001}\right)^{2}&\leq{}q_{0011}\left(1-q_{1010}\right)^{2}.
\end{mycases}
\end{split}\end{align}

Since $q_{1001}\leq{}q_{0101}$,
\begin{align}\begin{split}
\sqrt{q_{0011}}\left(q_{0101}-q_{1001}\right)&\geq{}0,
\end{split}\end{align}
as required. However, since
\begin{align}\begin{split}
\sqrt{q_{0011}}\left(q_{0101}-q_{1001}\right)&\leq{}\sqrt{q_{0101}}\left(q_{0011}-q_{1100}\right),
\end{split}\end{align}
it is implied that
\begin{align}\begin{split}
q_{0011}-q_{1100}&\geq{}0,
\end{split}\end{align}
and
\begin{align}\begin{split}
q_{0011}\left(q_{0101}-q_{1001}\right)^{2}&\leq{}q_{0101}\left(q_{0011}-q_{1100}\right)^{2}.
\end{split}\end{align}

We cannot take the square root of each side of this equality and preserve the inequality, however, since it must be assumed that $q_{0011}-q_{1100}\geq{}0$. Consequently, the original sets of constraints is the set of constraints with the minimum number of elements in it. In conclusion, the set of constraints for this convergence-divergence network is
\begin{align}\begin{split}
\begin{mycases}
q_{0101}&=q_{0110}, \\
q_{0011}^{2}q_{1100}^{2}-2q_{0011}q_{0101}q_{1001}q_{1100}-q_{0011}q_{0101}q_{1010}^{2}q_{1100}& \\
+q_{0011}q_{0101}q_{1010}q_{1100}+q_{0011}q_{1010}q_{1100}q_{1111}-q_{0011}q_{1100}q_{1111}& \\
+q_{0101}^{2}q_{1010}^{2}-q_{0101}^{2}q_{1001}q_{1010}+q_{0101}q_{1001}q_{1010}q_{1111}& \\
+q_{0101}q_{1001}q_{1111}-q_{1001}q_{1111}^{2}&=0, \\
q_{1001}&\leq{}q_{0101}, \\
q_{1001}&\leq{}q_{1010}, \\
q_{1001}&\leq{}q_{1100}, \\
\sqrt{q_{0011}}\left(q_{0101}-q_{1001}\right)&\leq{}\sqrt{q_{0101}}\left(q_{0011}-q_{1100}\right), \\
\left(q_{0011}-q_{1001}\right)^{2}&\leq{}q_{0011}\left(1-q_{1010}\right)^{2}.
\end{mycases}
\end{split}\end{align}
We will not proceed to find the intersections between this set of constraints and the sets of constraints for other trees and networks since we have already argued that this network is not an appropriate choice for a network. However, these intersections could be found in the same way that we found the intersections in Section~\ref{fourtaxondistinguish}.

\chapter{Conclusion \& Discussion}
\label{chapter7}

\section{Thesis Findings}
\label{section7.1}

\citet{sumner2012algebra} introduced a model for convergence on phylogenetic networks. We call these networks convergence-divergence networks. On these networks the assumption of independent evolution between edges is allowed to be broken and these edges converge with each other over a period of time. Some questions arise when we are attempting to determine whether the convergence-divergence networks are desirable choices for modelling. Three questions arise:
\begin{enumerate}
\item Are the trees and convergence-divergence networks \emph{identifiable}?
\item Are the trees and convergence-divergence networks \emph{network identifiable} from each other?
\item Are the trees and convergence-divergence networks \emph{distinguishable} from each other?
\end{enumerate}

For the convergence-divergence networks to be identifiable the numerical model parameters must be recoverable from the pattern frequencies. This is a matter of determining whether a set of solutions for the time parameters can be found from the expressions for the phylogenetic tensor for the network. For the convergence-divergence networks to be network identifiable the network itself must be recoverable from the pattern frequencies from all of the choices of trees and networks. In other words, can we find time parameters on one network that correspond to equivalent time parameters on another network? If we can do this then the convergence-divergence networks are not network identifiable. We used network identifiability to compare between taxon label permutations on a given tree or network. For any tree or network to be distinguishable from another tree or network there must be pattern frequencies that could have only arisen on one of the trees or networks. In other words, the probability spaces of the pattern frequencies for the two trees or networks must not be identical for them to be distinguishable.

In Chapter~\ref{chapter4} and Chapter~\ref{chapter6} we addressed the question of distinguishability by finding the equality and inequality constraints on the transformed phylogenetic tensors for various two-taxon, three-taxon and four-taxon trees and networks and comparing the intersections of the probability spaces. A simple consequence of set theory is that the intersection of two sets, $\Omega_{1}$ and $\Omega_{2}$, can take one of four possibilities:
\begin{align}\begin{split}
(i)&\text{	$\Omega{}_{1}=\Omega{}_{2}$. The sets are equal,} \\
(ii)&\text{	$\Omega{}_{1}\neq{}\Omega{}_{2}$ and $\Omega{}_{1}\cap{}\Omega{}_{2}=\Omega{}_{1}\Leftrightarrow{}\Omega{}_{1}\subset{}\Omega{}_{2}$. $\Omega{}_{1}$ is a proper subset of $\Omega{}_{2}$,} \\
(iii)&\text{	$\Omega{}_{1}\cap{}\Omega{}_{2}=\emptyset{}$. The sets are disjoint,} \\
(iv)&\text{	$\Omega{}_{1}\neq{}\Omega{}_{2}$, $\Omega{}_{1}\cap{}\Omega{}_{2}\subset{}\Omega{}_{1}$ and $\Omega{}_{1}\cap{}\Omega{}_{2}\subset{}\Omega{}_{2}$. The intersection of the sets is non-empty, but} \\
&\text{	neither set is a proper subset of the other.}
\end{split}\end{align}

If $\Omega_{1}$ and $\Omega_{2}$ are taken to be two sets of constraints for two different trees or networks then we say that the two trees or networks are distinguishable unless $\Omega_{1}=\Omega_{2}$.

All of the trees and networks that we examined for all number of taxa were modelled using the binary symmetric model since it is the simplest model to analyse. Being an Abelian group-based model it is diagonalisable, which allows us to find the constraints on the transformed phylogenetic tensors more easily. In Chapter~\ref{chapter2} we examined a number of Abelian group-based models, some of which can be found in \citet{sumner2014tensorial}, and found the diagonalising matrices for these models. 

We examined various two-taxon, three-taxon and four-taxon trees and networks. We made some restrictions on our convergence-divergence networks to simplify our analysis. For example, we did not consider any networks where there were alternating periods of convergence and divergence between the same group of taxa.

In Section \ref{twotaxon} we looked at the distinguishability of two-taxon trees and networks. For the two-taxon case, there were three trees and networks to examine: the non-clock-like tree, the clock-like tree and the convergence divergence network with convergence between the two taxa. We found that the non-clock-like tree and the clock-like convergence-divergence network were not identifiable or distinguishable from the clock-like tree. We also found that the clock-like convergence-divergence network was not network identifiable from the clock-like tree. We concluded that under the binary symmetric model this model of convergence is not desirable over the clock-like tree for two taxa. We also showed that if a two-taxon convergence-divergence network is embedded inside a larger network then that network will not be identifiable, nor will it be network identifiable nor distinguishable from the same network with the convergence period replaced by a divergence period.

We then looked at the three-taxon case in Section \ref{threetaxon}, for which there were four trees and networks to examine: the clock-like tree, the non-clock-like tree, the convergence-divergence network with ``sister'' taxa convergence and the convergence-divergence network with ``non-sister'' taxa convergence. We are using the term ``sister'' to refer to edges in the same time period that form a cluster with no other edges contained in it and the term ``non-sister'' to refer to all other pairs of edges in the same time period. From the result regarding embedded two-taxon convergence-divergence networks, we concluded that the three-taxon convergence-divergence network with sister taxa convergence was not identifiable, nor was it network identifiable nor distinguishable from the three-taxon clock-like tree. We found that the three-taxon convergence-divergence network with non-sister taxa convergence was identifiable and distinguishable from both the three-taxon non-clock-like tree and the three-taxon clock-like tree. We found that there were four regions in the probability space where the pattern frequencies could lie:
\begin{enumerate}
\item The pattern frequencies could lie on the convergence-divergence network with non-sister taxa convergence and on no other tree or network,
\item The pattern frequencies could lie on the non-clock-like tree and on no other tree or network,
\item The pattern frequencies could lie on either the convergence-divergence network with non-sister taxa convergence or the non-clock-like tree and on no other tree or network,
\item The pattern frequencies could lie on all three trees or networks: the convergence-divergence network with non-sister taxa convergence, the non-clock-like tree or the clock-like tree.
\end{enumerate}

In Chapter~\ref{chapter5} we examined the permutations of taxon labels for the three-taxon case. We first looked at the three-taxon clock-like tree. We showed that there are three taxon label permutations that are network identifiable from each other. Of the six possible taxon label permutations, there will be three pairs that have a taxon label permutation that is not network identifiable from the other taxon label permutation in the pair. Swapping the taxon labelling on the two sister taxa will not result in a network identifiable taxon label permutation from the original taxon label permutation.

We then looked at the convergence-divergence network with non-sister taxa convergence and compared the time parameters for different taxon label permutations. There was one taxon label permutation that was of particular interest to us, the permutation between the converging taxon not in the cluster of the two leaves sharing the most recent common ancestor and the non-converging taxon in that cluster. We found expressions for the time parameters on the second taxon label permutation in terms of the time parameters on the first taxon label permutation. Demanding the time parameters on the first taxon label permutation were positive, we found that for the transformed phylogenetic tensors to be equivalent one of the time parameters on the second taxon label permutation could be negative. From simulations we found that for $497$ out of $500$ samples all of the time parameters would be non-negative. For $3$ of the $500$ simulations one of the time parameters was negative. Since there are possible pattern frequencies that could have only arisen on one of the taxon label permutations, the two taxon label permutations are network identifiable. However, both taxon label permutations will usually have a set of non-negative time parameters that will give rise to the same phylogenetic tensors. There may therefore be a choice between the two taxon label permutations.

In Chapter~\ref{chapter6} we looked at the distinguishability of four-taxon trees and networks. For four-taxon convergence-divergence networks the problem became far more complicated. We could no longer find the constraints on the transformed phylogenetic tensors using elementary methods. We used methods from algebraic geometry to allow us to analyse these networks. As with the two-taxon and three-taxon cases, we first found the expressions for the transformed phylogenetic tensors for each tree and network. To find the equality and inequality constraints on the transformed phylogenetic tensors we first took the transformed phylogenetic tensor expressions and rearranged them to be identically zero. For example, suppose we had the transformed phylogenetic tensor expression, $q_{i}=f\left(y_{1},y_{2},\ldots{},y_{k}\right)$, where $q_{i}$ is an element of the transformed phylogenetic tensor and $y_{i}=e^{-\tau_{i}}$ are dependent on the time parameters. We then rearranged the expression to be $f\left(y_{1},y_{2},\ldots{},y_{k}\right)-q_{i}=0$. We then took the expression, $f\left(y_{1},y_{2},\ldots{},y_{k}\right)-q_{i}$, to be a generating polynomial of an ideal. After finding all of the generators of the ideal we transformed the basis of the ideal into the Gr\"obner basis. The generators of the Gr\"obner basis were then extracted and formed into expressions that were identical to zero. These expressions contained all of the equality constraints on the transformed phylogenetic tensor. The time parameters could be solved from the remaining expressions. The inequality constraints on the transformed phylogenetic tensors came from the expressions for the time parameters and the fact that the time parameters must be non-negative on trees and networks.

In Section \ref{fourtaxondistinguish} we compared the constraints on each of the four-taxon trees and networks in the same fashion as for the two-taxon and three-taxon cases. We found that the two structures of the four-taxon clock-like trees, the four-taxon non-clock-like tree and all of the four-taxon convergence-divergence networks with convergence between two non-sister leaves were identifiable and distinguishable from each other. We concluded that all of these convergence-divergence networks may be appropriate choices for particular pattern frequencies.

Finally, we showed that we could find the constraints on a very complicated four-taxon convergence-divergence network with multiple convergence periods using our method. Despite the Gr\"obner basis having $98$ generators, with the highest degree of the polynomials being $17$ and the largest number of terms in the generators being $266$, we were still able to find the set of constraints for this network.

We concluded that all three-taxon and four-taxon convergence-divergence networks with convergence between two non-sister leaves may be appropriate choices under the binary symmetric model when the pattern frequencies do not appear to have arisen on a tree. It must be noted, however, that our convergence-divergence networks may not always be an appropriate choice, even when it is clear that the pattern frequencies did not arise on a tree. If the pattern frequencies did not arise on a tree, then it cannot be assumed that the pattern frequencies arose on one of our convergence-divergence networks.

\section{Future Work}

We restricted our work to the binary symmetric model for simplicity. Being the simplest model, it is perhaps not surprising that the two-taxon convergence-divergence network was not identifiable, network identifiable or distinguishable from the clock-like tree. For both the two-taxon clock-like tree and the two-taxon convergence-divergence network there was only a single variable element of the transformed phylogenetic tensor. Hence, any two-taxon network with multiple time parameters will not be identifiable under the binary symmetric model.

The work we have done on determining identifiability of trees and networks and distinguishability between one tree or network and another could be done for any Abelian group-based model. For more parameter rich Abelian group-based models there will be more polynomial equations to solve, with more variables, more terms and higher degrees, which will make the sets of constraints more difficult to analyse. A more parameter rich model or a model with more than two states will allow for more freedom for the two-taxon convergence-divergence network to be identifiable.

For more parameter rich models we have a choice of fixing the rate parameters to be identical across all time periods or allowing for variable rate parameters across the time periods. For the binary symmetric model this choice is not available. For only one rate parameter, different rates across time periods will be equivalent to equal rates across time periods, with a different set of time parameters. With more parameter rich models than the binary symmetric model it may be possible to distinguish the two-taxon convergence-divergence network from the two-taxon clock-like tree if rates are allowed to vary across time periods since this allows for more flexibility.

The increase in parameters may allow for a two-taxon convergence-divergence network that is identifiable, as well as network identifiable and distinguishable from the two-taxon clock-like tree.

We didn't analyse any trees or networks beyond the four-taxon case, however there is no reason not to examine trees and networks with more taxa. If the trend from the three-taxon and four-taxon convergence-divergence networks holds for more taxa then any $n$-taxon convergence-divergence networks with convergence between two non-sister leaves will be distinguishable from $n$-taxon clock-like trees and $n$-taxon non-clock-like trees.

We only examined the taxon label permutations for the three-taxon case. It would be interesting to see if there are any taxon label permutations for the four-taxon or larger taxon cases which are not network identifiable from each other.

The least parameter rich of the trees and networks are the clock-like trees. \citet{steel2005should} noted that when developing models care must be taken to avoid overfitting the data. The more parameters that a tree or network has the more potential there is to overfit the data. \citet{steel2005should} also noted, however, that the models must be biologically relevant. In some circumstances a clock-like tree will not be biologically relevant and will not be the best fit to the data. If there is still variability in the data that is not explained by a clock-like tree then we can introduce more parameters to the model, provided these extra parameters significantly improve the fit. One option to introduce more parameters is to remove the molecular clock assumption and allow for a non-clock-like tree. Another option is to remove the assumption of independent evolution across edges in the same time period, allowing for convergence. Non-clock-like trees and convergence-divergence networks do not necessarily have the same number of parameters. Below in Table~\ref{numberparameters} is the number of parameters for clock-like trees, non-clock-like trees and convergence-divergence networks for different number of taxa. 

\begin{table}[H]
\centering
\begin{tabular}{c|c|c|c}
 & \multicolumn{3}{c}{Number of parameters} \\
\hline
Number of taxa & Clock-like tree & Non-clock-like tree & Convergence-divergence network \\
\hline
$2$ & $1$ & - & $\geq{}2$ \\
\hline
$3$ & $2$ & $3$ & $\geq{}3$ \\
\hline
$4$ & $3$ & $5$ & $\geq{}4$ \\
\hline
$\vdots{}$ & $\vdots{}$ & $\vdots{}$ & $\vdots{}$ \\
\hline
$n$ & $n-1$ & $2n-3$ & $\geq{}n$ \\
\hline
\end{tabular}
\caption{Number of parameters for clock-like trees, non-clock-like trees and clock-like convergence-divergence networks.}
\label{numberparameters}
\end{table}

The number of parameters for an $n$-taxon clock-like tree is $n-1$. If we wish to remove the molecular clock assumption, an $n$-taxon non-clock-like tree will have $2n-3$ parameters, $2n-3-\left(n-1\right)=n-2$ parameters more than the clock-like tree. If we have a large number of taxa then removing the molecular clock assumption introduces many more parameters, approximately doubling the number of parameters. Clearly this is a potential issue of overfitting. With convergence-divergence networks we have much more flexibility in our parameter number choice. Introducing one convergence period to a clock-like tree will only introduce one extra parameter. We have the option of introducing multiple convergence periods if there is still more variability in the data left to explain. Convergence-divergence networks could be an option when there is more variability in the data than can be explained by a clock-like tree, but when the data is overfit on a non-clock-like tree.

Suppose we decided on a convergence-divergence network, however we did not decide on where the taxa should be placed on the leaves. Recall from Chapter~\ref{chapter5} and earlier in Section~\ref{section7.1} that for the three-taxon convergence-divergence network with convergence between non-sister taxa all of the taxon label permutations were network identifiable. However, of the six taxon labellings, there were three pairs of permutations that almost always gave rise to equivalent transformed phylogenetic tensors with all parameters being non-negative. We found that for any given taxon labelling, if we swapped the labels between the converging edge having the root as the most recent common ancestor with the other two leaves and the edge which is not converging with another edge then the two taxon label permutations would almost always give rise to equivalent transformed phylogenetic tensors with all parameters being non-negative. Two of those taxon labellings are shown in Figure~\ref{twosimilarpermutations} below.
\begin{figure}[H]
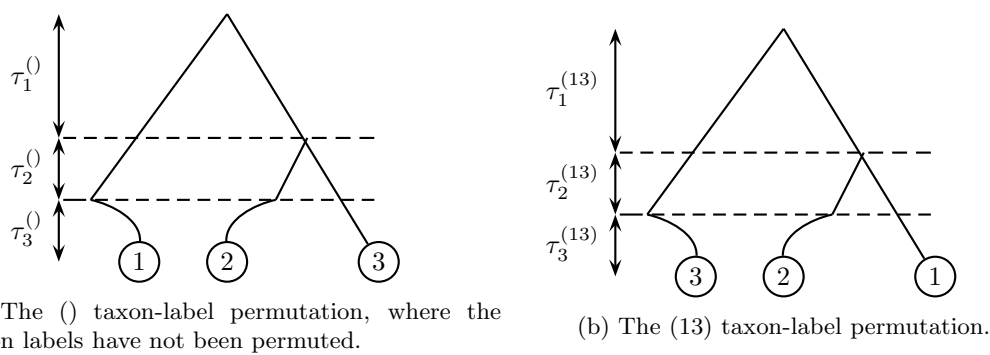

	\centering 
		\begin{subfigure}[h]{0.49\textwidth}
		\centering
				\psmatrix[colsep=.3cm,rowsep=.4cm,mnode=r]
				~ && && ~ \\
				~ \\
				~ && && && ~ && ~ \\
				~ & ~ && && ~ & && ~ \\
				~ && [mnode=circle] 1 && [mnode=circle] 2 && && [mnode=circle] 3
				\ncline{1,5}{4,2}
				\ncline{1,5}{5,9}
				\ncline{3,7}{4,6}
				\ncarc[arcangle=40]{4,2}{5,3}
				\ncarc[arcangle=-40]{4,6}{5,5}
				\ncline{<->,arrowscale=1.5}{1,1}{3,1}
				\tlput{$\tau_{1}^{\left(\right)}$}
				\ncline{<->,arrowscale=1.5}{3,1}{4,1}
				\tlput{$\tau_{2}^{\left(\right)}$}
				\ncline{<->,arrowscale=1.5}{4,1}{5,1}
				\tlput{$\tau_{3}^{\left(\right)}$}
				\psset{linestyle=dashed}\ncline{3,1}{3,7}
				\psset{linestyle=dashed}\ncline{3,7}{3,9}
				\psset{linestyle=dashed}\ncline{4,1}{4,2}
				\psset{linestyle=dashed}\ncline{4,2}{4,6}
				\psset{linestyle=dashed}\ncline{4,6}{4,9}
				\endpsmatrix
				\caption{The $()$ taxon-label permutation, where the taxon labels have not been permuted.}
		\end{subfigure}
		\begin{subfigure}[h]{0.49\textwidth}
		\centering
				\psmatrix[colsep=.3cm,rowsep=.4cm,mnode=r]
				~ && && ~ \\
				~ \\
				~ && && && ~ && ~ \\
				~ & ~ && && ~ & && ~ \\
				~ && [mnode=circle] 3 && [mnode=circle] 2 && && [mnode=circle] 1
				\ncline{1,5}{4,2}
				\ncline{1,5}{5,9}
				\ncline{3,7}{4,6}
				\ncarc[arcangle=40]{4,2}{5,3}
				\ncarc[arcangle=-40]{4,6}{5,5}
				\ncline{<->,arrowscale=1.5}{1,1}{3,1}
				\tlput{$\tau_{1}^{\left(13\right)}$}
				\ncline{<->,arrowscale=1.5}{3,1}{4,1}
				\tlput{$\tau_{2}^{\left(13\right)}$}
				\ncline{<->,arrowscale=1.5}{4,1}{5,1}
				\tlput{$\tau_{3}^{\left(13\right)}$}
				\psset{linestyle=dashed}\ncline{3,1}{3,7}
				\psset{linestyle=dashed}\ncline{3,7}{3,9}
				\psset{linestyle=dashed}\ncline{4,1}{4,2}
				\psset{linestyle=dashed}\ncline{4,2}{4,6}
				\psset{linestyle=dashed}\ncline{4,6}{4,9}
				\endpsmatrix
				\caption{The $(13)$ taxon-label permutation.}
		\end{subfigure}
		\caption{The two taxon-label permutations in question for the three-taxon convergence-divergence network.}
		\label{twosimilarpermutations}
\end{figure}

If we take $\tau_{1}^{\left(\right)}$, $\tau_{2}^{\left(\right)}$ and $\tau_{3}^{\left(\right)}$ to all be non-negative then $\tau_{1}^{\left(13\right)}$ and $\tau_{3}^{\left(13\right)}$ must always be non-negative, however $\tau_{2}^{\left(13\right)}$ will occasionally be negative. In most cases $\tau_{2}^{\left(13\right)}$ will also be non-negative, however. In these circumstances we will have a choice between the two taxon label permutations. This choice could be made by considering the biological implications of the two taxon label permutations. If it is known, or suspected, that a pair of taxa converged then we could choose the taxon label permutation that allowed that pair of taxa to converge. We could also choose the taxon label permutation that has the least convergence. In other words, we could compare $\tau_{3}^{\left(\right)}$ and $\tau_{3}^{\left(13\right)}$ and choose the taxon label permutation that had the smaller of the two parameters. Another choice that could be made is choosing the taxon label permutation that converges the two taxa with the least divergence or distance between them. In other words, we could compare $\tau_{1}^{\left(\right)}+\tau_{2}^{\left(\right)}$ to $\tau_{1}^{\left(13\right)}+\tau_{2}^{\left(13\right)}$ and choose the taxon label permutation with the smaller of the two sums.

It would be interesting to apply the results on identifiability and distinguishability of three-taxon and four-taxon trees and networks to some data sets for which a clock-like tree does not provide enough parameters and a non-clock-like tree provides too many parameters. The questions of identifiability and distinguishability could be used as a diagnostic tool for given data sets, particularly data sets for which there is known to be or suspected to be some form of convergence. An example for which convergence-divergence networks under the binary symmetric model could be used as a diagnostic tool is binary single nucleotide polymorphisms (SNPS).

\citet{holland2010identifying} analysed a binary morphological data set of cormorants and shags, with $30$ taxa and $137$ morphological characters. The taxa have been previously classified on morphological characteristics, which is reflected in their genus names and morphological consensus tree, with the nine genera representing the nine distinct morphological groupings. They showed that the molecular consensus tree differed considerably from the morphological consensus tree. The convergence-divergence networks could be applied to the morphological data set to further address the question of whether the species developed similar morphological traits due to inhabiting similar niches rather than sharing a recent common ancestor.

Finally, we have shown that the convergence-divergence networks introduced by \citet{sumner2012algebra} are a useful new way to model convergence. We have shown that these networks could be more appropriate models than trees when a pattern frequency does not appear to have arisen on a tree. These convergence-divergence networks could be used as a diagnostic tool when it is suspected that a set of taxa did not arise on a tree, with convergence having occurred between some of the taxa.

\begin{appendices}

\chapter{Time Parameters for the $\left(12\right)$ Taxon Label Permutation}\label{derivation1}

Equating the three pairwise distances for the two taxon label permutations for the three-taxon convergence-divergence network, the $\left(12\right)$ taxon label permutation and no taxon label permutation,
\begin{align}\begin{split}
\begin{mycases}
2\left(\tau_{2}^{\left(\right)}+\tau_{3}^{\left(\right)}\right)&=2\left(\tau_{1}^{\left(12\right)}+\tau_{2}^{\left(12\right)}+\tau_{3}^{\left(12\right)}\right), \\
2\left(\tau_{1}^{\left(\right)}+\tau_{2}^{\left(\right)}+\tau_{3}^{\left(\right)}\right)&=2\left(\tau_{2}^{\left(12\right)}+\tau_{3}^{\left(12\right)}\right), \\
-\ln{\left(1-e^{-\tau_{3}^{\left(\right)}}\left(1-e^{-2\left(\tau_{1}^{\left(\right)}+\tau_{2}^{\left(\right)}\right)}\right)\right)}&=-\ln{\left(1-e^{-\tau_{3}^{\left(12\right)}}\left(1-e^{-2\left(\tau_{1}^{\left(12\right)}+\tau_{2}^{\left(12\right)}\right)}\right)\right)}.
\end{mycases}
\end{split}\end{align}
Making the substitutions, $x_{i}=e^{-\tau_{i}^{\left(\right)}}$ and $y_{i}=e^{-\tau_{i}^{\left(12\right)}}$, we get the polynomial equations,
\begin{align}\begin{split}
\begin{mycases}
x_{2}^{2}x_{3}^{2}&=y_{1}^{2}y_{2}^{2}y_{3}^{2}, \\
x_{1}^{2}x_{2}^{2}x_{3}^{2}&=y_{2}^{2}y_{3}^{2}, \\
1-x_{3}\left(1-x_{1}^{2}x_{2}^{2}\right)&=1-y_{3}\left(1-y_{1}^{2}y_{2}^{2}\right).
\end{mycases}
\end{split}\end{align}
Making the further substitutions, $v_{1}=x_{1}^{2}$, $v_{2}=x_{2}^{2}x_{3}$, $v_{3}=x_{3}$, $w_{1}=y_{1}^{2}$, $w_{2}=y_{2}^{2}y_{3}$, $w_{3}=y_{3}$,
\begin{align}\begin{split}
\begin{mycases}
v_{2}v_{3}&=w_{1}w_{2}w_{3}, \\
v_{1}v_{2}v_{3}&=w_{2}w_{3}, \\
1-v_{3}+v_{1}v_{2}&=1-w_{3}+w_{1}w_{2}.
\end{mycases}
\end{split}\end{align}
Making the further substitutions, $v_{2}v_{3}=\alpha{}$, $v_{1}v_{2}v_{3}=\beta{}$, and $1-v_{3}+v_{1}v_{2}=\gamma{}$,
\begin{align}\begin{split}
\begin{mycases}
\alpha{}&=w_{1}w_{2}w_{3}, \\
\beta{}&=w_{2}w_{3}, \\
\gamma{}&=1-w_{3}+w_{1}w_{2}.
\end{mycases}
\end{split}\end{align}

We will now solve the system for the variables, $w_{1}$, $w_{2}$ and $w_{3}$, in terms of the variables, $\alpha{}$, $\beta{}$ and $\gamma{}$. We can immediately see that dividing the first equation by the second equation gives
\begin{align}\begin{split}
w_{1}=\frac{\alpha{}}{\beta{}}.
\end{split}\end{align}
Now rearranging the second equation to solve for $w_{2}$,
\begin{align}\begin{split}
w_{2}=\frac{\beta{}}{w_{3}}.
\end{split}\end{align}
Substituting in the expressions for $w_{1}$ and $w_{2}$ into the third equation,
\begin{align}\begin{split}
\gamma{}&=1-w_{3}+\frac{\alpha{}}{\beta{}}\frac{\beta{}}{w_{3}} \\
\Leftrightarrow{}\gamma{}&=1-w_{3}+\frac{\alpha{}}{w_{3}} \\
\Leftrightarrow{}-\left(1-\gamma{}\right)&=-w_{3}+\frac{\alpha{}}{w_{3}} \\
\Leftrightarrow{}w_{3}-\left(1-\gamma{}\right)-\frac{\alpha{}}{w_{3}}&=0 \\
\Leftrightarrow{}w_{3}^{2}-\left(1-\gamma{}\right)w_{3}-\alpha{}&=0 \\
\Leftrightarrow{}w_{3}&=\frac{1}{2}\left(\left(1-\gamma{}\right)\pm{}\sqrt{\left(1-\gamma{}\right)^{2}+4\alpha{}}\right).
\end{split}\end{align}

Demanding $w_{3}=y_{3}=e^{-\tau_{3}^{\left(12\right)}}>0$, the only possible solution is 
\begin{align}\begin{split}
w_{3}&=\frac{1}{2}\left(\left(1-\gamma{}\right)+\sqrt{\left(1-\gamma{}\right)^{2}+4\alpha{}}\right).
\end{split}\end{align}

We can now substitute in the expression for $w_{3}$ to find the expression for $w_{2}$,
\begin{align}\begin{split}
w_{2}&=\frac{\beta{}}{w_{3}} \\
&=\frac{\beta{}}{\frac{1}{2}\left(\left(1-\gamma{}\right)+\sqrt{\left(1-\gamma{}\right)^{2}+4\alpha{}}\right)} \\
&=\frac{2\beta{}}{\left(1-\gamma{}\right)+\sqrt{\left(1-\gamma{}\right)^{2}+4\alpha{}}} \\
&=\frac{2\beta{}}{\left(1-\gamma{}\right)+\sqrt{\left(1-\gamma{}\right)^{2}+4\alpha{}}}\cdot{}\frac{-\left(1-\gamma{}\right)+\sqrt{\left(1-\gamma{}\right)^{2}+4\alpha{}}}{-\left(1-\gamma{}\right)+\sqrt{\left(1-\gamma{}\right)^{2}+4\alpha{}}} \\
&=\frac{2\beta{}\left(-\left(1-\gamma{}\right)+\sqrt{\left(1-\gamma{}\right)^{2}+4\alpha{}}\right)}{4\alpha{}} \\
&=\frac{\beta{}}{2\alpha{}}\left(-\left(1-\gamma{}\right)+\sqrt{\left(1-\gamma{}\right)^{2}+4\alpha{}}\right).
\end{split}\end{align}

We now have the expressions for $w_{1}$, $w_{2}$ and $w_{3}$. Recall that
\begin{align}\begin{split}
w_{1}&=y_{1}^{2}=e^{-2\tau_{1}^{\left(12\right)}}
\end{split}\end{align}
and
\begin{align}\begin{split}
w_{1}&=\frac{\alpha{}}{\beta{}}.
\end{split}\end{align}

We can now use the appropriate back-substitutions to get expressions for the time parameters on the convergence-divergence network with the $\left(12\right)$ taxon label permutation in terms of the time parameters on the convergence-divergence network with no taxon label permutation.

The first time parameter will be
\begin{align}\begin{split}
\tau_{1}^{\left(12\right)}&=-\frac{1}{2}\ln{\left(\frac{\alpha{}}{\beta{}}\right)} \\
&=\ln{\sqrt{\frac{\beta{}}{\alpha{}}}} \\
&=\ln{\sqrt{\frac{v_{1}v_{2}v_{3}}{v_{2}v_{3}}}} \\
&=\ln{\sqrt{v_{1}}} \\
&=\ln{\sqrt{x_{1}^{2}}} \\
&=\ln{\left(x_{1}\right)} \\
&=\ln{\left(e^{-\tau_{1}^{\left(\right)}}\right)} \\
&=-\tau_{1}^{\left(\right)}.
\end{split}\end{align}

Recall that
\begin{align}\begin{split}
y_{2}^{2}&=e^{-2\tau_{2}^{\left(12\right)}}.
\end{split}\end{align}
Recall that
\begin{align}\begin{split}
w_{2}&=\frac{\beta{}}{w_{3}}=y_{2}^{2}y_{3}
\end{split}\end{align}
and
\begin{align}\begin{split}
w_{3}&=y_{3}.
\end{split}\end{align}
Consequently,
\begin{align}\begin{split}
y_{2}^{2}&=\frac{w_{2}}{w_{3}} \\
&=\frac{\beta{}}{w_{3}^{2}} \\
&=\frac{4\beta{}}{\left(\left(1-\gamma{}\right)+\sqrt{\left(1-\gamma{}\right)^{2}+4\alpha{}}\right)^{2}}.
\end{split}\end{align}
Again, using the appropriate substitutions,
\begin{align}\begin{split}
\tau_{2}^{\left(12\right)}&=-\frac{1}{2}\ln{\left(\frac{4\beta{}}{\left(\left(1-\gamma{}\right)+\sqrt{\left(1-\gamma{}\right)^{2}+4\alpha{}}\right)^{2}}\right)} \\
&=\ln{\left(\frac{\left(1-\gamma{}\right)+\sqrt{\left(1-\gamma{}\right)^{2}+4\alpha{}}}{2\sqrt{\beta{}}}\right)} \\
&=\ln{\left(\frac{\left(1-\left(1-v_{3}+v_{1}v_{2}\right)\right)+\sqrt{\left(1-\left(1-v_{3}+v_{1}v_{2}\right)\right)^{2}+4v_{2}v_{3}}}{2\sqrt{v_{1}v_{2}v_{3}}}\right)} \\
&=\ln{\left(\frac{\left(v_{3}-v_{1}v_{2}\right)+\sqrt{\left(v_{3}-v_{1}v_{2}\right)^{2}+4v_{2}v_{3}}}{2\sqrt{v_{1}v_{2}v_{3}}}\right)} \\
&=\ln{\left(\frac{\left(x_{3}-x_{1}^{2}x_{2}^{2}x_{3}\right)+\sqrt{\left(x_{3}-x_{1}^{2}x_{2}^{2}x_{3}\right)^{2}+4x_{2}^{2}x_{3}^{2}}}{2\sqrt{x_{1}^{2}x_{2}^{2}x_{3}^{2}}}\right)} \\
&=\ln{\left(\frac{x_{3}\left(1-x_{1}^{2}x_{2}^{2}\right)+\sqrt{x_{3}^{2}\left(1-x_{1}^{2}x_{2}^{2}\right)^{2}+4x_{2}^{2}x_{3}^{2}}}{2x_{1}x_{2}x_{3}}\right)} \\
&=\ln{\left(\frac{x_{3}\left(1-x_{1}^{2}x_{2}^{2}\right)+x_{3}\sqrt{\left(1-x_{1}^{2}x_{2}^{2}\right)^{2}+4x_{2}^{2}}}{2x_{1}x_{2}x_{3}}\right)} \\
&=\ln{\left(\frac{x_{3}\left(\left(1-x_{1}^{2}x_{2}^{2}\right)+\sqrt{\left(1-x_{1}^{2}x_{2}^{2}\right)^{2}+4x_{2}^{2}}\right)}{2x_{1}x_{2}x_{3}}\right)} \\
&=\ln{\left(\frac{\left(1-x_{1}^{2}x_{2}^{2}\right)+\sqrt{\left(1-x_{1}^{2}x_{2}^{2}\right)^{2}+4x_{2}^{2}}}{2x_{1}x_{2}}\right)} \\
&=\ln{\left(\frac{\left(1-e^{-2\left(\tau_{1}^{\left(\right)}+\tau_{2}^{\left(\right)}\right)}\right)+\sqrt{\left(1-e^{-2\left(\tau_{1}^{\left(\right)}+\tau_{2}^{\left(\right)}\right)}\right)^{2}+4e^{-2\tau_{2}^{\left(\right)}}}}{2e^{-\left(\tau_{1}^{\left(\right)}+\tau_{2}^{\left(\right)}\right)}}\right)}.
\end{split}\end{align}

Recall that
\begin{align}\begin{split}
w_{3}&=y_{3}=e^{-\tau_{3}^{\left(12\right)}}
\end{split}\end{align}
and
\begin{align}\begin{split}
w_{3}&=\frac{1}{2}\left(\left(1-\gamma{}\right)+\sqrt{\left(1-\gamma{}\right)^{2}+4\alpha{}}\right).
\end{split}\end{align}
Again, using the appropriate substitutions,
\begin{align}\begin{split}
\tau_{3}^{\left(12\right)}&=-\ln{\left(\frac{1}{2}\left(\left(1-\gamma{}\right)+\sqrt{\left(1-\gamma{}\right)^{2}+4\alpha{}}\right)\right)} \\
&=\ln{\left(\frac{2}{\left(1-\gamma{}\right)+\sqrt{\left(1-\gamma{}\right)^{2}+4\alpha{}}}\right)} \\
&=\ln{\left(\frac{2}{\left(1-\gamma{}\right)+\sqrt{\left(1-\gamma{}\right)^{2}+4\alpha{}}}\cdot{}\frac{-\left(1-\gamma{}\right)+\sqrt{\left(1-\gamma{}\right)^{2}+4\alpha{}}}{-\left(1-\gamma{}\right)+\sqrt{\left(1-\gamma{}\right)^{2}+4\alpha{}}}\right)} \\
&=\ln{\left(\frac{2\left(-\left(1-\gamma{}\right)+\sqrt{\left(1-\gamma{}\right)^{2}+4\alpha{}}\right)}{4\alpha{}}\right)} \\
&=\ln{\left(\frac{-\left(1-\gamma{}\right)+\sqrt{\left(1-\gamma{}\right)^{2}+4\alpha{}}}{2\alpha{}}\right)} \\
&=\ln{\left(\frac{-\left(1-\left(1-v_{3}+v_{1}v_{2}\right)\right)+\sqrt{\left(1-\left(1-v_{3}+v_{1}v_{2}\right)\right)^{2}+4v_{2}v_{3}}}{2v_{2}v_{3}}\right)} \\
&=\ln{\left(\frac{-\left(v_{3}-v_{1}v_{2}\right)+\sqrt{\left(v_{3}-v_{1}v_{2}\right)^{2}+4v_{2}v_{3}}}{2v_{2}v_{3}}\right)} \\
&=\ln{\left(\frac{-\left(x_{3}-x_{1}^{2}x_{2}^{2}x_{3}\right)+\sqrt{\left(x_{3}-x_{1}^{2}x_{2}^{2}x_{3}\right)^{2}+4x_{2}^{2}x_{3}^{2}}}{2x_{2}^{2}x_{3}^{2}}\right)} \\
&=\ln{\left(\frac{-x_{3}\left(1-x_{1}^{2}x_{2}^{2}\right)+\sqrt{x_{3}^{2}\left(1-x_{1}^{2}x_{2}^{2}\right)^{2}+4x_{2}^{2}x_{3}^{2}}}{2x_{2}^{2}x_{3}^{2}}\right)} \\
&=\ln{\left(\frac{-x_{3}\left(1-x_{1}^{2}x_{2}^{2}\right)+x_{3}\sqrt{\left(1-x_{1}^{2}x_{2}^{2}\right)^{2}+4x_{2}^{2}}}{2x_{2}^{2}x_{3}^{2}}\right)} \\
&=\ln{\left(\frac{x_{3}\left(-\left(1-x_{1}^{2}x_{2}^{2}\right)+\sqrt{\left(1-x_{1}^{2}x_{2}^{2}\right)^{2}+4x_{2}^{2}}\right)}{2x_{2}^{2}x_{3}^{2}}\right)} \\
&=\ln{\left(\frac{-\left(1-x_{1}^{2}x_{2}^{2}\right)+\sqrt{\left(1-x_{1}^{2}x_{2}^{2}\right)^{2}+4x_{2}^{2}}}{2x_{2}^{2}x_{3}}\right)} \\
&=\ln{\left(\frac{-\left(1-e^{-2\left(\tau_{1}^{\left(\right)}+\tau_{2}^{\left(\right)}\right)}\right)+\sqrt{\left(1-e^{-2\left(\tau_{1}^{\left(\right)}+\tau_{2}^{\left(\right)}\right)}\right)^{2}+4e^{-2\tau_{2}^{\left(\right)}}}}{2e^{-\left(2\tau_{2}^{\left(\right)}+\tau_{3}^{\left(\right)}\right)}}\right)}.
\end{split}\end{align}

In summary,
\begin{align}\begin{split}
\begin{mycases}
\tau_{1}^{\left(12\right)}&=-\tau_{1}^{\left(\right)}, \\
\tau_{2}^{\left(12\right)}&=\ln{\left(\frac{\left(1-e^{-2\left(\tau_{1}^{\left(\right)}+\tau_{2}^{\left(\right)}\right)}\right)+\sqrt{\left(1-e^{-2\left(\tau_{1}^{\left(\right)}+\tau_{2}^{\left(\right)}\right)}\right)^{2}+4e^{-2\tau_{2}^{\left(\right)}}}}{2e^{-\left(\tau_{1}^{\left(\right)}+\tau_{2}^{\left(\right)}\right)}}\right)}, \\
\tau_{3}^{\left(12\right)}&=\ln{\left(\frac{-\left(1-e^{-2\left(\tau_{1}^{\left(\right)}+\tau_{2}^{\left(\right)}\right)}\right)+\sqrt{\left(1-e^{-2\left(\tau_{1}^{\left(\right)}+\tau_{2}^{\left(\right)}\right)}\right)^{2}+4e^{-2\tau_{2}^{\left(\right)}}}}{2e^{-\left(2\tau_{2}^{\left(\right)}+\tau_{3}^{\left(\right)}\right)}}\right)}. \\
\end{mycases}
\end{split}\end{align}

Recall that $\tau_{2}^{\left(12\right)}$ can be expressed as
\begin{align}\begin{split}
\tau_{2}^{\left(12\right)}&=\ln{\left(\frac{\left(1-x_{1}^{2}x_{2}^{2}\right)+\sqrt{\left(1-x_{1}^{2}x_{2}^{2}\right)^{2}+4x_{2}^{2}}}{2x_{1}x_{2}}\right)}.
\end{split}\end{align}
Recall that $x_{i}=e^{-\tau_{i}^{\left(\right)}}$. Demanding $0<x_{1},x_{2}\leq{}1$,
\begin{align}\begin{split}
\left(1-x_{1}^{2}x_{2}^{2}\right)+\sqrt{\left(1-x_{1}^{2}x_{2}^{2}\right)^{2}+4x_{2}^{2}}&\geq{}\sqrt{\left(1-x_{1}^{2}x_{2}^{2}\right)^{2}+4x_{2}^{2}} \\
&\geq{}\sqrt{4x_{2}^{2}} \\
&=2x_{2} \\
&\geq{}2x_{1}x_{2}.
\end{split}\end{align}
We can conclude that $\tau_{2}^{\left(12\right)}\geq{}0$.

Recall that $\tau_{3}^{\left(12\right)}$ can be expressed as
\begin{align}\begin{split}
\tau_{3}^{\left(12\right)}&=\ln{\left(\frac{-\left(1-x_{1}^{2}x_{2}^{2}\right)+\sqrt{\left(1-x_{1}^{2}x_{2}^{2}\right)^{2}+4x_{2}^{2}}}{2x_{2}^{2}x_{3}}\right)}.
\end{split}\end{align}
Demanding $0<x_{1},x_{2},x_{3}\leq{}1$ and $\tau_{3}^{\left(12\right)}\geq{}0$,
\begin{align}
\begin{split}
-\left(1-x_{1}^{2}x_{2}^{2}\right)+\sqrt{\left(1-x_{1}^{2}x_{2}^{2}\right)^{2}+4x_{2}^{2}}&\geq{}2x_{2}^{2}x_{3} \\
\Leftrightarrow{}\sqrt{\left(1-x_{1}^{2}x_{2}^{2}\right)^{2}+4x_{2}^{2}}&\geq{}\left(1-x_{1}^{2}x_{2}^{2}\right)+2x_{2}^{2}x_{3} \\
\Leftrightarrow{}\left(1-x_{1}^{2}x_{2}^{2}\right)^{2}+4x_{2}^{2}&\geq{}\left(1-x_{1}^{2}x_{2}^{2}\right)^{2}+4x_{2}^{4}x_{3}^{2}+4x_{2}^{2}x_{3}\left(1-x_{1}^{2}x_{2}^{2}\right) \\
\Leftrightarrow{}4x_{2}^{2}&\geq{}4x_{2}^{4}x_{3}^{2}+4x_{2}^{2}x_{3}\left(1-x_{1}^{2}x_{2}^{2}\right) \\
\Leftrightarrow{}1&\geq{}x_{2}^{2}x_{3}^{2}+x_{3}\left(1-x_{1}^{2}x_{2}^{2}\right) \\
\Leftrightarrow{}1-x_{3}\left(1-x_{1}^{2}x_{2}^{2}\right)-x_{2}^{2}x_{3}^{2}&\geq{}0. \\
\Leftrightarrow{}1-x_{3}\left(1-x_{1}^{2}x_{2}^{2}+x_{2}^{2}x_{3}\right)&\geq{}0. \\
\Leftrightarrow{}1-x_{3}\left(1-x_{2}^{2}\left(x_{1}^{2}-x_{3}\right)\right)&\geq{}0.
\end{split}
\label{equation2}
\end{align}
It is not immediately clear whether this inequality is true and $\tau_{3}^{\left(12\right)}\geq{}0$. Suppose $x_{3}<x_{1}^{2}$. Then
\begin{align}\begin{split}
0&<x_{1}^{2}-x_{3} \\
\Leftrightarrow{}0&<x_{2}^{2}\left(x_{1}^{2}-x_{3}\right) \\
\Leftrightarrow{}-x_{2}^{2}\left(x_{1}^{2}-x_{3}\right)&<0 \\
\Leftrightarrow{}1-x_{2}^{2}\left(x_{1}^{2}-x_{3}\right)&<1 \\
\Leftrightarrow{}x_{3}\left(1-x_{2}^{2}\left(x_{1}^{2}-x_{3}\right)\right)&<1 \\
\Leftrightarrow{}-1&<-x_{3}\left(1-x_{2}^{2}\left(x_{1}^{2}-x_{3}\right)\right) \\
\Leftrightarrow{}0&<1-x_{3}\left(1-x_{2}^{2}\left(x_{1}^{2}-x_{3}\right)\right).
\end{split}\end{align}
We can conclude that the inequality will be satisfied when $x_{3}<x_{1}^{2}$. Clearly, when $x_{3}=x_{1}^{2}$ the inequality will be satisfied, since $1-x_{3}>0$. We can conclude that $\tau_{3}^{\left(12\right)}$ will sometimes be non-negative and sometimes be negative.

To get a rough estimate of how often the inequality is satisfied we decided to simulate some examples for $x_{1},x_{2},x_{3}\in{}U\left(0,1\right)$. We found that for $434$ out of $500$ samples $\tau_{3}^{\left(12\right)}\geq{}0$. Of the remaining $66$ out of $500$ samples $\tau_{3}^{\left(12\right)}<0$.

We will now see what happens when $\tau_{1}^{\left(\right)}=0$. When $\tau_{1}^{\left(\right)}=0$, $\tau_{1}^{\left(12\right)}=-\tau_{1}^{\left(\right)}=0$. We have already shown that $\tau_{2}^{\left(12\right)}\geq{}0$ for all possible $\tau_{1}^{\left(\right)}$, $\tau_{2}^{\left(\right)}$ and $\tau_{3}^{\left(\right)}$. Recall that
\begin{align}\begin{split}
\tau_{2}^{\left(12\right)}&=\ln{\left(\frac{\left(1-x_{1}^{2}x_{2}^{2}\right)+\sqrt{\left(1-x_{1}^{2}x_{2}^{2}\right)^{2}+4x_{2}^{2}}}{2x_{1}x_{2}}\right)}.
\end{split}\end{align}
When $\tau_{1}^{\left(\right)}=0$, $\tau_{2}^{\left(12\right)}$ becomes
\begin{align}\begin{split}
\tau_{2}^{\left(12\right)}&=\ln{\left(\frac{\left(1-x_{2}^{2}\right)+\sqrt{\left(1-x_{2}^{2}\right)^{2}+4x_{2}^{2}}}{2x_{2}}\right)} \\
&=\ln{\left(\frac{\left(1-x_{2}^{2}\right)+\sqrt{1-2x_{2}^{2}+x_{2}^{4}+4x_{2}^{2}}}{2x_{2}}\right)} \\
&=\ln{\left(\frac{\left(1-x_{2}^{2}\right)+\sqrt{1+2x_{2}^{2}+x_{2}^{4}}}{2x_{2}}\right)} \\
&=\ln{\left(\frac{\left(1-x_{2}^{2}\right)+\sqrt{\left(1+x_{2}^{2}\right)^{2}}}{2x_{2}}\right)} \\
&=\ln{\left(\frac{\left(1-x_{2}^{2}\right)+\sqrt{1+x_{2}^{2}}}{2x_{2}}\right)} \\
&=\ln{\left(\frac{2}{2x_{2}}\right)} \\
&=\ln{\left(\frac{1}{x_{2}}\right)} \\
&=\ln{\left(\frac{1}{e^{-\tau_{2}^{\left(\right)}}}\right)} \\
&=\ln{\left(e^{\tau_{2}^{\left(\right)}}\right)} \\
&=\tau_{2}^{\left(\right)}.
\end{split}\end{align}
We will now see what happens to $\tau_{3}^{\left(12\right)}$. Recall that
\begin{align}\begin{split}
\tau_{3}^{\left(12\right)}&=\ln{\left(\frac{-\left(1-x_{1}^{2}x_{2}^{2}\right)+\sqrt{\left(1-x_{1}^{2}x_{2}^{2}\right)^{2}+4x_{2}^{2}}}{2x_{2}^{2}x_{3}}\right)}.
\end{split}\end{align}
When $\tau_{1}^{\left(\right)}=0$, $\tau_{3}^{\left(12\right)}$ becomes
\begin{align}\begin{split}
\tau_{3}^{\left(12\right)}&=\ln{\left(\frac{-\left(1-x_{2}^{2}\right)+\sqrt{\left(1-x_{2}^{2}\right)^{2}+4x_{2}^{2}}}{2x_{2}^{2}x_{3}}\right)} \\
&=\ln{\left(\frac{-\left(1-x_{2}^{2}\right)+\sqrt{1-2x_{2}^{2}+x_{2}^{4}+4x_{2}^{2}}}{2x_{2}^{2}x_{3}}\right)} \\
&=\ln{\left(\frac{-\left(1-x_{2}^{2}\right)+\sqrt{1+2x_{2}^{2}+x_{2}^{4}}}{2x_{2}^{2}x_{3}}\right)} \\
&=\ln{\left(\frac{-\left(1-x_{2}^{2}\right)+\sqrt{\left(1+x_{2}^{2}\right)^{2}}}{2x_{2}^{2}x_{3}}\right)} \\
&=\ln{\left(\frac{-\left(1-x_{2}^{2}\right)+\left(1+x_{2}^{2}\right)}{2x_{2}^{2}x_{3}}\right)} \\
&=\ln{\left(\frac{2x_{2}^{2}}{2x_{2}^{2}x_{3}}\right)} \\
&=\ln{\left(\frac{1}{x_{3}}\right)} \\
&=\ln{\left(\frac{1}{e^{-\tau_{3}^{\left(\right)}}}\right)} \\
&=\ln\left({e^{\tau_{3}^{\left(\right)}}}\right) \\
&=\tau_{3}^{\left(\right)} \\
&\geq{}0.
\end{split}\end{align}

\chapter{Time Parameters for the $\left(13\right)$ Taxon Label Permutation}\label{derivation2}

Equating the three pairwise distances for the two taxon label permutations for the three-taxon convergence-divergence network, the $\left(13\right)$ taxon label permutation and no taxon label permutation,
\begin{align}\begin{split}
\begin{mycases}
e^{-2\left(\tau_{1}^{\left(\right)}+\tau_{2}^{\left(\right)}+\tau_{3}^{\left(\right)}\right)}&=e^{-2\left(\tau_{1}^{\left(13\right)}+\tau_{2}^{\left(13\right)}+\tau_{3}^{\left(13\right)}\right)}, \\
e^{-2\left(\tau_{2}^{\left(\right)}+\tau_{3}^{\left(\right)}\right)}&=1-e^{-\tau_{3}^{\left(13\right)}}\left(1-e^{-2\left(\tau_{1}^{\left(13\right)}+\tau_{2}^{\left(13\right)}\right)}\right), \\
1-e^{-\tau_{3}^{\left(\right)}}\left(1-e^{-2\left(\tau_{1}^{\left(\right)}+\tau_{2}^{\left(\right)}\right)}\right)&=e^{-2\left(\tau_{2}^{\left(13\right)}+\tau_{3}^{\left(13\right)}\right)}.
\end{mycases}
\end{split}\end{align}
Making the appropriate substitutions as before,
\begin{align}\begin{split}
\begin{mycases}
x_{1}^{2}x_{2}^{2}x_{3}^{2}&=y_{1}^{2}y_{2}^{2}y_{3}^{2}, \\
x_{2}^{2}x_{3}^{2}&=1-y_{3}\left(1-y_{1}^{2}y_{2}^{2}\right), \\
1-x_{3}\left(1-x_{1}^{2}x_{2}^{2}\right)&=y_{2}^{2}y_{3}^{2}.
\end{mycases}
\end{split}\end{align}
Making the further substitutions,
\begin{align}\begin{split}
\begin{mycases}
v_{1}v_{2}v_{3}&=w_{1}w_{2}w_{3}, \\
v_{2}v_{3}&=1-w_{3}+w_{1}w_{2}, \\
1-v_{3}+v_{1}v_{2}&=w_{2}w_{3}.
\end{mycases}
\end{split}\end{align}
Again, making further substitutions,
\begin{align}\begin{split}
\begin{mycases}
\alpha{}&=w_{1}w_{2}w_{3}, \\
\beta{}&=1-w_{3}+w_{1}w_{2}, \\
\gamma{}&=w_{2}w_{3}.
\end{mycases}
\end{split}\end{align}

We will now solve the system for the variables $w_{1}$, $w_{2}$ and $w_{3}$, in terms of the variables $\alpha{}$, $\beta{}$ and $\gamma{}$. We can immediately see that dividing the first equation by the third equation gives
\begin{align}\begin{split}
w_{1}=\frac{\alpha{}}{\gamma{}}.
\end{split}\end{align}
Now rearranging the second equation to solve for $w_{2}$,
\begin{align}\begin{split}
w_{1}w_{2}&=w_{3}-\left(1-\beta{}\right) \\
w_{2}&=\frac{w_{3}-\left(1-\beta{}\right)}{w_{1}}.
\end{split}\end{align}
Substituting in $w_{1}=\frac{\alpha{}}{\gamma{}}$,
\begin{align}\begin{split}
w_{2}&=\left(w_{3}-\left(1-\beta{}\right)\right)\cdot{}\frac{\gamma{}}{\alpha{}} \\
&=\frac{\gamma{}\left(w_{3}-\left(1-\beta{}\right)\right)}{\alpha{}}.
\end{split}\end{align}
Now rearranging the third equation to solve for $w_{2}$,
\begin{align}\begin{split}
w_{2}&=\frac{\gamma{}}{w_{3}}.
\end{split}\end{align}
Equating the two expressions for $w_{2}$,
\begin{align}\begin{split}
\frac{\gamma{}\left(w_{3}-\left(1-\beta{}\right)\right)}{\alpha{}}&=\frac{\gamma{}}{w_{3}} \\
w_{3}\left(w_{3}-\left(1-\beta{}\right)\right)&=\alpha{} \\
w_{3}^{2}-\left(1-\beta{}\right)w_{3}-\alpha{}&=0 \\
w_{3}&=\frac{1}{2}\left(\left(1-\beta{}\right)\pm{}\sqrt{\left(1-\beta{}\right)^{2}+4\alpha{}}\right).
\end{split}\end{align}

Demanding $w_{3}=y_{3}=e^{-\tau_{3}^{\left(13\right)}}>0$, the only possible solution is 
\begin{align}\begin{split}
w_{3}=\frac{1}{2}\left(\left(1-\beta{}\right)+\sqrt{\left(1-\beta{}\right)^{2}+4\alpha{}}\right).
\end{split}\end{align}

We can now substitute in the expression for $w_{3}$ to find the expression for $w_{2}$,
\begin{align}\begin{split}
w_{2}&=\frac{\gamma{}}{w_{3}} \\
&=\frac{\gamma{}}{\frac{1}{2}\left(\left(1-\beta{}\right)+\sqrt{\left(1-\beta{}\right)^{2}+4\alpha{}}\right)} \\
&=\frac{2\gamma{}}{\left(1-\beta{}\right)+\sqrt{\left(1-\beta{}\right)^{2}+4\alpha{}}} \\
&=\frac{2\gamma{}}{\left(1-\beta{}\right)+\sqrt{\left(1-\beta{}\right)^{2}+4\alpha{}}}\cdot{}\frac{-\left(1-\beta{}\right)+\sqrt{\left(1-\beta{}\right)^{2}+4\alpha{}}}{-\left(1-\beta{}\right)+\sqrt{\left(1-\beta{}\right)^{2}+4\alpha{}}} \\
&=\frac{2\gamma{}\left(-\left(1-\beta{}\right)+\sqrt{\left(1-\beta{}\right)^{2}+4\alpha{}}\right)}{4\alpha{}} \\
&=\frac{\gamma{}}{2\alpha{}}\left(-\left(1-\beta{}\right)+\sqrt{\left(1-\beta{}\right)^{2}+4\alpha{}}\right).
\end{split}\end{align}

We now have the expressions for $w_{1}$, $w_{2}$ and $w_{3}$. Recall that
\begin{align}\begin{split}
w_{1}&=y_{1}^{2}=e^{-2\tau_{1}^{\left(13\right)}}
\end{split}\end{align}
and
\begin{align}\begin{split}
w_{1}=\frac{\alpha{}}{\gamma{}}.
\end{split}\end{align}

We can now use the appropriate back-substitutions to get expressions for the time parameters on the convergence-divergence network with the $\left(23\right)$ taxon label permutation in terms of the time parameters on the convergence-divergence network with no taxon label permutation.

The first time parameter will be
\begin{align}\begin{split}
\tau_{1}^{\left(13\right)}&=-\frac{1}{2}\ln{\left(\frac{\alpha{}}{\gamma{}}\right)} \\
&=\ln{\sqrt{\frac{\gamma{}}{\alpha{}}}} \\
&=\ln{\sqrt{\frac{1-v_{3}+v_{1}v_{2}}{v_{1}v_{2}v_{3}}}} \\
&=\ln{\sqrt{\frac{1-x_{3}\left(1-x_{1}^{2}x_{2}^{2}\right)}{x_{1}^{2}x_{2}^{2}x_{3}^{2}}}} \\
&=\ln{\sqrt{\frac{1-e^{-\tau_{3}^{\left(\right)}}\left(1-e^{-2\left(\tau_{1}^{\left(\right)}+\tau_{2}^{\left(\right)}\right)}\right)}{e^{-2\left(\tau_{1}^{\left(\right)}+\tau_{2}^{\left(\right)}+\tau_{3}^{\left(\right)}\right)}}}}.
\end{split}\end{align}

Recall that
\begin{align}\begin{split}
y_{2}^{2}&=e^{-2\tau_{2}^{\left(13\right)}}.
\end{split}\end{align}
Recall that
\begin{align}\begin{split}
w_{2}&=\frac{\gamma{}}{w_{3}}=y_{2}^{2}y_{3}
\end{split}\end{align}
and
\begin{align}\begin{split}
w_{3}&=y_{3}.
\end{split}\end{align}
Consequently,
\begin{align}\begin{split}
y_{2}^{2}&=\frac{w_{2}}{w_{3}} \\
&=\frac{\gamma{}}{w_{3}^{2}} \\
&=\frac{4\gamma{}}{\left(\left(1-\beta{}\right)+\sqrt{\left(1-\beta{}\right)^{2}+4\alpha{}}\right)^{2}}.
\end{split}\end{align}
Again, using the appropriate substitutions,
\begin{align}\begin{split}
\tau_{2}^{\left(13\right)}&=-\frac{1}{2}\ln{\left(\frac{4\gamma{}}{\left(\left(1-\beta{}\right)+\sqrt{\left(1-\beta{}\right)^{2}+4\alpha{}}\right)^{2}}\right)} \\
&=\ln{\left(\frac{\left(1-\beta{}\right)+\sqrt{\left(1-\beta{}\right)^{2}+4\alpha{}}}{2\sqrt{\gamma{}}}\right)} \\
&=\ln{\left(\frac{\left(1-v_{2}v_{3}\right)+\sqrt{\left(1-v_{2}v_{3}\right)^{2}+4v_{1}v_{2}v_{3}}}{2\sqrt{1-v_{3}+v_{1}v_{2}}}\right)} \\
&=\ln{\left(\frac{\left(1-x_{2}^{2}x_{3}^{2}\right)+\sqrt{\left(1-x_{2}^{2}x_{3}^{2}\right)^{2}+4x_{1}^{2}x_{2}^{2}x_{3}^{2}}}{2\sqrt{1-x_{3}\left(1-x_{1}^{2}x_{2}^{2}\right)}}\right)} \\
&=\ln{\left(\frac{\left(1-e^{-2\left(\tau_{2}^{\left(\right)}+\tau_{3}^{\left(\right)}\right)}\right)+\sqrt{\left(1-e^{-2\left(\tau_{2}^{\left(\right)}+\tau_{3}^{\left(\right)}\right)}\right)^{2}+4e^{-2\left(\tau_{1}^{\left(\right)}+\tau_{2}^{\left(\right)}+\tau_{3}^{\left(\right)}\right)}}}{2\sqrt{1-e^{-\tau_{3}^{\left(\right)}}\left(1-e^{-2\left(\tau_{1}^{\left(\right)}+\tau_{2}^{\left(\right)}\right)}\right)}}\right)}.
\end{split}\end{align}

Recall that
\begin{align}\begin{split}
w_{3}&=y_{3}=e^{-\tau_{3}^{\left(13\right)}}
\end{split}\end{align}
and
\begin{align}\begin{split}
w_{3}&=\frac{1}{2}\left(\left(1-\beta{}\right)+\sqrt{\left(1-\beta{}\right)^{2}+4\alpha{}}\right).
\end{split}\end{align}
Again, using the appropriate substitutions,
\begin{align}\begin{split}
\tau_{3}^{\left(13\right)}&=-\ln{\left(\frac{1}{2}\left(\left(1-\beta{}\right)+\sqrt{\left(1-\beta{}\right)^{2}+4\alpha{}}\right)\right)} \\
&=\ln{\left(\frac{2}{\left(1-\beta{}\right)+\sqrt{\left(1-\beta{}\right)^{2}+4\alpha{}}}\right)} \\
&=\ln{\left(\frac{2}{\left(1-\beta{}\right)+\sqrt{\left(1-\beta{}\right)^{2}+4\alpha{}}}\cdot{}\frac{-\left(1-\beta{}\right)-\sqrt{\left(1-\beta{}\right)^{2}+4\alpha{}}}{-\left(1-\beta{}\right)-\sqrt{\left(1-\beta{}\right)^{2}+4\alpha{}}}\right)} \\
&=\ln{\left(\frac{2\left(-\left(1-\beta{}\right)-\sqrt{\left(1-\beta{}\right)^{2}+4\alpha{}}\right)}{4\alpha{}}\right)} \\
&=\ln{\left(\frac{-\left(1-\beta{}\right)+\sqrt{\left(1-\beta{}\right)^{2}+4\alpha{}}}{2\alpha{}}\right)} \\
&=\ln{\left(\frac{-\left(1-v_{2}v_{3}\right)+\sqrt{\left(1-v_{2}v_{3}\right)^{2}+4v_{1}v_{2}v_{3}}}{2v_{1}v_{2}v_{3}}\right)} \\
&=\ln{\left(\frac{-\left(1-x_{2}^{2}x_{3}^{2}\right)+\sqrt{\left(1-x_{2}^{2}x_{3}^{2}\right)^{2}+4x_{1}^{2}x_{2}^{2}x_{3}^{2}}}{2x_{1}^{2}x_{2}^{2}x_{3}^{2}}\right)} \\
&=\ln{\left(\frac{-\left(1-e^{-2\left(\tau_{2}^{\left(\right)}+\tau_{3}^{\left(\right)}\right)}\right)+\sqrt{\left(1-e^{-2\left(\tau_{2}^{\left(\right)}+\tau_{3}^{\left(\right)}\right)}\right)^{2}+4e^{-2\left(\tau_{1}^{\left(\right)}+\tau_{2}^{\left(\right)}+\tau_{3}^{\left(\right)}\right)}}}{2e^{-2\left(\tau_{1}^{\left(\right)}+\tau_{2}^{\left(\right)}+\tau_{3}^{\left(\right)}\right)}}\right)}.
\end{split}\end{align}

In summary, the time parameters for the $\left(13\right)$ permutation will be
\begin{align}\begin{split}
\begin{mycases}
\tau_{1}^{\left(13\right)}=&\ln{\sqrt{\frac{1-e^{-\tau_{3}^{\left(\right)}}\left(1-e^{-2\left(\tau_{1}^{\left(\right)}+\tau_{2}^{\left(\right)}\right)}\right)}{e^{-2\left(\tau_{1}^{\left(\right)}+\tau_{2}^{\left(\right)}+\tau_{3}^{\left(\right)}\right)}}}}, \\
\tau_{2}^{\left(13\right)}=&\ln{\left(\frac{\left(1-e^{-2\left(\tau_{2}^{\left(\right)}+\tau_{3}^{\left(\right)}\right)}\right)+\sqrt{\left(1-e^{-2\left(\tau_{2}^{\left(\right)}+\tau_{3}^{\left(\right)}\right)}\right)^{2}+4e^{-2\left(\tau_{1}^{\left(\right)}+\tau_{2}^{\left(\right)}+\tau_{3}^{\left(\right)}\right)}}}{2\sqrt{1-e^{-\tau_{3}^{\left(\right)}}\left(1-e^{-2\left(\tau_{1}^{\left(\right)}+\tau_{2}^{\left(\right)}\right)}\right)}}\right)}, \\
\tau_{3}^{\left(13\right)}=&\ln{\left(\frac{-\left(1-e^{-2\left(\tau_{2}^{\left(\right)}+\tau_{3}^{\left(\right)}\right)}\right)+\sqrt{\left(1-e^{-2\left(\tau_{2}^{\left(\right)}+\tau_{3}^{\left(\right)}\right)}\right)^{2}+4e^{-2\left(\tau_{1}^{\left(\right)}+\tau_{2}^{\left(\right)}+\tau_{3}^{\left(\right)}\right)}}}{2e^{-2\left(\tau_{1}^{\left(\right)}+\tau_{2}^{\left(\right)}+\tau_{3}^{\left(\right)}\right)}}\right)}.
\end{mycases}
\end{split}\end{align}

Recall that $\tau_{1}^{\left(13\right)}$ can be expressed as
\begin{align}\begin{split}
\tau_{1}^{\left(13\right)}=&\ln{\sqrt{\frac{1-x_{3}\left(1-x_{1}^{2}x_{2}^{2}\right)}{x_{1}^{2}x_{2}^{2}x_{3}^{2}}}}.
\end{split}\end{align}
Demanding $0<x_{1},x_{2},x_{3}\leq{}1$ and $\tau_{1}^{\left(13\right)}\geq{}0$,
\begin{align}\begin{split}
1-x_{3}\left(1-x_{1}^{2}x_{2}^{2}\right)&\geq{}x_{1}^{2}x_{2}^{2}x_{3}^{2} \\
\Leftrightarrow{}1-x_{3}\left(1-x_{1}^{2}x_{2}^{2}\right)-x_{1}^{2}x_{2}^{2}x_{3}^{2}&\geq{}0 \\
\Leftrightarrow{}1-x_{3}\left(1-x_{1}^{2}x_{2}^{2}+x_{1}^{2}x_{2}^{2}x_{3}\right)&\geq{}0 \\
\Leftrightarrow{}1-x_{3}\left(1-x_{1}^{2}x_{2}^{2}\left(1-x_{3}\right)\right)&\geq{}0.
\end{split}\end{align}
Since $0<x_{1},x_{2},x_{3}\leq{}1$,
\begin{align}
\begin{split}
0<x_{3}&\leq{}1 \\
\Leftrightarrow{}-1\leq{}-x_{3}&<0 \\
\Leftrightarrow{}0\leq{}1-x_{3}&<1 \\
\Leftrightarrow{}0\leq{}x_{1}^{2}x_{2}^{2}\left(1-x_{3}\right)&<1 \\
\Leftrightarrow{}-1<-x_{1}^{2}x_{2}^{2}\left(1-x_{3}\right)&\leq{}0 \\
\Leftrightarrow{}0<1-x_{1}^{2}x_{2}^{2}\left(1-x_{3}\right)&\leq{}1 \\
\Leftrightarrow{}0<x_{3}\left(1-x_{1}^{2}x_{2}^{2}\left(1-x_{3}\right)\right)&\leq{}1 \\
\Leftrightarrow{}-1\leq{}-x_{3}\left(1-x_{1}^{2}x_{2}^{2}\left(1-x_{3}\right)\right)&<0 \\
\Leftrightarrow{}0\leq{}1-x_{3}\left(1-x_{1}^{2}x_{2}^{2}\left(1-x_{3}\right)\right)&<1.
\end{split}
\label{equation1}
\end{align}
We can conclude that $\tau_{1}^{\left(13\right)}\geq{}0$.

Recall that $\tau_{2}^{\left(13\right)}$ can be expressed as
\begin{align}\begin{split}
\tau_{2}^{\left(13\right)}&=\ln{\left(\frac{\left(1-\beta{}\right)+\sqrt{\left(1-\beta{}\right)^{2}+4\alpha{}}}{2\sqrt{\gamma{}}}\right)} \\
&=\ln{\left(\frac{\left(1-\beta{}\right)+\sqrt{\left(1-\beta{}\right)^{2}+4\alpha{}}}{2\sqrt{\gamma{}}}\cdot{}\frac{-\left(1-\beta{}\right)+\sqrt{\left(1-\beta{}\right)^{2}+4\alpha{}}}{-\left(1-\beta{}\right)+\sqrt{\left(1-\beta{}\right)^{2}+4\alpha{}}}\right)} \\
&=\ln{\left(\frac{4\alpha{}}{2\sqrt{\gamma{}}\left(-\left(1-\beta{}\right)+\sqrt{\left(1-\beta{}\right)^{2}+4\alpha{}}\right)}\right)} \\
&=\ln{\left(\frac{2\alpha{}}{\sqrt{\gamma{}}\left(-\left(1-\beta{}\right)+\sqrt{\left(1-\beta{}\right)^{2}+4\alpha{}}\right)}\right)}.
\end{split}\end{align}
Demanding $0<\alpha{},\beta{},\gamma{}\leq{}1$ and $\tau_{2}^{\left(13\right)}\geq{}0$,
\begin{align}\begin{split}
2\alpha{}&\geq{}\sqrt{\gamma{}}\left(-\left(1-\beta{}\right)+\sqrt{\left(1-\beta{}\right)^{2}+4\alpha{}}\right) \\
\Leftrightarrow{}\frac{2\alpha{}}{\sqrt{\gamma{}}}&\geq{}-\left(1-\beta{}\right)+\sqrt{\left(1-\beta{}\right)^{2}+4\alpha{}} \\
\Leftrightarrow{}\frac{2\alpha{}}{\sqrt{\gamma{}}}+\left(1-\beta{}\right)&\geq{}\sqrt{\left(1-\beta{}\right)^{2}+4\alpha{}} \\
\Leftrightarrow{}\frac{4\alpha{}^{2}}{\gamma{}}+\left(1-\beta{}\right)^{2}+\frac{4\alpha{}\left(1-\beta{}\right)}{\sqrt{\gamma{}}}&\geq{}\left(1-\beta{}\right)^{2}+4\alpha{} \\
\Leftrightarrow{}\frac{4\alpha{}^{2}}{\gamma{}}+\frac{4\alpha{}\left(1-\beta{}\right)}{\sqrt{\gamma{}}}&\geq{}4\alpha{} \\
\Leftrightarrow{}\frac{\alpha{}}{\gamma{}}+\frac{\left(1-\beta{}\right)}{\sqrt{\gamma{}}}&\geq{}1 \\
\Leftrightarrow{}\alpha{}+\left(1-\beta{}\right)\sqrt{\gamma{}}&\geq{}\gamma{} \\
\Leftrightarrow{}\left(1-\beta{}\right)\sqrt{\gamma{}}&\geq{}\gamma{}-\alpha{}.
\end{split}\end{align}
Substituting in the expressions for $\alpha{}$, $\beta{}$ and $\gamma{}$ in terms of $x_{1}$, $x_{2}$ and $x_{3}$,
\begin{align}\begin{split}
\left(1-x_{2}^{2}x_{3}^{2}\right)\sqrt{1-x_{3}\left(1-x_{1}^{2}x_{2}^{2}\right)}&\geq{}1-x_{3}\left(1-x_{1}^{2}x_{2}^{2}\right)-x_{1}^{2}x_{2}^{2}x_{3}^{2} \\
\Leftrightarrow{}\left(1-x_{2}^{2}x_{3}^{2}\right)\sqrt{1-x_{3}\left(1-x_{1}^{2}x_{2}^{2}\right)}&\geq{}1-x_{3}+x_{1}^{2}x_{2}^{2}x_{3}-x_{1}^{2}x_{2}^{2}x_{3}^{2} \\
\Leftrightarrow{}\left(1-x_{2}^{2}x_{3}^{2}\right)\sqrt{1-x_{3}\left(1-x_{1}^{2}x_{2}^{2}\right)}&\geq{}1-x_{3}\left(1-x_{1}^{2}x_{2}^{2}+x_{1}^{2}x_{2}^{2}x_{3}\right) \\
\Leftrightarrow{}\left(1-x_{2}^{2}x_{3}^{2}\right)\sqrt{1-x_{3}\left(1-x_{1}^{2}x_{2}^{2}\right)}&\geq{}1-x_{3}\left(1-x_{1}^{2}x_{2}^{2}\left(1-x_{3}\right)\right).
\end{split}\end{align}
Since $0<x_{2},x_{3}\leq{}1$,
\begin{align}\begin{split}
0<x_{2}^{2}x_{3}^{2}&\leq{}1 \\
\Leftrightarrow{}-1\leq{}-x_{2}^{2}x_{3}^{2}&<0 \\
\Leftrightarrow{}0\leq{}1-x_{2}^{2}x_{3}^{2}&<1.
\end{split}\end{align}
Similarly, since $0<x_{1},x_{2},x_{3}\leq{}1$,
\begin{align}\begin{split}
0<x_{1}^{2}x_{2}^{2}&\leq{}1 \\
\Leftrightarrow{}-1\leq{}-x_{1}^{2}x_{2}^{2}&<0 \\
\Leftrightarrow{}0\leq{}1-x_{1}^{2}x_{2}^{2}&<1 \\
\Leftrightarrow{}0\leq{}x_{3}\left(1-x_{1}^{2}x_{2}^{2}\right)&<1 \\
\Leftrightarrow{}-1<-x_{3}\left(1-x_{1}^{2}x_{2}^{2}\right)&\leq{}0 \\
\Leftrightarrow{}0<1-x_{3}\left(1-x_{1}^{2}x_{2}^{2}\right)&\leq{}1 \\
\Leftrightarrow{}0<\sqrt{1-x_{3}\left(1-x_{1}^{2}x_{2}^{2}\right)}&\leq{}1.
\end{split}\end{align}
We can conclude that the left hand side of the inequality must be in the minimum length interval,
\begin{align}\begin{split}
0\leq{}\left(1-x_{2}^{2}x_{3}^{2}\right)\sqrt{1-x_{3}\left(1-x_{1}^{2}x_{2}^{2}\right)}&<1.
\end{split}\end{align}
Similarly, recall from (\ref{equation1})~on~page~\pageref{equation1} that the minimum length interval for the right hand side of the equation will be
\begin{align}\begin{split}
0\leq{}1-x_{3}\left(1-x_{1}^{2}x_{2}^{2}\left(1-x_{3}\right)\right)&<1.
\end{split}\end{align}
It is not immediately clear how $\tau_{2}^{\left(13\right)}$ is constrained. As a simple check we will substitute in some values for $x_{1}$, $x_{2}$ and $x_{3}$. First, suppose $x_{3}=1$. The inequality for $\tau_{2}^{\left(13\right)}\geq{}0$ then becomes
\begin{align}\begin{split}
\left(1-x_{2}^{2}\right)\sqrt{1-\left(1-x_{1}^{2}x_{2}^{2}\right)}&\geq{}1-\left(1-x_{1}^{2}x_{2}^{2}\left(1-x_{3}\right)\right) \\
\left(1-x_{2}^{2}\right)\sqrt{x_{1}^{2}x_{2}^{2}}&\geq{}0 \\
x_{1}x_{2}\left(1-x_{2}^{2}\right)&\geq{}0.
\end{split}\end{align}
Clearly this will be true for all $0<x_{1},x_{2}\leq{}1$. We can conclude that $\tau_{2}^{\left(13\right)}$ can at least sometimes be non-negative. As a second example, suppose $\epsilon{}$ is a ``small'' positive number and $x_{1}=\epsilon{}$, $x_{2}=1$ and $x_{3}=1-\epsilon{}$. The inequality for $\tau_{2}^{\left(13\right)}\geq{}0$ then becomes
\begin{align}\begin{split}
\left(1-\left(1-\epsilon{}\right)^{2}\right)\sqrt{1-\left(1-\epsilon{}\right)\left(1-\epsilon{}^{2}\right)}&\geq{}1-\left(1-\epsilon{}\right)\left(1-\epsilon{}^{2}\left(1-\left(1-\epsilon{}\right)\right)\right) \\
\Leftrightarrow{}\left(1-\left(1-2\epsilon{}+\epsilon{}^{2}\right)\right)\sqrt{1-\left(1-\epsilon{}-\epsilon{}^{2}+\epsilon{}^{3}\right)}&\geq{}1-\left(1-\epsilon{}\right)\left(1-\epsilon{}^{3}\right) \\
\Leftrightarrow{}\left(2\epsilon{}-\epsilon{}^{2}\right)\sqrt{\epsilon{}+\epsilon{}^{2}-\epsilon{}^{3}}&\geq{}1-\left(1-\epsilon{}-\epsilon{}^{3}+\epsilon{}^{4}\right) \\
\Leftrightarrow{}\epsilon{}\left(2-\epsilon{}\right)\sqrt{\epsilon{}\left(1+\epsilon{}-\epsilon{}^{2}\right)}&\geq{}\epsilon{}+\epsilon{}^{3}-\epsilon{}^{4} \\
\Leftrightarrow{}\sqrt{\epsilon{}}\left(2-\epsilon{}\right)\sqrt{1+\epsilon{}-\epsilon{}^{2}}&\geq{}1+\epsilon{}^{2}-\epsilon{}^{3} \\
\Leftrightarrow{}\sqrt{\epsilon{}}\left(2-\epsilon{}\right)\sqrt{1+\epsilon{}\left(1-\epsilon{}\right)}&\geq{}1+\epsilon{}^{2}\left(1-\epsilon{}\right).
\end{split}\end{align}
The left hand side of the inequality must be some ``small'' positive number, while the right hand side of the inequality is greater than $1$. We can conclude that the inequality is not always satisfied and $\tau_{2}^{\left(13\right)}$ can sometimes be non-negative and sometimes be negative.

Again, simulating some examples for $x_{1},x_{2},x_{3}\in{}U\left(0,1\right)$ we found that for $497$ out of $500$ samples $\tau_{3}^{\left(13\right)}\geq{}0$. Of the remaining $3$ out of $500$ samples $\tau_{3}^{\left(13\right)}<0$.

Recall that $\tau_{3}^{\left(13\right)}$ can be expressed as
\begin{align}\begin{split}
\tau_{3}^{\left(13\right)}&=\ln{\left(\frac{-\left(1-x_{2}^{2}x_{3}^{2}\right)+\sqrt{\left(1-x_{2}^{2}x_{3}^{2}\right)^{2}+4x_{1}^{2}x_{2}^{2}x_{3}^{2}}}{2x_{1}^{2}x_{2}^{2}x_{3}^{2}}\right)}.
\end{split}\end{align}
Demanding $0<x_{1},x_{2},x_{3}\leq{}1$ and $\tau_{3}^{\left(13\right)}\geq{}0$,
\begin{align}\begin{split}
-\left(1-x_{2}^{2}x_{3}^{2}\right)+\sqrt{\left(1-x_{2}^{2}x_{3}^{2}\right)^{2}+4x_{1}^{2}x_{2}^{2}x_{3}^{2}}&\geq{}2x_{1}^{2}x_{2}^{2}x_{3}^{2} \\
\Leftrightarrow{}\sqrt{\left(1-x_{2}^{2}x_{3}^{2}\right)^{2}+4x_{1}^{2}x_{2}^{2}x_{3}^{2}}&\geq{}\left(1-x_{2}^{2}x_{3}^{2}\right)+2x_{1}^{2}x_{2}^{2}x_{3}^{2} \\
\Leftrightarrow{}\left(1-x_{2}^{2}x_{3}^{2}\right)^{2}+4x_{1}^{2}x_{2}^{2}x_{3}^{2}&\geq{}\left(1-x_{2}^{2}x_{3}^{2}\right)^{2}+4x_{1}^{4}x_{2}^{4}x_{3}^{4}+4x_{1}^{2}x_{2}^{2}x_{3}^{2}\left(1-x_{2}^{2}x_{3}^{2}\right) \\
\Leftrightarrow{}4x_{1}^{2}x_{2}^{2}x_{3}^{2}&\geq{}4x_{1}^{4}x_{2}^{4}x_{3}^{4}+4x_{1}^{2}x_{2}^{2}x_{3}^{2}\left(1-x_{2}^{2}x_{3}^{2}\right) \\
\Leftrightarrow{}1&\geq{}x_{1}^{2}x_{2}^{2}x_{3}^{2}+\left(1-x_{2}^{2}x_{3}^{2}\right) \\
\Leftrightarrow{}0&\geq{}x_{1}^{2}x_{2}^{2}x_{3}^{2}-x_{2}^{2}x_{3}^{2} \\
\Leftrightarrow{}x_{2}^{2}x_{3}^{2}&\geq{}x_{1}^{2}x_{2}^{2}x_{3}^{2} \\
\Leftrightarrow{}1&\geq{}x_{1}^{2} \\
\Leftrightarrow{}1&\geq{}x_{1}.
\end{split}\end{align}
We can conclude that $\tau_{3}^{\left(13\right)}\geq{}0$.

\chapter{Time Parameters for the $\left(23\right)$ Taxon Label Permutation}\label{derivation3}

Equating the three pairwise distances for the two taxon label permutations for the three-taxon convergence-divergence network, the $\left(23\right)$ taxon label permutation and no taxon label permutation,
\begin{align}\begin{split}
\begin{mycases}
2\left(\tau_{2}^{\left(\right)}+\tau_{3}^{\left(\right)}\right)&=2\left(\tau_{2}^{\left(23\right)}+\tau_{3}^{\left(23\right)}\right), \\
2\left(\tau_{1}^{\left(\right)}+\tau_{2}^{\left(\right)}+\tau_{3}^{\left(\right)}\right)&=-\ln{\left(1-e^{-\tau_{3}^{\left(23\right)}}\left(1-e^{-2\left(\tau_{1}^{\left(23\right)}+\tau_{2}^{\left(23\right)}\right)}\right)\right)}, \\
-\ln{\left(1-e^{-\tau_{3}^{\left(\right)}}\left(1-e^{-2\left(\tau_{1}^{\left(\right)}+\tau_{2}^{\left(\right)}\right)}\right)\right)}&=2\left(\tau_{1}^{\left(23\right)}+\tau_{2}^{\left(23\right)}+\tau_{3}^{\left(23\right)}\right).
\end{mycases}
\end{split}\end{align}
Making the appropriate substitutions as before,
\begin{align}\begin{split}
\begin{mycases}
x_{2}^{2}x_{3}^{2}&=y_{2}^{2}y_{3}^{2}, \\
x_{1}^{2}x_{2}^{2}x_{3}^{2}&=1-y_{3}\left(1-y_{1}^{2}y_{2}^{2}\right), \\
1-x_{3}\left(1-x_{1}^{2}x_{2}^{2}\right)&=y_{1}^{2}y_{2}^{2}y_{3}^{2}.
\end{mycases}
\end{split}\end{align}
Making the further substitutions,
\begin{align}\begin{split}
\begin{mycases}
v_{2}v_{3}&=w_{2}w_{3}, \\
v_{1}v_{2}v_{3}&=1-w_{3}+w_{1}w_{2}, \\
1-v_{3}+v_{1}v_{2}&=w_{1}w_{2}w_{3}.
\end{mycases}
\end{split}\end{align}
Again, making the further substitutions,
\begin{align}\begin{split}
\begin{mycases}
\alpha{}&=w_{2}w_{3}, \\
\beta{}&=1-w_{3}+w_{1}w_{2}, \\
\gamma{}&=w_{1}w_{2}w_{3}.
\end{mycases}
\end{split}\end{align}

We will now solve the system for the variables, $w_{1}$, $w_{2}$ and $w_{3}$, in terms of the variables, $\alpha{}$, $\beta{}$ and $\gamma{}$. We can immediately see that dividing the third equation by the first equation gives
\begin{align}\begin{split}
w_{1}=\frac{\gamma{}}{\alpha{}}.
\end{split}\end{align}
Now rearranging the first equation to solve for $w_{2}$,
\begin{align}\begin{split}
w_{2}=\frac{\alpha{}}{w_{3}}.
\end{split}\end{align}
Substituting in the expressions for $w_{1}$ and $w_{2}$ into the second equation,
\begin{align}\begin{split}
\beta{}&=1-w_{3}+\frac{\gamma{}}{\alpha{}}\frac{\alpha{}}{w_{3}} \\
\Leftrightarrow{}\beta{}&=1-w_{3}+\frac{\gamma{}}{w_{3}} \\
\Leftrightarrow{}-\left(1-\beta{}\right)&=-w_{3}+\frac{\gamma{}}{w_{3}} \\
\Leftrightarrow{}w_{3}-\left(1-\beta{}\right)-\frac{\gamma{}}{w_{3}}&=0 \\
\Leftrightarrow{}w_{3}^{2}-\left(1-\beta{}\right)w_{3}-\gamma{}&=0 \\
\Leftrightarrow{}w_{3}&=\frac{1}{2}\left(\left(1-\beta{}\right)\pm{}\sqrt{\left(1-\beta{}\right)^{2}+4\gamma{}}\right).
\end{split}\end{align}

Demanding $w_{3}=y_{3}=e^{-\tau_{3}^{\left(23\right)}}>0$, the only possible solution is 
\begin{align}\begin{split}
w_{3}&=\frac{1}{2}\left(\left(1-\beta{}\right)+\sqrt{\left(1-\beta{}\right)^{2}+4\gamma{}}\right).
\end{split}\end{align}

We can now substitute in the expression for $w_{3}$ to find the expression for $w_{2}$,
\begin{align}\begin{split}
w_{2}&=\frac{\alpha{}}{w_{3}} \\
&=\frac{\alpha{}}{\frac{1}{2}\left(\left(1-\beta{}\right)+\sqrt{\left(1-\beta{}\right)^{2}+4\gamma{}}\right)} \\
&=\frac{2\alpha{}}{\left(1-\beta{}\right)+\sqrt{\left(1-\beta{}\right)^{2}+4\gamma{}}} \\
&=\frac{2\alpha{}}{\left(1-\beta{}\right)+\sqrt{\left(1-\beta{}\right)^{2}+4\gamma{}}}\cdot{}\frac{-\left(1-\beta{}\right)+\sqrt{\left(1-\beta{}\right)^{2}+4\gamma{}}}{-\left(1-\beta{}\right)+\sqrt{\left(1-\beta{}\right)^{2}+4\gamma{}}} \\
&=\frac{2\alpha{}\left(-\left(1-\beta{}\right)+\sqrt{\left(1-\beta{}\right)^{2}+4\gamma{}}\right)}{4\gamma{}} \\
&=\frac{\alpha{}}{2\gamma{}}\left(-\left(1-\beta{}\right)+\sqrt{\left(1-\beta{}\right)^{2}+4\gamma{}}\right).
\end{split}\end{align}

We now have the expressions for $w_{1}$, $w_{2}$ and $w_{3}$. Recall that
\begin{align}\begin{split}
w_{1}&=y_{1}^{2}=e^{-2\tau_{1}^{\left(23\right)}}
\end{split}\end{align}
and
\begin{align}\begin{split}
w_{1}=\frac{\gamma{}}{\alpha{}}.
\end{split}\end{align}

We can now use the appropriate back-substitutions to get expressions for the time parameters on the convergence-divergence network with the $\left(23\right)$ taxon label permutation in terms of the time parameters on the convergence-divergence network with no taxon label permutation.

The first time parameter will be
\begin{align}\begin{split}
\tau_{1}^{\left(23\right)}&=-\frac{1}{2}\ln{\left(\frac{\gamma{}}{\alpha{}}\right)} \\
&=\ln{\sqrt{\frac{\alpha{}}{\gamma{}}}} \\
&=\ln{\sqrt{\frac{v_{2}v_{3}}{1-v_{3}+v_{1}v_{2}}}} \\
&=\ln{\sqrt{\frac{x_{2}^{2}x_{3}^{2}}{1-x_{3}\left(1-x_{1}^{2}x_{2}^{2}\right)}}} \\
&=\ln{\sqrt{\frac{e^{-2\left(\tau_{2}^{\left(\right)}+\tau_{3}^{\left(\right)}\right)}}{1-e^{-\tau_{3}^{\left(\right)}}\left(1-e^{-2\left(\tau_{1}^{\left(\right)}+\tau_{2}^{\left(\right)}\right)}\right)}}}.
\end{split}\end{align}

Recall that
\begin{align}\begin{split}
y_{2}^{2}&=e^{-2\tau_{2}^{\left(23\right)}}.
\end{split}\end{align}
Recall that
\begin{align}\begin{split}
w_{2}&=\frac{\alpha{}}{w_{3}}=y_{2}^{2}y_{3}
\end{split}\end{align}
and
\begin{align}\begin{split}
w_{3}&=y_{3}.
\end{split}\end{align}
Consequently,
\begin{align}\begin{split}
y_{2}^{2}&=\frac{w_{2}}{w_{3}} \\
&=\frac{\alpha{}}{w_{3}^{2}} \\
&=\frac{4\alpha{}}{\left(\left(1-\beta{}\right)+\sqrt{\left(1-\beta{}\right)^{2}+4\gamma{}}\right)^{2}}.
\end{split}\end{align}
Again, using the appropriate substitutions,
\begin{align}\begin{split}
\tau_{2}^{\left(23\right)}&=-\frac{1}{2}\ln{\left(\frac{4\alpha{}}{\left(\left(1-\beta{}\right)+\sqrt{\left(1-\beta{}\right)^{2}+4\gamma{}}\right)^{2}}\right)} \\
&=\ln{\left(\frac{\left(1-\beta{}\right)+\sqrt{\left(1-\beta{}\right)^{2}+4\gamma{}}}{2\sqrt{\alpha{}}}\right)} \\
&=\ln{\left(\frac{\left(1-v_{1}v_{2}v_{3}\right)+\sqrt{\left(1-v_{1}v_{2}v_{3}\right)^{2}+4\left(1-v_{3}+v_{1}v_{2}\right)}}{2\sqrt{v_{2}v_{3}}}\right)} \\
&=\ln{\left(\frac{\left(1-x_{1}^{2}x_{2}^{2}x_{3}^{2}\right)+\sqrt{\left(1-x_{1}^{2}x_{2}^{2}x_{3}^{2}\right)^{2}+4\left(1-x_{3}\left(1-x_{1}^{2}x_{2}^{2}\right)\right)}}{2\sqrt{x_{2}^{2}x_{3}^{2}}}\right)} \\
&=\ln{\left(\frac{\left(1-x_{1}^{2}x_{2}^{2}x_{3}^{2}\right)+\sqrt{\left(1-x_{1}^{2}x_{2}^{2}x_{3}^{2}\right)^{2}+4\left(1-x_{3}\left(1-x_{1}^{2}x_{2}^{2}\right)\right)}}{2x_{2}x_{3}}\right)} \\
&=\scriptstyle{\ln{\left(\frac{\left(1-e^{-2\left(\tau_{1}^{\left(\right)}+\tau_{2}^{\left(\right)}+\tau_{3}^{\left(\right)}\right)}\right)+\sqrt{\left(1-e^{-2\left(\tau_{1}^{\left(\right)}+\tau_{2}^{\left(\right)}+\tau_{3}^{\left(\right)}\right)}\right)^{2}+4\left(1-e^{-\tau_{3}^{\left(\right)}}\left(1-e^{-2\left(\tau_{1}^{\left(\right)}+\tau_{2}^{\left(\right)}\right)}\right)\right)}}{2e^{-\left(\tau_{2}^{\left(\right)}+\tau_{3}^{\left(\right)}\right)}}\right)}}.
\end{split}\end{align}

Recall that
\begin{align}\begin{split}
w_{3}&=y_{3}=e^{-\tau_{3}^{\left(23\right)}}
\end{split}\end{align}
and
\begin{align}\begin{split}
w_{3}&=\frac{1}{2}\left(\left(1-\beta{}\right)+\sqrt{\left(1-\beta{}\right)^{2}+4\gamma{}}\right).
\end{split}\end{align}
Again, using the appropriate substitutions,
\begin{align}\begin{split}
\tau_{3}^{\left(23\right)}&=-\ln{\frac{1}{2}\left(\left(1-\beta{}\right)+\sqrt{\left(1-\beta{}\right)^{2}+4\gamma{}}\right)} \\
&=\ln{\left(\frac{2}{\left(1-\beta{}\right)+\sqrt{\left(1-\beta{}\right)^{2}+4\gamma{}}}\right)} \\
&=\ln{\left(\frac{2}{\left(1-\beta{}\right)+\sqrt{\left(1-\beta{}\right)^{2}+4\gamma{}}}\cdot{}\frac{-\left(1-\beta{}\right)+\sqrt{\left(1-\beta{}\right)^{2}+4\gamma{}}}{-\left(1-\beta{}\right)+\sqrt{\left(1-\beta{}\right)^{2}+4\gamma{}}}\right)} \\
&=\ln{\left(\frac{-2\left(1-\beta{}\right)+\sqrt{\left(1-\beta{}\right)^{2}+4\gamma{}}}{4\gamma{}}\right)} \\
&=\ln{\left(\frac{-\left(1-\beta{}\right)+\sqrt{\left(1-\beta{}\right)^{2}+4\gamma{}}}{2\gamma{}}\right)} \\
&=\ln{\left(\frac{-\left(1-v_{1}v_{2}v_{3}\right)+\sqrt{\left(1-v_{1}v_{2}v_{3}\right)^{2}+4\left(1-v_{3}+v_{1}v_{2}\right)}}{2\left(1-v_{3}+v_{1}v_{2}\right)}\right)} \\
&=\ln{\left(\frac{-\left(1-x_{1}^{2}x_{2}^{2}x_{3}^{2}\right)+\sqrt{\left(1-x_{1}^{2}x_{2}^{2}x_{3}^{2}\right)^{2}+4\left(1-x_{3}\left(1-x_{1}^{2}x_{2}^{2}\right)\right)}}{2\left(1-x_{3}\left(1-x_{1}^{2}x_{2}^{2}\right)\right)}\right)} \\
&=\scriptstyle{\ln{\left(\frac{-\left(1-e^{-2\left(\tau_{1}^{\left(\right)}+\tau_{2}^{\left(\right)}+\tau_{3}^{\left(\right)}\right)}\right)+\sqrt{\left(1-e^{-2\left(\tau_{1}^{\left(\right)}+\tau_{2}^{\left(\right)}+\tau_{3}^{\left(\right)}\right)}\right)^{2}+4\left(1-e^{-\tau_{3}^{\left(\right)}}\left(1-e^{-2\left(\tau_{1}^{\left(\right)}+\tau_{2}^{\left(\right)}\right)}\right)\right)}}{2\left(1-e^{-\tau_{3}^{\left(\right)}}\left(1-e^{-2\left(\tau_{1}^{\left(\right)}+\tau_{2}^{\left(\right)}\right)}\right)\right)}\right)}}.
\end{split}\end{align}

In summary,
\begin{align}\begin{split}
\begin{mycases}
\tau_{1}^{\left(23\right)}&=\ln{\sqrt{\frac{e^{-2\left(\tau_{2}^{\left(\right)}+\tau_{3}^{\left(\right)}\right)}}{1-e^{-\tau_{3}^{\left(\right)}}\left(1-e^{-2\left(\tau_{1}^{\left(\right)}+\tau_{2}^{\left(\right)}\right)}\right)}}}, \\
\tau_{2}^{\left(23\right)}&=\scriptstyle{\ln{\left(\frac{\left(1-e^{-2\left(\tau_{1}^{\left(\right)}+\tau_{2}^{\left(\right)}+\tau_{3}^{\left(\right)}\right)}\right)+\sqrt{\left(1-e^{-2\left(\tau_{1}^{\left(\right)}+\tau_{2}^{\left(\right)}+\tau_{3}^{\left(\right)}\right)}\right)^{2}+4\left(1-e^{-\tau_{3}^{\left(\right)}}\left(1-e^{-2\left(\tau_{1}^{\left(\right)}+\tau_{2}^{\left(\right)}\right)}\right)\right)}}{2e^{-\left(\tau_{2}^{\left(\right)}+\tau_{3}^{\left(\right)}\right)}}\right)}}, \\
\tau_{3}^{\left(23\right)}&=\scriptstyle{\ln{\left(\frac{-\left(1-e^{-2\left(\tau_{1}^{\left(\right)}+\tau_{2}^{\left(\right)}+\tau_{3}^{\left(\right)}\right)}\right)+\sqrt{\left(1-e^{-2\left(\tau_{1}^{\left(\right)}+\tau_{2}^{\left(\right)}+\tau_{3}^{\left(\right)}\right)}\right)^{2}+4\left(1-e^{-\tau_{3}^{\left(\right)}}\left(1-e^{-2\left(\tau_{1}^{\left(\right)}+\tau_{2}^{\left(\right)}\right)}\right)\right)}}{2\left(1-e^{-\tau_{3}^{\left(\right)}}\left(1-e^{-2\left(\tau_{1}^{\left(\right)}+\tau_{2}^{\left(\right)}\right)}\right)\right)}\right)}}. \\
\end{mycases}
\end{split}\end{align}

Recall that $\tau_{1}^{\left(23\right)}$ can be expressed as
\begin{align}\begin{split}
\ln{\sqrt{\frac{x_{2}^{2}x_{3}^{2}}{1-x_{3}\left(1-x_{1}^{2}x_{2}^{2}\right)}}}.
\end{split}\end{align}
Demanding $0<x_{1},x_{2},x_{3}\leq{}1$ and $\tau_{1}^{\left(23\right)}\geq{}0$,
\begin{align}\begin{split}
x_{2}^{2}x_{3}^{2}&\geq{}1-x_{3}\left(1-x_{1}^{2}x_{2}^{2}\right) \\
\Leftrightarrow{}x_{2}^{2}x_{3}^{2}&\geq{}1-x_{3}+x_{1}^{2}x_{2}^{2}x_{3} \\
\Leftrightarrow{}0&\geq{}1-x_{3}+x_{1}^{2}x_{2}^{2}x_{3}-x_{2}^{2}x_{3}^{2} \\
\Leftrightarrow{}0&\geq{}1-x_{3}\left(1-x_{1}^{2}x_{2}^{2}+x_{2}^{2}x_{3}\right) \\
\Leftrightarrow{}0&\geq{}1-x_{3}\left(1-x_{2}^{2}\left(x_{1}^{2}-x_{3}\right)\right).
\end{split}\end{align}
This is the same inequality as in (\ref{equation2})~on~page~\pageref{equation2}, except with the direction of the inequality swapped. We can conclude that the inequality will not be satisfied when $x_{3}<x_{1}^{2}$. Clearly, when $x_{3}=x_{1}^{2}$ the inequality will be satisfied, since $1-x_{3}>0$. We can conclude that $\tau_{1}^{\left(23\right)}$ can sometimes be non-negative and sometimes be negative.

To get a rough estimate of how often the inequality is satisfied we decided to simulate some samples for $x_{1},x_{2},x_{3}\in{}U\left(0,1\right)$. We found that for $52$ out of $500$ samples $\tau_{1}^{\left(23\right)}\geq{}0$. Of the remaining $448$ out of $500$ samples $\tau_{1}^{\left(23\right)}<0$.

Recall that
\begin{align}\begin{split}
\beta{}&=x_{1}^{2}x_{2}^{2}x_{3}^{2}
\end{split}\end{align}
and
\begin{align}\begin{split}
\gamma{}&=1-x_{3}\left(1-x_{1}^{2}x_{2}^{2}\right).
\end{split}\end{align}
To determine whether $\tau_{2}^{\left(23\right)}\geq{}0$ we first will see whether $\beta{}\leq{}\gamma{}$ is true. This is the statement,
\begin{align}\begin{split}
x_{1}^{2}x_{2}^{2}x_{3}^{2}&\leq{}1-x_{3}\left(1-x_{1}^{2}x_{2}^{2}\right) \\
\Leftrightarrow{}0&\leq{}1-x_{3}+x_{1}^{2}x_{2}^{2}-x_{1}^{2}x_{2}^{2}x_{3}^{2} \\
\Leftrightarrow{}0&\leq{}1-x_{3}\left(1-x_{1}^{2}x_{2}^{2}+x_{1}^{2}x_{2}^{2}x_{3}\right) \\
\Leftrightarrow{}0&\leq{}1-x_{3}\left(1-x_{1}^{2}x_{2}^{2}\left(1-x_{3}\right)\right).
\end{split}\end{align}
From (\ref{equation1})~on~page~\pageref{equation1} we can conclude that $\beta{}\leq{}\gamma{}$. Recall that
\begin{align}\begin{split}
\tau_{2}^{\left(23\right)}&=\ln{\left(\frac{\left(1-\beta{}\right)+\sqrt{\left(1-\beta{}\right)^{2}+4\gamma{}}}{2\sqrt{\alpha{}}}\right)}.
\end{split}\end{align}
Demanding $0<\alpha{},\beta{},\gamma{}\leq{}1$ and $\tau_{2}^{\left(23\right)}\geq{}0$,
\begin{align}\begin{split}
\left(1-\beta{}\right)+\sqrt{\left(1-\beta{}\right)^{2}+4\gamma{}}&\geq{}2\sqrt{\alpha{}}.
\end{split}\end{align}
Since $\beta{}\leq{}\gamma{}$,
\begin{align}\begin{split}
-\beta{}&\geq{}-\gamma{} \\
1-\beta{}&\leq{}1-\gamma{}.
\end{split}\end{align}
We can conclude that
\begin{align}\begin{split}
\left(1-\beta{}\right)+\sqrt{\left(1-\beta{}\right)^{2}+4\gamma{}}&\geq{}\left(1-\gamma{}\right)+\sqrt{\left(1-\gamma{}\right)^{2}+4\gamma{}} \\
&=\left(1-\gamma{}\right)+\sqrt{1-2\gamma{}+\gamma{}^{2}+4\gamma{}} \\
&=\left(1-\gamma{}\right)+\sqrt{1+2\gamma{}+\gamma{}^{2}} \\
&=\left(1-\gamma{}\right)+\sqrt{\left(1+\gamma{}\right)^{2}} \\
&=\left(1-\gamma{}\right)+\left(1+\gamma{}\right) \\
&=2 \\
&\geq{}2\sqrt{\alpha{}}.
\end{split}\end{align}
We can conclude that $\tau_{2}^{\left(23\right)}\geq{}0$.

Recall that $\tau_{3}^{\left(23\right)}$ can be expressed as
\begin{align}\begin{split}
\tau_{3}^{\left(23\right)}&=\ln{\left(\frac{-\left(1-x_{1}^{2}x_{2}^{2}x_{3}^{2}\right)+\sqrt{\left(1-x_{1}^{2}x_{2}^{2}x_{3}^{2}\right)^{2}+4\left(1-x_{3}\left(1-x_{1}^{2}x_{2}^{2}\right)\right)}}{2\left(1-x_{3}\left(1-x_{1}^{2}x_{2}^{2}\right)\right)}\right)}.
\end{split}\end{align}
Demanding $0<x_{1},x_{2},x_{3}\leq{}1$ and $\tau_{3}^{\left(23\right)}\geq{}0$,
\begin{align}\begin{split}
-\left(1-x_{1}^{2}x_{2}^{2}x_{3}^{2}\right)+\sqrt{\left(1-x_{1}^{2}x_{2}^{2}x_{3}^{2}\right)^{2}+4\left(1-x_{3}\left(1-x_{1}^{2}x_{2}^{2}\right)\right)}&\geq{}2\left(1-x_{3}\left(1-x_{1}^{2}x_{2}^{2}\right)\right) \\
\Leftrightarrow{}\sqrt{\left(1-x_{1}^{2}x_{2}^{2}x_{3}^{2}\right)^{2}+4\left(1-x_{3}\left(1-x_{1}^{2}x_{2}^{2}\right)\right)}&\geq{}\left(1-x_{1}^{2}x_{2}^{2}x_{3}^{2}\right) \\
&+2\left(1-x_{3}\left(1-x_{1}^{2}x_{2}^{2}\right)\right) \\
\Leftrightarrow{}\left(1-x_{1}^{2}x_{2}^{2}x_{3}^{2}\right)^{2}+4\left(1-x_{3}\left(1-x_{1}^{2}x_{2}^{2}\right)\right)&\geq{}\left(1-x_{1}^{2}x_{2}^{2}x_{3}^{2}\right)^{2} \\
&+4\left(1-x_{3}\left(1-x_{1}^{2}x_{2}^{2}\right)\right)^{2} \\
&+4\left(1-x_{1}^{2}x_{2}^{2}x_{3}^{2}\right)\left(1-x_{3}\left(1-x_{1}^{2}x_{2}^{2}\right)\right) \\
\Leftrightarrow{}4\left(1-x_{3}\left(1-x_{1}^{2}x_{2}^{2}\right)\right)&\geq{}4\left(1-x_{3}\left(1-x_{1}^{2}x_{2}^{2}\right)\right)^{2} \\
&+4\left(1-x_{1}^{2}x_{2}^{2}x_{3}^{2}\right)\left(1-x_{3}\left(1-x_{1}^{2}x_{2}^{2}\right)\right) \\
\Leftrightarrow{}1&\geq{}\left(1-x_{3}\left(1-x_{1}^{2}x_{2}^{2}\right)\right) \\
&+\left(1-x_{1}^{2}x_{2}^{2}x_{3}^{2}\right) \\
\Leftrightarrow{}0&\geq{}-x_{3}+x_{1}^{2}x_{2}^{2}x_{3}+1-x_{1}^{2}x_{2}^{2}x_{3}^{2} \\
\Leftrightarrow{}0&\geq{}1-x_{3}\left(1-x_{1}^{2}x_{2}^{2}+x_{1}^{2}x_{2}^{2}x_{3}\right) \\
\Leftrightarrow{}0&\geq{}1-x_{3}\left(1-x_{1}^{2}x_{2}^{2}\left(1-x_{3}\right)\right).
\end{split}\end{align}
This is the same inequality as in (\ref{equation1})~on~page~\pageref{equation1}, except with the direction of the inequality swapped. We can conclude that $\tau_{3}^{\left(23\right)}\leq{}0$.

\chapter{The $\left(23\right)$ Permutation in Terms of the $\left(12\right)$ Permutation}\label{derivation4}

Equating the three pairwise distances for the two taxon label permutations for the three-taxon convergence-divergence network, the $\left(23\right)$ taxon label permutation and the $\left(12\right)$ taxon label permutation,
\begin{align}\begin{split}
\begin{mycases}
2\left(\tau_{1}^{\left(12\right)}+\tau_{2}^{\left(12\right)}+\tau_{3}^{\left(12\right)}\right)&=2\left(\tau_{2}^{\left(23\right)}+\tau_{3}^{\left(23\right)}\right), \\
2\left(\tau_{2}^{\left(12\right)}+\tau_{3}^{\left(12\right)}\right)&=-\ln{\left(1-e^{-\tau_{3}^{\left(23\right)}}\left(1-e^{-2\left(\tau_{1}^{\left(23\right)}+\tau_{2}^{\left(23\right)}\right)}\right)\right)}, \\
-\ln{\left(1-e^{-\tau_{3}^{\left(12\right)}}\left(1-e^{-2\left(\tau_{1}^{\left(12\right)}+\tau_{2}^{\left(12\right)}\right)}\right)\right)}&=2\left(\tau_{1}^{\left(23\right)}+\tau_{2}^{\left(23\right)}+\tau_{3}^{\left(23\right)}\right).
\end{mycases}
\end{split}\end{align}
Making the appropriate substitutions as before,
\begin{align}\begin{split}
\begin{mycases}
x_{1}^{2}x_{2}^{2}x_{3}^{2}&=y_{2}^{2}y_{3}^{2}, \\
x_{2}^{2}x_{3}^{2}&=1-y_{3}\left(1-y_{1}^{2}y_{2}^{2}\right), \\
1-x_{3}\left(1-x_{1}^{2}x_{2}^{2}\right)&=y_{1}^{2}y_{2}^{2}y_{3}^{2}.
\end{mycases}
\end{split}\end{align}
Making the further substitutions,
\begin{align}\begin{split}
\begin{mycases}
v_{1}v_{2}v_{3}&=w_{2}w_{3}, \\
v_{2}v_{3}&=1-w_{3}+w_{1}w_{2}, \\
1-v_{3}+v_{1}v_{2}&=w_{1}w_{2}w_{3}.
\end{mycases}
\end{split}\end{align}
Again, making the further substitutions,
\begin{align}\begin{split}
\begin{mycases}
\alpha{}&=w_{2}w_{3}, \\
\beta{}&=1-w_{3}+w_{1}w_{2}, \\
\gamma{}&=w_{1}w_{2}w_{3}.
\end{mycases}
\end{split}\end{align}

We will now solve the system for the variables, $w_{1}$, $w_{2}$ and $w_{3}$, in terms of the variables, $\alpha{}$, $\beta{}$ and $\gamma{}$. We can immediately see that dividing the third equation by the first equation gives
\begin{align}\begin{split}
w_{1}=\frac{\gamma{}}{\alpha{}}.
\end{split}\end{align}
Now rearranging the first equation to solve for $w_{2}$,
\begin{align}\begin{split}
w_{2}=\frac{\alpha{}}{w_{3}}.
\end{split}\end{align}
Substituting in the expressions for $w_{1}$ and $w_{2}$ into the second equation,
\begin{align}\begin{split}
\beta{}&=1-w_{3}+\frac{\gamma{}}{\alpha{}}\frac{\alpha{}}{w_{3}} \\
\Leftrightarrow{}\beta{}&=1-w_{3}+\frac{\gamma{}}{w_{3}} \\
\Leftrightarrow{}-\left(1-\beta{}\right)&=-w_{3}+\frac{\gamma{}}{w_{3}} \\
\Leftrightarrow{}w_{3}-\left(1-\beta{}\right)-\frac{\gamma{}}{w_{3}}&=0 \\
\Leftrightarrow{}w_{3}^{2}-\left(1-\beta{}\right)w_{3}-\gamma{}&=0 \\
\Leftrightarrow{}w_{3}&=\frac{1}{2}\left(\left(1-\beta{}\right)\pm{}\sqrt{\left(1-\beta{}\right)^{2}+4\gamma{}}\right).
\end{split}\end{align}

Demanding $w_{3}=y_{3}=e^{-\tau_{3}^{\left(23\right)}}>0$, the only possible solution is 
\begin{align}\begin{split}
w_{3}&=\frac{1}{2}\left(\left(1-\beta{}\right)+\sqrt{\left(1-\beta{}\right)^{2}+4\gamma{}}\right).
\end{split}\end{align}

We can now substitute in the expression for $w_{3}$ to find the expression for $w_{2}$,
\begin{align}\begin{split}
w_{2}&=\frac{\alpha{}}{w_{3}} \\
&=\frac{\alpha{}}{\frac{1}{2}\left(\left(1-\beta{}\right)+\sqrt{\left(1-\beta{}\right)^{2}+4\gamma{}}\right)} \\
&=\frac{2\alpha{}}{\left(1-\beta{}\right)+\sqrt{\left(1-\beta{}\right)^{2}+4\gamma{}}} \\
&=\frac{2\alpha{}}{\left(1-\beta{}\right)+\sqrt{\left(1-\beta{}\right)^{2}+4\gamma{}}}\cdot{}\frac{-\left(1-\beta{}\right)+\sqrt{\left(1-\beta{}\right)^{2}+4\gamma{}}}{-\left(1-\beta{}\right)+\sqrt{\left(1-\beta{}\right)^{2}+4\gamma{}}} \\
&=\frac{2\alpha{}\left(-\left(1-\beta{}\right)+\sqrt{\left(1-\beta{}\right)^{2}+4\gamma{}}\right)}{4\gamma{}} \\
&=\frac{\alpha{}}{2\gamma{}}\left(-\left(1-\beta{}\right)+\sqrt{\left(1-\beta{}\right)^{2}+4\gamma{}}\right).
\end{split}\end{align}

We now have the expressions for $w_{1}$, $w_{2}$ and $w_{3}$. Recall that
\begin{align}\begin{split}
w_{1}&=y_{1}^{2}=e^{-2\tau_{1}^{\left(23\right)}}
\end{split}\end{align}
and
\begin{align}\begin{split}
w_{1}&=\frac{\gamma{}}{\alpha{}}.
\end{split}\end{align}

We can now use the appropriate back-substitutions to get expressions for the time parameters on the convergence-divergence network with the $\left(23\right)$ taxon label permutation in terms of the time parameters on the convergence-divergence network with the $\left(12\right)$ taxon label permutation.

The first time parameter will be
\begin{align}\begin{split}
\tau_{1}^{\left(23\right)}&=-\frac{1}{2}\ln{\left(\frac{\gamma{}}{\alpha{}}\right)} \\
&=\ln{\sqrt{\frac{\alpha{}}{\gamma{}}}} \\
&=\ln{\sqrt{\frac{v_{1}v_{2}v_{3}}{1-v_{3}+v_{1}v_{2}}}} \\
&=\ln{\sqrt{\frac{x_{1}^{2}x_{2}^{2}x_{3}^{2}}{1-x_{3}\left(1-x_{1}^{2}x_{2}^{2}\right)}}} \\
&=\ln{\sqrt{\frac{e^{-2\left(\tau_{1}^{\left(12\right)}+\tau_{2}^{\left(12\right)}+\tau_{3}^{\left(12\right)}\right)}}{1-e^{-\tau_{3}^{\left(12\right)}}\left(1-e^{-2\left(\tau_{1}^{\left(12\right)}+\tau_{2}^{\left(12\right)}\right)}\right)}}} \\
\end{split}\end{align}

Recall that
\begin{align}\begin{split}
y_{2}^{2}&=e^{-2\tau_{2}^{\left(23\right)}}.
\end{split}\end{align}
Recall that
\begin{align}\begin{split}
w_{2}&=\frac{\alpha{}}{w_{3}}=y_{2}^{2}y_{3}
\end{split}\end{align}
and
\begin{align}\begin{split}
w_{3}&=y_{3}.
\end{split}\end{align}
Consequently,
\begin{align}\begin{split}
y_{2}^{2}&=\frac{w_{2}}{w_{3}} \\
&=\frac{\alpha{}}{w_{3}^{2}} \\
&=\frac{4\alpha{}}{\left(\left(1-\beta{}\right)+\sqrt{\left(1-\beta{}\right)^{2}+4\gamma{}}\right)^{2}}.
\end{split}\end{align}
Again, using the appropriate substitutions,
\begin{align}\begin{split}
\tau_{2}^{\left(23\right)}&=-\frac{1}{2}\ln{\left(\frac{4\alpha{}}{\left(\left(1-\beta{}\right)+\sqrt{\left(1-\beta{}\right)^{2}+4\gamma{}}\right)^{2}}\right)} \\
&=\ln{\left(\frac{\left(1-\beta{}\right)+\sqrt{\left(1-\beta{}\right)^{2}+4\gamma{}}}{2\sqrt{\alpha{}}}\right)} \\
&=\ln{\left(\frac{\left(1-v_{2}v_{3}\right)+\sqrt{\left(1-v_{2}v_{3}\right)^{2}+4\left(1-v_{3}+v_{1}v_{2}\right)}}{2\sqrt{v_{1}v_{2}v_{3}}}\right)} \\
&=\ln{\left(\frac{\left(1-x_{2}^{2}x_{3}^{2}\right)+\sqrt{\left(1-x_{2}^{2}x_{3}^{2}\right)^{2}+4\left(1-x_{3}\left(1-x_{1}^{2}x_{2}^{2}\right)\right)}}{2\sqrt{x_{1}^{2}x_{2}^{2}x_{3}^{2}}}\right)} \\
&=\ln{\left(\frac{\left(1-x_{2}^{2}x_{3}^{2}\right)+\sqrt{\left(1-x_{2}^{2}x_{3}^{2}\right)^{2}+4\left(1-x_{3}\left(1-x_{1}^{2}x_{2}^{2}\right)\right)}}{2x_{1}x_{2}x_{3}}\right)} \\
&=\scriptstyle{\ln{\left(\frac{\left(1-e^{-2\left(\tau_{2}^{\left(12\right)}+\tau_{3}^{\left(12\right)}\right)}\right)+\sqrt{\left(1-e^{-2\left(\tau_{2}^{\left(12\right)}+\tau_{3}^{\left(12\right)}\right)}\right)^{2}+4\left(1-e^{-\tau_{3}^{\left(12\right)}}\left(1-e^{-2\left(\tau_{1}^{\left(12\right)}+\tau_{2}^{\left(12\right)}\right)}\right)\right)}}{2e^{-\left(\tau_{1}^{\left(12\right)}+\tau_{2}^{\left(12\right)}+\tau_{3}^{\left(12\right)}\right)}}\right)}}.
\end{split}\end{align}

Recall that
\begin{align}\begin{split}
w_{3}&=y_{3}=e^{-\tau_{3}^{\left(23\right)}}
\end{split}\end{align}
and
\begin{align}\begin{split}
w_{3}&=\frac{1}{2}\left(\left(1-\beta{}\right)+\sqrt{\left(1-\beta{}\right)^{2}+4\gamma{}}\right).
\end{split}\end{align}

Again, using the appropriate substitutions,
\begin{align}\begin{split}
\tau_{3}^{\left(23\right)}&=-\ln{\frac{1}{2}\left(\left(1-\beta{}\right)+\sqrt{\left(1-\beta{}\right)^{2}+4\gamma{}}\right)} \\
&=\ln{\left(\frac{2}{\left(1-\beta{}\right)+\sqrt{\left(1-\beta{}\right)^{2}+4\gamma{}}}\right)} \\
&=\ln{\left(\frac{2}{\left(1-\beta{}\right)+\sqrt{\left(1-\beta{}\right)^{2}+4\gamma{}}}\cdot{}\frac{-\left(1-\beta{}\right)+\sqrt{\left(1-\beta{}\right)^{2}+4\gamma{}}}{-\left(1-\beta{}\right)+\sqrt{\left(1-\beta{}\right)^{2}+4\gamma{}}}\right)} \\
&=\ln{\left(\frac{-2\left(1-\beta{}\right)+\sqrt{\left(1-\beta{}\right)^{2}+4\gamma{}}}{4\gamma{}}\right)} \\
&=\ln{\left(\frac{-\left(1-\beta{}\right)+\sqrt{\left(1-\beta{}\right)^{2}+4\gamma{}}}{2\gamma{}}\right)} \\
&=\ln{\left(\frac{-\left(1-v_{2}v_{3}\right)+\sqrt{\left(1-v_{2}v_{3}\right)^{2}+4\left(1-v_{3}+v_{1}v_{2}\right)}}{2\left(1-v_{3}+v_{1}v_{2}\right)}\right)} \\
&=\ln{\left(\frac{-\left(1-x_{2}^{2}x_{3}^{2}\right)+\sqrt{\left(1-x_{2}^{2}x_{3}^{2}\right)^{2}+4\left(1-x_{3}\left(1-x_{1}^{2}x_{2}^{2}\right)\right)}}{2\left(1-x_{3}\left(1-x_{1}^{2}x_{2}^{2}\right)\right)}\right)} \\
&=\scriptstyle{\ln{\left(\frac{-\left(1-e^{-2\left(\tau_{2}^{\left(12\right)}+\tau_{3}^{\left(12\right)}\right)}\right)+\sqrt{\left(1-e^{-2\left(\tau_{2}^{\left(12\right)}+\tau_{3}^{\left(12\right)}\right)}\right)^{2}+4\left(1-e^{-\tau_{3}^{\left(12\right)}}\left(1-e^{-2\left(\tau_{1}^{\left(12\right)}+\tau_{2}^{\left(12\right)}\right)}\right)\right)}}{2\left(1-e^{-\tau_{3}^{\left(12\right)}}\left(1-e^{-2\left(\tau_{1}^{\left(12\right)}+\tau_{2}^{\left(12\right)}\right)}\right)\right)}\right)}}.
\end{split}\end{align}

In summary,
\begin{align}\begin{split}
\begin{mycases}
\tau_{1}^{\left(23\right)}&=\ln{\sqrt{\frac{e^{-2\left(\tau_{1}^{\left(12\right)}+\tau_{2}^{\left(12\right)}+\tau_{3}^{\left(12\right)}\right)}}{1-e^{-\tau_{3}^{\left(12\right)}}\left(1-e^{-2\left(\tau_{1}^{\left(12\right)}+\tau_{2}^{\left(12\right)}\right)}\right)}}}, \\
\tau_{2}^{\left(23\right)}&=\scriptstyle{\ln{\left(\frac{\left(1-e^{-2\left(\tau_{2}^{\left(12\right)}+\tau_{3}^{\left(12\right)}\right)}\right)+\sqrt{\left(1-e^{-2\left(\tau_{2}^{\left(12\right)}+\tau_{3}^{\left(12\right)}\right)}\right)^{2}+4\left(1-e^{-\tau_{3}^{\left(12\right)}}\left(1-e^{-2\left(\tau_{1}^{\left(12\right)}+\tau_{2}^{\left(12\right)}\right)}\right)\right)}}{2e^{-\left(\tau_{1}^{\left(12\right)}+\tau_{2}^{\left(12\right)}+\tau_{3}^{\left(12\right)}\right)}}\right)}}, \\
\tau_{3}^{\left(23\right)}&=\scriptstyle{\ln{\left(\frac{-\left(1-e^{-2\left(\tau_{2}^{\left(12\right)}+\tau_{3}^{\left(12\right)}\right)}\right)+\sqrt{\left(1-e^{-2\left(\tau_{2}^{\left(12\right)}+\tau_{3}^{\left(12\right)}\right)}\right)^{2}+4\left(1-e^{-\tau_{3}^{\left(12\right)}}\left(1-e^{-2\left(\tau_{1}^{\left(12\right)}+\tau_{2}^{\left(12\right)}\right)}\right)\right)}}{2\left(1-e^{-\tau_{3}^{\left(12\right)}}\left(1-e^{-2\left(\tau_{1}^{\left(12\right)}+\tau_{2}^{\left(12\right)}\right)}\right)\right)}\right)}}. \\
\end{mycases}
\end{split}\end{align}

Recall that
\begin{align}\begin{split}
\tau_{1}^{\left(23\right)}&=\ln{\sqrt{\frac{x_{1}^{2}x_{2}^{2}x_{3}^{2}}{1-x_{3}\left(1-x_{1}^{2}x_{2}^{2}\right)}}}.
\end{split}\end{align}
Demanding $0<x_{1},x_{2},x_{3}\leq{}1$ and $\tau_{1}^{\left(23\right)}\geq{}0$,
\begin{align}\begin{split}
x_{1}^{2}x_{2}^{2}x_{3}^{2}&\geq{}1-x_{3}\left(1-x_{1}^{2}x_{2}^{2}\right) \\
\Leftrightarrow{}x_{1}^{2}x_{2}^{2}x_{3}^{2}&\geq{}1-x_{3}+x_{1}^{2}x_{2}^{2}x_{3} \\
\Leftrightarrow{}0&\geq{}1-x_{3}+x_{1}^{2}x_{2}^{2}x_{3}-x_{1}^{2}x_{2}^{2}x_{3}^{2} \\
\Leftrightarrow{}0&\geq{}1-x_{3}\left(1-x_{1}^{2}x_{2}^{2}+x_{1}^{2}x_{2}^{2}x_{3}\right) \\
\Leftrightarrow{}0&\geq{}1-x_{3}\left(1-x_{1}^{2}x_{2}^{2}\left(1-x_{3}\right)\right).
\end{split}\end{align}
This is the same inequality as in (\ref{equation1})~on~page~\pageref{equation1}, except with the direction of the inequality swapped. We can conclude that $\tau_{1}^{\left(23\right)}\leq{}0$.

Recall that
\begin{align}\begin{split}
\tau_{2}^{\left(23\right)}&=\ln{\left(\frac{\left(1-\beta{}\right)+\sqrt{\left(1-\beta{}\right)^{2}+4\gamma{}}}{2\sqrt{\alpha{}}}\right)} \\
&=\ln{\left(\frac{\left(1-\beta{}\right)+\sqrt{\left(1-\beta{}\right)^{2}+4\gamma{}}}{2\sqrt{\alpha{}}}\cdot{}\frac{-\left(1-\beta{}\right)+\sqrt{\left(1-\beta{}\right)^{2}+4\gamma{}}}{-\left(1-\beta{}\right)+\sqrt{\left(1-\beta{}\right)^{2}+4\gamma{}}}\right)} \\
&=\ln{\left(\frac{4\gamma{}}{2\sqrt{\alpha{}}\left(-\left(1-\beta{}\right)+\sqrt{\left(1-\beta{}\right)^{2}+4\gamma{}}\right)}\right)} \\
&=\ln{\left(\frac{2\gamma{}}{\sqrt{\alpha{}}\left(-\left(1-\beta{}\right)+\sqrt{\left(1-\beta{}\right)^{2}+4\gamma{}}\right)}\right)}.
\end{split}\end{align}

Demanding $0<\alpha{},\beta{},\gamma{}\leq{}1$ and $\tau_{2}^{\left(23\right)}\geq{}0$,
\begin{align}\begin{split}
2\gamma{}&\geq{}\sqrt{\alpha{}}\left(-\left(1-\beta{}\right)+\sqrt{\left(1-\beta{}\right)^{2}+4\gamma{}}\right) \\
\Leftrightarrow{}\frac{2\gamma{}}{\sqrt{\alpha{}}}&\geq{}-\left(1-\beta{}\right)+\sqrt{\left(1-\beta{}\right)^{2}+4\gamma{}} \\
\Leftrightarrow{}\left(1-\beta{}\right)+\frac{2\gamma{}}{\sqrt{\alpha{}}}&\geq{}\sqrt{\left(1-\beta{}\right)^{2}+4\gamma{}} \\
\Leftrightarrow{}\left(1-\beta{}\right)^{2}+\frac{4\gamma{}^{2}}{\alpha{}}+\frac{4\gamma{}\left(1-\beta{}\right)}{\sqrt{\alpha{}}}&\geq{}\left(1-\beta{}\right)^{2}+4\gamma{} \\
\Leftrightarrow{}\frac{4\gamma{}^{2}}{\alpha{}}+\frac{4\gamma{}\left(1-\beta{}\right)}{\sqrt{\alpha{}}}&\geq{}4\gamma{} \\
\Leftrightarrow{}\frac{\gamma{}}{\alpha{}}+\frac{\left(1-\beta{}\right)}{\sqrt{\alpha{}}}&\geq{}1 \\
\Leftrightarrow{}\gamma{}+\sqrt{\alpha{}}\left(1-\beta{}\right)&\geq{}\alpha{} \\
\Leftrightarrow{}\sqrt{\alpha{}}\left(1-\beta{}\right)&\geq{}\alpha{}-\gamma{} \\
\Leftrightarrow{}\sqrt{x_{1}^{2}x_{2}^{2}x_{3}^{2}}\left(1-x_{2}^{2}x_{3}^{2}\right)&\geq{}x_{1}^{2}x_{2}^{2}x_{3}^{2}-\left(1-x_{3}\left(1-x_{1}^{2}x_{2}^{2}\right)\right) \\
\Leftrightarrow{}x_{1}x_{2}x_{3}\left(1-x_{2}^{2}x_{3}^{2}\right)&\geq{}x_{1}^{2}x_{2}^{2}x_{3}^{2}-1+x_{3}-x_{1}^{2}x_{2}^{2}x_{3} \\
\Leftrightarrow{}x_{1}x_{2}x_{3}\left(1-x_{2}^{2}x_{3}^{2}\right)&\geq{}-\left(1-x_{3}+x_{1}^{2}x_{2}^{2}x_{3}-x_{1}^{2}x_{2}^{2}x_{3}^{2}\right) \\
\Leftrightarrow{}x_{1}x_{2}x_{3}\left(1-x_{2}^{2}x_{3}^{2}\right)&\geq{}-\left(1-x_{3}\left(1-x_{1}^{2}x_{2}^{2}+x_{1}^{2}x_{2}^{2}x_{3}\right)\right) \\
\Leftrightarrow{}x_{1}x_{2}x_{3}\left(1-x_{2}^{2}x_{3}^{2}\right)&\geq{}-\left(1-x_{3}\left(1-x_{1}^{2}x_{2}^{2}\left(1-x_{3}\right)\right)\right).
\end{split}\end{align}
Since $0<x_{1},x_{2},x_{3}\leq{}1$,
\begin{align}\begin{split}
0<x_{2}^{2}x_{3}^{2}&\leq{}1 \\
\Leftrightarrow{}-1\leq{}-x_{2}^{2}x_{3}^{2}&<0 \\
\Leftrightarrow{}0\leq{}1-x_{2}^{2}x_{3}^{2}&<1 \\
\Leftrightarrow{}0\leq{}x_{1}x_{2}x_{3}\left(1-x_{2}^{2}x_{3}^{2}\right)&<1.
\end{split}\end{align}
From (\ref{equation1})~on~page~\pageref{equation1}, we can conclude that $-1<-\left(1-x_{3}\left(1-x_{1}^{2}x_{2}^{2}\left(1-x_{3}\right)\right)\right)\leq{}0$. Since $0\leq{}x_{1}x_{2}x_{3}\left(1-x_{2}^{2}x_{3}^{2}\right)<1$, we can conclude that $\tau_{2}^{\left(23\right)}\geq{}0$.

Recall that
\begin{align}\begin{split}
\tau_{3}^{\left(23\right)}&=\ln{\left(\frac{-\left(1-x_{2}^{2}x_{3}^{2}\right)+\sqrt{\left(1-x_{2}^{2}x_{3}^{2}\right)^{2}+4\left(1-x_{3}\left(1-x_{1}^{2}x_{2}^{2}\right)\right)}}{2\left(1-x_{3}\left(1-x_{1}^{2}x_{2}^{2}\right)\right)}\right)}.
\end{split}\end{align}
Demanding $0<x_{1},x_{2},x_{3}\leq{}1$ and $\tau_{3}^{\left(23\right)}\geq{}0$,
\begin{align}\begin{split}
-\left(1-x_{2}^{2}x_{3}^{2}\right)+\sqrt{\left(1-x_{2}^{2}x_{3}^{2}\right)^{2}+4\left(1-x_{3}\left(1-x_{1}^{2}x_{2}^{2}\right)\right)}&\geq{}2\left(1-x_{3}\left(1-x_{1}^{2}x_{2}^{2}\right)\right) \\
\Leftrightarrow{}\sqrt{\left(1-x_{2}^{2}x_{3}^{2}\right)^{2}+4\left(1-x_{3}\left(1-x_{1}^{2}x_{2}^{2}\right)\right)}&\geq{}\left(1-x_{2}^{2}x_{3}^{2}\right)+2\left(1-x_{3}\left(1-x_{1}^{2}x_{2}^{2}\right)\right) \\
\Leftrightarrow{}\left(1-x_{2}^{2}x_{3}^{2}\right)^{2}+4\left(1-x_{3}\left(1-x_{1}^{2}x_{2}^{2}\right)\right)&\geq{}\left(1-x_{2}^{2}x_{3}^{2}\right)^{2}+4\left(1-x_{3}\left(1-x_{1}^{2}x_{2}^{2}\right)\right)^{2} \\
&+4\left(1-x_{2}^{2}x_{3}^{2}\right)\left(1-x_{3}\left(1-x_{1}^{2}x_{2}^{2}\right)\right) \\
\Leftrightarrow{}4\left(1-x_{3}\left(1-x_{1}^{2}x_{2}^{2}\right)\right)&\geq{}4\left(1-x_{3}\left(1-x_{1}^{2}x_{2}^{2}\right)\right)^{2} \\
&+4\left(1-x_{2}^{2}x_{3}^{2}\right)\left(1-x_{3}\left(1-x_{1}^{2}x_{2}^{2}\right)\right) \\
\Leftrightarrow{}1&\geq{}\left(1-x_{3}\left(1-x_{1}^{2}x_{2}^{2}\right)\right)+\left(1-x_{2}^{2}x_{3}^{2}\right) \\
\Leftrightarrow{}0&\geq{}-x_{3}+x_{1}^{2}x_{2}^{2}x_{3}+1-x_{2}^{2}x_{3}^{2} \\
\Leftrightarrow{}0&\geq{}1-x_{3}\left(1-x_{1}^{2}x_{2}^{2}+x_{2}^{2}x_{3}\right) \\
\Leftrightarrow{}0&\geq{}1-x_{3}\left(1-x_{2}^{2}\left(x_{1}^{2}-x_{3}\right)\right).
\end{split}\end{align}

This is the same inequality as in (\ref{equation2})~on~page~\pageref{equation2}, except with the direction of the inequality swapped. We can conclude that the inequality will not be satisfied when $x_{3}<x_{1}^{2}$. Clearly, when $x_{3}=x_{1}^{2}$ the inequality will be satisfied, since $1-x_{3}>0$. We can conclude that $\tau_{3}^{\left(23\right)}$ can sometimes be non-negative and sometimes be negative.

To get a rough estimate of how often the inequality is satisfied we decided to simulate some samples for $x_{1},x_{2},x_{3}\in{}U\left(0,1\right)$. We found that for $60$ out of $500$ samples $\tau_{3}^{\left(23\right)}\geq{}0$. Of the remaining $440$ out of $500$ samples $\tau_{3}^{\left(23\right)}<0$.

\chapter{Gr\"obner Basis for Network $1b$}\label{cha.Appendix1}

Using the \emph{Macaulay2} code from Section~\ref{macaulay2}~on~page~\pageref{macaulay2}, the Gr\"{o}bner basis of the ideal for Network $1b$ is
\begin{align*}
I=&\left<q_{1010}-q_{1100},\right. \\
&\qquad{}\left.q_{0101}-q_{0110},\right. \\
&\qquad{}\left.q_{0011}^{2}q_{1100}-q_{0011}q_{0110}q_{1001}^{2}+q_{0011}q_{0110}q_{1001}-2q_{0011}q_{0110}q_{1100}+q_{0011}q_{1001}q_{1111}\right. \\
&\qquad{}\left.-q_{0011}q_{1111}-q_{0110}^{2}q_{1001},\right. \\
&\qquad{}\left.y_{4}q_{0110}q_{1001}-y_{4}q_{1111}+q_{0011}q_{1100}-q_{0110}q_{1100}+q_{0110}^{2}q_{1100}+q_{0110}q_{1001}q_{1111}\right. \\
&\qquad{}\left.+q_{0110}q_{1111}-q_{1111}^{2},\right. \\
&\qquad{}\left.y_{4}q_{0011}-y_{4}q_{0110}+q_{0011}q_{1001}-q_{0011}+q_{0110}-q_{1111},\right. \\
&\qquad{}\left.y_{4}^{2}+y_{4}q_{1001}-y_{4}-q_{1100},\right. \\
&\qquad{}\left.y_{3}^{2}q_{1100}^{2}+y_{4}q_{0110}-y_{4}q_{1111}+q_{0110}q_{1001}-q_{0110}q_{1100}-q_{0110}-q_{1001}q_{1111}+q_{1111},\right. \\
&\qquad{}\left.y_{3}^{2}q_{0011}q_{1001}q_{1100}-y_{3}^{2}q_{0011}q_{1100}+y_{3}^{2}q_{0110}q_{1100}-y_{3}^{2}q_{1100}q_{1111}-q_{0011}q_{0110}q_{1001}\right. \\
&\qquad{}\left.+q_{0011}q_{1111},\right. \\
&\qquad{}\left.y_{3}^{2}q_{0011}q_{1001}^{2}-2y_{3}^{2}q_{0011}q_{1001}+y_{3}^{2}q_{0011}q_{1100}+y_{3}^{2}q_{0011}+y_{3}^{2}q_{0110}q_{1001}-y_{3}^{2}q_{0110}q_{1100}\right. \\
&\qquad{}\left.-y_{3}^{2}q_{0110}-y_{3}^{2}q_{1001}q_{1111}+y_{3}^{2}q_{1111}-q_{0011}^{2}+q_{0011}q_{0110},\right. \\
&\qquad{}\left.y_{3}^{2}y_{4}q_{1100}-y_{4}q_{0110}+q_{0110}-q_{1111},\right. \\
&\qquad{}\left.y_{3}^{2}y_{4}q_{1001}-y_{3}^{2}y_{4}-y_{3}^{2}q_{1100}+q_{0011},\right. \\
&\qquad{}\left.y_{3}^{2}y_{4}q_{0110}-y_{3}^{2}y_{4}q_{1111}+y_{3}^{2}q_{0011}q_{1100}-q_{0011}q_{0110},\right. \\
&\qquad{}\left.y_{2}^{2}q_{0110}^{2}q_{1001}-y_{2}^{2}q_{0110}^{2}q_{1100}-y_{2}^{2}q_{0110}q_{1001}q_{1111}-y_{2}^{2}q_{0110}q_{1111}+y_{2}^{2}q_{1111}^{2}\right. \\
&\qquad{}\left.-q_{0011}q_{0110}q_{1100}+q_{0110}^{2}q_{1001}^{2}-q_{0110}^{2}q_{1001}+2q_{0110}^{2}q_{1100}-q_{0110}q_{1001}q_{1111}+q_{0110}q_{1111},\right. \\
&\qquad{}\left.y_{2}^{2}q_{0011}-q_{0110},\right. \\
&\qquad{}\left.y_{2}^{2}y_{4}q_{1111}-y_{2}^{2}q_{0110}q_{1001}+y_{2}^{2}q_{0110}q_{1100}+y_{2}^{2}q_{1001}q_{1111}-y_{4}q_{1111}+q_{0011}q_{1100}\right. \\
&\qquad{}\left.-q_{0110}q_{1001}^{2}+q_{0110}q_{1001}-2q_{0110}q_{1100},\right. \\
&\qquad{}\left.y_{2}^{2}y_{4}q_{0110}-y_{2}^{2}q_{0110}+y_{2}^{2}q_{1111}-y_{4}q_{0110}-q_{0110}q_{1001}+q_{0110},\right. \\
&\qquad{}\left.y_{2}^{2}y_{3}^{2}q_{0110}q_{1100}-y_{2}^{2}y_{3}^{2}q_{1100}q_{1111}+y_{3}^{2}q_{0110}q_{1001}q_{1100}-y_{3}^{2}q_{0110}q_{1100}-q_{0110}^{2}q_{1001}\right. \\
&\qquad{}\left.+q_{0110}q_{1111},\right. \\
&\qquad{}\left.y_{2}^{2}y_{3}^{2}q_{0110}q_{1001}-y_{2}^{2}y_{3}^{2}q_{0110}-y_{2}^{2}y_{3}^{2}q_{1001}q_{1111}-y_{2}^{2}y_{3}^{2}q_{1100}q_{1111}+y_{2}^{2}y_{3}^{2}q_{1111}+y_{3}^{2}q_{0110}q_{1001}^{2}\right. \\
&\qquad{}\left.+y_{3}^{2}q_{0110}q_{1001}q_{1100}-2y_{3}^{2}q_{0110}q_{1001}+y_{3}^{2}q_{0110}-q_{0011}q_{0110}-q_{0110}^{2}q_{1001}+q_{0110}^{2}\right. \\
&\qquad{}\left.+q_{0110}q_{1111},\right. \\
&\qquad{}\left.y_{2}^{2}y_{3}^{2}q_{0110}^{2}-2y_{2}^{2}y_{3}^{2}q_{0110}q_{1111}+y_{2}^{2}y_{3}^{2}q_{1111}^{2}-y_{3}^{2}q_{0011}q_{0110}q_{1100}+y_{3}^{2}q_{0110}^{2}q_{1001}\right. \\
&\qquad{}\left.+y_{3}^{2}q_{0110}^{2}q_{1100}-y_{3}^{2}q_{0110}^{2}-y_{3}^{2}q_{0110}q_{1001}q_{1111}+y_{3}^{2}q_{0110}q_{1111}+q_{0011}q_{0110}^{2}-q_{0110}^{3},\right. \\
&\qquad{}\left.y_{1}^{2}q_{0110}-q_{1100},\right. \\
&\qquad{}\left.y_{1}^{2}y_{4}q_{1111}-y_{1}^{2}q_{0011}q_{1100}-y_{4}q_{1001}q_{1100}+q_{1100}^{2},\right. \\
&\qquad{}\left.y_{1}^{2}y_{2}^{2}q_{1111}+y_{2}^{2}y_{4}q_{1100}-y_{2}^{2}q_{1100}-y_{4}q_{1100}-q_{1001}q_{1100}+q_{1100},\right. \\
&\qquad{}\left.y_{1}^{2}y_{2}^{2}y_{3}^{2}q_{1100}-y_{4}q_{1001}+y_{4}-q_{1001}^{2}+2q_{1001}-q_{1100}-1,\right. \\
&\qquad{}\left.y_{1}^{2}y_{2}^{2}y_{3}^{2}y_{4}-y_{4}-q_{1001}+1\right>.
\end{align*}

\chapter{Gr\"obner Basis for Network $1c$}\label{cha.Appendix2}

Using the \emph{Macaulay2} code from Section~\ref{macaulay2}~on~page~\pageref{macaulay2}, the Gr\"{o}bner basis of the ideal for Network $1c$ is
\begin{align*}
I=&\left<q_{1010}-q_{1100},\right. \\
&\qquad{}\left.q_{1001}-q_{1100},\right. \\
&\qquad{}\left.q_{0011}^{2}q_{1100}^{2}-q_{0011}q_{0101}^{2}q_{1100}^{2}+q_{0011}q_{0101}q_{1100}^{2}+q_{0011}q_{0101}q_{1100}q_{1111}-2q_{0011}q_{0110}q_{1100}^{2}\right. \\
&\qquad{}\left.-q_{0011}q_{1100}q_{1111}-q_{0101}q_{0110}q_{1100}^{2}+q_{0101}q_{0110}q_{1100}q_{1111}+q_{0110}^{2}q_{1100}^{2}+q_{0110}q_{1100}q_{1111}\right. \\
&\qquad{}\left.-q_{0110}q_{1111}^{2},\right. \\
&\qquad{}\left.y_{4}q_{0101}q_{1100}-y_{4}q_{1111}+q_{0011}q_{1100}-q_{0110}q_{1100},\right. \\
&\qquad{}\left.y_{4}q_{0011}q_{1111}-y_{4}q_{0110}q_{1111}-q_{0011}^{2}q_{1100}+q_{0011}q_{0101}^{2}q_{1100}-q_{0011}q_{0101}q_{1100}\right. \\
&\qquad{}\left.+2q_{0011}q_{0110}q_{1100}+q_{0101}q_{0110}q_{1100}-q_{0101}q_{0110}q_{1111}-q_{0110}^{2}q_{1100},\right. \\
&\qquad{}\left.y_{4}q_{0011}q_{1100}-y_{4}q_{0110}q_{1100}+q_{0011}q_{0101}q_{1100}-q_{0011}q_{1100}+q_{0110}q_{1100}-q_{0110}q_{1111},\right. \\
&\qquad{}\left.y_{4}^{2}+y_{4}q_{0101}-y_{4}-q_{0110},\right. \\
&\qquad{}\left.y_{3}^{2}q_{0110}q_{1100}+y_{4}q_{1100}-y_{4}q_{1111}+q_{0101}q_{1100}-q_{0101}q_{1111}-q_{0110}q_{1100}-q_{1100}+q_{1111},\right. \\
&\qquad{}\left.y_{3}^{2}q_{0110}^{2}-y_{4}q_{0011}q_{0101}+y_{4}q_{0011}-q_{0011}q_{0101}^{2}+2q_{0011}q_{0101}-q_{0011}q_{0110}-q_{0011},\right. \\
&\qquad{}\left.y_{3}^{2}q_{0011}q_{1100}^{2}+y_{4}q_{1100}^{2}-y_{4}q_{1100}q_{1111}-q_{0011}q_{1100}^{2}-q_{1100}^{2}+2q_{1100}q_{1111}-q_{1111}^{2},\right. \\
&\qquad{}\left.y_{3}^{2}q_{0011}q_{0101}q_{1100}-y_{3}^{2}q_{0011}q_{1100}-y_{3}^{2}q_{0110}q_{1111}-y_{4}q_{1100}+y_{4}q_{1111}-q_{0011}q_{0101}q_{1100}\right. \\
&\qquad{}\left.+q_{0011}q_{1111}-q_{0101}q_{1100}+q_{0101}q_{1111}+q_{0110}q_{1100}+q_{1100}-q_{1111},\right. \\
&\qquad{}\left.y_{3}^{2}y_{4}q_{1111}-y_{3}^{2}q_{0011}q_{1100}-y_{4}q_{1100}+q_{0011}q_{1100}+q_{1100}-q_{1111},\right. \\
&\qquad{}\left.y_{3}^{2}y_{4}q_{1100}-y_{4}q_{1100}+q_{1100}-q_{1111},\right. \\
&\qquad{}\left.y_{3}^{2}y_{4}q_{0110}-y_{4}q_{0011}-q_{0011}q_{0101}+q_{0011},\right. \\
&\qquad{}\left.y_{3}^{2}y_{4}q_{0101}-y_{3}^{2}y_{4}-y_{3}^{2}q_{0110}+q_{0011},\right. \\
&\qquad{}\left.y_{2}^{2}q_{0101}q_{1100}^{2}-y_{2}^{2}q_{0101}q_{1100}q_{1111}-y_{2}^{2}q_{0110}q_{1100}^{2}-y_{2}^{2}q_{1100}q_{1111}+y_{2}^{2}q_{1111}^{2}-q_{0011}q_{1100}^{2}\right. \\
&\qquad{}\left.+q_{0101}^{2}q_{1100}^{2}-q_{0101}q_{1100}^{2}-q_{0101}q_{1100}q_{1111}+2q_{0110}q_{1100}^{2}+q_{1100}q_{1111},\right. \\
&\qquad{}\left.y_{2}^{2}q_{0011}-q_{0110},\right. \\
&\qquad{}\left.y_{2}^{2}y_{4}q_{1111}-y_{2}^{2}q_{0101}q_{1100}+y_{2}^{2}q_{0101}q_{1111}+y_{2}^{2}q_{0110}q_{1100}-y_{4}q_{1111}+q_{0011}q_{1100}\right. \\
&\qquad{}\left.-q_{0101}^{2}q_{1100}+q_{0101}q_{1100}-2q_{0110}q_{1100},\right. \\
&\qquad{}\left.y_{2}^{2}y_{4}q_{1100}-y_{2}^{2}q_{1100}+y_{2}^{2}q_{1111}-y_{4}q_{1100}-q_{0101}q_{1100}+q_{1100},\right. \\
&\qquad{}\left.y_{2}^{2}y_{3}^{2}q_{1100}-y_{2}^{2}y_{3}^{2}q_{1111}+y_{3}^{2}q_{0101}q_{1100}-y_{3}^{2}q_{1100}-q_{0101}q_{1100}+q_{1111},\right. \\
&\qquad{}\left.y_{2}^{2}y_{3}^{2}q_{0110}-y_{4}q_{0101}+y_{4}-q_{0101}^{2}+2q_{0101}-q_{0110}-1,\right. \\
&\qquad{}\left.y_{2}^{2}y_{3}^{2}y_{4}-y_{4}-q_{0101}+1,\right. \\
&\qquad{}\left.y_{1}^{2}q_{0110}-q_{1100},\right. \\
&\qquad{}\left.y_{1}^{2}q_{0011}q_{1100}-y_{1}^{2}q_{0101}^{2}q_{1100}+y_{1}^{2}q_{0101}q_{1100}+y_{1}^{2}q_{0101}q_{1111}-y_{1}^{2}q_{1111}+y_{2}^{2}y_{3}^{2}q_{1111}^{2}\right. \\
&\qquad{}\left.-y_{3}^{2}q_{0101}q_{1100}^{2}-y_{3}^{2}q_{0101}q_{1100}q_{1111}+y_{3}^{2}q_{1100}q_{1111}+q_{0101}q_{1100}^{2}+q_{0101}q_{1100}q_{1111}-q_{1100}^{2}\right. \\
&\qquad{}\left.-q_{1100}q_{1111}-q_{1111}^{2},\right. \\
&\qquad{}\left.y_{1}^{2}y_{4}q_{1100}-y_{1}^{2}y_{4}q_{1111}+y_{1}^{2}q_{0101}q_{1100}-y_{1}^{2}q_{0101}q_{1111}-y_{1}^{2}q_{1100}+y_{1}^{2}q_{1111}\right. \\
&\qquad{}\left.+y_{3}^{2}q_{1100}^{2}-q_{1100}^{2},\right. \\
&\qquad{}\left.y_{1}^{2}y_{4}q_{0101}-y_{1}^{2}y_{4}+y_{1}^{2}q_{0101}^{2}-2y_{1}^{2}q_{0101}+y_{1}^{2}-y_{2}^{2}y_{3}^{2}q_{1111}+y_{3}^{2}q_{0101}q_{1100}-y_{3}^{2}q_{1100}\right. \\
&\qquad{}\left.-q_{0101}q_{1100}+q_{1100}+q_{1111},\right. \\
&\qquad{}\left.y_{1}^{2}y_{4}q_{0011}+y_{1}^{2}q_{0011}q_{0101}-y_{1}^{2}q_{0011}-y_{4}q_{1100}+q_{1100}-q_{1111},\right. \\
&\qquad{}\left.y_{1}^{2}y_{3}^{2}q_{0101}q_{1100}^{2}-y_{1}^{2}y_{3}^{2}q_{0101}q_{1100}q_{1111}-y_{1}^{2}y_{3}^{2}q_{1100}^{2}+y_{1}^{2}y_{3}^{2}q_{1100}q_{1111}-y_{1}^{2}q_{0101}q_{1100}^{2}\right. \\
&\qquad{}\left.+y_{1}^{2}q_{0101}q_{1100}q_{1111}+y_{1}^{2}q_{1100}q_{1111}-y_{1}^{2}q_{1111}^{2}+y_{3}^{4}q_{1100}^{3}-2y_{3}^{2}q_{1100}^{3}+q_{1100}^{3},\right. \\
&\qquad{}\left.y_{1}^{2}y_{3}^{2}q_{0101}^{2}-2y_{1}^{2}y_{3}^{2}q_{0101}+y_{1}^{2}y_{3}^{2}-y_{1}^{2}q_{0011}-y_{2}^{2}y_{3}^{4}q_{1111}+y_{3}^{4}q_{0101}q_{1100}-y_{3}^{4}q_{1100}\right. \\
&\qquad{}\left.-y_{3}^{2}q_{0101}q_{1100}+2y_{3}^{2}q_{1100}+y_{3}^{2}q_{1111},\right. \\
&\qquad{}\left.y_{1}^{2}y_{2}^{2}q_{0101}^{2}q_{1100}-y_{1}^{2}y_{2}^{2}q_{0101}q_{1100}-y_{1}^{2}y_{2}^{2}q_{0101}q_{1111}+y_{1}^{2}y_{2}^{2}q_{1111}-y_{2}^{4}y_{3}^{2}q_{1111}^{2}\right. \\
&\qquad{}\left.+2y_{2}^{2}y_{3}^{2}q_{0101}q_{1111}^{2}-y_{2}^{2}y_{3}^{2}q_{1111}^{2}-2y_{2}^{2}q_{0101}q_{1100}q_{1111}-y_{2}^{2}q_{0110}q_{1100}^{2}+y_{2}^{2}q_{1100}^{2}+2y_{2}^{2}q_{1111}^{2}\right. \\
&\qquad{}\left.-y_{3}^{2}q_{0101}^{2}q_{1100}^{2}-2y_{3}^{2}q_{0101}^{2}q_{1100}q_{1111}+y_{3}^{2}q_{0101}q_{1100}^{2}+3y_{3}^{2}q_{0101}q_{1100}q_{1111}-y_{3}^{2}q_{1100}q_{1111}\right. \\
&\qquad{}\left.-q_{0011}q_{1100}^{2}+2q_{0101}^{2}q_{1100}^{2}+2q_{0101}^{2}q_{1100}q_{1111}-q_{0101}q_{1100}^{2}-3q_{0101}q_{1100}q_{1111}\right. \\
&\qquad{}\left.-2q_{0101}q_{1111}^{2}+2q_{0110}q_{1100}^{2}-q_{1100}^{2}+q_{1100}q_{1111}+q_{1111}^{2}\right>.
\end{align*}

\chapter{Gr\"obner Basis for Network $2b$}\label{cha.Appendix3}

Using the \emph{Macaulay2} code from Section~\ref{macaulay2}~on~page~\pageref{macaulay2}, the Gr\"{o}bner basis of the ideal for Network $2b$ is
\begin{align*}
I=&\left<q_{0110}-q_{1001},\right. \\
&\qquad{}\left.q_{0101}-q_{1001},\right. \\
&\qquad{}\left.q_{0011}^{2}q_{1100}^{2}-2q_{0011}q_{1001}^{2}q_{1100}-q_{0011}q_{1001}q_{1010}^{2}q_{1100}+q_{0011}q_{1001}q_{1010}q_{1100}\right. \\
&\qquad{}\left.+q_{0011}q_{1010}q_{1100}q_{1111}-q_{0011}q_{1100}q_{1111}+q_{1001}^{4}-q_{1001}^{3}q_{1010}+q_{1001}^{2}q_{1010}q_{1111}\right. \\
&\qquad{}\left.+q_{1001}^{2}q_{1111}-q_{1001}q_{1111}^{2},\right. \\
&\qquad{}\left.y_{4}q_{1001}q_{1010}-y_{4}q_{1111}+q_{0011}q_{1100}-q_{1001}^{2},\right. \\
&\qquad{}\left.y_{4}q_{0011}q_{1100}-y_{4}q_{1001}^{2}+q_{0011}q_{1010}q_{1100}-q_{0011}q_{1100}+q_{1001}^{2}-q_{1001}q_{1111},\right. \\
&\qquad{}\left.y_{4}^{2}+y_{4}q_{1010}-y_{4}-q_{1001},\right. \\
&\qquad{}\left.y_{3}^{2}q_{1001}q_{1100}+y_{4}q_{1001}-y_{4}q_{1111}-q_{1001}^{2}+q_{1001}q_{1010}-q_{1001}-q_{1010}q_{1111}+q_{1111},\right. \\
&\qquad{}\left.y_{3}^{2}q_{1001}^{2}-y_{4}q_{0011}q_{1010}+y_{4}q_{0011}-q_{0011}q_{1001}-q_{0011}q_{1010}^{2}+2q_{0011}q_{1010}-q_{0011},\right. \\
&\qquad{}\left.y_{3}^{2}q_{0011}q_{1100}^{2}+y_{4}q_{1001}^{2}-y_{4}q_{1001}q_{1111}-q_{0011}q_{1001}q_{1100}-q_{1001}^{2}+2q_{1001}q_{1111}-q_{1111}^{2},\right. \\
&\qquad{}\left.y_{3}^{2}q_{0011}q_{1010}q_{1100}-y_{3}^{2}q_{0011}q_{1100}-y_{3}^{2}q_{1001}q_{1111}+y_{4}q_{0011}q_{1010}-y_{4}q_{0011}\right. \\
&\qquad{}\left.-q_{0011}q_{1001}q_{1010}+q_{0011}q_{1001}+q_{0011}q_{1010}^{2}-2q_{0011}q_{1010}+q_{0011}q_{1111}+q_{0011},\right. \\
&\qquad{}\left.y_{3}^{2}y_{4}q_{1111}-y_{3}^{2}q_{0011}q_{1100}-y_{4}q_{0011}+q_{0011}q_{1001}-q_{0011}q_{1010}+q_{0011},\right. \\
&\qquad{}\left.y_{3}^{2}y_{4}q_{1100}-y_{4}q_{1001}+q_{1001}-q_{1111},\right. \\
&\qquad{}\left.y_{3}^{2}y_{4}q_{1010}-y_{3}^{2}y_{4}-y_{3}^{2}q_{1001}+q_{0011},\right. \\
&\qquad{}\left.y_{3}^{2}y_{4}q_{1001}-y_{4}q_{0011}-q_{0011}q_{1010}+q_{0011},\right. \\
&\qquad{}\left.y_{2}^{2}q_{1001}^{3}-y_{2}^{2}q_{1001}^{2}q_{1010}+y_{2}^{2}q_{1001}q_{1010}q_{1111}+y_{2}^{2}q_{1001}q_{1111}-y_{2}^{2}q_{1111}^{2}+y_{3}^{2}q_{1100}^{3}\right. \\
&\qquad{}\left.-y_{4}q_{1010}q_{1100}^{2}+y_{4}q_{1100}^{2}-2q_{1001}q_{1100}^{2}-q_{1010}^{2}q_{1100}^{2}+2q_{1010}q_{1100}^{2}-q_{1100}^{2},\right. \\
&\qquad{}\left.y_{2}^{2}q_{0011}-q_{1100},\right. \\
&\qquad{}\left.y_{2}^{2}y_{4}q_{1111}^{2}+y_{2}^{2}q_{1001}^{2}q_{1111}-y_{2}^{2}q_{1001}q_{1010}q_{1111}+y_{2}^{2}q_{1010}q_{1111}^{2}-y_{3}^{2}q_{1010}q_{1100}^{3}\right. \\
&\qquad{}\left.-y_{4}q_{1010}q_{1100}^{2}+q_{1001}q_{1010}q_{1100}^{2}-q_{1010}^{2}q_{1100}^{2}+q_{1010}q_{1100}^{2}-q_{1100}^{2}q_{1111},\right. \\
&\qquad{}\left.y_{2}^{2}y_{4}q_{1001}q_{1111}-y_{2}^{2}q_{1001}q_{1111}+y_{2}^{2}q_{1111}^{2}-y_{3}^{2}q_{1100}^{3}-y_{4}q_{1100}^{2}+q_{1001}q_{1100}^{2}-q_{1010}q_{1100}^{2}\right. \\
&\qquad{}\left.+q_{1100}^{2},\right. \\
&\qquad{}\left.y_{2}^{2}y_{4}q_{1001}^{2}-y_{2}^{2}q_{1001}^{2}+y_{2}^{2}q_{1001}q_{1111}-y_{4}q_{1100}^{2}-q_{1010}q_{1100}^{2}+q_{1100}^{2},\right. \\
&\qquad{}\left.y_{2}^{2}y_{3}^{2}q_{1111}^{2}-y_{3}^{4}q_{1100}^{3}-2y_{3}^{2}q_{1010}q_{1100}^{2}+2y_{3}^{2}q_{1100}^{2}-2y_{4}q_{1001}q_{1100}-y_{4}q_{1010}q_{1100}\right. \\
&\qquad{}\left.+2y_{4}q_{1100}q_{1111}+y_{4}q_{1100}+q_{1001}^{2}q_{1100}+q_{1001}q_{1100}-q_{1010}^{2}q_{1100}+2q_{1010}q_{1100}q_{1111}\right. \\
&\qquad{}\left.+2q_{1010}q_{1100}-4q_{1100}q_{1111}-q_{1100},\right. \\
&\qquad{}\left.y_{2}^{2}y_{3}^{2}q_{1001}q_{1111}-y_{3}^{2}q_{1010}q_{1100}^{2}+y_{3}^{2}q_{1100}^{2}-y_{4}q_{1010}q_{1100}+y_{4}q_{1100}+q_{1001}q_{1010}q_{1100}\right. \\
&\qquad{}\left.-q_{1001}q_{1100}-q_{1010}^{2}q_{1100}+2q_{1010}q_{1100}-q_{1100}q_{1111}-q_{1100},\right. \\
&\qquad{}\left.y_{1}^{2}q_{1100}-q_{1001},\right. \\
&\qquad{}\left.y_{1}^{2}q_{1001}^{3}-y_{1}^{2}q_{1001}^{2}q_{1010}+y_{1}^{2}q_{1001}q_{1010}q_{1111}+y_{1}^{2}q_{1001}q_{1111}-y_{1}^{2}q_{1111}^{2}+q_{0011}^{2}q_{1100}\right. \\
&\qquad{}\left.-2q_{0011}q_{1001}^{2}-q_{0011}q_{1001}q_{1010}^{2}+q_{0011}q_{1001}q_{1010}+q_{0011}q_{1010}q_{1111}-q_{0011}q_{1111},\right. \\
&\qquad{}\left.y_{1}^{2}y_{4}q_{1111}+y_{1}^{2}q_{1001}^{2}-y_{1}^{2}q_{1001}q_{1010}+y_{1}^{2}q_{1010}q_{1111}-y_{4}q_{0011}q_{1010}-q_{0011}q_{1001}\right. \\
&\qquad{}\left.-q_{0011}q_{1010}^{2}+q_{0011}q_{1010},\right. \\
&\qquad{}\left.y_{1}^{2}y_{4}q_{1001}-y_{1}^{2}q_{1001}+y_{1}^{2}q_{1111}-y_{4}q_{0011}-q_{0011}q_{1010}+q_{0011},\right. \\
&\qquad{}\left.y_{1}^{2}y_{3}^{2}q_{1111}^{2}-y_{1}^{2}y_{4}q_{0011}q_{1010}+y_{1}^{2}y_{4}q_{0011}-y_{1}^{2}q_{0011}q_{1001}^{2}+2y_{1}^{2}q_{0011}q_{1001}q_{1010}\right. \\
&\qquad{}\left.-y_{1}^{2}q_{0011}q_{1001}-y_{1}^{2}q_{0011}q_{1010}^{2}+2y_{1}^{2}q_{0011}q_{1010}-2y_{1}^{2}q_{0011}q_{1111}-y_{1}^{2}q_{0011}-y_{3}^{2}q_{0011}^{2}q_{1100}\right. \\
&\qquad{}\left.-y_{3}^{2}q_{0011}q_{1001}q_{1010}+y_{3}^{2}q_{0011}q_{1001}-y_{3}^{2}q_{0011}q_{1010}q_{1111}+y_{3}^{2}q_{0011}q_{1111}+y_{4}q_{0011}^{2}q_{1010}\right. \\
&\qquad{}\left.-y_{4}q_{0011}^{2}+2q_{0011}^{2}q_{1001}+q_{0011}^{2}q_{1010}^{2}-2q_{0011}^{2}q_{1010}+q_{0011}^{2},\right. \\
&\qquad{}\left.y_{1}^{2}y_{3}^{2}q_{1001}-y_{1}^{2}y_{3}^{2}q_{1111}-y_{1}^{2}y_{4}q_{0011}-y_{1}^{2}q_{0011}q_{1010}+y_{1}^{2}q_{0011}+y_{3}^{2}y_{4}q_{0011}+y_{3}^{2}q_{0011}q_{1010}\right. \\
&\qquad{}\left.-y_{3}^{2}q_{0011},\right. \\
&\qquad{}\left.y_{1}^{2}y_{2}^{2}y_{3}^{2}q_{1111}-y_{3}^{2}q_{1010}q_{1100}+y_{3}^{2}q_{1100}-y_{4}q_{1010}+y_{4}+q_{1001}q_{1010}-q_{1001}-q_{1010}^{2}\right. \\
&\qquad{}\left.+2q_{1010}-q_{1111}-1,\right. \\
&\qquad{}\left.y_{1}^{2}y_{2}^{2}y_{3}^{2}y_{4}-y_{4}-q_{1010}+1\right>.
\end{align*}

\chapter{Gr\"obner Basis for Network $3$}\label{cha.Appendix4}

Using the \emph{Macaulay2} code from Section~\ref{macaulay2}~on~page~\pageref{macaulay2}, the Gr\"{o}bner basis of the ideal for Network $3$ is
\begin{align*}
I=&\left<q_{0101}q_{1010}-q_{0110}q_{1001},\right. \\
&\qquad{}\left.q_{0011}q_{1100}-q_{1111},\right. \\
&\qquad{}\left.y_{1234}^2q_{1111}-q_{0110}q_{1001},\right. \\
&\qquad{}\left.y_{4}^2q_{1010}-q_{0011}q_{1001},\right. \\
&\qquad{}\left.y_{4}^2q_{0110}-q_{0011}q_{0101},\right. \\
&\qquad{}\left.y_{4}^2y_{1234}^2q_{1100}-q_{0101}q_{1001},\right. \\
&\qquad{}\left.y_{3}q_{1001}-y_{4}q_{1010},\right. \\
&\qquad{}\left.y_{3}q_{0101}-y_{4}q_{0110},\right. \\
&\qquad{}\left.y_{3}y_{4}-q_{0011},\right. \\
&\qquad{}\left.y_{3}^2y_{1234}^2q_{1100}-q_{0110}q_{1010},\right. \\
&\qquad{}\left.y_{2}q_{1010}-y_{3}y_{1234}q_{1100},\right. \\
&\qquad{}\left.y_{2}q_{1001}-y_{4}y_{1234}q_{1100},\right. \\
&\qquad{}\left.y_{2}y_{1234}q_{1111}-y_{4}q_{0110}q_{1100},\right. \\
&\qquad{}\left.y_{2}y_{1234}q_{0011}-y_{4}q_{0110},\right. \\
&\qquad{}\left.y_{2}y_{4}y_{1234}-q_{0101},\right. \\
&\qquad{}\left.y_{2}y_{3}y_{1234}-q_{0110},\right. \\
&\qquad{}\left.y_{1}q_{0110}-y_{3}y_{1234}q_{1100},\right. \\
&\qquad{}\left.y_{1}q_{0101}-y_{4}y_{1234}q_{1100},\right. \\
&\qquad{}\left.y_{1}y_{1234}q_{1111}-y_{4}q_{1010}q_{1100},\right. \\
&\qquad{}\left.y_{1}y_{1234}q_{0011}-y_{4}q_{1010},\right. \\
&\qquad{}\left.y_{1}y_{4}y_{1234}-q_{1001},\right. \\
&\qquad{}\left.y_{1}y_{3}y_{1234}-q_{1010},\right. \\
&\qquad{}\left.y_{1}y_{2}-q_{1100}\right>.
\end{align*}

\chapter{Regions in the Probability Space}\label{cha.Appendix5}

\section*{Region $R_{1}\cap{}R_{2}\cap{}R_{3}\cap{}R_{4}\cap{}R_{5}$}

From (\ref{regions}),
\begin{align}\begin{split}
R_{1}\cap{}R_{2}\cap{}R_{3}\cap{}R_{4}\cap{}R_{5}&=\Omega_{1}\cap{}\left(\Omega_{1}\cap{}\Omega_{2}\right)\cap{}\left(\Omega_{1a}\cap{}\Omega_{3}\right)\cap{}\left(\Omega_{1a}\cap{}\Omega_{2}\right)\cap{}\Omega_{2} \\
&=\Omega_{1}\cap{}\Omega_{1a}\cap{}\Omega_{2}\cap{}\Omega_{3}.
\end{split}\end{align}
From (\ref{regions2}),
\begin{align}\begin{split}
\Omega{}_{1}&=\Omega{}_{1}\cap{}\Omega{}_{1a}\cap{}\Omega{}_{1b}\cap{}\Omega{}_{1c}\cap{}\Omega{}_{3}.
\end{split}\end{align}
From (\ref{regions3}),
\begin{align}\begin{split}
\Omega{}_{2}&=\Omega{}_{2}\cap{}\Omega{}_{2b}\cap{}\Omega{}_{3}.
\end{split}\end{align}
We can conclude that
\begin{align}\begin{split}
R_{1}\cap{}R_{2}\cap{}R_{3}\cap{}R_{4}\cap{}R_{5}&=\left(\Omega{}_{1}\cap{}\Omega{}_{1a}\cap{}\Omega{}_{1b}\cap{}\Omega{}_{1c}\cap{}\Omega{}_{3}\right)\cap{}\left(\Omega{}_{2}\cap{}\Omega{}_{2b}\cap{}\Omega{}_{3}\right)\cap{}\Omega{}_{3} \\
&=\Omega{}_{1}\cap{}\Omega{}_{1a}\cap{}\Omega{}_{1b}\cap{}\Omega{}_{1c}\cap{}\Omega{}_{2}\cap{}\Omega{}_{2b}\cap{}\Omega{}_{3}.
\end{split}\end{align}

\section*{Region $R_{1}\cap{}R_{2}^{C}\cap{}R_{3}\cap{}R_{4}^{C}\cap{}R_{5}^{C}$}

From (\ref{regions}),
\begin{align}\begin{split}
R_{1}\cap{}R_{2}^{C}\cap{}R_{3}\cap{}R_{4}^{C}\cap{}R_{5}^{C}&=\Omega_{1}\cap{}\left(\Omega_{1}\cap{}\Omega_{2b}\right)^{C}\cap{}\left(\Omega_{1a}\cap{}\Omega_{3}\right)\cap{}\left(\Omega_{1a}\cap{}\Omega_{2b}\right)^{C}\cap{}\Omega_{2}^{C} \\
&=\Omega_{1}\cap{}\Omega_{1a}\cap{}\Omega_{2}^{C}\cap{}\Omega_{3}\cap{}\left(\Omega_{1}\cap{}\Omega_{2b}\right)^{C}\cap{}\left(\Omega_{1a}\cap{}\Omega_{2b}\right)^{C}.
\end{split}\end{align}
For any two sets, $A$ and $B$,
\begin{align}
\begin{split}
\left(A\cap{}B\right)^{C}&=A^{C}\cup{}B^{C}.
\end{split}
\label{setcomplements}
\end{align}
From (\ref{setcomplements}),
\begin{align}\begin{split}
R_{1}\cap{}R_{2}^{C}\cap{}R_{3}\cap{}R_{4}^{C}\cap{}R_{5}^{C}&=\Omega_{1}\cap{}\Omega_{1a}\cap{}\Omega_{2}^{C}\cap{}\Omega_{3}\cap{}\left(\Omega_{1}^{C}\cup{}\Omega_{2b}^{C}\right)\cap{}\left(\Omega_{1a}^{C}\cup{}\Omega_{2b}^{C}\right).
\end{split}\end{align}
For any three sets, $A$, $B$ and $C$,
\begin{align}
\begin{split}
\left(A\cup{}B\right)\cap{}\left(A\cup{}C\right)&=A\cup{}\left(B\cap{}C\right).
\end{split}
\label{setcomplements2}
\end{align}
From (\ref{setcomplements2}),
\begin{align}\begin{split}
R_{1}\cap{}R_{2}^{C}\cap{}R_{3}\cap{}R_{4}^{C}\cap{}R_{5}^{C}&=\Omega_{1}\cap{}\Omega_{1a}\cap{}\Omega_{2}^{C}\cap{}\Omega_{3}\cap{}\left(\Omega_{2b}^{C}\cup{}\left(\Omega_{1}^{C}\cap{}\Omega_{1a}^{C}\right)\right).
\end{split}\end{align}
For any three sets, $A$, $B$ and $C$,
\begin{align}
\begin{split}
A\cap{}\left(B\cup{}C\right)&=\left(A\cap{}B\right)\cup{}\left(A\cap{}C\right).
\end{split}
\label{setcomplements3}
\end{align}
From (\ref{setcomplements3}),
\begin{align}\begin{split}
R_{1}\cap{}R_{2}^{C}\cap{}R_{3}\cap{}R_{4}^{C}\cap{}R_{5}^{C}&=\left(\Omega_{1}\cap{}\Omega_{1a}\cap{}\Omega_{2}^{C}\cap{}\Omega_{3}\cap{}\Omega_{2b}^{C}\right)\cup{}\left(\Omega_{1}\cap{}\Omega_{1a}\cap{}\Omega_{2}^{C}\cap{}\Omega_{3}\cap{}\Omega_{1}^{C}\cap{}\Omega_{1a}^{C}\right).
\end{split}\end{align}
Since $\Omega_{1a}\cap{}\Omega_{1a}^{C}=\emptyset{}$,
\begin{align}\begin{split}
R_{1}\cap{}R_{2}^{C}\cap{}R_{3}\cap{}R_{4}^{C}\cap{}R_{5}^{C}&=\left(\Omega_{1}\cap{}\Omega_{1a}\cap{}\Omega_{2}^{C}\cap{}\Omega_{3}\cap{}\Omega_{2b}^{C}\right)\cup{}\emptyset{} \\
&=\Omega_{1}\cap{}\Omega_{1a}\cap{}\Omega_{2}^{C}\cap{}\Omega_{2b}^{C}\cap{}\Omega_{3}.
\end{split}\end{align}
From (\ref{regions2}),
\begin{align}\begin{split}
R_{1}\cap{}R_{2}^{C}\cap{}R_{3}\cap{}R_{4}^{C}\cap{}R_{5}^{C}&=\left(\Omega{}_{1}\cap{}\Omega{}_{1a}\cap{}\Omega{}_{1b}\cap{}\Omega{}_{1c}\cap{}\Omega{}_{3}\right)\cap{}\Omega_{1a}\cap{}\Omega_{2}^{C}\cap{}\Omega_{2b}^{C}\cap{}\Omega_{3} \\
&=\Omega{}_{1}\cap{}\Omega{}_{1a}\cap{}\Omega{}_{1b}\cap{}\Omega{}_{1c}\cap{}\Omega_{2}^{C}\cap{}\Omega_{2b}^{C}\cap{}\Omega_{3}.
\end{split}\end{align}

\section*{Region $R_{1}^{C}\cap{}R_{2}^{C}\cap{}R_{3}\cap{}R_{4}\cap{}R_{5}$}

From (\ref{regions}),
\begin{align}\begin{split}
R_{1}^{C}\cap{}R_{2}^{C}\cap{}R_{3}\cap{}R_{4}\cap{}R_{5}&=\Omega{}_{1}^{C}\cap{}\left(\Omega{}_{1b}\cap{}\Omega{}_{2}\right)^{C}\cap{}\left(\Omega{}_{1a}\cap{}\Omega{}_{3}\right)\cap{}\left(\Omega{}_{1a}\cap{}\Omega{}_{2b}\right)\cap{}\Omega{}_{2} \\
&=\Omega{}_{1}^{C}\cap{}\Omega{}_{1a}\cap{}\Omega{}_{2}\cap{}\Omega{}_{2b}\cap{}\Omega{}_{3}\cap{}\left(\Omega{}_{1b}\cap{}\Omega{}_{2}\right)^{C}.
\end{split}\end{align}
From (\ref{setcomplements}),
\begin{align}\begin{split}
R_{1}^{C}\cap{}R_{2}^{C}\cap{}R_{3}\cap{}R_{4}\cap{}R_{5}&=\Omega{}_{1}^{C}\cap{}\Omega{}_{1a}\cap{}\Omega{}_{2}\cap{}\Omega{}_{2b}\cap{}\Omega{}_{3}\cap{}\left(\Omega{}_{1b}^{C}\cup{}\Omega{}_{2}^{C}\right).
\end{split}\end{align}
From (\ref{setcomplements3}),
\begin{align}\begin{split}
&R_{1}^{C}\cap{}R_{2}^{C}\cap{}R_{3}\cap{}R_{4}\cap{}R_{5} \\
&=\left(\Omega{}_{1}^{C}\cap{}\Omega{}_{1a}\cap{}\Omega{}_{2}\cap{}\Omega{}_{2b}\cap{}\Omega{}_{3}\cap{}\Omega{}_{1b}^{C}\right)\cup{}\left(\Omega{}_{1}^{C}\cap{}\Omega{}_{1a}\cap{}\Omega{}_{2}\cap{}\Omega{}_{2b}\cap{}\Omega{}_{3}\cap{}\Omega{}_{2}^{C}\right).
\end{split}\end{align}
Since $\Omega_{2}\cap{}\Omega_{2}^{C}=\emptyset{}$,
\begin{align}\begin{split}
R_{1}^{C}\cap{}R_{2}^{C}\cap{}R_{3}\cap{}R_{4}\cap{}R_{5}&=\left(\Omega{}_{1}^{C}\cap{}\Omega{}_{1a}\cap{}\Omega{}_{1b}^{C}\cap{}\Omega{}_{2}\cap{}\Omega{}_{2b}\cap{}\Omega{}_{3}\right)\cup{}\emptyset{} \\
&=\Omega{}_{1}^{C}\cap{}\Omega{}_{1a}\cap{}\Omega{}_{1b}^{C}\cap{}\Omega{}_{2}\cap{}\Omega{}_{2b}\cap{}\Omega{}_{3}.
\end{split}\end{align}
Since $\left(R_{1}^{C}\cap{}R_{2}^{C}\cap{}R_{3}\cap{}R_{4}\cap{}R_{5}\right)\cap{}R_{2}^{C}=R_{1}^{C}\cap{}R_{2}^{C}\cap{}R_{3}\cap{}R_{4}\cap{}R_{5}$, from (\ref{regions}),
\begin{align}\begin{split}
R_{1}^{C}\cap{}R_{2}^{C}\cap{}R_{3}\cap{}R_{4}\cap{}R_{5}&=\left(\Omega{}_{1}^{C}\cap{}\Omega{}_{1a}\cap{}\Omega{}_{1b}^{C}\cap{}\Omega{}_{2}\cap{}\Omega{}_{2b}\cap{}\Omega{}_{3}\right)\cap{}\left(\Omega{}_{1c}\cap{}\Omega{}_{2}\right)^{C}.
\end{split}\end{align}
From (\ref{setcomplements}),
\begin{align}\begin{split}
R_{1}^{C}\cap{}R_{2}^{C}\cap{}R_{3}\cap{}R_{4}\cap{}R_{5}&=\left(\Omega{}_{1}^{C}\cap{}\Omega{}_{1a}\cap{}\Omega{}_{1b}^{C}\cap{}\Omega{}_{2}\cap{}\Omega{}_{2b}\cap{}\Omega{}_{3}\right)\cap{}\left(\Omega{}_{1c}^{C}\cup{}\Omega{}_{2}^{C}\right).
\end{split}\end{align}
From (\ref{setcomplements3}),
\begin{align}\begin{split}
&R_{1}^{C}\cap{}R_{2}^{C}\cap{}R_{3}\cap{}R_{4}\cap{}R_{5} \\
&=\left(\Omega{}_{1}^{C}\cap{}\Omega{}_{1a}\cap{}\Omega{}_{1b}^{C}\cap{}\Omega{}_{2}\cap{}\Omega{}_{2b}\cap{}\Omega{}_{3}\cap{}\Omega{}_{1c}^{C}\right)\cup{}\left(\Omega{}_{1}^{C}\cap{}\Omega{}_{1a}\cap{}\Omega{}_{1b}^{C}\cap{}\Omega{}_{2}\cap{}\Omega{}_{2b}\cap{}\Omega{}_{3}\cap{}\Omega{}_{2}^{C}\right).
\end{split}\end{align}
Since $\Omega_{2}\cap{}\Omega_{2}^{C}=\emptyset{}$,
\begin{align}\begin{split}
R_{1}^{C}\cap{}R_{2}^{C}\cap{}R_{3}\cap{}R_{4}\cap{}R_{5}&=\left(\Omega{}_{1}^{C}\cap{}\Omega{}_{1a}\cap{}\Omega{}_{1b}^{C}\cap{}\Omega{}_{1c}^{C}\cap{}\Omega{}_{2}\cap{}\Omega{}_{2b}\cap{}\Omega{}_{3}\right)\cup{}\emptyset{} \\
&=\Omega{}_{1}^{C}\cap{}\Omega{}_{1a}\cap{}\Omega{}_{1b}^{C}\cap{}\Omega{}_{1c}^{C}\cap{}\Omega{}_{2}\cap{}\Omega{}_{2b}\cap{}\Omega{}_{3}.
\end{split}\end{align}

\section*{Region $R_{1}^{C}\cap{}R_{2}^{C}\cap{}R_{3}\cap{}R_{4}^{C}\cap{}R_{5}^{C}$}

From (\ref{regions}),
\begin{align}\begin{split}
R_{1}^{C}\cap{}R_{2}^{C}\cap{}R_{3}\cap{}R_{4}^{C}\cap{}R_{5}^{C}&=\Omega{}_{1}^{C}\cap{}\left(\Omega{}_{1b}\cap{}\Omega{}_{2b}\right)^{C}\cap{}\left(\Omega{}_{1a}\cap{}\Omega{}_{3}\right)\cap{}\left(\Omega{}_{1a}\cap{}\Omega{}_{2b}\right)^{C}\cap{}\Omega{}_{2}^{C} \\
&=\Omega{}_{1}^{C}\cap{}\Omega{}_{1a}\cap{}\Omega{}_{2}^{C}\cap{}\Omega{}_{3}\cap{}\left(\Omega{}_{1b}\cap{}\Omega{}_{2b}\right)^{C}\cap{}\left(\Omega{}_{1a}\cap{}\Omega{}_{2b}\right)^{C}.
\end{split}\end{align}
From (\ref{setcomplements}),
\begin{align}\begin{split}
R_{1}^{C}\cap{}R_{2}^{C}\cap{}R_{3}\cap{}R_{4}^{C}\cap{}R_{5}^{C}&=\Omega{}_{1}^{C}\cap{}\Omega{}_{1a}\cap{}\Omega{}_{2}^{C}\cap{}\Omega{}_{3}\cap{}\left(\Omega{}_{1b}^{C}\cup{}\Omega{}_{2b}^{C}\right)\cap{}\left(\Omega{}_{1a}^{C}\cup{}\Omega{}_{2b}^{C}\right).
\end{split}\end{align}
From (\ref{setcomplements2}),
\begin{align}\begin{split}
R_{1}^{C}\cap{}R_{2}^{C}\cap{}R_{3}\cap{}R_{4}^{C}\cap{}R_{5}^{C}&=\Omega{}_{1}^{C}\cap{}\Omega{}_{1a}\cap{}\Omega{}_{2}^{C}\cap{}\Omega{}_{3}\cap{}\left(\Omega{}_{2b}^{C}\cup{}\left(\Omega{}_{1b}^{C}\cap{}\Omega{}_{1a}^{C}\right)\right).
\end{split}\end{align}
From (\ref{setcomplements3}),
\begin{align}\begin{split}
&R_{1}^{C}\cap{}R_{2}^{C}\cap{}R_{3}\cap{}R_{4}^{C}\cap{}R_{5}^{C} \\
&=\left(\Omega{}_{1}^{C}\cap{}\Omega{}_{1a}\cap{}\Omega{}_{2}^{C}\cap{}\Omega{}_{3}\cap{}\Omega{}_{2b}^{C}\right)\cup{}\left(\Omega{}_{1}^{C}\cap{}\Omega{}_{1a}\cap{}\Omega{}_{2}^{C}\cap{}\Omega{}_{3}\cap{}\Omega{}_{1b}^{C}\cup{}\Omega{}_{1a}^{C}\right).
\end{split}\end{align}
Since $\Omega_{1a}\cap{}\Omega_{1a}^{C}=\emptyset{}$,
\begin{align}\begin{split}
R_{1}^{C}\cap{}R_{2}^{C}\cap{}R_{3}\cap{}R_{4}^{C}\cap{}R_{5}^{C}&=\left(\Omega{}_{1}^{C}\cap{}\Omega{}_{1a}\cap{}\Omega{}_{2}^{C}\cap{}\Omega{}_{2b}^{C}\cap{}\Omega{}_{3}\right)\cup{}\emptyset{} \\
&=\Omega{}_{1}^{C}\cap{}\Omega{}_{1a}\cap{}\Omega{}_{2}^{C}\cap{}\Omega{}_{2b}^{C}\cap{}\Omega{}_{3}.
\end{split}\end{align}
Since $\left(R_{1}^{C}\cap{}R_{2}^{C}\cap{}R_{3}\cap{}R_{4}^{C}\cap{}R_{5}^{C}\right)\cap{}R_{1}^{C}=R_{1}^{C}\cap{}R_{2}^{C}\cap{}R_{3}\cap{}R_{4}^{C}\cap{}R_{5}^{C}$, from (\ref{regions}),
\begin{align}\begin{split}
R_{1}^{C}\cap{}R_{2}^{C}\cap{}R_{3}\cap{}R_{4}^{C}\cap{}R_{5}^{C}&=\left(\Omega{}_{1}^{C}\cap{}\Omega{}_{1a}\cap{}\Omega{}_{2}^{C}\cap{}\Omega{}_{2b}^{C}\cap{}\Omega{}_{3}\right)\cap{}\left(\Omega{}_{1b}\cap{}\Omega{}_{3}\right)^{C}.
\end{split}\end{align}
From (\ref{setcomplements}),
\begin{align}\begin{split}
R_{1}^{C}\cap{}R_{2}^{C}\cap{}R_{3}\cap{}R_{4}^{C}\cap{}R_{5}^{C}&=\left(\Omega{}_{1}^{C}\cap{}\Omega{}_{1a}\cap{}\Omega{}_{2}^{C}\cap{}\Omega{}_{2b}^{C}\cap{}\Omega{}_{3}\right)\cap{}\left(\Omega{}_{1b}^{C}\cup{}\Omega{}_{3}^{C}\right).
\end{split}\end{align}
From (\ref{setcomplements3}),
\begin{align}\begin{split}
&R_{1}^{C}\cap{}R_{2}^{C}\cap{}R_{3}\cap{}R_{4}^{C}\cap{}R_{5}^{C} \\
&=\left(\Omega{}_{1}^{C}\cap{}\Omega{}_{1a}\cap{}\Omega{}_{2}^{C}\cap{}\Omega{}_{2b}^{C}\cap{}\Omega{}_{3}\cap{}\Omega{}_{1b}^{C}\right)\cup{}\left(\Omega{}_{1}^{C}\cap{}\Omega{}_{1a}\cap{}\Omega{}_{2}^{C}\cap{}\Omega{}_{2b}^{C}\cap{}\Omega{}_{3}\cap{}\Omega{}_{3}^{C}\right).
\end{split}\end{align}
Since $\Omega_{3}\cap{}\Omega_{3}^{C}=\emptyset{}$,
\begin{align}\begin{split}
R_{1}^{C}\cap{}R_{2}^{C}\cap{}R_{3}\cap{}R_{4}^{C}\cap{}R_{5}^{C}&=\left(\Omega{}_{1}^{C}\cap{}\Omega{}_{1a}\cap{}\Omega{}_{1b}^{C}\cap{}\Omega{}_{2}^{C}\cap{}\Omega{}_{2b}^{C}\cap{}\Omega{}_{3}\right)\cup{}\emptyset{} \\
&=\Omega{}_{1}^{C}\cap{}\Omega{}_{1a}\cap{}\Omega{}_{1b}^{C}\cap{}\Omega{}_{2}^{C}\cap{}\Omega{}_{2b}^{C}\cap{}\Omega{}_{3}.
\end{split}\end{align}
Since $\left(R_{1}^{C}\cap{}R_{2}^{C}\cap{}R_{3}\cap{}R_{4}^{C}\cap{}R_{5}^{C}\right)\cap{}R_{1}^{C}=R_{1}^{C}\cap{}R_{2}^{C}\cap{}R_{3}\cap{}R_{4}^{C}\cap{}R_{5}^{C}$, from (\ref{regions}),
\begin{align}\begin{split}
R_{1}^{C}\cap{}R_{2}^{C}\cap{}R_{3}\cap{}R_{4}^{C}\cap{}R_{5}^{C}&=\left(\Omega{}_{1}^{C}\cap{}\Omega{}_{1a}\cap{}\Omega{}_{1b}^{C}\cap{}\Omega{}_{2}^{C}\cap{}\Omega{}_{2b}^{C}\cap{}\Omega{}_{3}\right)\cap{}\left(\Omega{}_{1c}\cap{}\Omega{}_{3}\right)^{C}.
\end{split}\end{align}
From (\ref{setcomplements}),
\begin{align}\begin{split}
R_{1}^{C}\cap{}R_{2}^{C}\cap{}R_{3}\cap{}R_{4}^{C}\cap{}R_{5}^{C}&=\left(\Omega{}_{1}^{C}\cap{}\Omega{}_{1a}\cap{}\Omega{}_{1b}^{C}\cap{}\Omega{}_{2}^{C}\cap{}\Omega{}_{2b}^{C}\cap{}\Omega{}_{3}\right)\cap{}\left(\Omega{}_{1c}^{C}\cup{}\Omega{}_{3}^{C}\right).
\end{split}\end{align}
From (\ref{setcomplements3}),
\begin{align}\begin{split}
&R_{1}^{C}\cap{}R_{2}^{C}\cap{}R_{3}\cap{}R_{4}^{C}\cap{}R_{5}^{C} \\
&=\left(\Omega{}_{1}^{C}\cap{}\Omega{}_{1a}\cap{}\Omega{}_{1b}^{C}\cap{}\Omega{}_{2}^{C}\cap{}\Omega{}_{2b}^{C}\cap{}\Omega{}_{3}\cap{}\Omega{}_{1c}^{C}\right)\cup{}\left(\Omega{}_{1}^{C}\cap{}\Omega{}_{1a}\cap{}\Omega{}_{1b}^{C}\cap{}\Omega{}_{2}^{C}\cap{}\Omega{}_{2b}^{C}\cap{}\Omega{}_{3}\cap{}\Omega{}_{3}^{C}\right).
\end{split}\end{align}
Since $\Omega_{3}\cap{}\Omega_{3}^{C}=\emptyset{}$,
\begin{align}\begin{split}
R_{1}^{C}\cap{}R_{2}^{C}\cap{}R_{3}\cap{}R_{4}^{C}\cap{}R_{5}^{C}&=\left(\Omega{}_{1}^{C}\cap{}\Omega{}_{1a}\cap{}\Omega{}_{1b}^{C}\cap{}\Omega{}_{1c}^{C}\cap{}\Omega{}_{2}^{C}\cap{}\Omega{}_{2b}^{C}\cap{}\Omega{}_{3}\right)\cup{}\emptyset{} \\
&=\Omega{}_{1}^{C}\cap{}\Omega{}_{1a}\cap{}\Omega{}_{1b}^{C}\cap{}\Omega{}_{1c}^{C}\cap{}\Omega{}_{2}^{C}\cap{}\Omega{}_{2b}^{C}\cap{}\Omega{}_{3}.
\end{split}\end{align}

\section*{Region $R_{1}^{C}\cap{}R_{2}^{C}\cap{}R_{3}^{C}\cap{}R_{4}^{C}\cap{}R_{5}$}

From (\ref{regions}),
\begin{align}\begin{split}
R_{1}^{C}\cap{}R_{2}^{C}\cap{}R_{3}^{C}\cap{}R_{4}^{C}\cap{}R_{5}&=\Omega{}_{1}^{C}\cap{}\left(\Omega{}_{1b}\cap{}\Omega{}_{2}\right)^{C}\cap{}\left(\Omega{}_{1a}\cap{}\Omega{}_{3}\right)^{C}\cap{}\left(\Omega{}_{1a}\cap{}\Omega{}_{2}\right)^{C}\cap{}\Omega{}_{2} \\
&=\Omega{}_{1}^{C}\cap{}\Omega{}_{2}\cap{}\left(\Omega{}_{1b}\cap{}\Omega{}_{2}\right)^{C}\cap{}\left(\Omega{}_{1a}\cap{}\Omega{}_{3}\right)^{C}\cap{}\left(\Omega{}_{1a}\cap{}\Omega{}_{2}\right)^{C}.
\end{split}\end{align}
From (\ref{setcomplements}),
\begin{align}\begin{split}
R_{1}^{C}\cap{}R_{2}^{C}\cap{}R_{3}^{C}\cap{}R_{4}^{C}\cap{}R_{5}&=\Omega{}_{1}^{C}\cap{}\Omega{}_{2}\cap{}\left(\Omega{}_{1a}^{C}\cup{}\Omega{}_{3}^{C}\right)\cap{}\left(\Omega{}_{1a}^{C}\cup{}\Omega{}_{2}^{C}\right)\cap{}\left(\Omega{}_{1b}^{C}\cup{}\Omega{}_{2}^{C}\right).
\end{split}\end{align}
From (\ref{regions3}),
\begin{align}\begin{split}
&R_{1}^{C}\cap{}R_{2}^{C}\cap{}R_{3}^{C}\cap{}R_{4}^{C}\cap{}R_{5}=\Omega{}_{1}^{C}\cap{}\left(\Omega{}_{2}\cap{}\Omega{}_{2b}\cap{}\Omega{}_{3}\right)\cap{}\left(\Omega{}_{1a}^{C}\cup{}\Omega{}_{3}^{C}\right)\cap{}\left(\Omega{}_{1a}^{C}\cup{}\Omega{}_{2}^{C}\right)\cap{}\left(\Omega{}_{1b}^{C}\cup{}\Omega{}_{2}^{C}\right) \\
&=\left(\left(\Omega{}_{1}^{C}\cap{}\Omega{}_{2}\cap{}\Omega{}_{2b}\cap{}\Omega{}_{3}\right)\cap{}\left(\Omega{}_{1a}^{C}\cup{}\Omega{}_{3}^{C}\right)\right)\cap{}\left(\Omega{}_{1a}^{C}\cup{}\Omega{}_{2}^{C}\right)\cap{}\left(\Omega{}_{1b}^{C}\cup{}\Omega{}_{2}^{C}\right).
\end{split}\end{align}
From (\ref{setcomplements3}),
\begin{align}\begin{split}
&R_{1}^{C}\cap{}R_{2}^{C}\cap{}R_{3}^{C}\cap{}R_{4}^{C}\cap{}R_{5}=\left(\left(\Omega{}_{1}^{C}\cap{}\Omega{}_{2}\cap{}\Omega{}_{2b}\cap{}\Omega{}_{3}\cap{}\Omega{}_{1a}^{C}\right)\cup{}\left(\Omega{}_{1}^{C}\cap{}\Omega{}_{2}\cap{}\Omega{}_{2b}\cap{}\Omega{}_{3}\cap{}\cup{}\Omega{}_{3}^{C}\right)\right) \\
&\cap{}\left(\Omega{}_{1a}^{C}\cup{}\Omega{}_{2}^{C}\right)\cap{}\left(\Omega{}_{1b}^{C}\cup{}\Omega{}_{2}^{C}\right).
\end{split}\end{align}
Since $\Omega_{3}\cap{}\Omega_{3}^{C}=\emptyset{}$,
\begin{align}\begin{split}
R_{1}^{C}\cap{}R_{2}^{C}\cap{}R_{3}^{C}\cap{}R_{4}^{C}\cap{}R_{5}&=\left(\left(\Omega{}_{1}^{C}\cap{}\Omega{}_{1a}^{C}\cap{}\Omega{}_{2}\cap{}\Omega{}_{2b}\cap{}\Omega{}_{3}\right)\cup{}\emptyset{}\right)\cap{}\left(\Omega{}_{1a}^{C}\cup{}\Omega{}_{2}^{C}\right)\cap{}\left(\Omega{}_{1b}^{C}\cup{}\Omega{}_{2}^{C}\right) \\
&=\left(\Omega{}_{1}^{C}\cap{}\Omega{}_{1a}^{C}\cap{}\Omega{}_{2}\cap{}\Omega{}_{2b}\cap{}\Omega{}_{3}\cap{}\left(\Omega{}_{1a}^{C}\cup{}\Omega{}_{2}^{C}\right)\right)\cap{}\left(\Omega{}_{1b}^{C}\cup{}\Omega{}_{2}^{C}\right).
\end{split}\end{align}
From (\ref{setcomplements3}),
\begin{align}\begin{split}
&R_{1}^{C}\cap{}R_{2}^{C}\cap{}R_{3}^{C}\cap{}R_{4}^{C}\cap{}R_{5} \\
&=\left(\left(\Omega{}_{1}^{C}\cap{}\Omega{}_{1a}^{C}\cap{}\Omega{}_{2}\cap{}\Omega{}_{2b}\cap{}\Omega{}_{3}\cap{}\Omega{}_{1a}^{C}\right)\cup{}\left(\Omega{}_{1}^{C}\cap{}\Omega{}_{1a}^{C}\cap{}\Omega{}_{2}\cap{}\Omega{}_{2b}\cap{}\Omega{}_{3}\cap{}\Omega{}_{2}^{C}\right)\right)\cap{}\left(\Omega{}_{1b}^{C}\cup{}\Omega{}_{2}^{C}\right).
\end{split}\end{align}
Since $\Omega_{2}\cap{}\Omega_{2}^{C}=\emptyset{}$,
\begin{align}\begin{split}
R_{1}^{C}\cap{}R_{2}^{C}\cap{}R_{3}^{C}\cap{}R_{4}^{C}\cap{}R_{5}&=\left(\left(\Omega{}_{1}^{C}\cap{}\Omega{}_{1a}^{C}\cap{}\Omega{}_{2}\cap{}\Omega{}_{2b}\cap{}\Omega{}_{3}\right)\cup{}\emptyset{}\right)\cap{}\left(\Omega{}_{1b}^{C}\cup{}\Omega{}_{2}^{C}\right) \\
&=\Omega{}_{1}^{C}\cap{}\Omega{}_{1a}^{C}\cap{}\Omega{}_{2}\cap{}\Omega{}_{2b}\cap{}\Omega{}_{3}\cap{}\left(\Omega{}_{1b}^{C}\cup{}\Omega{}_{2}^{C}\right).
\end{split}\end{align}
From (\ref{setcomplements3}),
\begin{align}\begin{split}
&R_{1}^{C}\cap{}R_{2}^{C}\cap{}R_{3}^{C}\cap{}R_{4}^{C}\cap{}R_{5} \\
&=\left(\Omega{}_{1}^{C}\cap{}\Omega{}_{1a}^{C}\cap{}\Omega{}_{2}\cap{}\Omega{}_{2b}\cap{}\Omega{}_{3}\cap{}\Omega{}_{1b}^{C}\right)\cup{}\left(\Omega{}_{1}^{C}\cap{}\Omega{}_{1a}^{C}\cap{}\Omega{}_{2}\cap{}\Omega{}_{2b}\cap{}\Omega{}_{3}\cap{}\Omega{}_{2}^{C}\right).
\end{split}\end{align}
Since $\Omega_{2}\cap{}\Omega_{2}^{C}=\emptyset{}$,
\begin{align}\begin{split}
R_{1}^{C}\cap{}R_{2}^{C}\cap{}R_{3}^{C}\cap{}R_{4}^{C}\cap{}R_{5}&=\left(\Omega{}_{1}^{C}\cap{}\Omega{}_{1a}^{C}\cap{}\Omega{}_{1b}^{C}\cap{}\Omega{}_{2}\cap{}\Omega{}_{2b}\cap{}\Omega{}_{3}\right)\cup{}\emptyset{} \\
&=\Omega{}_{1}^{C}\cap{}\Omega{}_{1a}^{C}\cap{}\Omega{}_{1b}^{C}\cap{}\Omega{}_{2}\cap{}\Omega{}_{2b}\cap{}\Omega{}_{3}.
\end{split}\end{align}
Since $\left(R_{1}^{C}\cap{}R_{2}^{C}\cap{}R_{3}^{C}\cap{}R_{4}^{C}\cap{}R_{5}\right)\cap{}R_{2}^{C}=R_{1}^{C}\cap{}R_{2}^{C}\cap{}R_{3}^{C}\cap{}R_{4}^{C}\cap{}R_{5}$, from (\ref{regions}),
\begin{align}\begin{split}
R_{1}^{C}\cap{}R_{2}^{C}\cap{}R_{3}^{C}\cap{}R_{4}^{C}\cap{}R_{5}&=\left(\Omega{}_{1}^{C}\cap{}\Omega{}_{1a}^{C}\cap{}\Omega{}_{1b}^{C}\cap{}\Omega{}_{2}\cap{}\Omega{}_{2b}\cap{}\Omega{}_{3}\right)\cap{}\left(\Omega{}_{1c}\cap{}\Omega{}_{2}\right)^{C}.
\end{split}\end{align}
From (\ref{setcomplements}),
\begin{align}\begin{split}
R_{1}^{C}\cap{}R_{2}^{C}\cap{}R_{3}^{C}\cap{}R_{4}^{C}\cap{}R_{5}&=\left(\Omega{}_{1}^{C}\cap{}\Omega{}_{1a}^{C}\cap{}\Omega{}_{1b}^{C}\cap{}\Omega{}_{2}\cap{}\Omega{}_{2b}\cap{}\Omega{}_{3}\right)\cap{}\left(\Omega{}_{1c}^{C}\cup{}\Omega{}_{2}^{C}\right).
\end{split}\end{align}
From (\ref{setcomplements3}),
\begin{align}\begin{split}
&R_{1}^{C}\cap{}R_{2}^{C}\cap{}R_{3}^{C}\cap{}R_{4}^{C}\cap{}R_{5} \\
&=\left(\Omega{}_{1}^{C}\cap{}\Omega{}_{1a}^{C}\cap{}\Omega{}_{1b}^{C}\cap{}\Omega{}_{2}\cap{}\Omega{}_{2b}\cap{}\Omega{}_{3}\cap{}\Omega{}_{1c}^{C}\right)\cup{}\left(\Omega{}_{1}^{C}\cap{}\Omega{}_{1a}^{C}\cap{}\Omega{}_{1b}^{C}\cap{}\Omega{}_{2}\cap{}\Omega{}_{2b}\cap{}\Omega{}_{3}\cap{}\Omega{}_{2}^{C}\right).
\end{split}\end{align}
Since $\Omega_{2}\cap{}\Omega_{2}^{C}=\emptyset{}$,
\begin{align}\begin{split}
R_{1}^{C}\cap{}R_{2}^{C}\cap{}R_{3}^{C}\cap{}R_{4}^{C}\cap{}R_{5}&=\left(\Omega{}_{1}^{C}\cap{}\Omega{}_{1a}^{C}\cap{}\Omega{}_{1b}^{C}\cap{}\Omega{}_{1c}^{C}\cap{}\Omega{}_{2}\cap{}\Omega{}_{2b}\cap{}\Omega{}_{3}\right)\cup{}\emptyset{} \\
&=\Omega{}_{1}^{C}\cap{}\Omega{}_{1a}^{C}\cap{}\Omega{}_{1b}^{C}\cap{}\Omega{}_{1c}^{C}\cap{}\Omega{}_{2}\cap{}\Omega{}_{2b}\cap{}\Omega{}_{3}.
\end{split}\end{align}

\section*{Region $R_{1}^{C}\cap{}R_{2}^{C}\cap{}R_{3}^{C}\cap{}R_{4}^{C}\cap{}R_{5}^{C}$}

This is the region where only a single tree or network can be found. Since Network $1$ and Network $2$ can only be found in regions where other trees and networks can be found, neither of these trees will be found in this region. There will be five non-overlapping subregions in this region. They are
\begin{align}\begin{split}
\begin{mycases}
S_{1a}&=\Omega{}_{1}^{C}\cap{}\Omega{}_{1a}\cap{}\Omega{}_{1b}^{C}\cap{}\Omega{}_{1c}^{C}\cap{}\Omega{}_{2}^{C}\cap{}\Omega{}_{2b}^{C}\cap{}\Omega{}_{3}^{C}, \\
S_{1b}&=\Omega{}_{1}^{C}\cap{}\Omega{}_{1a}^{C}\cap{}\Omega{}_{1b}\cap{}\Omega{}_{1c}^{C}\cap{}\Omega{}_{2}^{C}\cap{}\Omega{}_{2b}^{C}\cap{}\Omega{}_{3}^{C}, \\
S_{1c}&=\Omega{}_{1}^{C}\cap{}\Omega{}_{1a}^{C}\cap{}\Omega{}_{1b}^{C}\cap{}\Omega{}_{1c}\cap{}\Omega{}_{2}^{C}\cap{}\Omega{}_{2b}^{C}\cap{}\Omega{}_{3}^{C}, \\
S_{2b}&=\Omega{}_{1}^{C}\cap{}\Omega{}_{1a}^{C}\cap{}\Omega{}_{1b}^{C}\cap{}\Omega{}_{1c}^{C}\cap{}\Omega{}_{2}^{C}\cap{}\Omega{}_{2b}\cap{}\Omega{}_{3}^{C}, \\
S_{3}&=\Omega{}_{1}^{C}\cap{}\Omega{}_{1a}^{C}\cap{}\Omega{}_{1b}^{C}\cap{}\Omega{}_{1c}^{C}\cap{}\Omega{}_{2}^{C}\cap{}\Omega{}_{2b}^{C}\cap{}\Omega{}_{3}.
\end{mycases}
\end{split}\end{align}

\end{appendices}

\backmatter

\renewcommand{\bibname}{References}
\bibliographystyle{plainnat}
\bibliography{bibliography}

\end{document}